\documentclass[11pt,letterpaper]{article}
\pdfoutput=1
\usepackage{jheppub}
\usepackage{bbm}
\usepackage{mathrsfs}
\usepackage{slashed}
\usepackage{caption}
\usepackage{epstopdf}
\usepackage[normalem]{ulem}
\usepackage[bottom]{footmisc}
\usepackage{subcaption}
\usepackage{bbold}
\usepackage{titlesec}
\usepackage{threeparttable}
\usepackage{booktabs}
\usepackage{changepage}
\usepackage[utf8]{inputenc}

\usepackage{grffile}

\usepackage{graphicx}  
\usepackage{dcolumn}   
\usepackage{bm}        
\usepackage{amssymb}   
\usepackage{setspace}
\usepackage{amsmath, amssymb, setspace}
\usepackage{array}
\usepackage{booktabs}
\usepackage{caption}
\usepackage{indentfirst}
\usepackage{float}
\usepackage{lmodern}
\usepackage{multirow}
\usepackage{soul}
\usepackage[normalem]{ulem}

\usepackage{braket}
\usepackage{comment}

\usepackage[draft]{pgf}

\usepackage{adjustbox} 
\newcommand{\ER}{E_{\rm R}}
\newcommand{\vmin}{v_{\rm min}}

\newcommand{\DoBox}[1]{\begin{center}
\color{red}\fbox{
\begin{minipage}{0.9\textwidth}

\end{minipage}}
\end{center}}

\newcommand{\Fig}[1]{Fig.~\ref{#1}}
\newcommand{\Eq}[1]{Eq.~(\ref{#1})}
\newcommand{\Sec}[1]{Sec.~\ref{#1}}
\newcommand{\Tab}[1]{Table~\ref{#1}}

\newlength{\myimageoversize}
\newsavebox{\myimage}
\newcommand{\mycenter}[1]{%
\savebox{\myimage}{#1}
\settowidth{\myimageoversize}{\usebox{\myimage}}
\addtolength{\myimageoversize}{-\textwidth}
\setlength{\leftskip}{-0.5\myimageoversize}
\noindent
\usebox{\myimage}}

\newcommand{\triplt}{0.38\textwidth}

\titleformat{\subsubsection}
 {\normalfont\fontsize{12}{17}\itshape}{\thesubsubsection}{1em}{}

\newcommand{\AddrIFIC}{Instituto de F\'{i}sica Corpuscular (IFIC), CSIC-Universitat de Val\`{e}ncia, Apartado de Correos 22085, E-46071 Valencia, Spain}

\title{ \huge{Casting a Wide Signal Net  with Future Direct Dark Matter Detection Experiments}}

\author[a]{Graciela B. Gelmini}
\author[a]{,~Volodymyr Takhistov}
\author[a,b]{,~Samuel J. Witte}

 \affiliation[a]{Department of Physics and Astronomy, University of California, Los Angeles\\
 Los Angeles, CA 90095-1547, USA}
 \affiliation[b]{\AddrIFIC}

 \emailAdd{gelmini@physics.ucla.edu}
 \emailAdd{vtakhist@physics.ucla.edu}
 \emailAdd{sam.witte@ific.uv.es}
 
\abstract{
As dark matter (DM) direct detection experiments continue to improve their sensitivity they will inevitably encounter an irreducible background arising from coherent neutrino scattering. This so-called “neutrino floor” may significantly reduce the sensitivity of an experiment to DM-nuclei interactions, particularly if the recoil spectrum of the neutrino background is approximately degenerate with the DM signal. This occurs for the conventionally considered spin-independent (SI) or spin-dependent (SD) interactions.
In such case, an increase in the experiment's exposure by multiple orders of magnitude may not yield any significant increase in sensitivity. The typically considered SI and SD interactions, however, do not adequately reflect the whole landscape of the well-motivated DM models, which includes other interactions. Since particle DM has not been detected yet in laboratories, it is essential to understand and maximize the detection capabilities for a broad variety of possible models and signatures.
In this work we explore the impact of the background arising from various neutrino sources on the discovery potential of a DM signal for a large class of viable DM-nucleus interactions and several potential futuristic experimental settings, with different target elements. For some momentum suppressed cross sections, large DM particle masses and heavier targets, we find that there is no suppression of the discovery limits due to neutrino backgrounds.
Further, we explicitly demonstrate that inelastic scattering, which could appear in models with multicomponent dark sectors, would help to lift the signal degeneracy associated with the neutrino floor.
This study could assist with mapping out the optimal DM detection strategy for the next generation of experiments.
 }

\begin{document}
\preprint{}
 \maketitle
\flushbottom


\section{Introduction}
\label{sec:intro}

For the past several decades, direct dark matter (DM) detection experiments have attempted to detect  the energy imparted to nuclei in underground laboratories by collisions with DM particles that are gravitationally bound to the Galactic halo. In particular, direct detection experiments are ideal for probing DM particle candidates with weak-scale interactions and masses in the range of $\sim 1-10^4$ GeV, which interact coherently with nuclei, referred to as weakly interacting massive particles (WIMPs). These experiments have made tremendous strides over this time period, increasing both in size and detection efficiency. 
The next multi-ton scale generation of direct DM detection experiments are currently being planned. 
Thus, it is worthwhile to consider what is the ultimate capabilities and reach of such experiments. 

It is well-known that direct detection experiments will soon encounter an irreducible background due to  coherent nuclear interactions of neutrinos (e.g. \cite{Cabrera:1984rr, Monroe:2007xp, Strigari:2009bq, Billard:2013qya}). At lower recoil energies, near the detection threshold, the background is predominantly due to solar neutrinos. At larger nuclear recoil energies it comes from the diffuse supernovae neutrino background and atmospheric neutrinos. It will be difficult to claim a discovery of DM if the signal lies below the neutrino background. Early studies of this so-called ``neutrino floor" concentrated on the usual elastic DM spin-independent (SI) and spin-dependent (SD) interactions and showed that the degeneracy between recoil spectra  due to coherent neutrino and WIMP scattering for particular DM masses would significantly limit the ultimate sensitivity of direct detection experiments~ \cite{Billard:2013qya,Ruppin:2014bra}. For the usual SI or SD interactions, solar neutrinos mimic a WIMP signal for a DM mass close to 6 GeV. On the other hand, atmospheric neutrinos mimic a WIMP signal for a DM mass of about  100 GeV. It has been shown (e.g. \cite{Ruppin:2014bra}) that combining data from different target materials can enhance the subtraction of the neutrino background in direct detection experiments for DM particles with SD interaction and masses below 10 GeV. However, target complementarity will not alleviate the problem for SI interactions, since the cross-sections of both DM and neutrinos have the same scaling with the nuclear mass number (e.g. \cite{Ruppin:2014bra}). Proposals to distinguish DM and neutrino signals include searching for an annual modulation~\cite{Davis:2014ama, OHare:2016pjy} or measuring the resulting recoil momentum~\cite{Grothaus:2014hja, OHare:2015utx} in directional direct detection experiments~\cite{Mayet:2016zxu}. However, both of these could be very challenging for a DM signal below the neutrino background.

In this paper we present for the first time a look at the neutrino floor for a large number of viable DM-nucleus interactions beyond the standard SI and SD, defined by a fully relativistic Lagrangian formalism, for several target nuclei. 
Recent studies of the neutrino floor have used non-relativistic effective field theory (EFT) DM-nucleus couplings~\cite{Dent:2016iht, Dent:2016wor, Dutta:2017nht}  to show that the degeneracy between neutrino and DM recoil spectra  may not be present for less conventional interactions. Ref.~\cite{Dent:2016iht} determined that for 11 out of the 14 possible non-relativistic  EFT operators considered, the predicted recoil spectra can be cleanly distinguished from the corresponding neutrino-induced recoil spectra with moderate size detectors (exposure of few ton$\cdot$years), for low mass WIMPs (i.e. with masses $\lesssim 10$ GeV).  
Non-relativistic EFT provides a theoretical framework for determining different nuclear responses to DM scattering events, thus yielding some insight into different viable DM couplings.  However, most of the DM-nuclei interactions defined in terms of a field theoretical Lagrangian formalism involve complex linear combinations of EFT operators, with the relative importance of each EFT operator weighted by nuclide-specific factors. Thus, for a particular interaction, different experiments may expect a varying degree of degeneracy of the DM and neutrino recoil spectra. Here we consider interactions, expressed in a Lagrangian formalism, due to the exchange of a single mediator that can be heavy or light with respect to the typical momentum being exchanged. Let us remark that although this is the right formalism to describe realistic interactions, it is still limited with respect to a possible complete theory of the dark and visible sectors. In a complete model of DM interactions, scattering amplitude terms arising from various mediators would be summed to obtain the total amplitude. They could thus interfere with each other, and different terms could dominate the cross section for different energy regimes.

Some of DM interactions produce recoil spectra that appear nearly degenerate with neutrinos in particular target elements, but not in others (e.g. \cite{Ruppin:2014bra}). For example, the absence of spin in argon results in a high level of degeneracy between the SI and anapole interactions (see \Eq{eq:diffxsecsi} and \Eq{eq:anapolediff} below), which is not necessarily present in other target elements. Hence, complementarity of a variety of target nuclei needs to be considered in order to maximize the detection sensitivity to DM in future experiments.

 The main questions we are going to address in this paper are the following: \emph{(i)} beyond the conventional SI and SD interactions, what types of DM candidates should one expect to give rise to a high level of degeneracy with the neutrino background, and for which elements can such a degeneracy be evaded, \emph{(ii)} for which combinations of elements and interactions can one exploit target complementarity to help ameliorate the neutrino background, \emph{(iii)} how much exposure is required to distinguish a particular DM candidate below the neutrino background, and \emph{(iv)} what is required for experiments to maintain a broad sensitivity to the largest possible number of DM interactions.

Recent detection  of coherent elastic neutrino-nucleus scattering by the COHERENT experiment \cite{Akimov:2017ade} further strengthens the case for exploiting the sensitivity of next-generation direct detection experiments also for exploring neutrino physics. Coherent neutrino-nucleus and elastic electron-neutrino scatterings allow one to study a range of physics topics, including sterile neutrinos \cite{Pospelov:2011ha,Billard:2014yka}, non-standard neutrino interactions (NSI) (e.g. \cite{Harnik:2012ni,Dutta:2017nht}) and supernovae \cite{Chakraborty:2013zua}. While we do not investigate physics related specifically to neutrinos in this work, we stress the richness of research opportunities available to future large direct detection experiments in this context.

The format of this paper is as follows. In \Sec{sec:darkmatter} we review the fundamentals of direct DM detection and discuss the DM-nucleus interactions which can result from the exchange of a single mediator in a Lagrangian formalism. Both elastic as well as inelastic scatterings are discussed.  In \Sec{sec:neutrinos} we introduce the fluxes and recoil spectra arising from neutrinos in direct detection experiments. \Sec{sec:detection} introduces our statistical analysis and presents the discovery limits for the various DM interactions considered. We conclude in \Sec{sec:summary}.


\section{Dark Matter Signal \label{sec:darkmatter}}


\subsection{Flux and recoil spectrum}

Direct DM detection experiments attempt to observe the recoils of nuclei due to collisions with DM particles gravitationally bound to the halo of the Milky Way.
The differential scattering rate per unit time and target mass of a WIMP $\chi$, assumed to account for the bulk of the DM, off a target nuclide $T$ as a function of nuclear recoil energy $\ER$ is given by
\begin{equation}\label{eq:dm_diff_rate}
\frac{d R_T}{d \ER} = \frac{\rho_\chi}{m_\chi}\frac{C_T}{m_T}\int_{v \geqslant \vmin(\ER)} \, d^3 \, v \, f(\vec{v},t) \, v \, \frac{d \sigma_T}{d \ER}(\ER, \vec{v}) \, ,
\end{equation}
where $\rho_\chi$ is the local DM density, $m_\chi$ is the DM particle mass, $m_T$ and $C_T$ are the mass and mass fraction of $T$ in the detector, $f(\vec{v}, t)$ is the distribution of the DM velocity $\vec{v}$ in the Earth's rest frame, and $d \sigma_T/d \ER$ is the DM-nucleus differential cross section. When multiple nuclides are present in the detector, \Eq{eq:dm_diff_rate} is summed over the element and/or isotopic composition, taking into account the respective element-dependence of the differential cross-section including the nuclear form factors, to obtain the total DM rate  
\begin{equation} \label{eq:totdmrate}
\frac{d R_{\rm DM}}{d \ER} =  \sum_T \frac{d R_T}{d \ER}~.
\end{equation}
The velocity integral in \Eq{eq:dm_diff_rate} is restricted to speeds larger than $\vmin$, the minimum DM particle speed required to impart the particular recoil energy to the target nucleus. For elastic scattering it is given by
\begin{equation}\label{eq:vmin_el}
\vmin = \dfrac{|\vec{q}|}{2 \mu_T} = \sqrt{\frac{m_T \ER}{2 \mu_T^2}} ~,
\end{equation}
where $\vec{q}$ is momentum transfer and $\mu_T = m_{\chi} m_T/ (m_{\chi} + m_T)$ is the reduced DM-nucleus mass. 

We assume that the local characteristics of the DM halo of the Galaxy are described by the Standard Halo Model (SHM), with the local DM density  $\rho_{\chi} = 0.3$ GeV/cm$^3$ and the distribution of the WIMP velocity $\vec{u}$ in the Galactic frame given by a truncated Maxwell-Boltzmann distribution
\begin{equation}
f_{\rm G}(\vec{u}) = \dfrac{1}{N_{\rm esc} (v_0 \sqrt{\pi})^3} \text{exp}(-u^2/v_0^2) \theta(v_{\rm esc} - u)~.
\end{equation}
Here, $v_0$ is the velocity dispersion taken to be the speed of the Local Standard of Rest $v_0 = 220$ km/s. The Galactic escape velocity $v_{\rm esc}$ is taken to be $v_{\rm esc} = 533$ km/s, following measurements of \cite{Piffl:2013mla}. The normalization factor 
\begin{equation}
N_{\rm esc} = \text{erf}(v_{\rm esc}/v_0) - 2(v_{\rm esc}/v_0)\text{exp}(-v_{\rm esc}^2/v_0^2)/\sqrt{\pi}~,
\end{equation}
by construction gives $\int d^3 u f_{\rm G} (\vec{u}) = 1$.
In the Earth's frame, neglecting the gravitational DM lensing by the Sun, the velocity $f(\vec{v},t)$ of \Eq{eq:totdmrate} is obtained from a Galilean transformation
\begin{equation} \label{eq:dmrelvel}
f(\vec{v},t) = f_{\rm G}(\vec{v}_{\odot} + \vec{v}_{\oplus}(t) + \vec{v})~,
\end{equation}
where $\vec{v}_{\odot} = 232$ km/s and $\vec{v}_{\oplus} = 30$ km/s denote the velocity of the Sun with respect to the Galaxy and the time-dependent velocity of Earth with respect to the Sun, taken in an orbit inclined at $60^{\circ}$ with respect to the Galactic plane, respectively \cite{Schoenrich:2009bx}.
Since we do not consider here the time variation of the expected DM signal we take $\vec{v}_{\oplus} = 0$. We do not expect that alternative realistic halo models would significantly alter the results of the analysis presented here~\cite{Morgan:2004ys,Billard:2011zj}. 
In practice, the differential rate in \Eq{eq:dm_diff_rate} must also be convolved with the efficiency and energy resolution of the particular experiment to obtain the observable recoil spectrum. Throughout this paper we optimistically assume that future direct detection experiments have perfect resolution and efficiency.

For realistic astrophysical distributions of DM, the rotation of the Earth around the Sun produces an annual modulation in the scattering rate, following \Eq{eq:dmrelvel} and \Eq{eq:totdmrate}. The recoil spectrum arising from neutrinos may also modulate, e.g. in the case of solar neutrinos with the proximity of the Sun. The modulations may enhance the ability of direct detection experiments to differentiate neutrinos from a DM candidate as phases and amplitudes of the modulation will differ. A recent analysis investigating DM-nucleus interactions (without including a neutrino signal) showed that using the annual modulation as few as $\mathcal{O}(500)$ events may be sufficient to significantly increase model identification~\cite{Witte:2016ydc}. This effect may be enhanced for a DM differential cross section containing a non-standard dependence on the DM velocity (e.g. due to a magnetic dipole coupling). In such scenario, the properties of the annual modulation (e.g. the phases and amplitudes) are nuclide-dependent~\cite{DelNobile:2015tza,DelNobile:2015rmp}. This may help distinguishing DM and neutrino signals, since the modulation from neutrinos is target-universal.

In this work we do not consider time-dependence of the signal. However, a combined analysis using both time and recoil energies of observed events will likely show that the discriminatory power obtained in the following sections to be conservative.

\subsection{Interactions}

Conventional analyses of direct detection data assume WIMP-nucleus scattering proceeds through an SI or SD interaction. These are the types of interactions that arise from the exchange of heavy scalar and vector (in the case of a SI interaction), or axial-vector (in the case of a SD interaction), mediators in the non-relativistic limit to zeroth order in $|\vec{q}|$ and $v$. These interactions, however, hardly constitute an exhaustive list of the viable DM-nucleus interactions, and fail to characterize the richer phenomenology that may arise in direct detection experiments. 

Recently, an effective field theory (EFT) analysis for direct detection experiments has been developed~\cite{Fan:2010gt,Fitzpatrick:2012ix}. The EFT approach provides an exhaustive list of non-relativistic interactions that are capable of characterizing the various types of DM signals that could appear in direct detection experiments. This is a valuable study that may provide additional insight that could otherwise be lost in conventional analyses, however, viable UV-complete models often produce non-trivial linear combinations of such operators. While this work does not focus on the neutrino floor within the context of UV complete models, we will consider DM-nuclei interactions that have been shown to arise from UV-complete models. Within such models, the coefficients determining the relative importance of each EFT term in the interaction often depends on the target material. Hence, well-motivated signatures are likely to be overlooked in a conventional EFT analysis, in particular when signals are detected in multiple experiments employing different target elements. 

In this section, we will motivate a large list of viable DM-nucleus interactions that capture complex phenomenology that may arise in these experiments, but are also motivated with some UV completion. 
We first provide a brief overview of the EFT approach for direct detection, as we use tools developed for EFT analyses to streamline the calculation of the differential cross sections studied in this paper.

\subsubsection{Effective field theory approach}

In the EFT method, one starts by specifying the DM couplings to nucleons and then computes the corresponding interactions with nuclei.
The non-relativistic nature of DM-nucleon interactions in direct detection experiments allows for interactions to be decomposed into a  non-relativistic EFT basis. A complete set of Galilean-invariant Hermitian operators can be constructed from the following dimensionless three-vectors \cite{Fitzpatrick:2012ix}: the momentum transfer divided by the nucleon mass $\vec{q}/m_N$, the relative DM-nucleon velocity in the direction perpendicular to the momentum exchange $\vec{v}^\perp \equiv \vec{v} - \vec{q}/2\mu_N$, where $\mu_N = m_{\chi} m_N/ (m_{\chi} + m_N) $ is the DM-nucleon reduced mass, as well as the DM $\vec{S}_\chi$ and the nucleon $\vec{S}_N$ spins. There exist a total of 15 Hermitian operators that can be constructed from these variables, enumerated in \cite{Fitzpatrick:2012ix,Anand:2013yka}.

For any given interaction, such as those listed in the second column of \Tab{tab:fermop}, one can take the non-relativistic limit and write the interaction in terms of the EFT operators. Here, we present a brief example, outlined in \cite{Anand:2013yka}, of an axial vector mediated contact interaction (see SD in \Tab{tab:fermop} and also assuming a heavy mediator) given by
\begin{equation}\label{eq:egoper}
\mathcal{A}_{\rm int} = c \bar{\chi}\gamma^\mu\gamma^5\chi\bar{N}\gamma_\mu\gamma^5N ~,
\end{equation}
where $\chi$ and $N$ are the DM and nucleon spinors, respectively. Using the Bjorken and Drell gamma matrix conventions with spinor normalization, one can write the non-relativistic limit of a spinor as
\begin{equation}\label{eq:nonrel}
\chi(p) = \sqrt{\dfrac{E+m_{\chi}}{2m_{\chi}}}\left(
\begin{array}{c}
\xi \\
\dfrac{\vec{\sigma}\cdot \vec{p}}{E+m_{\chi}}\xi
\end{array}\right) \rightarrow 
\left(
\begin{array}{c} 
\xi \\
\dfrac{\vec{\sigma}\cdot \vec{p}}{2m_{\chi}}\xi
\end{array}\right) \, ,
\end{equation}
where $\xi$ is a two component spinor (and likewise for the nucleon spinor $N$). Using Eq.~\eqref{eq:nonrel} with the interaction provided in Eq.~\eqref{eq:egoper}, and recalling that the $\vec{S} = \xi^{\prime\dagger}(\sigma/2)\xi$, one can rewrite the interaction as 
\begin{equation}
\mathcal{A}_{\rm int} = -4c \, \vec{S}_\chi \cdot \vec{S}_N \, \equiv -4c \,\mathcal{O}_4 \, , 
\end{equation}
where the operator $\mathcal{O}_4$ follows the conventions of ~\cite{Fitzpatrick:2012ix}. In general, interactions will be summed over multiple operators $\mathcal{O}_i$ for both neutrons and protons. The spin-averaged DM-nucleus scattering amplitude can then be computed using
\begin{equation}
\frac{1}{2J_\chi+1}\frac{1}{2J_T+1}\sum_{\rm spins}|\mathcal{M}|^2 \equiv \frac{m_T^2}{m_N^2}\sum_{i,j} \, c_i c_j\,F_{ij}^T(v^2,q^2) \, ,
\end{equation}
where $\mathcal{M}$ si the DM-nucleon scattering amplitude summed over nucleons and computer between nuclear states, $J_\chi$ and $J_T$ are the total angular momentum of the DM and nucleus respectively, the $c_i$ are the coefficients multiplying the operator $i$ (e.g. $-4c$ in the example provided above), and $F_{ij}^T$ is a nuclide-dependent response factor (these can be found in ~\cite{Fitzpatrick:2012ix,Fitzpatrick:2012ib,Anand:2013yka}). The differential cross section as a function of scattering angle is then obtained using the spin-averaged scattering amplitude:
\begin{equation}
\dfrac{d\sigma}{d \cos{\theta}} = \dfrac{1}{(2 J_{\chi} + 1)} \dfrac{1}{(2 J_T + 1)} \sum_{\rm spins} \dfrac{1}{32 \pi} \dfrac{|\mathcal{M}|^2}{(m_{\chi} + m_T)^2} \, .
\end{equation}
Finally, for elastic DM scattering one can change variables using $d \cos{\theta} = m_T/\mu_T^2 v^2$ (with $\theta$ defined in the center of mass frame) and obtain the differential cross-section $d\sigma/dE_R$  needed for the interaction in \Eq{eq:dm_diff_rate}.

The calculations using the above formalism can be expedited with a publicly available code, described in~\cite{Anand:2013yka}, which allows for a straightforward computation of the cross sections considered using the form factors specific to particular isotopes (see \Tab{tab:experiments}) and that we employ for our analysis.

It is worth noting at this point that some of the interactions we consider are dominated in the non-relativistic limit by one EFT operator, but some consist of a combination of several operators.  Using the interaction names in the first column of \Tab{tab:fermop} and the usual names of EFT operators as introduced in~\cite{Fitzpatrick:2012ix}, the scattering amplitude for SI is dominated by $\mathcal{O}_1$, for SD by $\mathcal{O}_4$, for PS-S by $\mathcal{O}_{11}$, for S-PS by $\mathcal{O}_{10}$ and PS-PS by $\mathcal{O}_6$, as can be seen in \cite{Dent:2015zpa}, and the amplitude for ED is dominated by $\mathcal{O}_{11}$. Using dimensionless versions of the EFT operators  (as e.g. in \cite{Dent:2015zpa}, in which $\vec{q}/m_N$ is used instead of  $\vec{q}$ for the operators defined in~\cite{Fitzpatrick:2012ix}) the scattering amplitudes with nucleons $N$ for Ana (the anapole coupling) and AV-V are proportional to the combination~\cite{Dent:2015zpa}
\begin{equation}
\mathcal{A}_{\rm Ana} \propto  Q_N \mathcal{O}_8 - \frac{\lambda}{e/2m_N} \mathcal{O}_9~,
\end{equation}
and the amplitude for MD (magnetic dipole couplings) is proportional to (excluding the propagator)
\begin{equation}
\mathcal{A}_{\rm MD} \propto  m_N Q_N ( |\vec{q}|^2 \mathcal{O}_1 + 4 m_\chi m_N \mathcal{O}_5) + \frac{\lambda}{e/2m_N} 4 m_\chi( |\vec{q}|^2 \mathcal{O}_4 - m_N^2 \mathcal{O}_6)~,
\end{equation}
where $m_N$, $Q_N$ and $\lambda$ are the mass, charge and magnetic moment of the nucleon  $N$ ($e/2 m_N$ is the nuclear magneton). Notice that when several EFT operators appear in the amplitude, their  coefficients are target element dependent and, thus, different operators may be dominant in the amplitude for different targets in a particular energy range.

 \subsubsection{Interaction models}
 \label{sssec:models}

We  now proceed to describe specific DM interaction models considered in our study, as specified for a DM fermion $\chi$ in \Tab{tab:fermop}, a DM scalar $\phi$ in \Tab{tab:scaleop} and DM vector $X^{\mu}$ in \Tab{tab:vecop}. In this section we also provide the differential cross-sections for fermionic DM candidates, assuming elastic DM-nuclei scattering.
For momentum-independent interactions, the differential cross-section is related to the full cross-section $\sigma_T$ as
\begin{equation}
\dfrac{d \sigma_T}{d E_R} = \dfrac{\sigma_T m_T}{2 \mu_T^2 v^2}~.
\end{equation}

For each interaction we consider the case of either a heavy or a light mediator, where `heavy' and `light' refer to the mass of the mediator as compared to the typical momentum transfer $|\vec{q}|$. We only specify the full differential cross-sections for the light (massless) mediator case, i.e. when the mediator mass $M$ obeys $|\vec{q}| \gg M$. For the case of heavy mediator, i.e. $M \gg |\vec{q}|$, the results are readily obtainable using the relation
\begin{equation}
\dfrac{d \sigma}{d E_R} = \dfrac{ |\vec{q}|^4}{(|\vec{q}|^2 + M^2)^2} \left(\dfrac{d \sigma_{\rm light}}{d E_R} \right)~
\end{equation} 
and $|\vec{q}|^2 + M^2 \simeq M^2$.

For a DM scalar and vector field, many of the interactions produce cross sections which, at leading order, have a dependence on the momentum transfer and DM speed identical to the interactions produced by a fermionic DM candidate. Consequently, the resulting degeneracy of the neutrino and DM recoil spectra in these models is qualitatively similar, although one should bear in mind that other nuclide-specific factors may favor particular experimental targets. In this case we simply specify the relevant $|\vec{q}|$, $v$ dependence of the interaction and the fermionic model they mimic, given that the qualitative features of the recoil spectrum are determined by these factors (as will be shown in the following sections). The shape of the recoil spectrum is determined by the $|\vec{q}|$ dependence, while the annual modulation and the relative importance of various terms in the cross section may be altered by the $v$-dependent terms. From each cross-section we extract an arbitrary constant ``reference cross-section'' $\sigma_{\rm ref}$, to indicate its magnitude (sometimes it is chosen to be the DM-proton total cross-section). It is a normalization factor, stated in terms of reference momentum $|\vec{q}_{\rm ref}|  =  100$ MeV, for the DM differential spectrum. We now list the form of the relativistic scattering amplitudes and the resulting differential cross-sections for all the interactions we study, which we assume all describe a single-mediator-exchange scattering. This a rather complete but not necessarily exhaustive list. 
Throughout this work we assume that the couplings of DM to neutrons and protons are the same. In principle these couplings could be different which would additionally not subsequently enhance or suppress the overall scattering rate in a particular target nuclei \cite{Feng:2011vu,Kurylov:2003ra}.

\begin{table}[tb]
\begin{centering}
 \renewcommand{\arraystretch}{2.25}
   \begin{threeparttable}
\begin{tabular}{|c|c|c|c|c|c|} \hline
  &   & \multicolumn{2}{c|}{  Dependence ($|\vec q|$, $v$)} & \multicolumn{2}{c|}{$\sigma_{\rm ref}$} 
\\  
Model & {\rm Interaction} 	& \text{Heavy} & \text{Light} & \text{Heavy} & \text{Light}
\\ \hline \hline
 \multirow{2}{*}{SI} 		& $\bar{\chi} \chi \bar{N} N $ &  \multirow{2}{*}{$\dfrac{1}{v^2}$} &  \multirow{2}{*}{$\dfrac{1}{v^2 |\vec{q}|^4}$} &  \multirow{2}{*}{$\dfrac{\mu_N^2}{\pi}\left(\dfrac{f_p^{\rm SI}}{M^2} \right)^2$} &
  \multirow{2}{*}{$\dfrac{\mu_N^2}{\pi}\left(\dfrac{f_p^{\rm SI}}{|\vec{q}_{\rm ref}|^2} \right)^2$}
\\  
   & $ \bar{\chi} \gamma^{\mu} \chi \bar{N} \gamma_{\mu} N $ &   &   &   &
\\ \hline 
SD  & $   \bar{\chi} \gamma^{\mu}\gamma^5 \chi \bar{N} \gamma_{\mu}\gamma_5 N $ & $\dfrac{1}{v^2}$ &  $\dfrac{1}{v^2 |\vec{q}|^4}$ & $\dfrac{3 \mu_N^2}{\pi}\left(\dfrac{a_p^{\rm SD}}{ M ^2} \right)^2$  & $\dfrac{3 \mu_N^2}{\pi}\left(\dfrac{a_p^{\rm SD}}{|\vec{q}_{\rm ref}|^2} \right)^2$
\\ \hline 
 Ana   &   $    \bar{\chi} \gamma^{\mu}\gamma_5 \chi  \partial^{\nu} F_{\mu\nu} $  &   $|\vec{q}|^4~,~\dfrac{|\vec{q}|^6}{v^2}$   &    1~,~$\dfrac{|\vec{q} |^2}{v^2 }$  &  $ \dfrac{\varepsilon^2 \mu_N^2}{4 \pi}\left(\dfrac{e g_{\chi}^{\rm A} |\vec{q}_{\rm ref}|^2}{\Lambda^2 M^2} \right)^2$ &  $ \dfrac{\varepsilon^2 \mu_N^2}{4 \pi}\left(\dfrac{e g_{\chi}^{\rm A}}{\Lambda^2} \right)^2$ \\    \hline 
MD  & $\bar{\chi} \sigma^{\mu \nu} \chi F_{\mu \nu}  $ & $|\vec{q}|^2~,~\dfrac{|\vec{q}|^4}{ v^2}$  &  $\dfrac{1}{|\vec{q}|^2}~,~\dfrac{1}{v^2}$   & $\dfrac{ \mu_N^2}{\pi} \left(\dfrac{e g_{\chi}^{\rm MD} |\vec{q}_{\rm ref}|}{ \Lambda M^2}\right)^2$  & $\dfrac{ \mu_N^2}{\pi} \left(\dfrac{e g_{\chi}^{\rm MD}}{ \Lambda |\vec{q}_{\rm ref}|}\right)^2$
\\ \hline 
ED  & $ \bar{\chi} \sigma^{\mu \nu} \gamma^5 \chi F_{\mu \nu}$ &      $\dfrac{|\vec{q}|^2}{v^2}$ & $ \dfrac{1}{v^2 |\vec{q}|^2}$  & $\dfrac{ \mu_N^2}{\pi} \left(\dfrac{e g_{\chi}^{\rm ED} |\vec{q}_{\rm ref}|}{ \Lambda M^2}\right)^2$  & $\dfrac{ \mu_N^2}{\pi} \left(\dfrac{e g_{\chi}^{\rm ED}}{ \Lambda |\vec{q}_{\rm ref}|}\right)^2$
\\ \hline 
  mC\tnote{1}  & $ \bar{\chi} \gamma^{\mu} \chi \bar{N} \gamma_{\mu} N$ &   $\dfrac{1}{  v^2}$ & $ \dfrac{1}{v^2 |\vec{q}|^4 }$    & $\dfrac{\varepsilon^2 \mu_N^2}{\pi} \left(\dfrac{ e g_{\chi}^{\rm mC} }{ M^2}\right)^2$  & $\dfrac{\varepsilon^2 \mu_N^2}{\pi} \left(\dfrac{ e g_{\chi}^{\rm mC} }{|\vec{q}_{\rm ref}|^2}\right)^2$
\\ \hline 
PS-S  & $ \bar{\chi} \gamma^5 \chi \bar{N} N$ & $\dfrac{|\vec{q}|^2}{v^2}$  & $\dfrac{1}{v^2 |\vec{q}|^2}$   & $\dfrac{ \mu_N^2}{4 \pi}  \left(\dfrac{g_{\chi}^{\rm PS} f_p^{\rm S}|\vec{q}_{\rm ref}|}{M^2 m_{\chi}}\right)^2 $  & $\dfrac{  \mu_N^2}{4 \pi m_{\chi}^2} \left(\dfrac{g_{\chi}^{\rm PS} f_p^{\rm S}}{|\vec{q}_{\rm ref}|}\right)^2 $
\\ \hline 
S-PS  & $  \bar{\chi} \chi \bar{N} \gamma^5  N $ & $\dfrac{|\vec{q}|^2}{v^2}$   &  $\dfrac{1}{v^2 |\vec{q}|^2}$  & $\dfrac{  \mu_N^2}{4 \pi} \left(\dfrac{g_{\chi}^{\rm S} f_p^{\rm PS} |\vec{q}_{\rm ref}|}{M^2 m_p}\right)^2 $  & $\dfrac{  \mu_N^2}{4 \pi  m_p^2} \left(\dfrac{g_{\chi}^{\rm S} f_p^{\rm PS}}{|\vec{q}_{\rm ref}|}\right)^2 $
\\ \hline 
PS-PS & $ \bar{\chi}  \gamma^5 \chi \bar{N} \gamma^5  N $ & $\dfrac{|\vec{q}|^4}{v^2}$  &  $\dfrac{1}{v^2}$  & $\dfrac{  \mu_N^2}{16 \pi} \left(\dfrac{g_{\chi}^{\rm PS} f_p^{\rm PS} |\vec{q}_{\rm ref}|^2}{M^2 m_p  m_{\chi}}\right)^2 $  & $\dfrac{  \mu_N^2}{16 \pi} \left(\dfrac{g_{\chi}^{\rm PS} f_p^{\rm PS}}{  m_p  m_{\chi}}\right)^2 $
\\ \hline 
AV-V & $\bar{\chi} \gamma^{\mu} \gamma_5 \chi \bar{N}\gamma_\mu N $  & $1~,~\dfrac{|\vec{q}|^2}{v^2}$ & $\dfrac{1}{|\vec{q}|^4}~,~\dfrac{1}{v^2 |\vec{q}|^2}$ & $\dfrac{ \varepsilon^2 \mu_N^2}{4\pi}  \left(\dfrac{g_{\chi}^{\rm AV-V}}{M^2}\right)^2 $ & $\dfrac{ \varepsilon^2 \mu_N^2}{4\pi}  \left(\dfrac{g_{\chi}^{\rm AV-V}}{|\vec{q}_{\rm ref}|^2}\right)^2 $ \\ \hline 
\end{tabular}
\begin{tablenotes}
\item[1] This model is equivalent to SI. 
\end{tablenotes}
\caption{Interaction models with fermionic DM particles. Model name, interaction, dependence on $|\vec{q}|$ and $v$, as well as the definition of the reference cross-section $\sigma_{\rm ref}$ are shown. Values for both heavy and light mediators are included.}
\label{tab:fermop} 
  \end{threeparttable}
\end{centering}
\end{table}

\begin{table}[tb]
\begin{centering}
\renewcommand{\arraystretch}{2.1}
\begin{tabular}{|c|c|c|c|c| } \hline
  &   & \multicolumn{2}{c|}{  Dependence ($|\vec q|$, $v$)} &  Similar
\\  
Model & {\rm Interaction} 	& \text{Heavy} & \text{Light} &  Model \\ 
\hline 
\hline
S1 & $\phi^\dagger \phi \bar{N}N$ & $\dfrac{1}{v^2}$  &  $\dfrac{1}{v^2 |\vec{q}|^4}$ & SI \\ \hline 
S2 & $\phi^\dagger \phi \bar{N}\gamma^5 N $ &  $\dfrac{|\vec{q}|^2}{v^2}$  & $\dfrac{1}{v^2 |\vec{q}|^2}$   &  S-PS \\ \hline 
S3 & $\phi^\dagger \overset\leftrightarrow{\partial_\mu}\phi \bar{N}\gamma^\mu N $ & $\dfrac{1}{v^2}$  &  $\dfrac{1}{v^2 |\vec{q}|^4}$ & SI  \\ \hline 
S4 & $\phi^\dagger \overset\leftrightarrow{\partial_\mu}\phi \bar{N}\gamma^\mu\gamma^5 N $ &  $1~,~\dfrac{|\vec{q}|^2}{v^2}$ &  $\dfrac{1}{|\vec{q}|^4}~,~\dfrac{1}{v^2|\vec{q}|^2}$ &  AV-V \\ \hline 
\end{tabular}
\caption{Interaction models with scalar DM particles. Model name, interaction, dependence on $|\vec{q}|$ and $v$, as well as the fermionic DM particle model to which the behavior is similar are shown.}
\label{tab:scaleop} 
\end{centering}
\end{table}
 
\begin{table}[tb]
\begin{centering}
\renewcommand{\arraystretch}{2.1}
\begin{tabular}{|c|c|c|c|c| } \hline
  &   & \multicolumn{2}{c|}{  Dependence ($|\vec q|$, $v$)} &  Similar
\\  
Model & {\rm Interaction} 	& \text{Heavy} & \text{Light} &  Model \\ 
\hline 
\hline
V1 & $X^\mu X_\mu^\dagger \bar{N}N $ & $\dfrac{1}{v^2}$  &  $\dfrac{1}{v^2 |\vec{q}|^4}$ & SI \\ \hline 
V2 & $X^\mu X_\mu^\dagger \bar{N}\gamma^5 N $ & $\dfrac{|\vec{q}|^2}{v^2}$   &  $\dfrac{1}{v^2 |\vec{q}|^2}$ &  S-PS \\ \hline 
V3 & $X^\mu \partial_\nu X_\mu^\dagger \bar{N}\gamma^\nu N $ & $\dfrac{1}{v^2}$  &  $\dfrac{1}{v^2 |\vec{q}|^4}$ & SI  \\ \hline 
V4 & $X^\mu \partial_\nu X_\mu^\dagger \bar{N}\gamma^\nu\gamma^5 N $ &  $1~,~\dfrac{|\vec{q}|^2}{v^2}$ &  $\dfrac{1}{|\vec{q}|^4}~,~\dfrac{1}{v^2|\vec{q}|^2}$ &  AV-V \\ \hline 
V5 & $(X_\mu^\dagger X_\nu -X_\nu^\dagger X_\mu)\bar{N}\sigma^{\mu\nu}N$ & $\dfrac{1}{v^2}$ &  $\dfrac{1}{v^2 |\vec{q}|^4}$ & SD \\ \hline 
V6 & $(X_\mu^\dagger X_\nu -X_\nu^\dagger X_\mu)\bar{N}\sigma^{\mu\nu}\gamma^5N$ & $\dfrac{|\vec{q}|^2}{v^2}$  & $\dfrac{1}{v^2 |\vec{q}|^2}$ &  PS-S \\ \hline 
V7 & $X_\nu^\dagger \partial^\nu X_\mu \bar{N}\gamma^\mu\gamma^5 N$ & $\dfrac{|\vec{q}|^2}{v^2}$   &  $\dfrac{1}{v^2 |\vec{q}|^2}$ & S-PS \\ \hline 
V8 & $\epsilon^{\mu\nu\rho\sigma}X_\nu^\dagger \partial_\rho X_\sigma \bar{N}\gamma_\mu N$ &  $\dfrac{|\vec{q}|^2}{v^2}~, 1$ &  $\dfrac{1}{v^2|\vec{q}|^2}~,~\dfrac{1}{|\vec{q}|^4}$  & AV-V \\ \hline 
V9 & $\epsilon^{\mu\nu\rho\sigma}X_\nu^\dagger \partial_\rho X_\sigma \bar{N}\gamma_\mu\gamma^5 N$ & $\dfrac{1}{v^2}$ &  $\dfrac{1}{v^2 |\vec{q}|^4}$ & SD \\ \hline 
\end{tabular}
\caption{Interaction models with vectorial DM particles.  Model name, interaction, dependence on $|\vec{q}|$ and $v$, as well as the fermionic DM particle model to which the behavior is similar are shown.}
\label{tab:vecop} 
\end{centering}
\end{table}

\paragraph{\underline{Spin--1/2 fermionic DM $\chi$}} 

~\newline

The standard SI and SD are the most generic DM interactions. They are due to the exchange of a scalar or vector boson mediator (for SI), or an axial vector boson mediator (for SD).

\begin{itemize}
\item \emph{Spin-independent  } (SI):

The SI scattering amplitude is given by
\begin{equation} 
\mathcal{A}_{\rm SI} \propto  \dfrac{f_N^{\rm SI}}{(|\vec{q}|^2 + M^2)}  \bar{\chi} \chi \bar{N} N  ~~~~ \text{or} ~~~ \mathcal{A}_{\rm SI} \propto \dfrac{f_N^{\rm SI}}{(|\vec{q}|^2 + M^2)}  \bar{\chi} \gamma^{\mu} \chi \bar{N} \gamma_{\mu} N ~.
\end{equation}
For a light ($M = 0$) mediator the differential cross-section is given by (e.g. \cite{DelNobile:2013gba}): 
\begin{equation} \label{eq:diffxsecsi}
\dfrac{d \sigma_T^{\rm SI}}{d E_R} = \sigma_{\rm ref}^{\rm SI}  \dfrac{|\vec{q}_{\rm ref}|^4}{|\vec{q}|^4} \dfrac{m_T }{2 \mu_N^2 v^2} \Big[ Z_T + \dfrac{f_n^{\rm SI}}{f_p^{\rm SI}} (A_T - Z_T) \Big]^2 F_{{\rm SI}, T}^2~,
\end{equation}
where $Z_T$ and $A_T$ are the atomic and mass numbers of the target nucleus $T$, $F_{{\rm SI},T}$ is the charge nuclear form factor (note that there is an implicit assumption here that neutrons $n$ have the same form factor as protons $p$), taken to be the Helm form factor \cite{Helm:1956zz}, $f_n$ and $f_p$ are the effective couplings of the DM particle to neutrons and protons, respectively, $\sigma_{\rm ref}^{\rm SI}$ is the DM-proton cross-section (i.e. $\sigma_{\rm ref}^{\rm SI} = \sigma_{\rm p}^{\rm SI}$), given in the 1$^{\rm st}$ row of \Tab{tab:fermop} (assumed to be the same for $n$) and $\mu_N$ is the DM-nucleon reduced mass. For small enough $|\vec{q}|$ values the SI scattering acts coherently with the entire nucleus, leading to the $\sim A_T^2$ enhancement for $f_n^{\rm SI} = f_p^{\rm SI}$.
~\newline

\item \emph{Spin-dependent } 
(SD):

The conventional SD interaction arises from an axial-vector (AV) mediated interaction with an amplitude given by
\begin{equation}
\mathcal{A}_{\rm SD} \propto \dfrac{a_N^{\rm SD}}{(|\vec{q}|^2 + M^2)} \bar{\chi} \gamma^{\mu}\gamma^5 \chi \bar{N} \gamma_{\mu}\gamma_5 N~.
\end{equation}
The differential cross-section is given by (e.g. \cite{DelNobile:2015lxa})
\begin{equation}
\dfrac{d \sigma_T^{\rm SD}}{d E_R} = \sigma_{\rm ref}^{\rm SD} \dfrac{|\vec{q}_{\rm ref}|^4}{|\vec{q}|^4} \dfrac{m_T }{8 \mu_N^2 v^2}   \dfrac{4}{3} \dfrac{J_T+1}{J_T}
\Big[ \langle S_{p, T} \rangle + \dfrac{a_n^{\rm SD}}{a_p^{\rm SD}} \langle S_{n, T} \rangle \Big]^2 F_{{\rm SD}, T}^2~,
\end{equation}
where $J_T$ is the spin of the target nucleus, while $\langle S_{p, T} \rangle$, $\langle S_{n, T} \rangle$ represent the proton and the neutron spin of the target nucleus and $a_p^{\rm SD}, a_n^{\rm SD}$ the DM coupling to protons and neutrons, respectively, with $F_{{\rm SD},T}^2$ being the spin form factor of the nucleus.
\end{itemize}

Neutral DM particle candidates can interact with photons through higher electromagnetic moments. 
Such interactions can naturally occur in models of composite DM formed from charged components (e.g. \cite{Bagnasco:1993st}), e.g. a ``dark neutron'', or in models where the DM couples to an intermediate sector that leads to an effective photon coupling (e.g. through heavy charged messengers \cite{Weiner:2012gm} or kinetic mixing with a dark photon \cite{Holdom:1985ag}), as occurs with neutrinos in the Standard Model. 
The effective low energy non-renormalizable operators describing these interactions are suppressed by the scale of new physics $\Lambda$, such as the messenger particle mass or the compositeness scale of the ultraviolet theory. Here, for the electromagnetic moment interactions, we focus explicitly only on vector mediators which couple to the electromagnetic current of the nucleus.

For fermionic DM, the most studied candidates are WIMPs with the lowest order electromagnetic moments (e.g.~\cite{Pospelov:2000bq,An:2010kc,Sigurdson:2004zp,Barger:2010gv,Chang:2010en,Cho:2010br,Heo:2009vt,Gardner:2008yn,Masso:2009mu,Banks:2010eh,Fortin:2011hv,Kumar:2011iy,Barger:2012pf,DelNobile:2012tx,Cline:2012is,Weiner:2012cb,Tulin:2012uq,Cline:2012bz}), the magnetic and the electric dipole moments, given by dimension five effective operators and thus proportional to the inverse of a large scale of new physics $\Lambda$. For Majorana fermions the magnetic and electric dipole moments vanish (although non-diagonal couplings are possible). In this case the only possible electromagnetic moment is the anapole, with the respective interaction described by a dimension-six effective operator proportional to $1/\Lambda^2$. The anapole moment DM has been studied in various contexts, including  direct detection \cite{Pospelov:2000bq,Ho:2012bg,Fitzpatrick:2010br,Frandsen:2013cna,Gresham:2013mua} and colliders \cite{Gao:2013vfa}. 

~\newline
\begin{itemize}
\item \emph{Anapole} (Ana) \emph{moment}:

The anapole amplitude is given by
\begin{equation}
\mathcal{A}_{\rm anapole} \propto \dfrac{\varepsilon g_{\chi}^{\rm A}}{ \left( |\vec{q}|^2 + M^2\right) } \dfrac{|\vec{q}|^2}{ \Lambda^2} \bar{\chi} \gamma^{\mu}\gamma_5 \chi J_{\mu}~.
\end{equation}
This interaction is mediated by a vector boson that couples to the electromagnetic current $J_{\mu}$ (e.g. via a kinetic mixing $\varepsilon$). For $M = 0$ the mediator can be the photon.
With $F_{\mu\nu} = \partial_{\mu} A_{\nu} - \partial_{\mu} A_{\nu}$, for a massless mediator (e.g. $A_{\mu}$ is the SM photon) the equations of motions imply $J_{\mu} = \partial^{\nu} F_{\mu \nu}$.
Unlike magnetic and electric dipole moments, the anapole \cite{ZelDovich:1958} is not part of the pure electro-magnetic multipole expansion (however, see discussion in \cite{fernandez-corbaton:natsci2017}). The anapole moment violates charge conjugation $C$ and parity $P$, but preserves $C P$.  It can arise in models of Majorana fermion DM coupling to a photon. The anapole can couple diagonally to Majorana fermions since it is CPT self-conjugate, which is not possible for magnetic or electric CPT-odd dipole moments.  For a light mediator the anapole differential cross-section is given by (e.g. \cite{DelNobile:2014eta})
\begin{equation} \label{eq:anapolediff}
\dfrac{d \sigma_T^{\rm A}}{d E_R} =  \sigma_{\rm ref}^{\rm A} \dfrac{2 m_T }{\mu_N^2 v^2} \Big[ Z_T^2 \left(v^2 - \dfrac{|\vec{q}|^2}{4 \mu_T^2} \right) F_{{\rm E}, T}^2 
 + \dfrac{\lambda_T^2}{\lambda_N^2} \dfrac{ |\vec{q}|^2}{2 m_N^2} \left(\dfrac{J_T + 1}{3 J_T}\right)  F_{{\rm M},T} \Big] ~,
\end{equation}
where $\sigma_{\rm ref}^{\rm A}$ is given in \Tab{tab:fermop}, $F_{{\rm E}, T}, F_{{\rm M}, T}$ are the electric and magnetic form factors, with $F_{{\rm E}, T} = F_{{\rm SI}, T}$, $\lambda_T$ is the nuclear magnetic moment (see \Tab{tab:experiments}) and $\lambda_N = e/2 m_N$ is the nuclear magneton. 
A distinct characteristic of this cross-section is that it contains two different terms with different dependencies on the DM particle speed.
\newline

\item \emph{Magnetic dipole} (MD):

The MD amplitude is given by
\begin{equation}
\mathcal{A}_{\rm MD} \propto \dfrac{ g_{\chi}^{\rm MD}}{\Lambda (|\vec{q}|^2 + M^2)}  \bar{\chi} \sigma^{\mu \nu} \chi~q_\mu J_\nu~.
\end{equation}
where $J_\nu$ is the electromagnetic current. For $M = 0$ this can be a photon-mediated scattering. In general, it can be due to the exchange of a vector boson that couples to the electromagnetic field tensor $F_{\mu \nu}$.
In this case, the differential cross-section for MD is given by \cite{DelNobile:2014eta}
\begin{equation}
\dfrac{d \sigma_T^{\rm MD}}{d E_R} = \sigma_{\rm ref}^{\rm MD} \dfrac{|\vec{q}_{\rm ref}|^2}{|\vec{q}|^2} \dfrac{m_T^2}{4 v^2 \mu_N^2} 
\Big[ Z_T^2 \left( 4  v^2  - |\vec{q}|^2 \Big\{ \dfrac{1}{\mu_T^2} - \dfrac{1}{m_{\chi}^2}\Big\} \right) F_{{\rm E}, T}^2 
 + 2 \dfrac{|\vec{q}|^2}{ m_N^2}  \dfrac{\lambda_T^2}{\lambda_N^2} \left(\dfrac{J_T + 1}{3 J_T}\right)   F_{{\rm M},T} \Big] ~,
\end{equation}
where $m_{\chi}$ is the mass of the DM fermion and $\sigma_{\rm ref}^{\rm MD}$ is an arbitrary factor extracted from the cross-section and defined in \Tab{tab:fermop}. Here again $F_{{\rm E},T}^2 = F_{{\rm SI},T}^2$ is the charge nuclear form factor and $F_{{\rm M},T}^2$ the magnetic nuclear form factor, with other variables defined as before. We note that this cross-section also contains two different terms, as is the case with an anapole interaction, with different dependencies on the DM particle speed. This could lead to distinct annual modulation signals in different target materials~\cite{DelNobile:2015nua,DelNobile:2015tza}.
~\newline

\item \emph{Electric dipole} (ED):

The ED amplitude is given by
\begin{equation}
\mathcal{A}_{\rm ED} \propto \dfrac{g_{\chi}^{\rm ED}}{\Lambda (|\vec{q}|^2 + M^2 )} \bar{\chi} \sigma^{\mu \nu} \gamma^5 \chi ~q_\mu J_\nu~.
\end{equation}
where $J_\nu$ is the electromagnetic current. For a light mediator, the ED differential cross-section is given by  (e.g. \cite{Barger:2010gv})
\begin{equation}
\dfrac{d \sigma_T^{\rm ED}}{d E_R} = \sigma_{\rm ref}^{\rm ED} \dfrac{|\vec{q}_{\rm ref}|^2}{|\vec{q} |^2  }  \dfrac{m_T}{2 v^2 \mu_N^2} \Big[ Z_T^2  F_{{\rm E},T}^2 \Big] ~.
\end{equation} 
We note that ED has no dependence on the spin and the magnetic form factor $F_{\rm M, T}$ like the anapole or MD cross-sections, allowing it to test the charge $Z_T$ signal component separately from other contributions.
~\newline

\item \emph{Milli-charge} (mC):

The mC amplitude is given by
\begin{equation}
\mathcal{A}_{\rm mC} \propto \dfrac{\varepsilon g_{\chi}^{\rm mC}}{\Lambda (|\vec{q}|^2 + M^2 )} \bar{\chi} \gamma^\mu \chi  J_\mu~.
\end{equation}
Like ED and MD for $M = 0$ can be a photon-mediated scattering, but with a DM particle electric charge suppressed by $\varepsilon$. In general, this interaction is due to exchange of a vector boson coupled to the electromagnetic current $J_{\mu}$ with a small coupling $\varepsilon$. DM with a small electric charge can arise when the SM photon
kinetically mixes with a photon of a dark sector $U(1)$ \cite{Holdom:1985ag}, which can be massive through the Stuckelberg
mechanism \cite{Kors:2004dx}.
The mC differential cross-section is given by (e.g. \cite{Gluscevic:2015sqa})  
\begin{equation}
\dfrac{d \sigma_T^{\rm mC}}{d E_R} = \sigma_{\rm ref}^{\rm mC}  \dfrac{|\vec{q}_{\rm ref}|^4}{|\vec{q}|^4} \dfrac{m_T }{2 \mu_N^2 v^2} \Big[ Z_T\Big]^2 F_{{\rm SI}, T}^2 ~.
\end{equation}
Since it is so similar to the SI interaction, we do not treat the mC interaction separately in the rest of the paper.
\end{itemize}

Pseudo-scalar couplings lead to non-standard spin-dependent DM interactions in the non-relativistic limit. While the PS-S does depend on the spin of the DM particle, this is a SI interaction because it does not depend on the spin of the nucleus. 
The $\gamma^5$ coupling is $CP$-odd. Such interactions can arise in models with an extended Higgs sector  (e.g.~Two-Higgs Doublet Models \cite{Gunion:1989we}) and allow for a rich phenomenology (e.g.~\cite{Arina:2014yna}). We consider all of the possible $\gamma^5$ interaction combinations, i.e. ($\mathbb{1} \cdot\gamma_5$), ($\gamma_5\cdot\mathbb{1}$), ($\gamma_5\cdot\gamma_5$) in the vertices. For universal flavor-diagonal quark couplings to the pseudo-scalar mediator the WIMP couples primarily to protons. Rare meson decays already strongly constrain the coupling of light pseudo-scalar bosons (M $< 7$ GeV) to quarks \cite{Hiller:2004ii,Andreas:2010ms,Dolan:2014ska}. If the quark couplings are non-universal and are instead proportional to the quark mass, the flavor constraints are less stringent \cite{Dolan:2014ska}. 

\begin{itemize}
\item \emph{Pseudo-scalar-scalar} (PS-S):

The PS-S amplitude is given by
\begin{equation}
\mathcal{A}_{\rm PS-S} \propto \dfrac{  g_{\chi}^{\rm PS} f_N^{\rm S} }{(|\vec{q} |^2 + M^2)} \bar{\chi} \gamma^5 \chi \bar{N} N~.
\end{equation}
It arises from a scalar mediator with a CP-odd vertex with the DM.
For a light mediator, the  differential cross-section is given by (e.g. \cite{Gluscevic:2015sqa}) 
\begin{equation}
\dfrac{d \sigma_T^{\rm PS-S}}{d E_R} = \sigma_{\rm ref}^{\rm PS-S}  \dfrac{|\vec{q}_{\rm ref}|^2}{|\vec{q}|^2} \dfrac{m_T }{16 m_{\chi}^2 \mu_N^2 v^2} \Big[ Z_T + \dfrac{f_n^{\rm S}}{f_p^{\rm S}} (A_T - Z_T) \Big]^2 F_{{\rm SI}, T}^2 ~.
\end{equation}
 ~\newline

\item \emph{Scalar-pseudo-scalar} (S-PS):

The S-PS amplitude is given by
\begin{equation}
\mathcal{A}_{\rm S-PS} \propto \dfrac{  g_{\chi}^{\rm S} f_n^{\rm PS} }{(|\vec{q} |^2 + M^2)} \bar{\chi} \chi \bar{N} \gamma^5  N~.
\end{equation}

The  differential cross-section for a light mediator is given by (e.g. \cite{Gluscevic:2015sqa}) 
\begin{equation}
\dfrac{d \sigma_T^{\rm S-PS}}{d E_R} = \sigma_{\rm ref}^{\rm S-PS} \dfrac{|\vec{q}_{\rm ref}|^2}{|\vec{q}|^2} \dfrac{4 m_{\chi}^2 m_T }{\mu_T^2 v^2} \dfrac{\mu_T^2}{\mu_N^2}  \dfrac{4}{3} \dfrac{J_T+1}{J_T}
\Big[ \langle S_{p, T} \rangle + \dfrac{f_n^{\rm PS}}{f_p^{\rm PS}} \langle S_{n, T} \rangle \Big]^2 F_{{\rm SD}, T}^2~.
\end{equation}
~\newline

\item \emph{Pseudo-scalar-pseudo-scalar} (PS-PS):

The PS-PS amplitude is given by
\begin{equation}
\mathcal{A}_{\rm PS-PS} \propto \dfrac{  g_{\chi}^{\rm PS} f_n^{\rm PS} }{(|\vec{q} |^2 + M^2)} \bar{\chi}  \gamma^5 \chi \bar{N} \gamma^5  N~.
\end{equation}
Arises from pseudo-scalar mediator. Both of vertices include a $\gamma_5$ and are CP-violating.
For $M = 0$ the  differential cross-section is given by (e.g. \cite{Gluscevic:2015sqa}) 
\begin{equation}
\dfrac{d \sigma_T^{\rm PS-PS}}{d E_R} = \sigma_{\rm ref}^{\rm PS-PS} \dfrac{m_T }{2 \mu_T^2 v^2} \dfrac{\mu_T^2}{\mu_N^2}  \dfrac{4}{3} \dfrac{J_T+1}{J_T}
\Big[ \langle S_{p, T} \rangle + \dfrac{f_n^{\rm S}}{f_p^{\rm S}} \langle S_{n, T} \rangle \Big]^2 F_{{\rm SD}, T}^2~.
\end{equation}
 ~\newline
 
 ~~~~~ In \cite{Fitzpatrick:2010br} it was suggested that 
a coupling $\bar{\chi} \gamma^{\mu} \gamma_5 \chi A_{\mu}^{\prime}$, with $A_{\mu}^{\prime}$ denoting a vector boson, also constitutes an anapole-DM interaction if $A_{\mu}^{\prime}$ couples with nucleons with a vector coupling $\overline{N} \gamma^{\mu} N$. This AV-V coupling also violates $C$ and $P$ symmetries, but preserves $CP$. It was shown in \cite{Ho:2012bg}, however, that $A_{\mu}^{\prime}$ cannot be the SM photon and this operator must arise from another mechanism (e.g. kinetic mixing with a dark photon, suppressed by $\varepsilon$). We treat this operator separately, as the AV-V interaction. The differential cross-section for the AV-V interaction for heavy mediator \cite{Fitzpatrick:2010br} leads to the same differential cross-section as the anapole interaction with a light mediator, i.e. \eqref{eq:anapolediff}.

\item \emph{Axial-vector-vector} (AV-V):

The AV-V amplitude is given by
\begin{equation}
\mathcal{A}_{\rm AV-V} \propto \dfrac{  \varepsilon g_{\chi}^{\rm AV-V} }{(|\vec{q}|^2 + M^2)}  \bar{\chi}  \gamma^{\mu} \gamma_5 \chi  \bar{N}\gamma_\mu N ~.
\end{equation}
We treat here the anapole-like couplings $\bar{\chi} \gamma^{\mu} \gamma_5 \chi A_{\mu}^{\prime}$ and $A^{\prime \mu} J_{\mu}$ (see anapole above), where $J_{\mu}$ is the electro-magnetic current, as suggested in \cite{Fitzpatrick:2010br}. While $A_{\mu}^{\prime}$ cannot be the SM photon, it can readily appear as a dark photon, with a kinetic mixing $\varepsilon$ with the SM electro-magnetic photon. For a light mediator the cross-section is given by
\begin{equation}
\dfrac{d \sigma_T^{\rm AV-V}}{d E_R} = \sigma_{\rm ref}^{\rm AV-V} \dfrac{|\vec{q}_{\rm ref}|^4}{|\vec{q}|^4} \dfrac{2 m_T }{v^2 \mu_N^2} \Big[ Z_T^2 \left(v^2 - \dfrac{|\vec{q}|^2}{4 \mu_T^2} \right) F_{{\rm E}, T}^2 
 + \dfrac{\lambda_T^2}{\lambda_N^2} \dfrac{ |\vec{q}|^2}{2 m_N^2} \left(\dfrac{J_T + 1}{3 J_T}\right)  F_{{\rm M},T} \Big]~.
\end{equation}

\end{itemize}

In addition to the listed fermion DM interactions above, it was suggested in~\cite{Fitzpatrick:2012ix} that when the nucleon angular momentum orbitals are not completely filled one also can expect from very specific DM interactions a response proportional to the product of nucleon angular momentum and spin (i.e. $\vec{L}_N \cdot \vec{S}_N$). We do not consider this interaction in this work.
~\newline 

\paragraph{\underline{Spin--0 scalar DM $\phi$}} 
~\newline

A list of scalar DM interactions from a general matrix element analysis can be found in \cite{Kumar:2013iva}. While all four interactions in \cite{Kumar:2013iva} have cross-sections with an identical dependence on the momentum transfer and the DM velocity to the already considered fermionic ones, we reproduce the cross section below for the only interaction (S4) for which the similarity is perhaps not immediately obvious.

\begin{itemize}
\item \emph{S4}:

The S4 amplitude is given by
\begin{equation}
\mathcal{A}_{\rm S4} \propto \dfrac{  g_{\chi}^{\rm S4} }{(|\vec{q} |^2 + M^2)} \phi^\dagger \overset\leftrightarrow{\partial_\mu}\phi \bar{N}\gamma^\mu\gamma^5 N~.
\end{equation}
We obtain the following differential cross-section
\begin{equation}
\dfrac{d \sigma_T^{\rm S4}}{d E_R} = \sigma_{\rm ref}^{\rm S4} \dfrac{|\vec{q}_{\rm ref}|^4}{|\vec{q}|^4} \dfrac{m_T \mu_T^2 }{4 \mu_N^2 m_N^2 v^2}   \dfrac{1}{3} \dfrac{J_T+1}{J_T} \left( v^2 - \dfrac{|\vec{q}|^2}{4 \mu_T^2}\right) 
\Big[ \langle S_{p, T} \rangle + \dfrac{a_n^{\rm SD}}{a_p^{\rm SD}} \langle S_{n, T} \rangle \Big]^2 F_{{\rm SD}, T}^2~,
\end{equation}
where $\sigma_{\rm ref}^{S4} = \sigma_{\rm ref}^{SD}$.
\end{itemize}
~\newline 

\paragraph{\underline{Spin--1 vectorial DM $X^{\mu}$}}
~\newline
 
A list of vectorial DM interactions from a general matrix element analysis can be found in \cite{Kumar:2013iva}. Nine out of the ten interactions in \cite{Kumar:2013iva} give similar responses to the already considered fermionic DM ones. The only vector interaction with a novel response,  $X_\nu^\dagger \partial^\nu X_\mu \bar{N}\gamma^\mu N$, contains terms in the differential cross section proportional to $|\vec{q}|^4$ and $|\vec{q}|^2v^2$. This interaction is difficult to observe due to the highly suppressed nature of the cross section for small $|\vec{q}|$ and $v$. Further, it will produce no degeneracy with the neutrino spectrum, as will become evident in later sections. Therefore the discovery limits for this vector DM interaction will be similar to those of the PS-PS interaction, and thus we do not consider this interaction independently.

\subsubsection{Inelastic scattering}

So far we have only considered DM to consist of a single component. However, multi-component dark sectors can also appear in well-motivated scenarios and allows for rich phenomenology when the component mass is distinct (e.g.~\cite{Feng:2008ya,Zurek:2008qg,Feldman:2010wy,Aoki:2012ub,Chen:2015jkt,DeSimone:2010tf}). In some of them it is possible that the dominant DM-nucleus scattering is inelastic. This happens when the elastic scattering is suppressed or forbidden by the particular couplings of the mediator~\cite{TuckerSmith:2001hy,ArkaniHamed:2008qn, Cui:2009xq,Graham:2010ca,Batell:2009vb,Essig:2010ye}.

In an inelastic scattering, the initial DM particle of mass $m_\chi$ scatters dominantly into a different mass state $m_\chi^\prime = m_\chi + \delta$, where $|\delta| \ll m_\chi$.  Here $\delta > 0$  describes ``endothermic'' inelastic scattering~\cite{TuckerSmith:2001hy,ArkaniHamed:2008qn, Cui:2009xq}  and $\delta < 0$ describes ``exothermic'' inelastic  scattering~\cite{Graham:2010ca,Batell:2009vb,Essig:2010ye}. The case of elastic scattering is recovered when $\delta = 0$. In the limit $\mu_T |\delta|/m_\chi^2 \ll 1$, $\vmin(\ER)$ is given by 
\begin{equation} \label{eq:vmin}
\vmin(\ER) = \frac{1}{\sqrt{2 m_T \ER}} \left| \frac{m_T \ER}{\mu_T} +\delta \right| \, .
\end{equation}
\Eq{eq:vmin} can be used to obtain the range of possible recoil energies, $[\ER^{T,-}(v),\ER^{T,+}(v)]$, that can be imparted to a target nucleus by a DM particle traveling at speed $v$ in Earth's frame, given by
\begin{equation}\label{eq:Ebranch}
\ER^{T,\pm} (v) = \frac{\mu_T^2 v^2}{2 m_T} \left( 1 \pm \sqrt{1-\frac{2\delta}{\mu_T v^2}} \right)^2 \, ,
\end{equation}
with $\ER^{T,-} (v) \leq E_R \leq \ER^{T,+} (v)$.
\Eq{eq:Ebranch} shows that for endothermic scattering there exists a non-zero kinematic endpoint in DM speed  $v_\delta^T = \sqrt{2 \delta / \mu_T}$, below which incoming DM particles cannot induce nuclear recoils (this endpoint is 0 for elastic and inelastic exothermic scattering). The maximum and minimum possible recoil energies $\ER^{T,-} (v)$ and $\ER^{T,+} (v)$ become equal at $E_{\delta} = \ER^{T,-} (v_{\delta}) = \ER^{T,+} (v_{\delta}) = \mu_T |\delta|/m_T$, where $v_{\delta}$ is the minimum possible value of $v_{\rm min}$.  
For an exothermic scattering the recoiling nucleus energy is close to  $E_{\delta}$, which is proportional
to the splitting between the DM states and is inversely proportional to the nuclear
mass. Thus, the nuclear recoils originating from exothermic interactions are more visible in experiments
with light nuclei and low thresholds. On the other hand, for endothermic DM, only high velocity DM particles have enough energy to up-scatter and the minimum necessary speed decreases with increasing target mass. Thus, high mass targets are favored. Hence, different target materials can act as sensitive probes of multi-component dark sector models in which inelastic scattering dominates over elastic.


\section{Neutrino Background \label{sec:neutrinos}}

\begin{table*}[tbp]
  \setlength{\extrarowheight}{2pt}
  \setlength{\tabcolsep}{10pt}
  \begin{center}

  \begin{threeparttable}
	\begin{tabular}{|l|l|c|}  \hline
	 Lab Site &  Location & Depth (m.w.e)\\
	\hline
	SNOLab 			&  ~46$^\circ$28$'$19$''$\,N 	&   \multirow{2}{*}{$6010$}  \\
    (Sudbury, CA) 	& ~81$^\circ$11$'$12$''$\,W  		&      \\ \hline
    LNGS &   ~42$^\circ$28$'$09$''$\,N  &   \multirow{2}{*}{$3400$}  \\ 
    (Gran Sasso, IT) & ~13$^\circ$33$'$56$''$\,E &      \\ \hline
    SURF & ~44$^\circ$21$'$07$''$\,N &   \multirow{2}{*}{$4400$}  \\ 
    (Homestake, US) & 103$^\circ$45$'$50$''$\,W  &      \\ \hline
 	\end{tabular}
\caption{\label{tab:labs} Experimental laboratories that are likely to host new generation of direct detection experiments considered in this work. Exact location as well as depth (meter water equivalent) are displayed.}
\end{threeparttable}
  \end{center}
\end{table*}

\begin{figure}
\centering
\includegraphics[width=.6\textwidth]{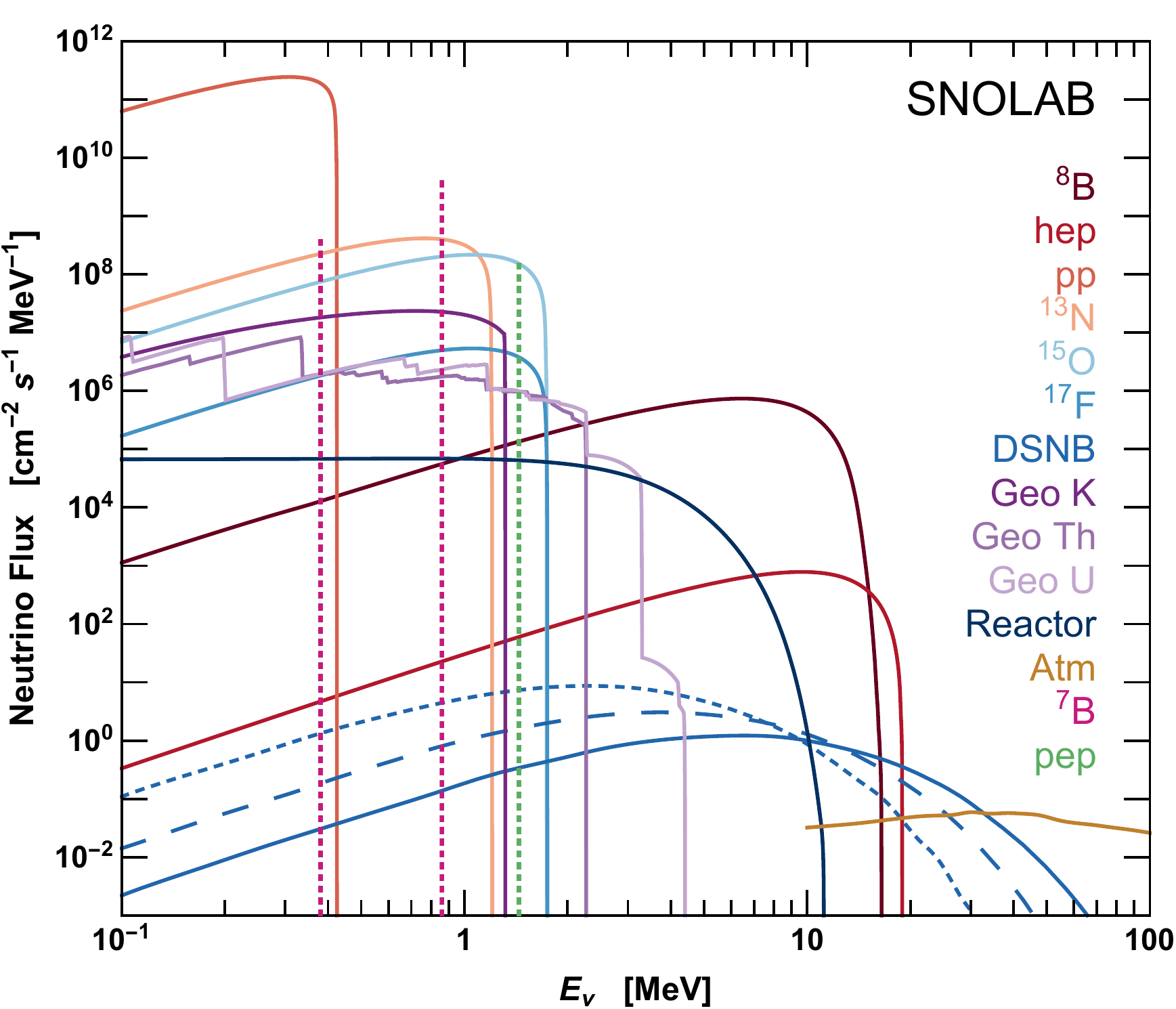}
\caption{\label{fig:neutrino_flux} Neutrino flux components comprising the ``neutrino floor'' at the SNOLAB location. Details of the specific flux components can be found in \Tab{tab:nucomponents}. The atmospheric neutrino contribution has been summed over all the neutrino flavors.}
\end{figure}

Neutrino coherent scattering gives origin to an irreducible background (i.e. ``neutrino floor'') for direct detection experiments, with contributions coming from solar, reactor, geo-, diffusive supernovae background as well as  atmospheric neutrinos. Below we discuss each specific source and the resulting flux. 
Since neutrino fluxes depend on location, we consider likely future direct detection experiment sites \cite{Cushman:2013zza} at SNOLAB, SURF and LGNS laboratories (see \Tab{tab:labs}). 
As an example, we display the combined contribution of neutrinos at the SNOLAB location in \Fig{fig:neutrino_flux}. A detailed description of the specific flux components can be found in \Tab{tab:nucomponents}. Results for other laboratories are not significantly different.
For a neutrino flux $\phi_{\nu} (E_{\nu})$ given as a function of the neutrino energy $E_{\nu}$ (see \Fig{fig:neutrino_flux}) and originating from a particular source, the resulting differential event rate as a function of the nuclear recoil energy $E_R$, per unit time and detector mass off a target nuclide $T$ in a detector is given by
\begin{equation} \label{eq:nu_diff_rate}
\dfrac{d R_{\nu, T}}{d E_R} = \dfrac{C_T}{m_T} \int_{E_{\nu}^{\rm min}} \phi_{\nu} (E_{\nu}) \dfrac{d \sigma^T(E_{\nu}, E_R)}{d E_R} d E_{\nu}~,
\end{equation}
where $d \sigma^T(E_{\nu}, E_R) / d E_R$ is the  coherent neutrino-nucleus scattering cross-section. 
Thus, summing over all nuclides in a detector we obtain the differential rate for each type of neutrino flux $\phi_{\nu}(E_{\nu})$,
\begin{equation} \label{eq:totnurate}
\dfrac{d R_{\nu}}{d E_R} = \sum_T \dfrac{d R_{\nu, T}}{d E_R}~.
\end{equation}
From kinematics, the maximum recoil energy is given by
\begin{equation}
E_R^{\rm max} = \dfrac{2 E_{\nu}^2}{m_T + 2 E_{\nu}}~.
\end{equation}

Neutrinos penetrate the rock surrounding the laboratory sites nearly unimpeded. Their oscillation effects \cite{Fukuda:1998mi,Ahmad:2002jz}, however, will result in a varying depth-dependent flavor composition of the neutrino flux. Since the coherent cross-section is nearly identical for various neutrino species and we are not concerned with the exact  composition, throughout this work we neglect the oscillation effects\footnote{They are important, however, for studies focusing on detecting neutrinos of a specific flavor (such as in probing non-standard neutrino interactions, e.g. \cite{Harnik:2012ni,Dutta:2017nht}).}. 

\subsection{Flux sources}

\subsubsection{Solar neutrinos}

 Solar electron neutrinos $\nu_e$ are produced as a byproduct of nuclear fusion reactions in the Sun (see  \cite{Robertson:2012ib} for review). 
They vary in flux and energy, depending on the specific step in the reaction chain that they originate from. Around $98\%$ of the Sun's energy is produced via the proton-proton cycle, starting with $p + p \rightarrow ~^{2}{\rm H} + e^+ + \nu_e$ (yielding $pp$, $hep$, $pep$, $^{7}$Be, $^{8}$B neutrinos). The remaining energy is released in the Carbon-Nitrogen-Oxygen cycle (yielding $^{13} $N, $^{15}$O, $^{17}$F neutrinos). Solar neutrinos provide the dominant background for direct detection experiments for  energies $E_{\nu} \lesssim 20$ MeV. 

For our analysis, we take solar neutrino fluxes from a high metallicity Standard Solar Model GS98 of \cite{Serenelli:2011py} (Table 2), which shows good agreement with the helioseismological studies. Since for a range of parameters $^{8}$B neutrinos provide the dominant background contribution, some of the earlier neutrino floor-related studies have focused solely on this single solar neutrino component (e.g. the target complementarity analysis of \cite{Ruppin:2014bra}). Throughout this work, for completeness, we consider contributions from all of the solar neutrinos. 

\subsubsection{Atmospheric neutrinos}

Atmospheric neutrinos (for a review see \cite{Gaisser:2002jj}) provide the dominant direct detection background at energies $E_{\nu} \gtrsim 50$ MeV, although the spectrum extends to lower energies. They are produced by cosmic ray (primarily proton) collisions with the nuclei in the atmosphere. The resulting hadronic showers, mostly composed of pions, produce copious amounts of neutrinos from decays. Pion decay $\pi^+ \rightarrow \nu_{\mu} + \mu^+$, followed by $\mu^+ \rightarrow e^+ + \nu_e + \overline{\nu}_{\mu}$, allows for a simple prediction of 2-to-1 muon-to-electron neutrino ratio within the atmospheric flux.

\begin{table*}[tbp]
\setlength{\extrarowheight}{2pt}
  \setlength{\tabcolsep}{10pt}
  \begin{center}
  \begin{threeparttable}
	\begin{tabular}{|l|l|c|l|}  \hline
	~~~~~~~~Neutrino Flux  & ~~~~~~~~Total Flux & Maximum Energy & ~~~~Reference \\ 
    ~~~~~~~~~~Component   &   ~~~~~~~~(cm$^{-2}$ s$^{-1}$) &  $E_{\nu}$, (MeV) & ~~~~~(model) \\ \hline
	\hline
	Solar ($\nu_e$, pp) & $5.98 (1 \pm  0.006) \times 10^{10}$ & ~~~$0.42$ & \cite{Serenelli:2011py} (GS98) \\ \hline
	Solar ($\nu_e$, pep [line]) & $1.44 (1 \pm  0.012) \times 10^{8}$ & ~~~$1.45$ & \cite{Serenelli:2011py} (GS98)\\ \hline
    Solar ($\nu_e$, hep) & $8.04 (1 \pm  0.300) \times 10^{3}$ & ~~$18.77$ & \cite{Serenelli:2011py} (GS98) \\ \hline
    Solar ($\nu_e$, $^{7}$Be [line-1]) & $5.00 (1 \pm  0.070) \times 10^{8}$ & ~~~$0.39$ & \cite{Serenelli:2011py} (GS98) \\ \hline
    Solar ($\nu_e$, $^{7}$Be [line-2]) & $4.50 (1 \pm  0.070) \times 10^{9}$ & ~~~$0.87$ & \cite{Serenelli:2011py} (GS98)  \\ \hline
    Solar ($\nu_e$, $^{8}$B) & $5.58 (1 \pm  0.140) \times 10^{6}$ & ~~$16.80$ & \cite{Serenelli:2011py} (GS98) \\ \hline
    Solar ($\nu_e$, $^{13}$N) & $2.96 (1 \pm  0.140) \times 10^{8}$ & ~~~$1.20$ & \cite{Serenelli:2011py} (GS98) \\ \hline
    Solar ($\nu_e$, $^{15}$O) & $2.23 (1 \pm  0.150) \times 10^{8}$ & ~~~$1.73$ & \cite{Serenelli:2011py} (GS98) \\ \hline
    Solar ($\nu_e$, $^{17}$F) & $5.52 (1 \pm  0.170) \times 10^{6}$ & ~~~$1.74$ & \cite{Serenelli:2011py} (GS98) \\ \hline
    \hline
    Atm. ($\nu_e$)  & $1.27 (1 \pm  0.500) \times 10^{1}$ & $944.00$ & \cite{Battistoni:2005pd} (FLUKA) \\ \hline   
    Atm. ($\overline{\nu}_e$)  & $1.17 (1 \pm  0.500) \times 10^{1}$ & $944.00$ & \cite{Battistoni:2005pd} (FLUKA) \\ \hline   
    Atm. ($\nu_{\mu}$)  & $2.46 (1 \pm  0.500) \times 10^{1}$ & $944.00$ & \cite{Battistoni:2005pd} (FLUKA) \\ \hline   
    Atm. ($\overline{\nu}_{\mu}$)\tnote{1} & $2.45 (1 \pm  0.500) \times 10^{1}$ & $944.00$ & \cite{Battistoni:2005pd} (FLUKA) \\ \hline
    \hline
DSNB ($\nu_e$, $T_{\nu} = 3$ MeV) & $4.55 (1 \pm  0.500) \times 10^{1}$ & ~~$36.90$ & \cite{Horiuchi:2008jz} (th.~avrg.)\tnote{1} \\ \hline
DSNB ($\overline{\nu}_e$, $T_{\nu} = 5$ MeV) & $2.73 (1 \pm  0.500) \times 10^{1}$ & ~~$57.01$ & \cite{Horiuchi:2008jz} (th.~avrg.)\tnote{1} \\ \hline
DSNB ($\nu_x$, $T_{\nu} = 8$ MeV)\tnote{2} & $1.75 (1 \pm  0.500) \times 10^{1}$ & ~~$81.91$ & \cite{Horiuchi:2008jz} (th.~avrg.)\tnote{1} \\ \hline
\hline
Reactor ($\overline{\nu}_e$, $^{235}$U)\tnote{3} & $1.88(1 \pm 0.080) \times 10^5$  & ~~10.00  & ~~(combined)\tnote{4} \\
\hline
\hline
Geo. ($\overline{\nu}_e$, $^{40}$K)  & $2.19 (1 \pm  0.168) \times 10^{7}$  &  ~~~1.32  &  \cite{huang:2013geomodel} (global)\tnote{5} \\
\hline
Geo. ($\overline{\nu}_e$, $^{238}$U)  &  $4.90 (1 \pm  0.200) \times 10^{6}$ &  ~~~3.99  &  \cite{huang:2013geomodel} (global)\tnote{5} \\
\hline
Geo. ($\overline{\nu}_e$, $^{232}$Th)  & $4.55 (1 \pm  0.257) \times 10^{6}$  &  ~~~2.26  &   \cite{huang:2013geomodel} (global)\tnote{5} \\
\hline
 	\end{tabular}
\begin{tablenotes}
\item[1] Average of several theoretical models.
\item[2] $\nu_x$ is the total contribution from all other neutrinos and antineutrinos.
\item[3] Only the most dominant element is considered. 
\item[4] Combined result from multiple nearby reactors.
\item[5] Global Earth model, incorporates several theoretical models.
\end{tablenotes}
\caption{\label{tab:nucomponents} Neutrino flux components that contribute to the coherent neutrino scattering background in direct detection experiments at the  SNOLab location.~Contributions from solar, atmospheric, diffuse supernovae, reactor as well as geo--neutrinos are shown.}
\label{tab:nuflux}
\end{threeparttable}
  \end{center}
\end{table*}

The atmospheric neutrino flux depends on the laboratory location. This is primarily due to the effect of the location-dependent Earth's geo-magnetic field, which results in a rigidity cut-off for the flux of charged parent cosmic ray and secondary particles. Atmospheric neutrino flux predictions from several models have been employed in the neutrino-oscillation experiments \cite{Abe:2011ph}, known colloquially as ``Bartol'' \cite{Barr:2004br}, ``Honda'' \cite{Honda:2006qj,Honda:2015fha} and ``FLUKA'' \cite{Battistoni:2007zzb} fluxes. The only available atmospheric flux predictions for $E_{\nu} < 100$ MeV are the tabulated results of FLUKA (see Appendix A, Table 2-4, of \cite{Battistoni:2005pd}), and  we employ the values for the Kamioka site in this work\footnote{While we do not consider the Kamioka site in this work this provides us with a reasonable estimate for other laboratory locations since we are only interested in the total flux and not specific components.}. Due to large modeling uncertainties, especially at lower energies, we conservatively take the systematic error on the predicted fluxes to be $50\%$. 

\subsubsection{Diffuse supernova neutrino background}

The diffuse supernova neutrino background (DSNB)\footnote{Also known as ``supernovae relic neutrinos'' in the older literature.} refers to neutrinos and anti-neutrinos originating from all of the past core-collapse supernovae (for a review see~\cite{Beacom:2010kk}). The signal, which includes red-shifted contributions from various epochs, is effectively isotropic in space and time-independent for experiments. 
 The DSNB flux, over $4 \pi$, is given by \cite{Horiuchi:2008jz}
\begin{equation}
 \phi_{\nu} (E)  = \int R_{\rm CCSN}(z) \dfrac{dN(E^{\prime})}{dE^{\prime}} (1 + z)\Big|\dfrac{dt}{dz}\Big| dz~,
\end{equation}
where $z$ is the redshift, $E^{\prime} = E(1 + z)$, $R_{\rm CCSN}(z)$ is the historical rate of core-collapse supernovae, $dN/dE^{\prime}$ is the time-integrated neutrino spectrum per supernova and $t$ is the cosmological time that is related to redshift as $|dz/dt| =  H_0 (1 + z) [\Omega_m (1+z)^3 + \Omega_{\Lambda}]^{1/2}$, where the adopted approximate parameter values are for the present day $\Lambda$CDM cosmology, with $H_0 = 70$ km s$^{-1}$ Mpc$^{-1}$ for the Hubble constant as well as $\Omega_m = 0.3$ and $\Omega_{\Lambda} = 0.7$ for the matter and the cosmological constant density fractions, respectively. Provided the historical supernovae rate, the DSNB flux depends only on the effective neutrino temperature $T_{\nu}$ of the respective neutrino-sphere. The neutrino spectrum, which can be well approximated by the Fermi-Dirac
distribution with zero chemical potential \cite{Kotake:2005zn}, depends on $T_{\nu}$ as
\begin{equation}
\dfrac{dN}{dE^{\prime}_{\nu}} = E_{\nu}^{\rm tot} \dfrac{20}{7 \pi^4} \dfrac{(E_{\nu}^{\prime})^2 }{T_{\nu}^4} \dfrac{1}{\left(e^{E_{\nu}^{\prime}/T_{\nu}} + 1\right)}~,
\end{equation}
where $E_{\rm tot} \simeq 3 \times 10^{53}$ erg is the approximate total energy released from a supernova explosion.
 
The DSNB component is expected to significantly contribute to the neutrino background of direct detection experiments in the $20~{\rm MeV} \lesssim E_{\nu} \lesssim 50$ MeV energy range. In this work we employ DSNB fluxes from \cite{Strigari:2009bq} (Figure~1), which were obtained using the above formalism assuming the following neutrino temperatures: $T_{\nu_e} = 3$ MeV for $\nu_e$, $T_{\overline{\nu}_e} = 5$ MeV for $\overline{\nu}_e$ and $T_{\nu_x} = 8$ MeV for the combined contribution from all other neutrinos and anti-neutrinos, denoted as $\nu_x$. These temperature values represent an approximate average of different theoretical models found in the literature, as summarized in Table 3 of \cite{Horiuchi:2008jz}. Due to large model uncertainties the systematic errors on DSNB fluxes are taken to be 50\%.

\subsubsection{Reactor neutrinos}

Reactor anti-neutrinos $\bar{\nu}_e$ (for a review see \cite{Hayes:2016qnu}) originate from the $\beta$-decay of unstable isotopes from reactor fuel fissions. Since the isotopes are short-lived, the corresponding neutrino flux directly follows the reactor operation.
The core elements include $^{235}$U, $^{238}$U, $^{239}$Pu and $^{241}$Pu. A typical reactor core  (e.g. \cite{Declais:1994ma}) contains these fuel elements in an approximate ratio of 0.6 : 0.08 : 0.3 : 0.05. As the reactor operates, nuclear processes change the element composition. In our work we only focus on the dominant $^{235}$U contribution and take it to be approximately constant throughout the reactor operation, neglecting element recomposition that depends on each reactor's specifications. An example of analysis including these effects can be found in \cite{Murayama:2000iq}.

The flux of reactor anti-electron neutrinos is given by 
\begin{equation}
\phi_k(E) = \frac{R_{\bar{\nu}_e}}{4\pi d^2}S_k (E) \, ,
\end{equation}
where $R_{\bar{\nu}_e}$ is the emitted rate of reactor neutrinos, $d$ is the distance from a given reactor to the laboratory  and $S_k(E)$ is the neutrino spectrum for isotope $k$. Approximate analytic expressions for $S_k (E)$ have been developed in \cite{Vogel:1989iv,Mueller:2011nm,Huber:2011wv,Murayama:2000iq}. We employ the model of \cite{Mueller:2011nm}, which is based on a phenomenological fit to data with an exponentiated polynomial, with the resulting spectrum being
\begin{equation}\label{eq:reactor_nuspec}
 S_k (E_\nu) = \dfrac{dN_{\nu}}{dE_{\nu}} = \exp \left(\sum_{i=1}^{6} \alpha_{i,k}E_\nu^{i-1} \right) ~,
\end{equation}
where $\alpha_{i,k}$ is the respective fit coefficient of order $i$. \Tab{tab:reactor_fit} displays the values of the best fit coefficients as obtained by \cite{Mueller:2011nm}\footnote{Strictly, this is only valid for energies $\gtrsim 1.8$ MeV. However, calculations of \cite{Vogel:1989iv} do not show substantial deviations for energies above $0.5$ MeV, allowing us to truncate the distributions at that point. The presented results are insensitive to this choice.}.
\begin{table*}[tbp]
  \setlength{\extrarowheight}{2pt}
  \setlength{\tabcolsep}{10pt}
  \begin{center}
	\begin{tabular}{|c|c|c|c|c|}  \hline
	$k$  & $^{235}U$ & $^{238}U$ & $^{239}P$ & $^{241}P$ \\ \hline
	\hline
	1 & 3.217 & 4.833 & 6.413 & 3.251\\  \hline
	2 & -3.111 & 1.927 & -7.432 & -3.204 \\  \hline
    3 & 1.395 & -1.283 & 3.535 & 1.428 \\  \hline
    4 & -3.690 & -6.762 & -8.820 & -3.675 \\  \hline
    5 & 4.445 & 2.233 & 1.025 & 4.254 \\  \hline
	6 &  -2.053 & -1.536 & -4.550 & -1.896 \\  \hline

	\end{tabular}
  \end{center}
\caption{\label{tab:reactor_fit} Fitted values of the reactor neutrino spectrum $\alpha_{i, k}$ coefficients, used in \Eq{eq:reactor_nuspec}, for dominant nuclear isotopes, from \cite{Mueller:2011nm}.}

\end{table*}
The neutrino emission rate is given by
\begin{equation}
R_{\bar{\nu}_e} = N_{\nu, {\rm fiss}} \, \frac{P_{th}}{E_{\rm fiss}} e ~,
\end{equation}
where $N_{\nu, {\rm fiss}} = 6$ is the average number of anti-neutrinos produced per fission, $P_{th}$ is the power output of the reactor, $E_{\rm fiss}$ is the fission energy that is around 200 MeV for all major isotopes and $e = 0.75 \, (1 \pm 0.080) $ is the average reactor operation efficiency that includes shut-downs \cite{usenergy}. The uncertainty on the efficiency reflects an $\sim 8\%$ uncertainty on the reactor anti-neutrino spectrum (see Table 3 of \cite{Kopeikin:2012zz}), which is also the source of the reactor flux uncertainty as specified in \Tab{tab:nuflux}.
In \Tab{tab:reactors} we list the reactors  considered in this work, which constitute the dominant reactor-neutrino sources for the SNOLab, the LNGS and the SURF laboratories \footnote{Note that the Pickering Nuclear Generating Station is not included in \Tab{tab:reactors}. The reason is that the experiments we are interested in studying, those likely to reach the neutrino floor, are unlikely to be taking data by 2024, while this Station will still operate. After running for nearly 40 years, the reactor operations are expected to cease in August 2018. However, there is a strong movement to extend the operations until 2024, at which time it will be decommissioned.}. For the reactor neutrino flux we have calculated their combined near-by reactor contribution at the laboratory site using the above formalism.

\begin{table}[tbp]
  \setlength{\extrarowheight}{2pt}
  \setlength{\tabcolsep}{10pt}
  \begin{center}
	\begin{tabular}{|l|m{2.3cm}|c|c|c|}  \hline
	Nuclear Reactor Name & Location & Nearest Lab & Distance (km) & Output (MW) \\ \hline \hline
    Cooper  & 40$^\circ$21$'$43$''$\,N 95$^\circ$38$'$29$''$\,W & SURF & 801 & 830  \\ \hline
    Monticello  & 45$^\circ$20$'$01$''$\,N 93$^\circ$50$'$57$''$\,W& SURF & 788 & 671 \\ \hline
    Prarie Island  & 44$^\circ$37$'$18$''$\,N 92$^\circ$73$'$59$''$\,W & SURF & 835 & 1096 \\ \hline
    Nine Mile Point  & 43$^\circ$31$'$15$''$\,N 76$^\circ$24$'$25$''$\,W & SNOLAB & 498 & 1761 \\ \hline
    R.E. Ginna  & 43$^\circ$16$'$40$''$\,N 77$^\circ$18$'$36$''$\,W & SNOLAB & 468 & 610 \\ \hline
    James A. Fitzpatrick  & 43$^\circ$31$'$04$''$\,N 76$^\circ$23$'$09$''$\,W & SNOLAB & 500 & 838 \\ \hline
    Point Beach  & 44$^\circ$16$'$52$''$\,N 87$^\circ$32$'$12$''$\,W & SNOLAB & 552 & 1200 \\ \hline
    Enrico Fermi  & 41$^\circ$57$'$46$''$\,N 83$^\circ$15$'$27$''$\,W & SNOLAB & 527 & 1198 \\ \hline
    Davis Besse & 41$^\circ$35$'$48$''$\,N 83$^\circ$05$'$11$''$\,W & SNOLAB & 563 & 889 \\ \hline
    Perry  & 41$^\circ$48$'$03$''$\,N 81$^\circ$08$'$36$''$\,W & SNOLAB & 519 & 1261 \\ \hline    
    Bruce  & 44$^\circ$19$'$31$''$\,N 81$^\circ$35$'$58$''$ W & SNOLAB & 240 & 6384 \\ \hline
    Darlington  & 43$^\circ$55$'$22$''$\,N 78$^\circ$43$'$11$''$\,W & SNOLAB & 343 & 3512 \\ \hline
    Tricastin  & 44$^\circ$19$'$47$''$\,N 04$^\circ$43$'$56$''$\,E & LNGS & 744 & 3820 \\ \hline
    Cruas & 44$^\circ$37$'$59$''$\,N 04$^\circ$45$'$29$''$\,E & LNGS & 750 & 3842 \\ \hline
    Saint-Alban & 45$^\circ$24$'$16$''$\,N 04$^\circ$45$'$19$''$\,E & LNGS & 778 & 2600 \\ \hline
    Bugey  & 45$^\circ$47$'$54$''$\,N 05$^\circ$16$'$15$''$\,E & LNGS & 760 & 3724 \\ \hline
	\end{tabular}
  \end{center}
\caption{\label{tab:reactors}  List of most relevant nuclear reactors for SURF, SNOLAB, and LNGS. Columns contain, from left to right, the name of the reactor, the GPS location, the laboratory for which the reactor is relevant, the distance to the laboratory in kilometers, and the output of the reactor in MW~\cite{USNuclear,CanadaNuclear,EUNuclear}}  
\end{table}

\subsubsection{Geo-neutrinos}

Geo-neutrinos are predominantly electron anti--neutrinos $\overline{\nu}_e$ originating from the $\beta$-decay branches of the major Earth's heat-producing nuclear reactions, involving isotopes of potassium $^{40}$K, thorium $^{232}$Th and uranium $^{238}$U. Recently, KamLAND~\cite{Araki:2005qa} as well as Borexino~\cite{Bellini:2010hy} have definitively observed a geoneutrino flux.
The spectrum for each of these elements is taken from \cite{Araki:2005qa} (Figure 1). The respective location-dependent total flux is predicted from a geophysically-based three-dimensional global Earth model of heat-producing element distribution \cite{huang:2013geomodel} (Table 1). The systematic uncertainties are taken to be the larger of the quoted $\pm 1\sigma$ fluctuations in these values.
 
\subsection{Coherent neutrino scattering interactions}

If the neutrino energy is not sufficient to discern individual quarks or nucleons, neutrinos have coherent elastic  scattering off the whole nucleus through the weak neutral current \cite{Freedman:1977xn,Drukier:1983gj}.
At low momentum transfer $|\vec{q}|$ the coherence condition for target nucleons to interact in phase is $|\vec{q}| R \ll 1$, where $R$ is the nuclear radius. The total cross-section for coherent interaction scales as the square of the number of participating nucleons. While heavier targets thus enjoy a dramatically enhanced neutrino interaction cross-section, they are  also penalized with a smaller maximum recoil energy.
Recent observations by the  COHERENT experiment have definitively confirmed this process \cite{Akimov:2017ade}. At energies above $E_{\nu} \sim 50$ MeV, other channels, such as quasi-elastic scattering (QE) and deep-inelastic-scattering (DIS), start to dominate  (for review see  \cite{Formaggio:2013kya}). 
The Standard Model coherent-scattering neutrino-nucleus cross-section is given by 
\begin{equation}
\dfrac{d \sigma^T (E_{\nu}, E_R)}{d E_R} = \dfrac{G_f^2}{4 \pi} Q_w^2 m_{T} \left(1 - \dfrac{m_T E_R}{2 E_{\nu}^2}\right) F_{{\rm SI},T}^2 (E_R)~,
\end{equation}
where $m_T$ is target nucleus mass, $G_f$ is Fermi coupling constant, $F_{{\rm SI},T}(E_R)$ is the form factor (as before, we take this to be the Helm form factor \cite{Helm:1956zz}),$Q_w = (1 - 4 \sin^2 \theta_{\rm W}) Z-N$ is the weak nuclear charge, $N$ is the number of neutrons, $Z$ is the number of protons and $\theta_{\rm W}$ is the Weinberg angle. Since $\sin^2 \theta_{\rm W} = 0.223$ \cite{Patrignani:2016xqp}, the coherent neutrino-nucleus scattering cross-section follows an approximate $N^2$ scaling.

Assuming good electron-tagging in future direct detection experiments, we neglect in this work  $\nu_e + e^- \rightarrow \nu_e + e^-$ neutrino-electron scattering\footnote{This scattering has been also studied in the context of direct detection experiments before (e.g. \cite{Billard:2013qya}). See \cite{Marciano:2003eq} for a list of possible new physics topics associated with this process, and~\cite{Essig:2018tss,Wyenberg:2018eyv} for a recent discussion of solar neutrinos as a background for electron recoil analyses in direct detection experiments.}.


\begin{table*}[tb]
  \setlength{\extrarowheight}{2pt}
  \small
  \begin{center}
    \begin{threeparttable}
	\begin{tabular}{|l|c|c|c|c|c|c|c|}  \hline
	 \multirow{2}{*}{Target Material} &  \multirow{2}{*}{ $A (Z)$} &  Isotope & \multirow{2}{*}{  $J$} &  \multirow{2}{*}{  $\langle S_p \rangle$} &  \multirow{2}{*}{  $\langle S_n \rangle$} &  \multirow{2}{*}{ $\lambda/\lambda_N$} & Energy     \\  
& & Fraction  & & & &  & Range (keVnr) \\ \hline
	\hline
	Xenon (Xe) & 124 (54) 	& 0.001 	& 3/2 	& -0.009 	& -0.227	& 0.692 	& 0.1--50 (10--300)\tnote{2} \\  \hline
	 & 126 (54)	& 0.001 	& 0 	& 0.0 		& 0.0 		&	 0.0 	& 0.1--50 (10--300)\tnote{2}\\  \hline
     & 128 (54)	& 0.019 	& 0 	& 0.0	 	& 0.0 		& 	 0.0 	& 0.1--50 (10--300)\tnote{2}\\  \hline
	 & 129 (54)	& 0.264 	& 1/2 	& 0.028 	& 0.359 	& -0.778 	& 0.1--50 (10--300)\tnote{2}\\  \hline
     & 130 (54)	& 0.041 	& 0 	& 0.0	 	& 0.0 		& 0.0 	& 0.1--50 (10--300)\tnote{2}\\  \hline
     & 131 (54) & 0.212 	& 3/2 	& -0.009 	& -0.227	& 0.692 & 0.1--50 (10--300)\tnote{2}\\  \hline
     & 132 (54)	& 0.269 	& 0 	& 0.0	 	& 0.0 		& 0.0 	& 0.1--50 (10--300)\tnote{2}\\  \hline
     & 134 (54)	& 0.104 	& 0 	& 0.0	 	& 0.0 		& 0.0 	& 0.1--50 (10--300)\tnote{2}\\  \hline
     & 136 (54)	& 0.089 	& 0 	& 0.0	 	& 0.0 		& 0.0 	& 0.1--50(10--300)\tnote{2} \\  \hline
    Germanium (Ge) & 70 (32)  	& 0.208 	& 0 	& 0.0 	& 0.0	& 0.0 	& 0.04--50	(10--300)\tnote{2} \\  \hline
     	& 72 (32)  	& 0.275 	& 0 	& 0.0 	& 0.0	& 0.0 	& 0.04--50	 (10--300)\tnote{2}\\  \hline
        & 73 (32)  	& 0.077 	& 9/2 	& 0.038 	& 0.37	& 	-0.879 	& 0.04--50	(10--300)\tnote{2} \\  \hline
        & 74 (32)  	& 0.363 	& 0 	& 0.0 	& 0.0	& 0.0 	& 0.04--50	(10--300)\tnote{2} \\  \hline
        & 76 (32)  	& 0.076 	& 0 	& 0.0 	& 0.0	& 0.0 	& 0.04--50	(10--300)\tnote{2} \\  \hline
   	Argon (Ar) 			& 40 (18)    	& 0.996 	& 0 	& -- 		& -- 		& 0 		& 1--50 (10--300)\tnote{2} \\  \hline
    Sodium (Na) 			& 11 (23)    	& 1  		& 3/2 	& 0.248 	& 0.020 	& 2.218 	& 1--50 (10--300)\tnote{2}\\  \hline
    Iodine (I) 				& 127 (53)    	& 1  		& 5/2 	& 0.309 	& 0.075	& 2.813 	& 1--50 (10--300)\tnote{2}\\  \hline
	Fluorine (F)\tnote{1} 			& 19 (9)  		& 1 		& 1/2 	& 0.477 	& -0.004	& 2.629 	& -- \\  \hline
    Silicon (Si)\tnote{1} 	& 28 (14)  		& 0.922 		& 0 	& 0.0 	& 0.0	& 0.0 	& -- \\  \hline 
     	& 29 (14)  		& 0.047 		& 1/2 	& -0.002 	& 0.130	& -0.555 	& -- \\  \hline 
       	& 30 (14)  		& 0.031 		& 0 	& 0.0 	& 0.0	& 0.0 	& -- \\  \hline
    Helium (He)\tnote{1} 	& 4 (2)  		& 1 		& 0 	& 0.0 	& 0.0	& 0.0 	& -- \\  \hline 
	\end{tabular}
    \begin{tablenotes}
\item[1] This element is not considered in our work and is shown for completeness. 
\item[2] For momentum suppressed interactions (i.e. those whose differential cross sections are proportional to $q^{b}$ with $b > 0$), the DM nuclear recoil spectrum can extend to larger recoil energies than for other interactions (see e.g.~\cite{Aprile:2017aas}). Thus, for these interactions our analyses are performed over a low energy range and a high energy range (the latter indicated in parenthesis), and the result is taken to be the stronger of the two.
\end{tablenotes}
\caption{\label{tab:experiments} Experimental configurations considered in this work (except fluorine, silicon and helium, which are included for completeness). Shown in columns from left to right are: nucleon number and the nuclear charge, isotope fraction (rounded to three decimal places), total nuclear spin, the expectation values of the proton and neutron spin content, the expected nuclear neutron spin, the nuclear magnetic moment $\lambda$ (in terms of the nuclear magneton $\lambda_B$)  and the range of recoil energies that a particular experiment is sensitive to (a perfect detection efficiency is assumed throughout the entire range). Nuclear properties are taken from \cite{Tovey:2000mm, webelements}. }
\label{tab:experiments}
  \end{threeparttable}
    \end{center}
\end{table*}

\section{Detection\label{sec:detection}}

\subsection{Considered experimental configurations}
Making definitive statements about the scientific capabilities of future direct detection experiments requires explicit assumptions about their size, composition, energy resolution and detection threshold. Various proposals for the next multi-ton scale generation of experiments have been put forth \cite{Cushman:2013zza}, but which experiments will be constructed and what their respective ultimate characteristics will be remains uncertain. Thus, we study the future scientific reach of a variety of potential experimental configurations that could reside near the optimistic edge of such realizations. 

Specifically, \Tab{tab:experiments} lists the five experimental configurations we consider, each with a different target element (xenon, argon, sodium, germanium, iodine). Some proposals have considered a fluorine target, but these are energy threshold experiments that cannot measure the recoil spectrum, and we thus do not study them\footnote{Fluorine will assist in breaking model degeneracy if detection is already made with other experiments~\cite{Gluscevic:2015sqa}.}. It has been shown in~\cite{Ruppin:2014bra} that fluorine could be extremely useful to help disentangle various DM models, should a detection be made in another target elements, as its nuclear properties differ strongly from other direct detection target elements. 

We optimistically assume that experiments have perfect detection efficiency and resolution in the energy range provided. Furthermore, the background is assumed to arise exclusively from neutrinos. While for sodium and iodine this assumption is at the present time not realistic, it allows us to analyze the intrinsic properties of all the target elements on the same footing.

Before continuing we briefly comment on the choice of assumptions for the energy thresholds adopted in this work, which can be extremely import in determining sensitivity to low mass WIMPs. Xenon experiments have historically been able to probe recoils down to $\sim 1$ keV (see Figure 1 of~\cite{Aprile:2017iyp}), albeit not with perfect efficiency. However, xenon experiments do have the ability to probe lower energies via specialized analyses (see e.g.~\cite{Aprile:2016wwo}), and thus we consider an optimistic threshold of 0.1 keV. The projected threshold for the next generation of germanium detectors from SuperCDMS SNOLAB reaches energy values as low as 40 eV~\cite{Agnese:2016cpb}\footnote{We note that the 40 eV value chosen by the SuperCDMS collaboration as the cutoff of their ionization yield is somewhat optimistic because the Lindhard model on which this choice is based has not be tested below $\mathcal{O}(100)$ eV. Also this threshold only applies to the HV detectors which are not capable of discriminating nuclear and electronic recoils, thus the zero background assumption adopted in this analysis may not be realistic.}, which is the value adopted here. Typically, argon based experiments have projected higher thresholds for future experiments like DarkSide-20k than the value of 1 keV adopted here (typically quoted values are not below  $\sim 10$ keV, see e.g.~\cite{Aalseth:2017fik}), however  a recent analysis by the DarkSide-50 experiment has demonstrated that a specialized analysis~\cite{Agnes:2018ves} can be performed that extends the reach of argon-based experiments down to 0.6 keV while maintaining a low background, potentially making our analysis slightly conservative. Finally, the energy threshold in sodium and iodine experiments is typically only slightly larger than the 1 keV value adopted here (see e.g.~\cite{Thompson:2017yvq}), implying our adopted values represent reasonable benchmarks.

Finally, we have taken into account that momentum suppressed interactions (i.e. those with differential cross sections containing positive powers of the momentum transfer) can produce significant scattering rates at recoil energies larger than 50 keV, which we have chosen as the upper limit of our analysis range (see e.g.~\cite{Aprile:2017aas}). Typically, experiments place an upper limit on the recoil energy to the search window assuming the conventional SI and SD interactions which predict the bulk of the DM recoils to be below $\sim 50$ keV (although this depends on the target and DM particle mass); however experiments can analyze larger recoil energies should their be a sufficient reason to look in this energy range (see e.g.~\cite{Aprile:2017aas}). Thus, for momentum suppressed interactions, we perform our analysis over two energy regions, one extended from the aforementioned thresholds to 50 keV, and the other running from 10 keV to 300 keV. The derived experimental sensitivity is then taken to be the stronger of the two.

\begin{figure}
\mycenter{
\includegraphics[width=.41\textwidth]{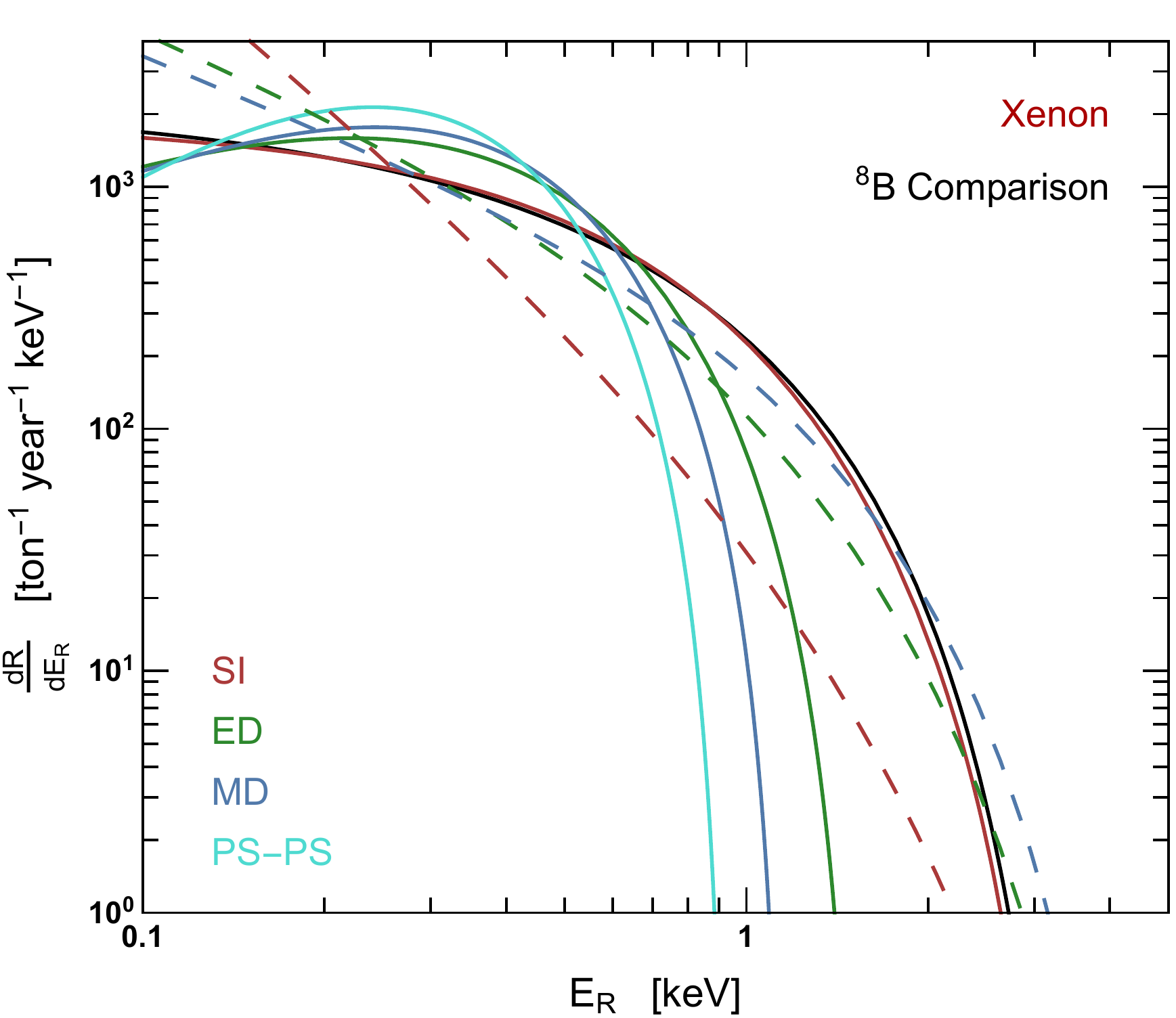}
\includegraphics[trim={12.5mm 0 0 0},clip,width=.3875\textwidth]{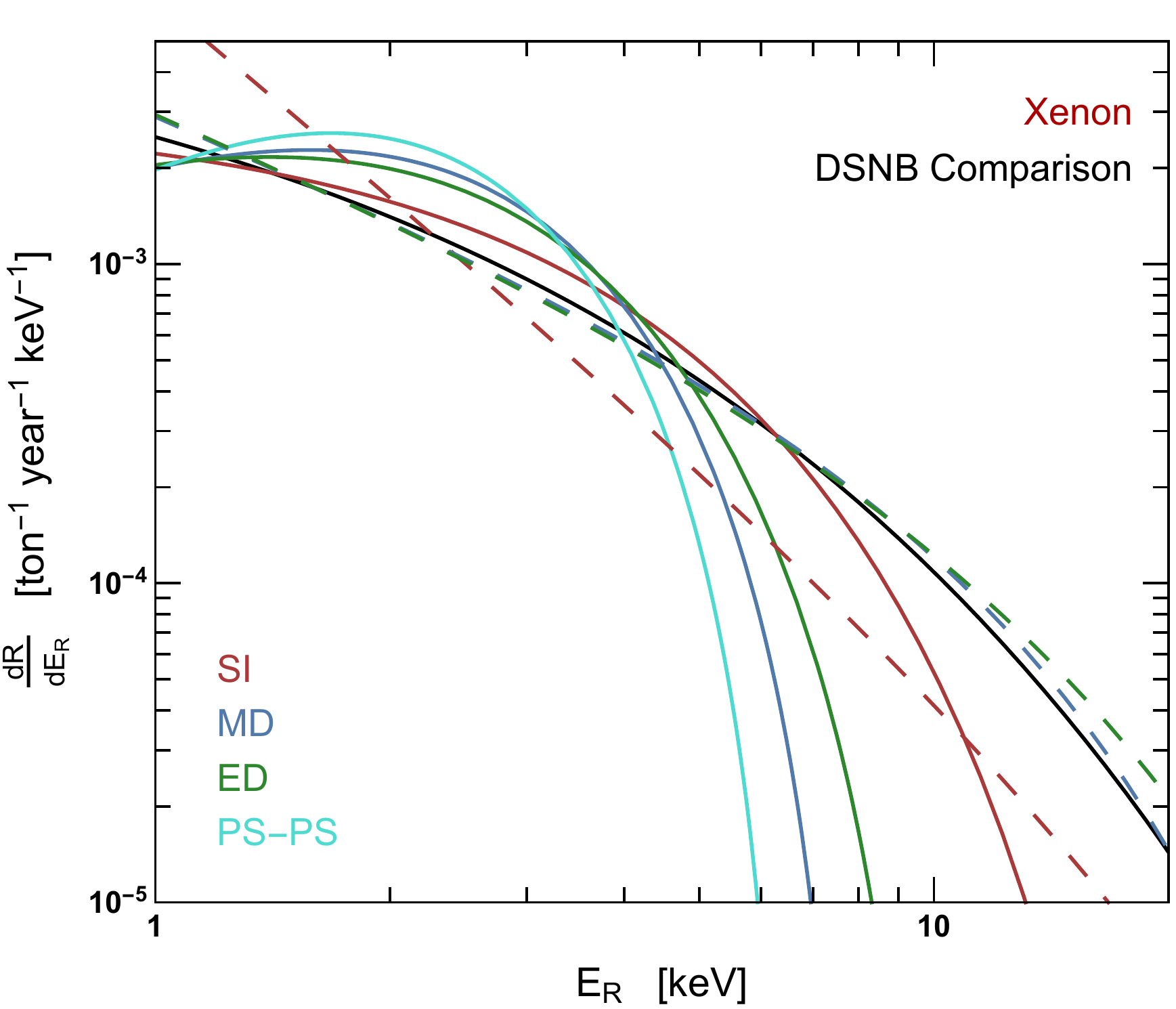}
\includegraphics[trim={12.5mm 0 0 0},clip,width=.3875\textwidth]{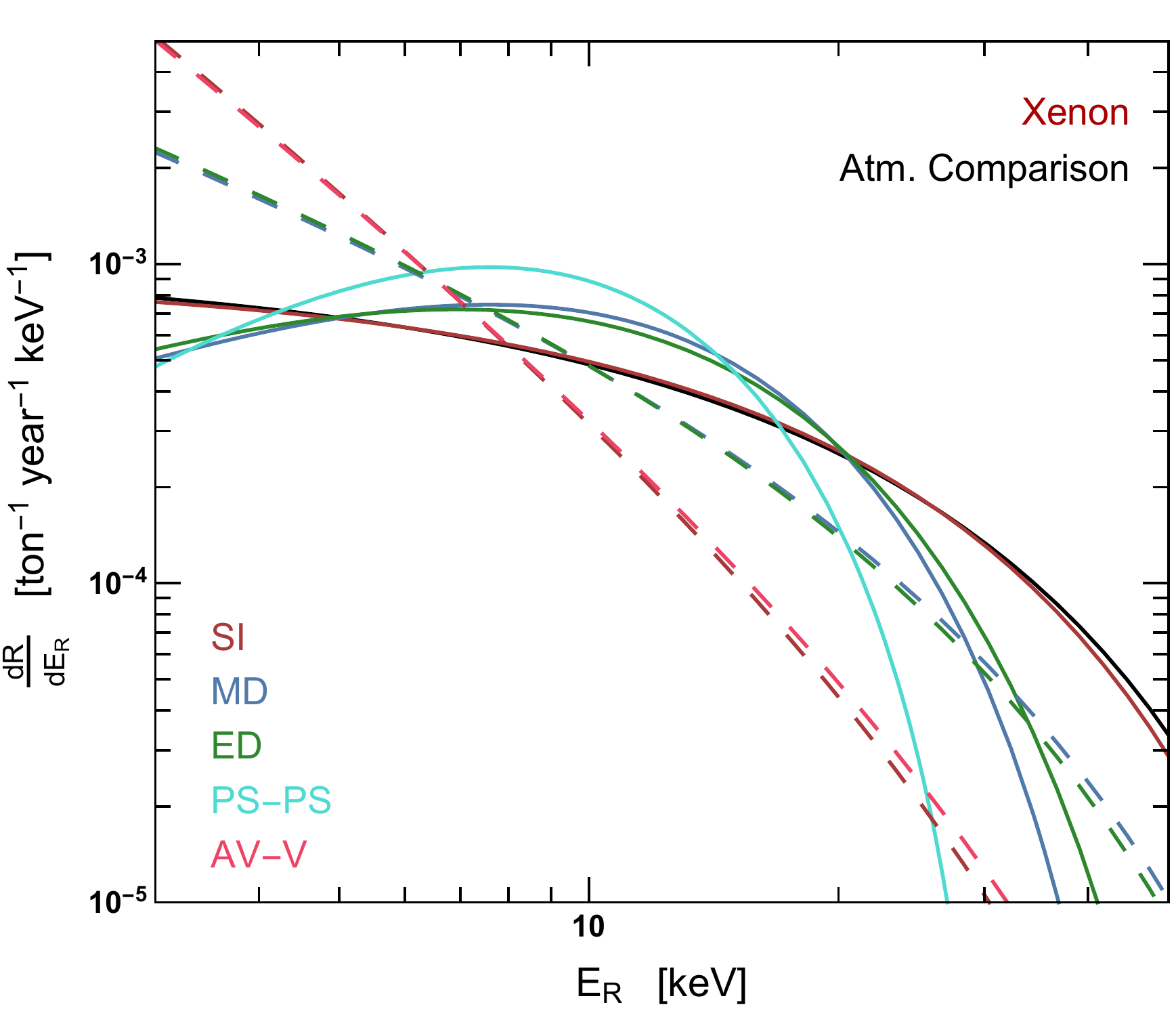}}
\caption{\label{fig:Dark_Matter_Degeneracy} Comparison of recoil spectra induced in a xenon target from $^8 {\rm B}$ (left), DSNB (middle), and atmospheric (right) neutrinos with the recoil spectra arising in various DM interaction models. Dashed lines are used to represent the light mediator limit of a particular interaction. Interactions not shown are understood to be approximately degenerate with one of the shown DM models for the particular target element and energy range considered. The best fit mass values are given in Table 10.}
\end{figure}

\subsection{Fitting DM and neutrino recoil spectra}

In the following sections we study the extent to which future direct detection experiments will be sensitive to a variety of DM candidates. Provided an experiment is capable of probing background neutrinos, the projected sensitivity will intimately depend on the shapes of the DM and background neutrino recoil spectra. Thus, in order to gain intuition about the sensitivity to a particular DM model, we perform here a brief analysis of the extent to which an observed recoil spectrum arising from background neutrinos could be misinterpreted in terms of various DM models.

Specifically, for a particular experimental configuration (defined by a target element and energy range), we divide the experimental energy range into 1000 log-spaced intervals and calculate the differential recoil rate arising from a particular source of background neutrinos in each interval (in particular we focus on either $^8$B, DSNB, or atmospheric neutrinos). For each DM model, we minimize the statistic
\begin{equation}\label{eq:gof}
\chi_{\rm weighted}^2 = \sum_i \frac{(R_{{\rm DM},i} - R_{\nu,i})^2}{R_{\nu,i}}\, ,
\end{equation}
where $R_{{\rm DM},i}$ and $R_{\nu,i}$ are the DM and neutrino predicted rates in bin $i$, allowing both the WIMP mass as well as the normalization of the DM recoil spectrum (given by $\sigma_{\rm eff}$) to vary. Starting with the usual definition of the $\chi^2$-statistic, we obtain the statistic of \Eq{eq:gof} by inserting the corresponding number of events of DM and neutrinos in each bin, $N_{{\rm DM}, i} = M T R_{{\rm DM},i}$ and $N_{\nu,i}= MT R_{\nu,i}$, with the uncertainty taken to be $N_{\nu, i}$ to be $\sigma_{\nu,i} = \sqrt{MT R_{\nu,i}}$. Dividing out the $MT$ factor gives a weighted $\chi^2$ statistic \cite{Baker:1983tu}: $\chi^2/ MT = \chi_{\rm weighted}^2$. Note that minimizing $\chi_{\rm weighted}^2$ does not provide a $\chi^2$-goodness of fit test (although multiplying the obtained minimum by $MT$ would). We are only interested in finding the best fit, not in evaluating how good the fit actually is. Given a particular DM model, this procedure identifies the DM mass maximally degenerate with the particular neutrino source.  

Comparison of the ${}^8$B  (left), DSNB (middle), and atmospheric (right) neutrino recoil spectra in a xenon-target experiment with the best-fit recoil spectrum derived for a variety of DM models is shown in \Fig{fig:Dark_Matter_Degeneracy}. In the fit we did not allow the DM mass to exceed 1~TeV~\footnote{Fit results for various interactions show a slight preference for very massive DM candidates. However, the change in the quality of the fit for a $\sim 10 {\, {\rm TeV}}$ DM candidate is only marginally better than that of a 1 TeV candidate.}.
Comparisons with other elements are shown in \Fig{fig:recoilspec_full} (App.~\ref{app:recoilspec}). The resulting best-fit DM masses  for each particle model and element are shown in Appendix~\ref{app:recoilspec}  \Tab{tab:bestfitDM_elastic_xenon} for xenon, \Tab{tab:bestfitDM_elastic_germanium} for germanium, \Tab{tab:bestfitDM_elastic_argon} for argon, \Tab{tab:bestfitDM_elastic_iodine} for iodine and \Tab{tab:bestfitDM_elastic_sodium} for sodium. For each plot in \Fig{fig:Dark_Matter_Degeneracy} and \Fig{fig:recoilspec_full}, recoil spectra for interactions not explicitly shown are understood to be degenerate with one of the plotted DM spectra. An interaction that produces a degenerate recoil spectra in all three neutrino plots (for a particular target element) to one of the shown interactions, is referred to as `similar to' the shown degenerate interaction throughout the remainder of the paper, as the qualitative features of these interactions with respect to the neutrino degeneracy will be identical.

As can be seen in \Fig{fig:Dark_Matter_Degeneracy}, it is the momentum dependence of the cross section which determines the physical shape of the recoil spectrum. The SI massless mediator interaction has a $1/|\vec{q}|^4$ dependence and is thus extremely peaked at low energies. The ED and MD massless mediator interaction have a $1/|\vec{q}|^2$ dependence and are thus still peaked, however not to the same extent as the SI massless mediator interaction. The SI massive mediator interaction, being independent of the momentum transfer, is much flatter at lower energies. It is actually \emph{only} for these momentum-independent interactions (as pointed out in ~\cite{Dent:2016iht} and \cite{Dent:2016wor}  using EFT operators) that the shape of the recoil spectra can match the recoil spectra of background neutrinos.  Spectra arising from $d \sigma_{\chi}/dE_R \sim |\vec{q}|^2, |\vec{q}|^4, |\vec{q}|^6$ are also physically distinct from the neutrino spectrum, as can be readily seen for ED, MD and PS-PS with a heavy mediator (denoted with green, blue and cyan solid lines), and have the unique feature of producing a maximum scattering rate at some non-zero value of $\ER$. 

Thus, it is clear that \emph{only} DM-nucleus interactions whose differential cross-sections are independent of $|\vec{q}|$ lead to spectral degeneracy with $^8$B neutrinos. These neutrinos constitute the most important background for experiments with realistic exposures (i.e. MT $\lesssim 100$ ton $\cdot$ yrs) since they provide the largest contribution to the background, and hence they are the most relevant to analyze. Similar considerations relating the $|\vec{q}|$ dependence of the DM scattering cross-section to spectral degeneracy can be made for the other neutrino backgrounds. However, the degeneracies with these backgrounds are not as detrimental for the exposures considered in this work.

\subsection{Discovery Limit Statistical Analysis \label{sec:like_fit}}

To establish the discovery sensitivity of experiments we use a frequentist analysis based on the profile likelihood ratio test \cite{Cowan:2010js,Rolke:2004mj}, the power-constrained limits \cite{Cowan:2011an}. This has been utilized in recent direct detection studies \cite{Billard:2011zj,Billard:2013qya,Aprile:2011hx,Ruppin:2014bra}. This test is performed by generating simulated datasets for each experiment, assuming for each dataset a particular DM interaction, WIMP mass $m_{\chi}$, reference DM cross section $\sigma_{\rm ref}^{\ast}$, and a normalization (i.e. an energy integrated neutrino flux) for each of the considered neutrino fluxes ${\phi}_{\nu_k}$ ($k = 1, 2 \dots n_{\nu}$, with $n_{\nu} = 14$), which we take here to be the average theoretical predicted value $\overline{\phi}_{\nu_k}$. More specifically, for each of the aforementioned parameters and experimental configurations we generate fake data consisting of a total number of ``observed'' events $N_{\rm o}$ at particular recoil energies $E_j$ with $j = 1, 2, \dots N_{\rm o}$ that we use to define the following likelihood function
\begin{equation} \label{eq:likemain}
\mathcal{L} = \Bigg\{\frac{e^{-N_{\rm E}}}{N_{\rm o}!}  \prod_{j=1}^{N_{\rm o}}{\rm MT} \Big(\dfrac{d R_{\rm tot}}{d E_R}\biggr|_{\ER=E_j}  \Big) \Bigg\} \prod_{k=1}^{n_{\nu}}\dfrac{1}{\sqrt{2 \pi}\sigma_{\nu_k}}{\rm exp}\biggr[-\left(\frac{\phi_{\nu_k} - \bar{\phi}_{\nu_k}}{\sqrt{2}\sigma_{\nu_k}}\right)^2 \biggr] .
\end{equation}
Here, the total differential rate is the sum of the signal (due to DM) and the background (due to all types of neutrinos) contributions
\begin{equation} \label{eq:drtot}
\dfrac{d R_{\rm tot}}{d E_R} =  \frac{d R_\chi}{d\ER} (\sigma_{\rm ref}, m_{\chi})  + \sum_{k}\frac{d R_{\nu_k}}{d\ER}(\phi_{\nu_k})~,
\end{equation}
as defined in \Eq{eq:totdmrate} and \Eq{eq:totnurate}, respectively. The total number of predicted events $N_{\rm E}$ is obtained by integrating \Eq{eq:drtot} over the energy range of observation of the particular experiment, and multiplying by the exposure MT. In \Eq{eq:likemain} the extended likelihood in the curly brackets for the DM and neutrino events
is multiplied by a Gaussian product of likelihoods, centered around the mean predicted flux normalization $\overline{\phi}_{\nu_k}$, for each neutrino species $\nu_k$ ($k = 1, 2, \dots n_{\nu} = 14$) to take into account the systematic uncertainty in the flux $\phi_{\nu_k}$.
 In the Gaussian likelihoods the $\sigma_{\nu_k}$ is the 1-$\sigma$ uncertainty in the particular flux (see \Tab{tab:nuflux}, where $\sigma_{\nu_k}$ is taken to be the largest of the two asymmetric 1-$\sigma$ uncertainties).
 
The procedure to obtain each set of simulated data involves two steps: 1) finding the total number of events of each type $t$ (i.e. $t = 1$ for the DM signal events and $t = k+1$ for the neutrino background events, with $k$ running as before), and 2) finding the corresponding recoil energy for each of the events. The number of events of a specific type, $n_{t}$, is found from a Poisson distribution $P_{t}$ as
\begin{equation}
 P_{t} = \dfrac{\mu_{t}^{n_{t}} \, e^{- \mu_{t}}}{n_{t}!}~, 
\end{equation}
where the mean $\mu_{t}$ is the number of events predicted by the model being tested, defined by a particular set of values $(\sigma_{\rm ref}^{\ast}, m_{\chi})$ for $t$ = 1, and the total neutrino flux $\bar{\phi}_{\nu_k}$ for $t = k + 1$. Choosing a random number for the cumulative probability distribution (CDF) of each $P_t$ one value of $n_{t}$ is randomly generated (inverse transform sampling).
The number $N_0$ of ``observed'' events is then $N_0 = \sum_{t = 1}^{15} n_t$.

To determine the energy of the $n_t$ events of each type we use as probability density function (PDF) the corresponding differential recoil rate $d R_{t}/d E_{R}$ normalized by the total rate (i.e. the rate integrated over the specified energy range for each experiment $R_{t}$) as
\begin{equation}
\Big(\text{PDF}\Big)_{t} = \dfrac{1}{R_{t}} \dfrac{d R_{t}}{d E_R}~.
\end{equation}
The corresponding $n_{t}$ recoil energies are again obtained with inverse transform sampling. With the above procedure, we simulate 250 - 500 datasets for each particle model, as specified by the choice of $(\sigma_{\rm ref}^{\ast}, m_{\chi})$, and experimental configuration. 

For each simulated data set we define a test statistic $q_0$ that  allows  to reject the background only hypothesis $H_0$ (in which $\sigma_{\rm ref} = 0$) if it is true, with a probability not larger than some value $\alpha$ that denotes  the significance level of the test. We further impose that the probability of not rejecting $H_0$ when the DM alternative hypothesis $H_{\sigma}$ (with $\sigma_{\rm ref} \neq 0$) is true, is less than some value $\beta$. For our analysis we chose $\alpha$ to correspond to 3$\sigma$ ($\alpha = 0.0135$) and $(1-\beta)$, which denotes the ``power'' of the test of $H_0$ with respect to the alternative hypothesis $H_{\sigma}$ (e.g. \cite{Patrignani:2016xqp}, Sec. 40), to be 90\%. The test statistic $q_0$ for each simulated dataset is the profile likelihood ratio, defined as  
\begin{equation} \label{eq:likeratio}
q_0 = 
\begin{cases}
   - 2 \,\text{ln} \left( \dfrac{\mathcal{L}(\sigma_{\rm ref} = 0, \hat{\hat{\vec{\phi}}}_{\nu_k})}{\mathcal{L}(\hat{\sigma}_p, \hat{\vec{\phi}}_{\nu_k})} \right) ,& \hat{\sigma}_{\rm ref} \geq 0 \\
    0 ,& \hat{\sigma}_{\rm ref} < 0 \, ,
\end{cases}
\end{equation}
where the $\phi_{\nu_k}$ are treated as nuisance parameters. For numerical calculations we parametrize $\sigma_{\rm ref} = 10^x$, which ensures that $\sigma_{\rm ref} > 0$\footnote{It has previously been verified that enforcing semi-positive values of the cross section does not affect the distribution of the test statistic, see e.g.~\cite{Billard:2011zj}.}. The hats refer to values that maximize the likelihood, with double-hats referring to values maximizing the likelihood subject to the constraint $\sigma_{\rm ref} = 0$. Note that by definition $q_0 \geq 0$, with larger values of $q_0$ indicating greater incompatibility of the simulated data with the background only hypothesis $H_0$. Thus, we require that the probability $p_0$ of having a $q_0$ value larger (i.e. more incompatible with the data if due to background only) than the ``observed'' (i.e. than the $q_0$ of the simulated dataset) $q_0^{\rm obs}$ 
 \begin{equation}\label{eq:pval0}
p_0 = \int_{q_0^{\rm obs}}^\infty dq_0 \, f(q_0|H_0) \, , 
\end{equation}
is not larger than $\alpha$, $p_0 \leq \alpha$. Here, $f(q_0|H_0)$ is the PDF of obtaining $q_0$ under the background-only hypothesis $H_0$. In the large sample limit, Wilks' theorem ensures that $f(q_0|H_0)$ is given by a $\chi^2$ distribution with one degree of freedom (d.o.f.). This implies that the value of $\alpha$ corresponding to a $Z\sigma$ significance can be obtained by requiring $q_0^{\rm obs} \geq Z^2$. In the analyses detailed below, we have required a value of $\alpha$ corresponding to a $3\sigma$ significance (i.e. $Z = 3$). If at least 90\% of the simulated datasets generated with a particular set ($\sigma_{\rm ref}^{\ast} , m_{\chi}$) produce $q_0^{\rm obs} > 9$, then we accept this value of $\sigma_{\rm ref}^{\ast}$ as `discoverable'. That is to say that we require the probability $p_{\sigma}$ of having a value of $q_0^{\rm obs}$ larger than $Z^2$
 \begin{equation}\label{eq:pval}
p_{\sigma} = \int_{Z^2}^\infty dq_0^{\rm obs} \, f(q_0^{\rm obs}|H_\sigma) \, , 
\end{equation}
to be $p_{\sigma} \geq 90\%$. The discovery limit of $\sigma_{\rm ref}$ of each model and for each DM mass is thus set by the value of $\sigma_{\rm ref}^{\ast}$ for which the aforementioned condition, $p_{\sigma} \geq (1 - \beta) = 90\%$, is strictly equal, i.e. $p_{\sigma} = 90\%$.
The procedure described above is widely used in modern high energy experiment analyses, for example in a typical LHC combined Higgs search \cite{ATLAS:2012ae} one assumes $(1-\beta) = 50\%$ and $\alpha$ corresponding to $5 \sigma$.
 
In practice we determine the fraction of simulated data sets that produce p-values $p_0^i \leq \alpha = 0.00135$ (i.e. those producing a $\geq 3\sigma$ detection of DM), by computing 
\begin{equation}
 f^{\rm 90}(\sigma_{\rm ref}^{\ast}, m_\chi) \equiv \sum_{i=1}^{N_{\rm sim}}\frac{1}{N_{\rm sim}} \begin{cases} 1 &\mbox{if } p_0^i \leq 0.00135 \\ 0 &\mbox{if } p_0^i > 0.00135 \end{cases} \, , 
\end{equation}
 where $N_{\rm sim}$ is the number of simulated data sets with fixed parameters $(\sigma_{\rm ref}^{\ast}, m_{\chi})$. The value of $\sigma_{\rm ref}^{\ast}$ is varied until $f^{90}(\sigma_{\rm ref}^{\ast}, m_\chi) = 0.9$,  thus identifying the cross section at which $90\%$ of experimental realizations are expected to obtain a $\geq 3\sigma$ detection of DM. For a fixed DM particle mass $m_\chi$, it is this value of $\sigma_{\rm ref}$ that defines the discovery limit.  The full discovery limit for a particular experiment and interaction is then obtained by scanning over $m_{\chi}$.  

\begin{figure}[h]
\mycenter{
\includegraphics[trim={0mm 15.5mm 0 0},clip,width=.4\textwidth]{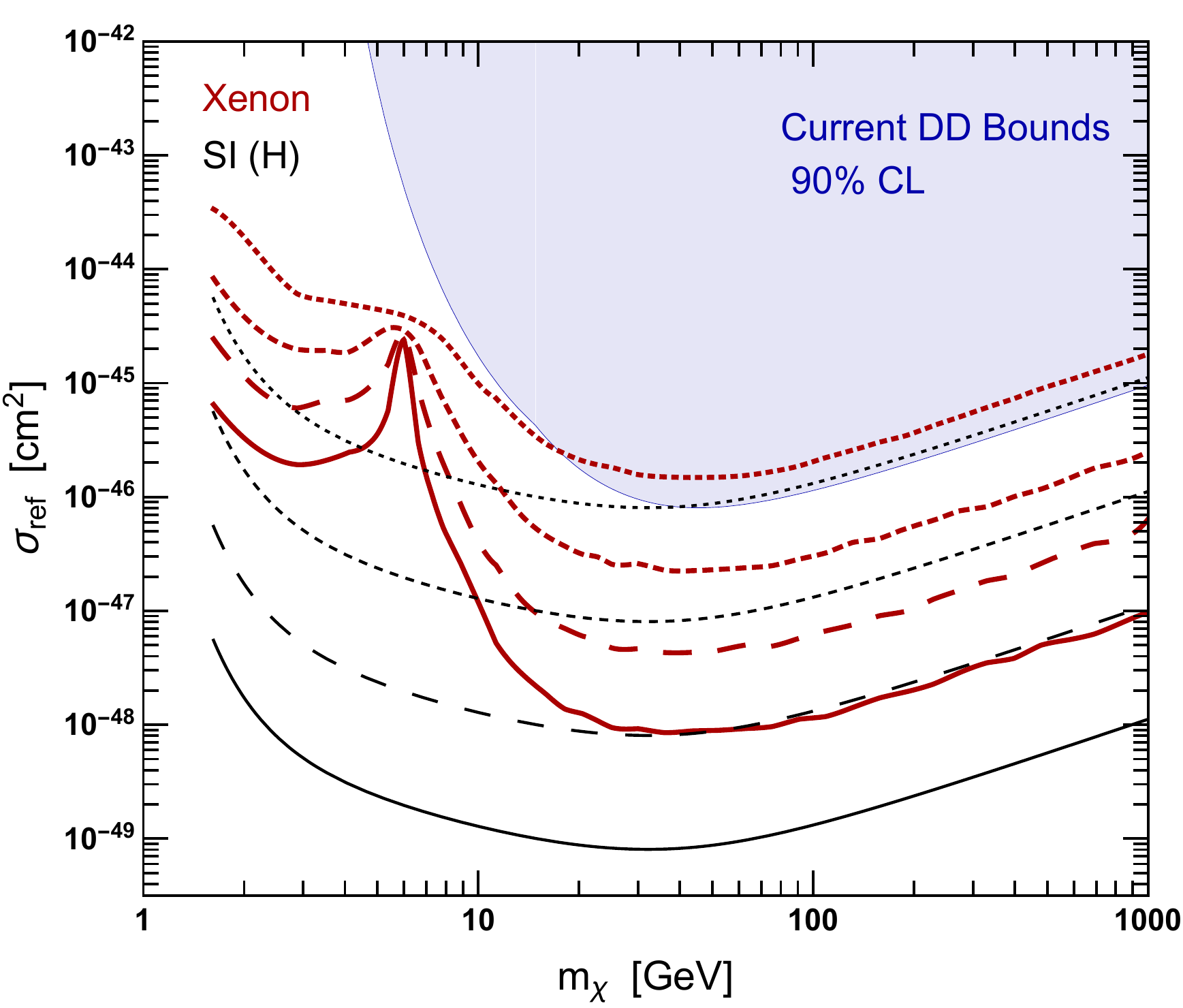}
\includegraphics[trim={9mm 15.5mm 0 0},clip,width=.38\textwidth]{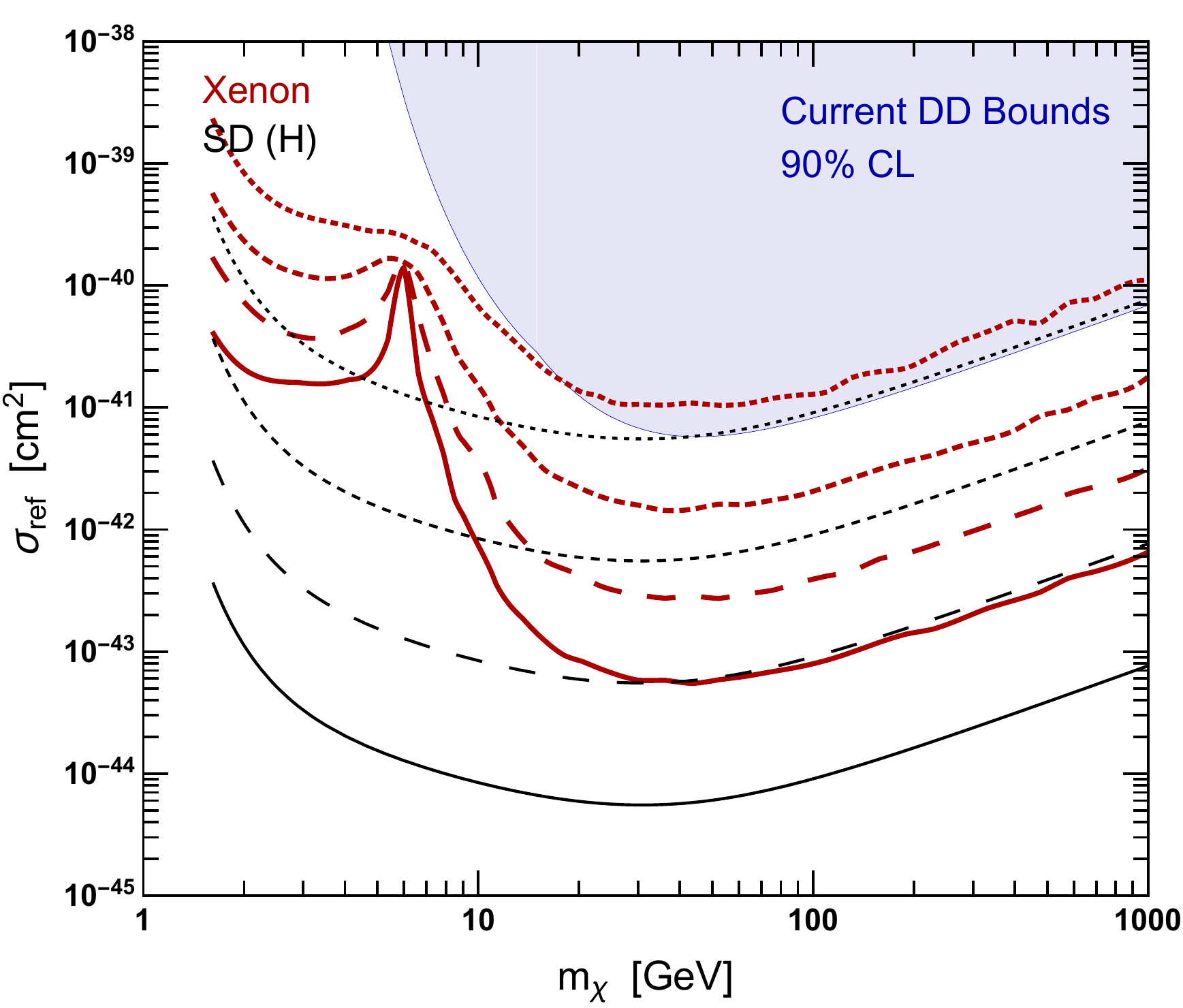}
\includegraphics[trim={9mm 15.5mm 0 0},clip,width=.38\textwidth]{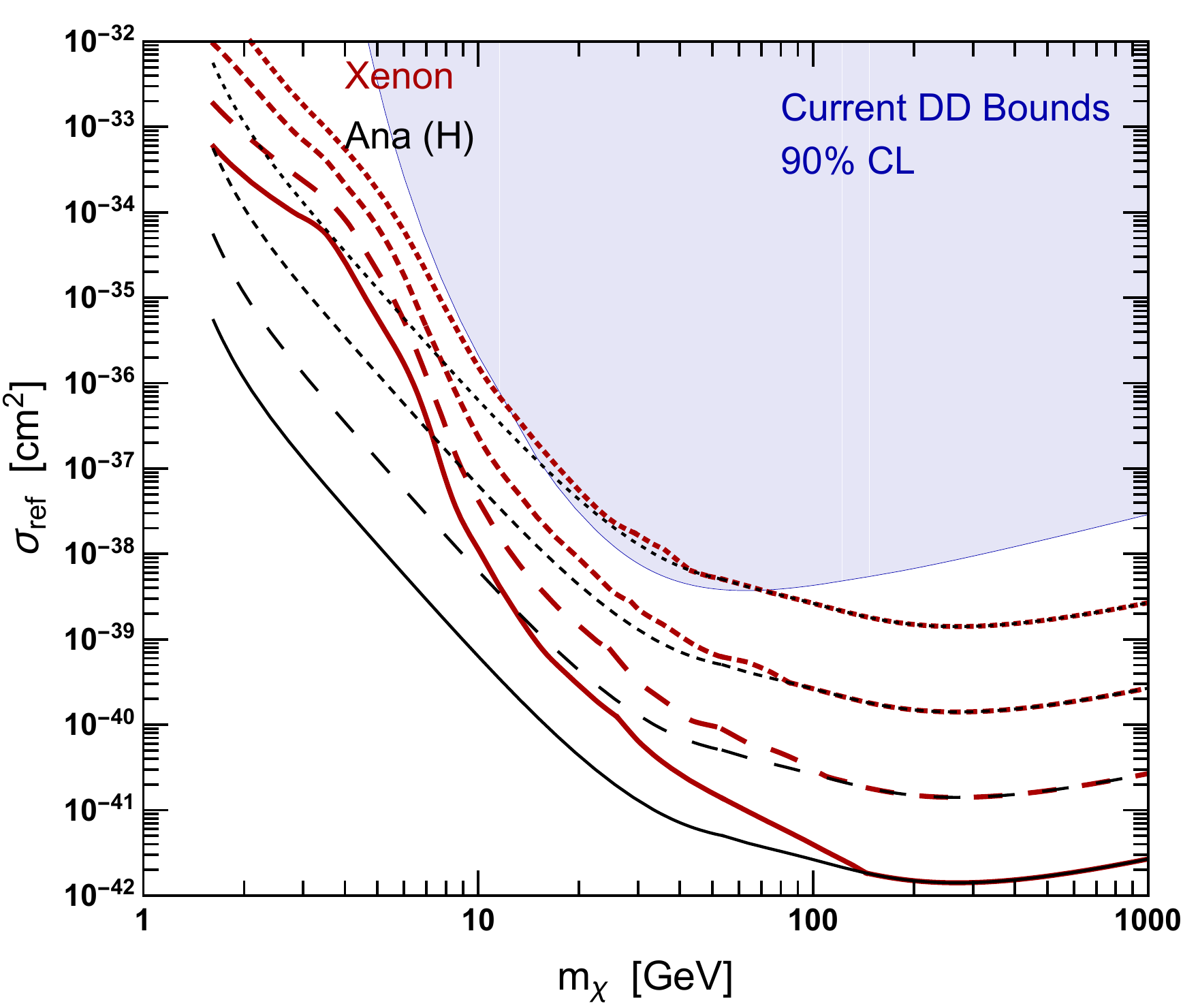}}
\mycenter{
\includegraphics[trim={0mm 15.5mm 0 0},clip,width=.4\textwidth]{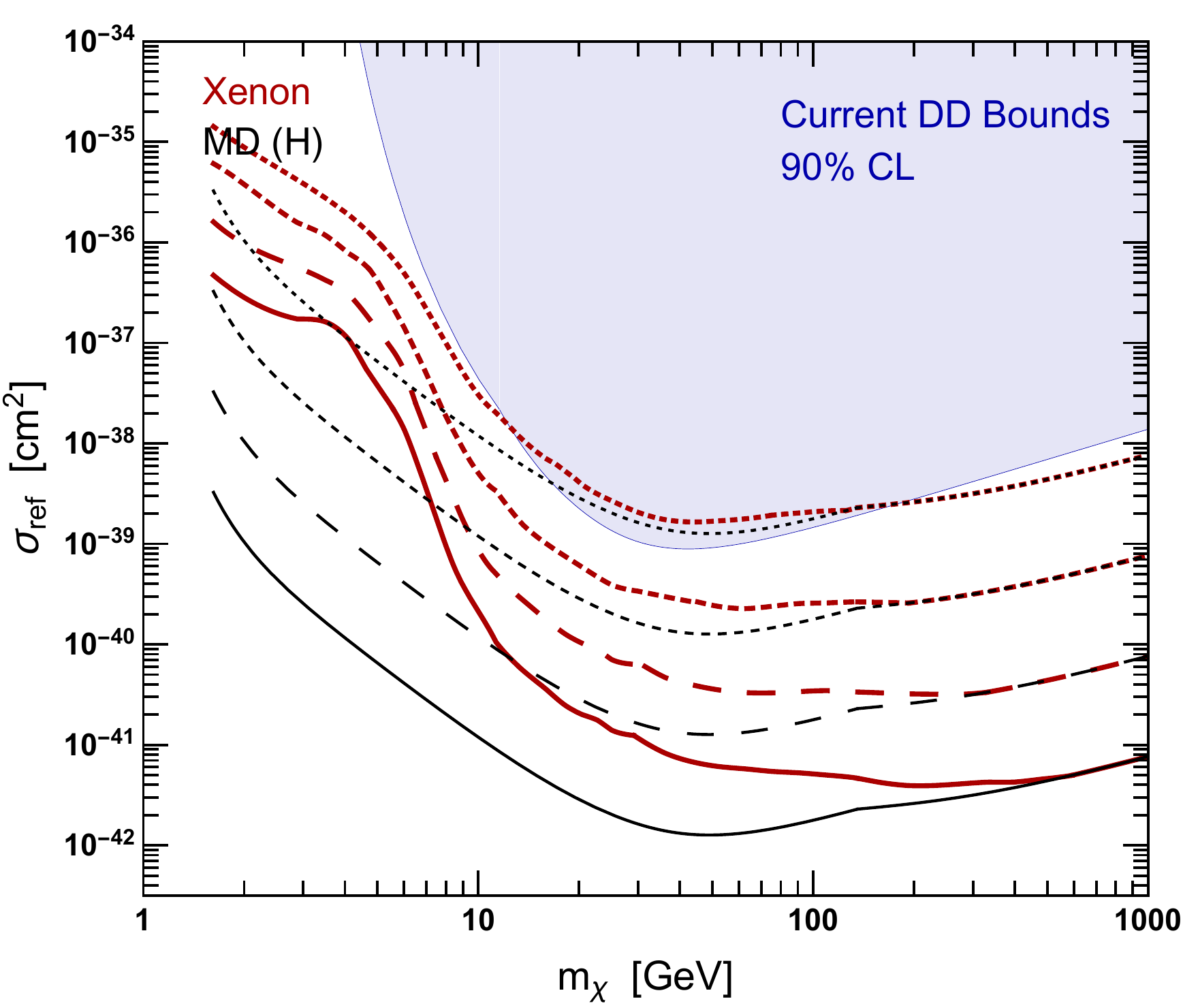}
\includegraphics[trim={9mm 15.5mm 0 0},clip,width=.38\textwidth]{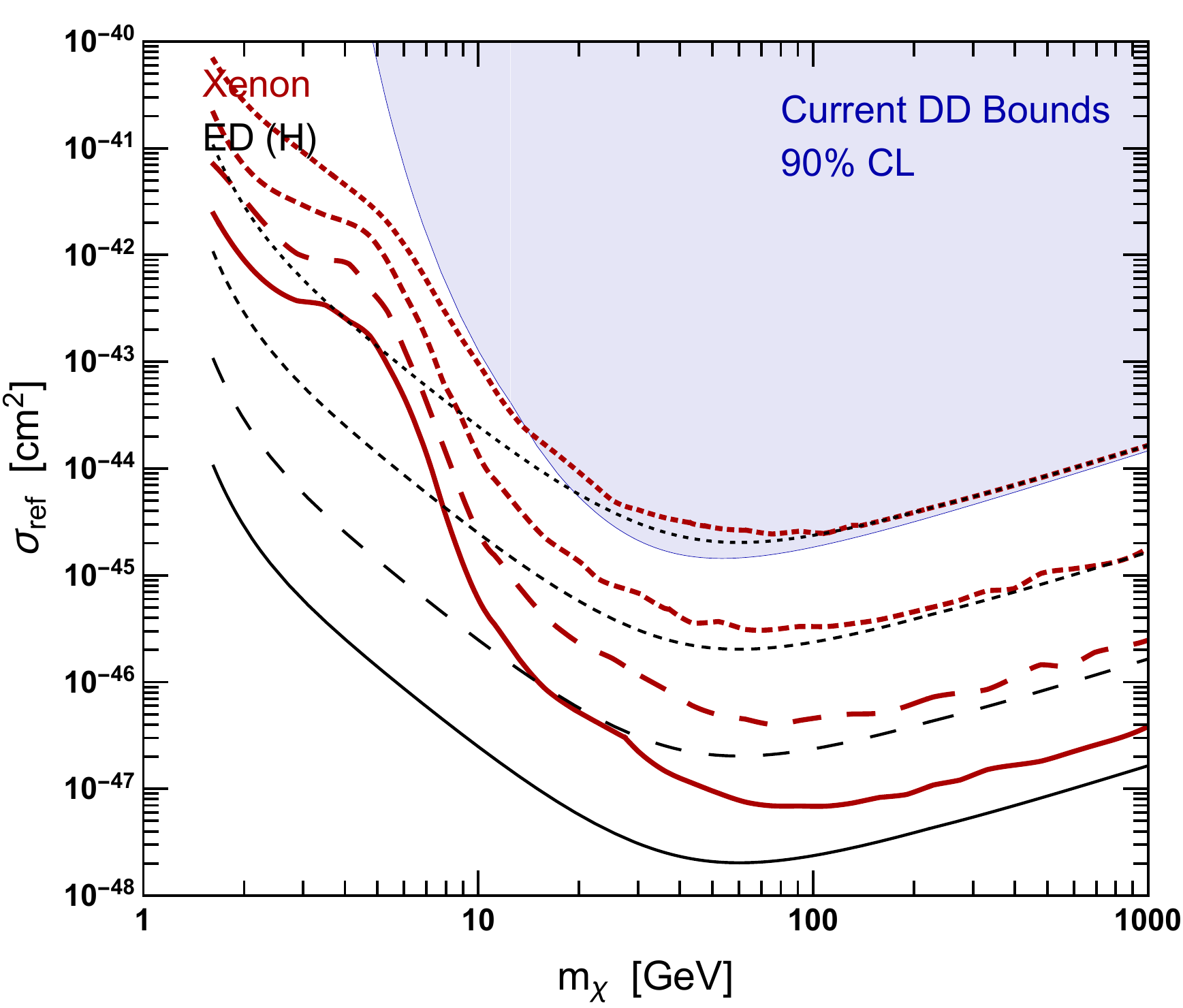}
\includegraphics[trim={9mm 15.5mm 0 0},clip,width=.38\textwidth]{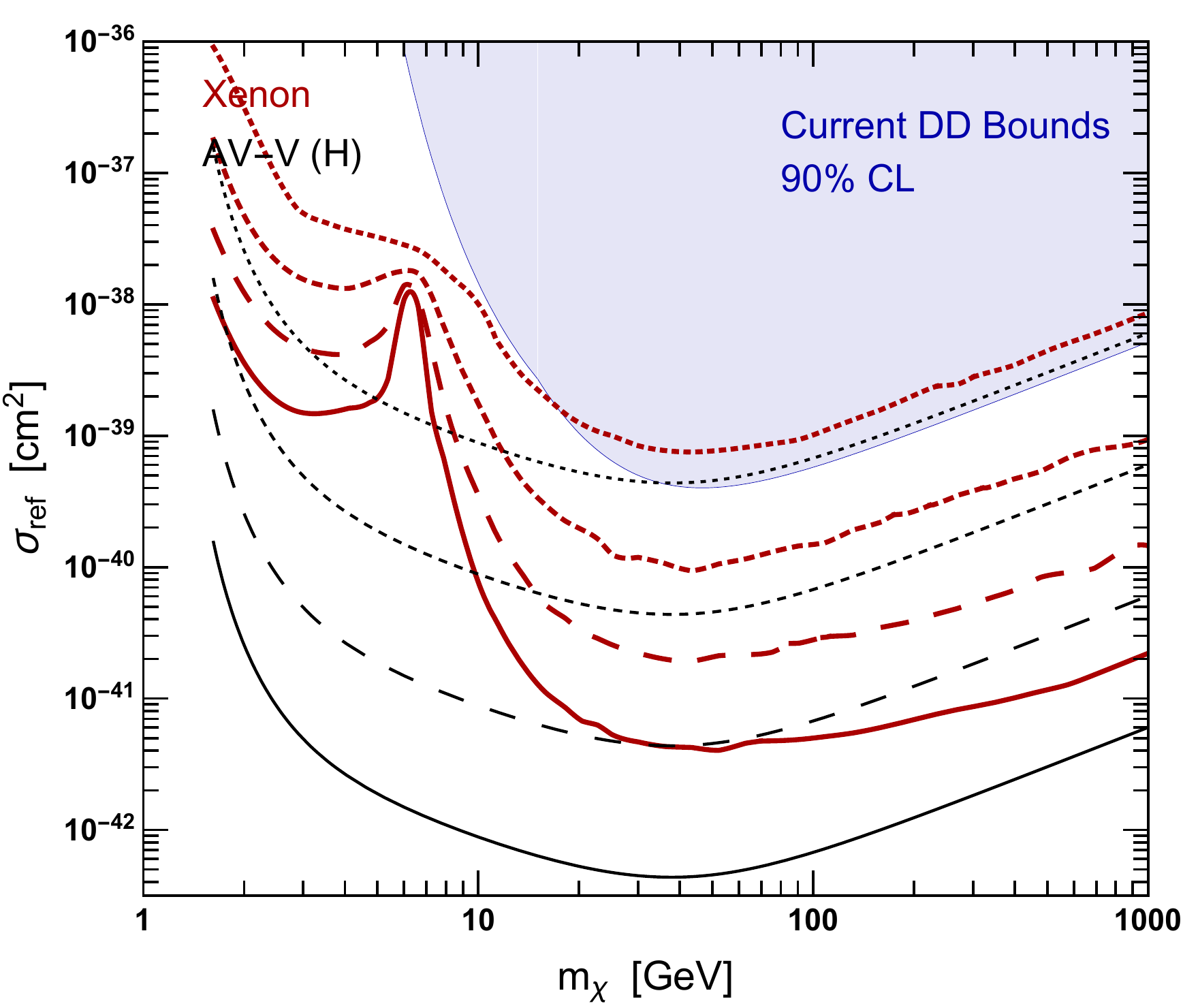}
}
\mycenter{
\includegraphics[trim={0mm 0mm 0 0},clip,width=.4\textwidth]{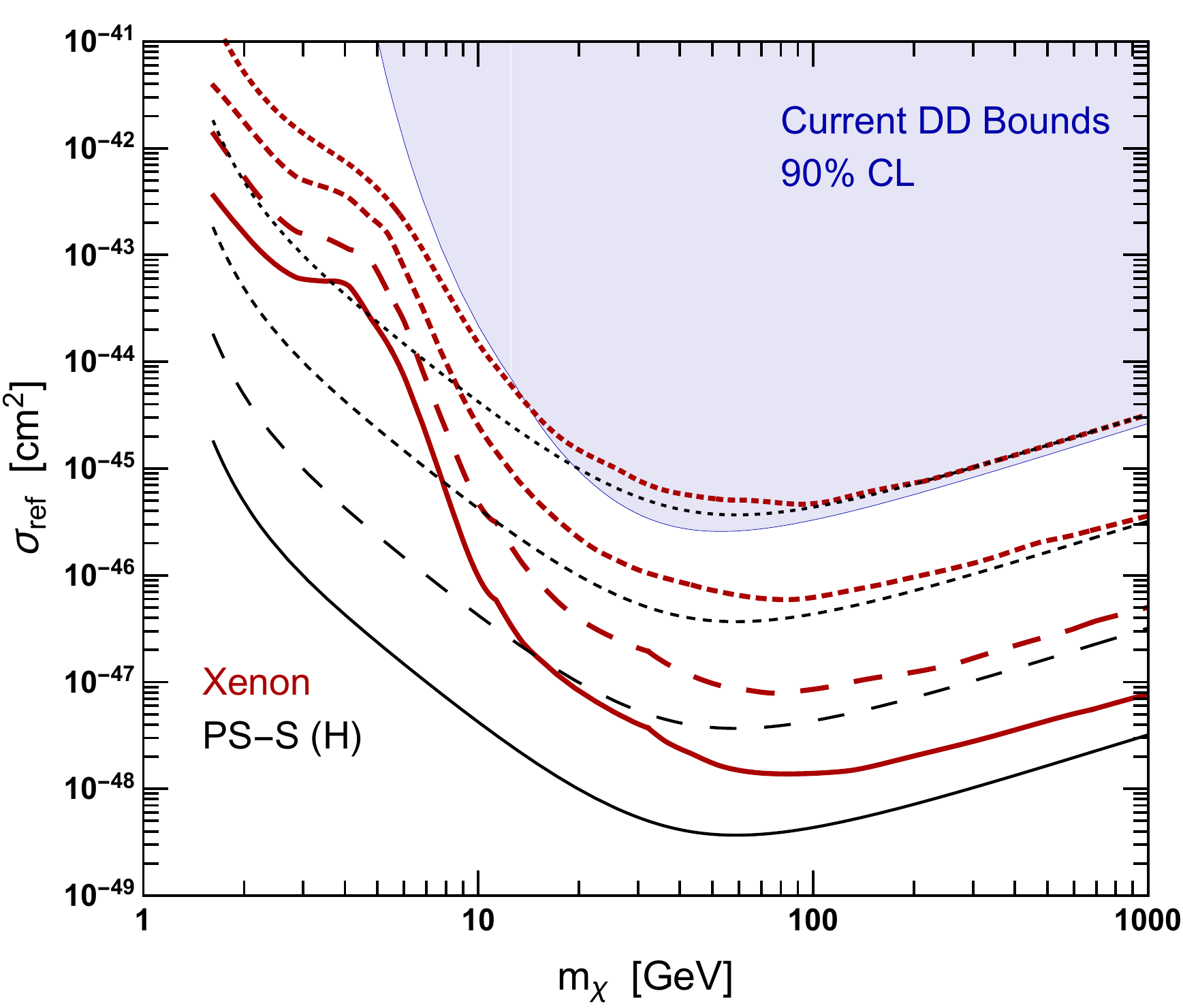}

\includegraphics[trim={9mm 0mm 0 0cm},clip,width=.38\textwidth]{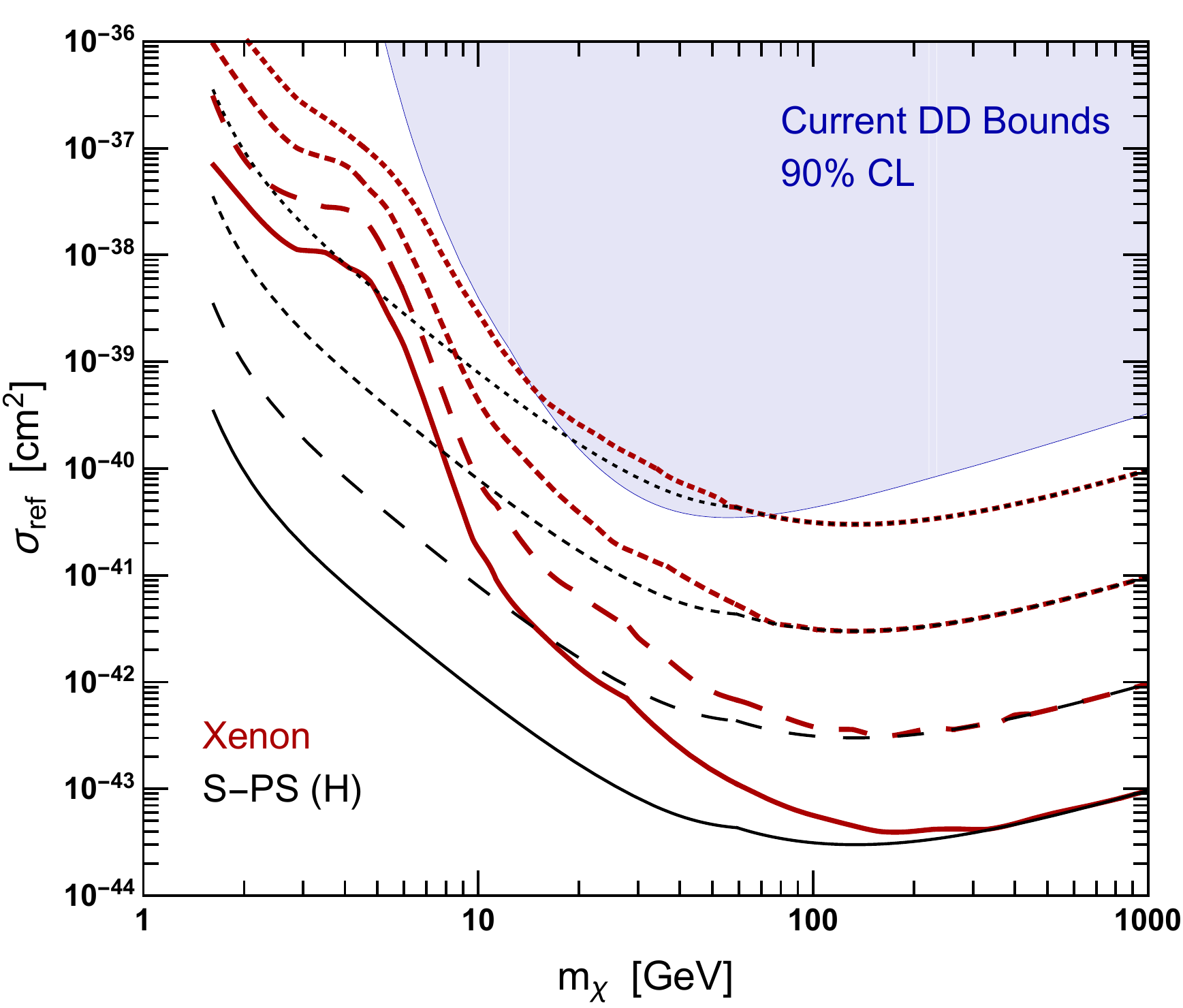}
\includegraphics[trim={9mm 0mm 0 0},clip,width=.38\textwidth]{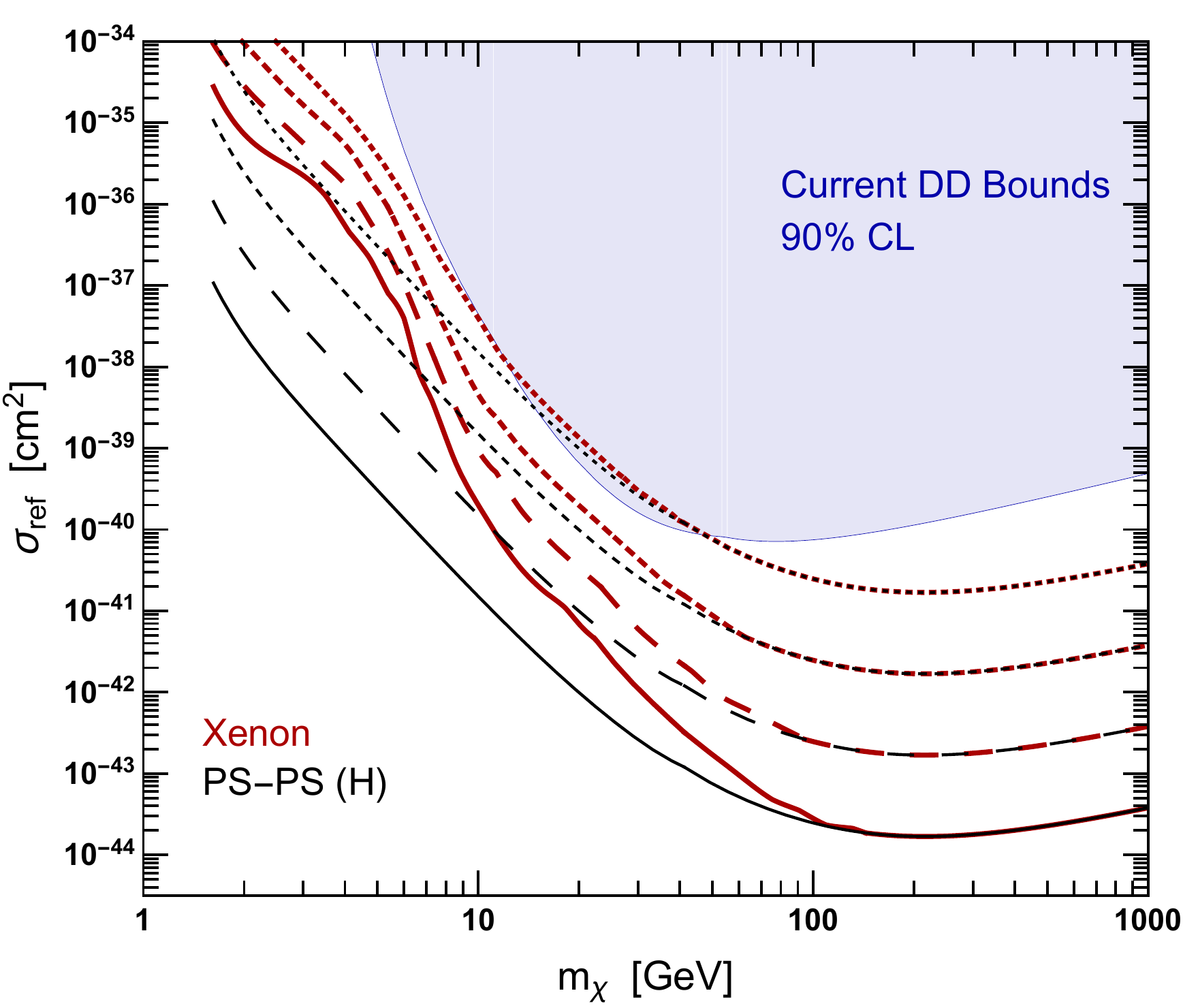}}

\caption{\label{fig:xenon_floors_massive} 3$\sigma$ discovery limit for a xenon-based experiment (see \Tab{tab:experiments}) for a 0.1 ton-year exposure (dotted), 1 ton-year exposure (short dashed), 10 ton-year exposure (long dashed), and 100 ton-year exposure (solid), including (red) and neglecting (black) the neutrino background. Results assume the mediator mass is much larger than the momentum transfer. Shown for comparison is the current combined 90\% upper limits from XENON1T, PandaX-II, LUX, and PICO (shaded blue).}
\end{figure}

\begin{figure}[h]
 \mycenter{
  \includegraphics[trim={0mm 15.5mm 0 0},clip,width=.4\textwidth]{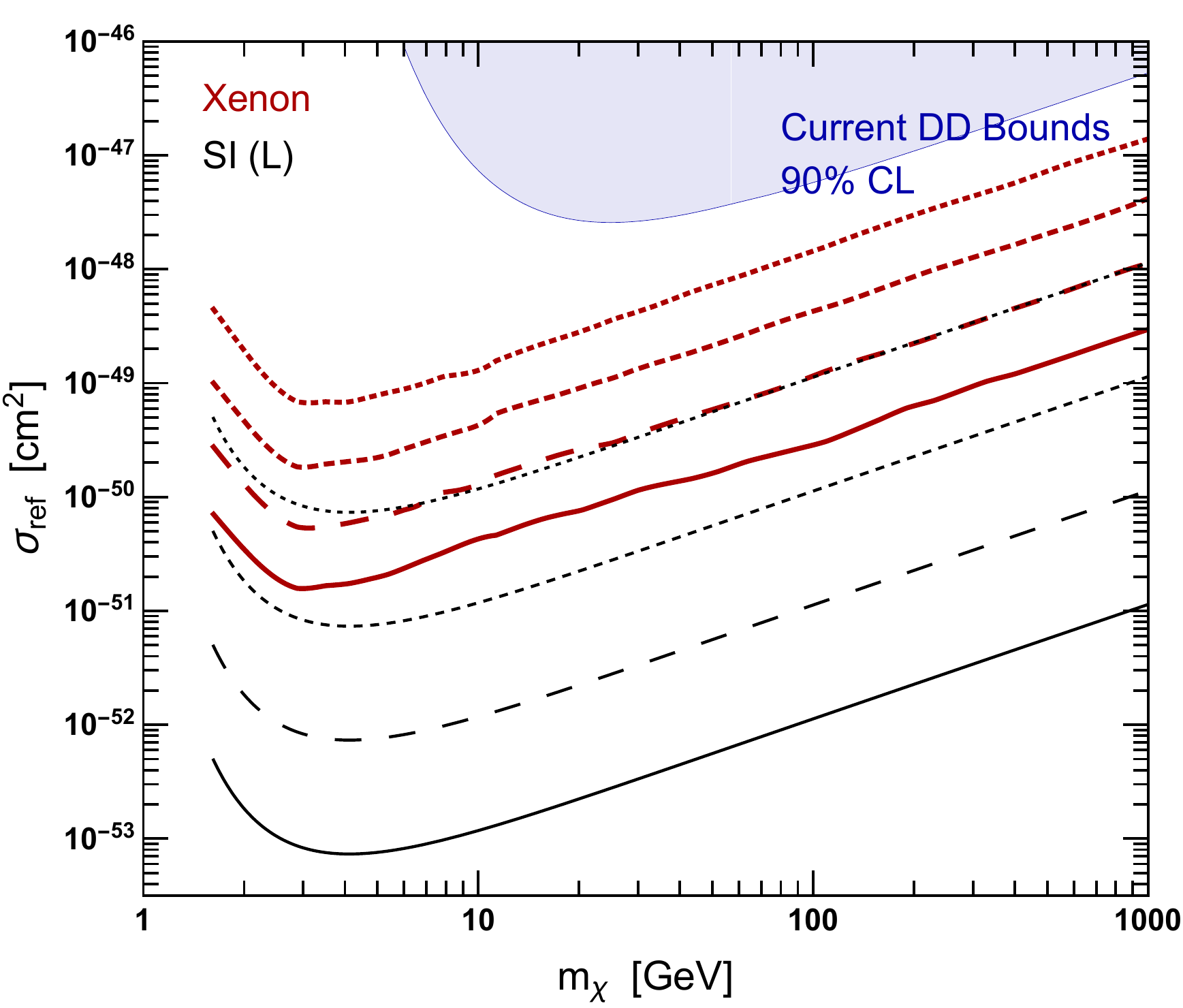}
 \includegraphics[trim={9mm 15.5mm 0 0},clip,width=.38\textwidth]{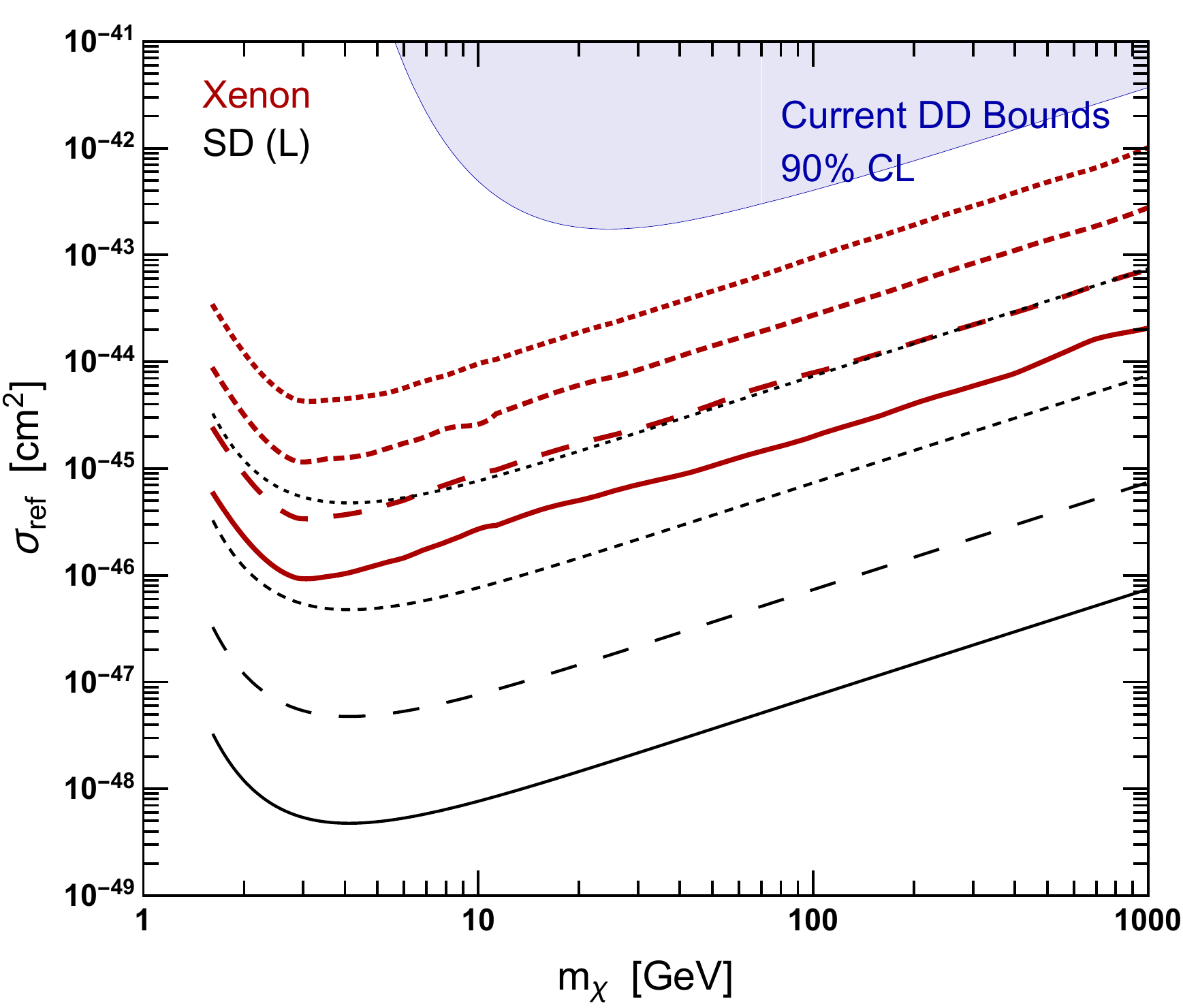}
 \includegraphics[trim={9mm 15.5mm 0 0},clip,width=.38\textwidth]{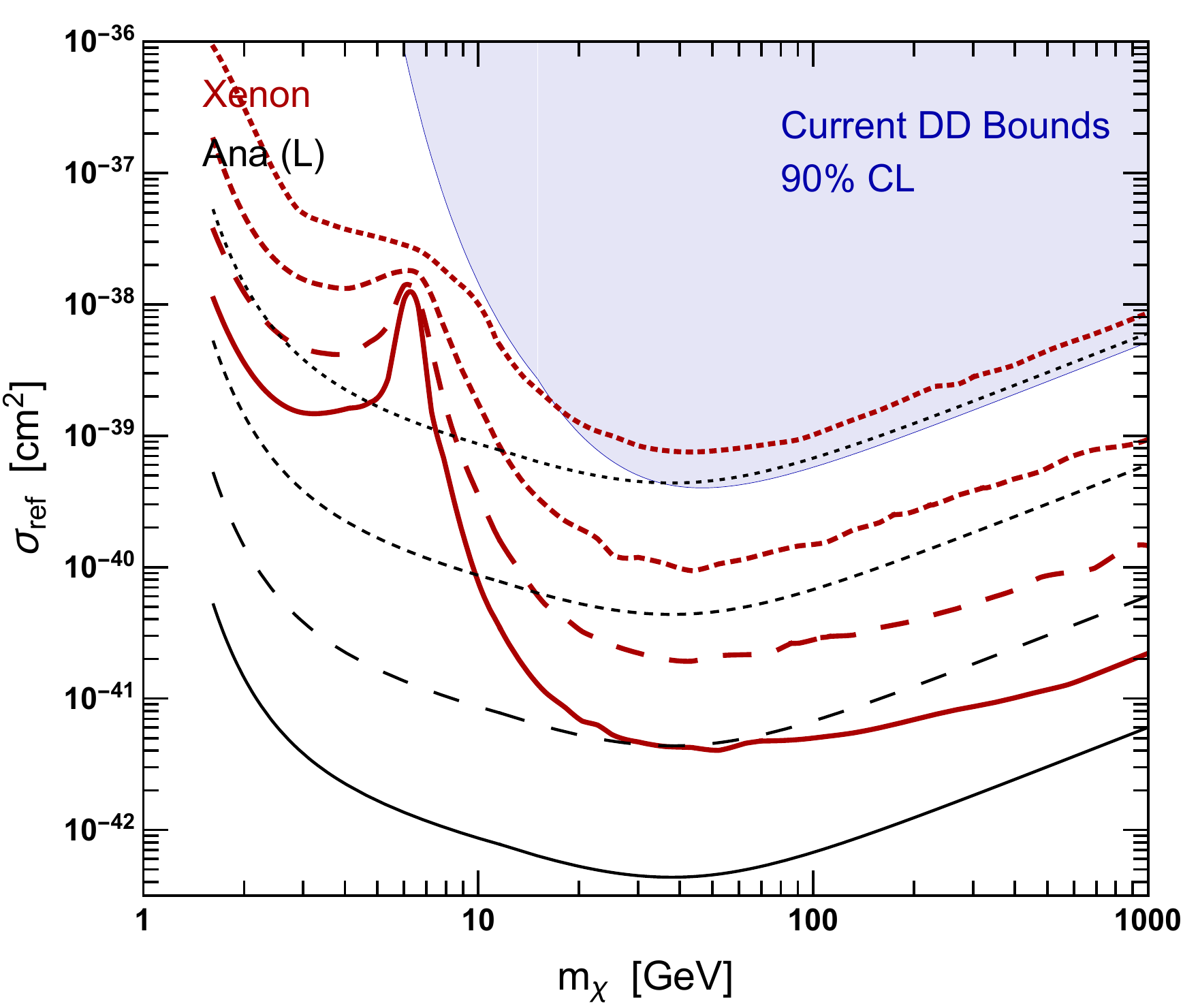}}
 \mycenter{
 \includegraphics[trim={0mm 15.5mm 0 0},clip,width=.4\textwidth]{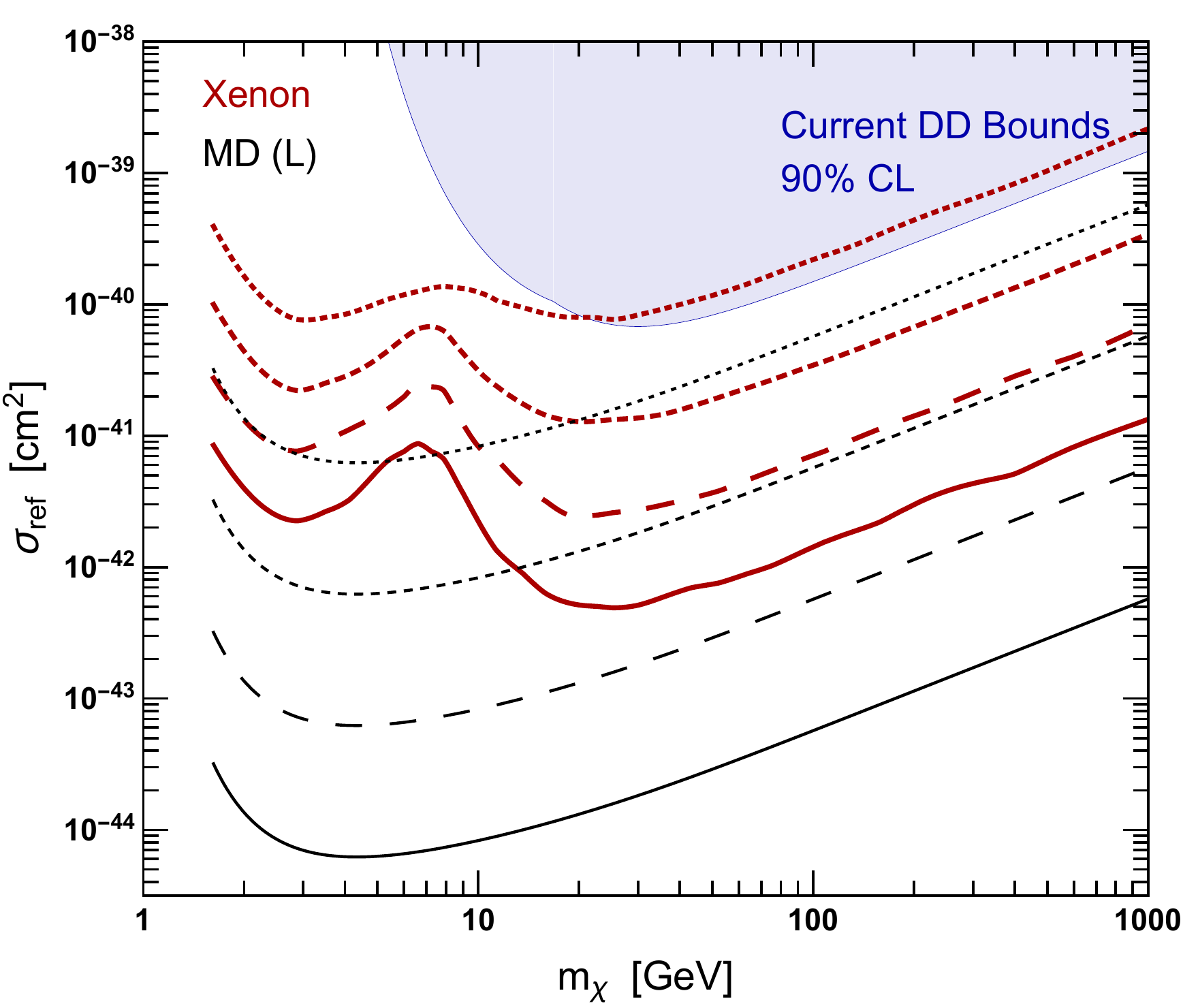}
 \includegraphics[trim={9mm 15.5mm 0 0},clip,width=.38\textwidth]{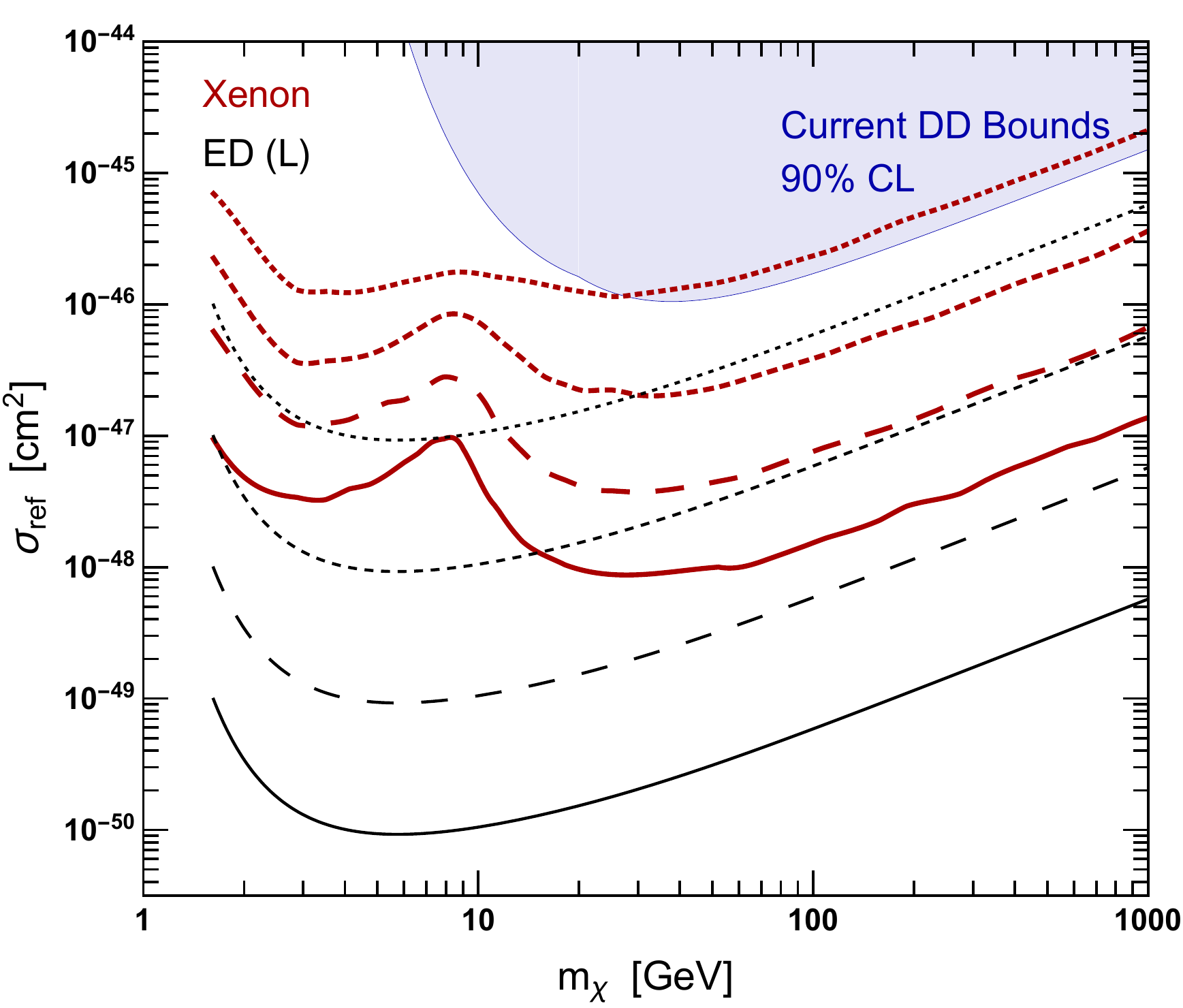}
 \includegraphics[trim={9mm 15.5mm 0 0},clip,width=.38\textwidth]{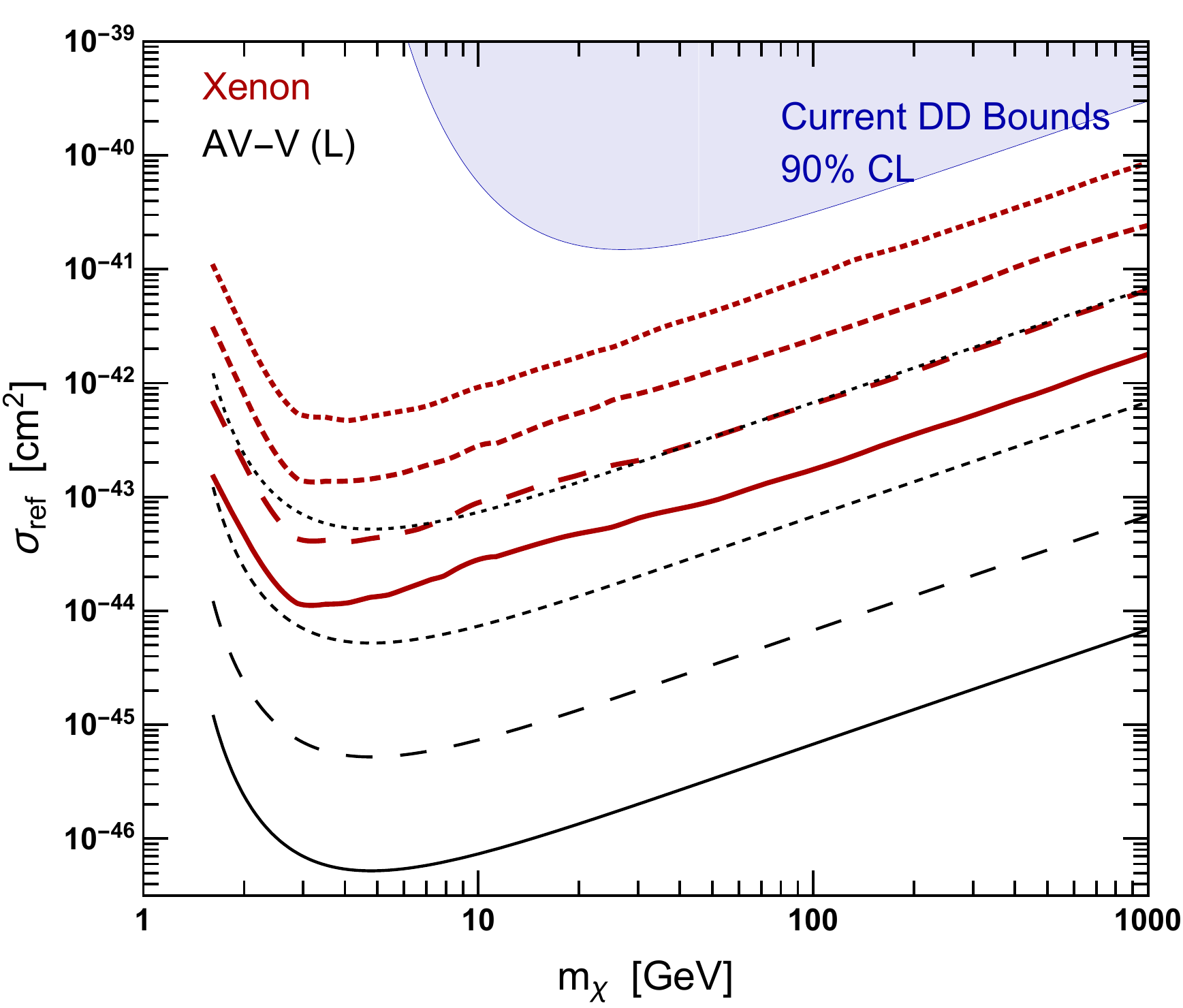}}
 \mycenter{
 \includegraphics[trim={0mm 0mm 0 0},clip,width=.4\textwidth]{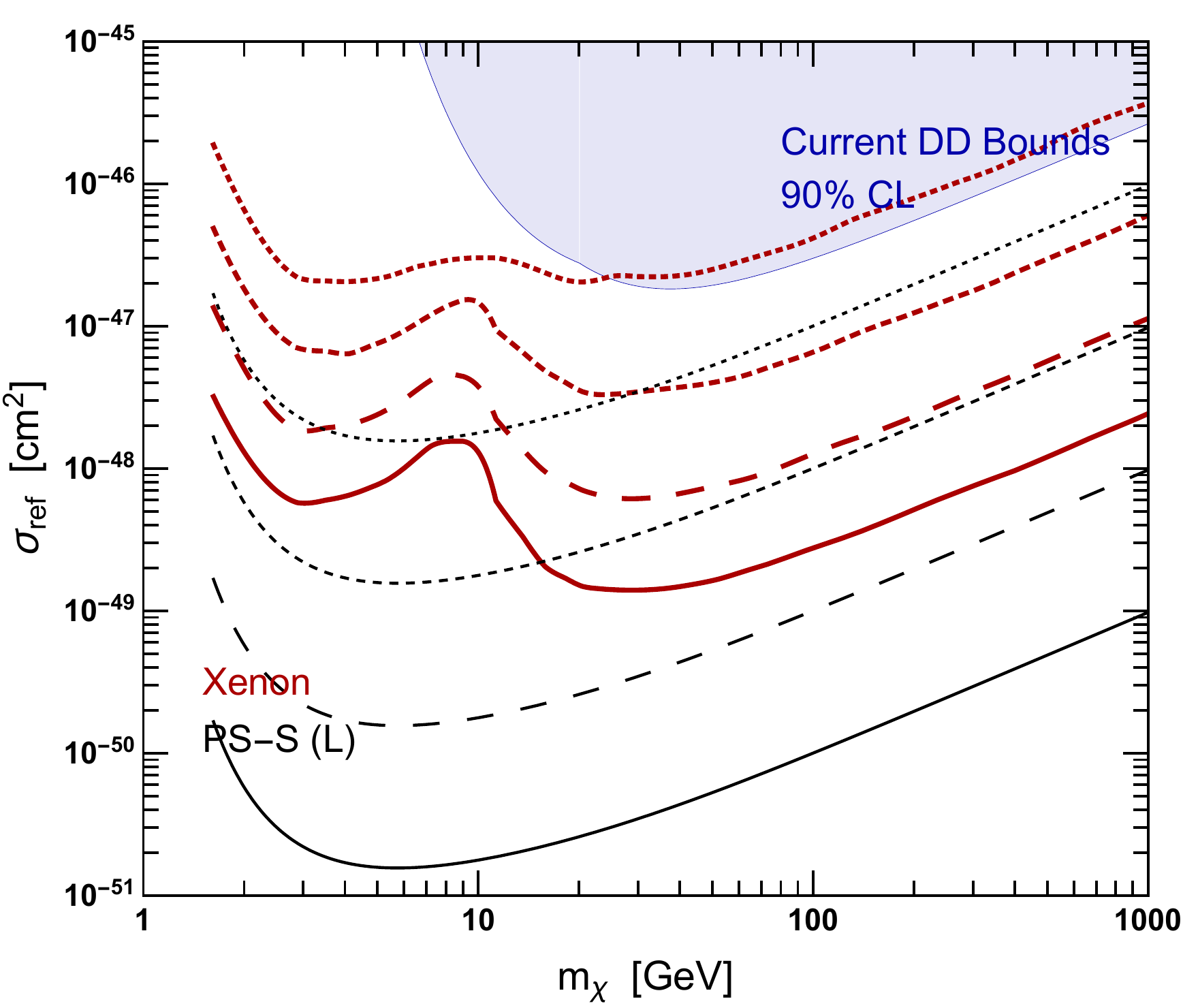}
 \includegraphics[trim={9mm 0mm 0 0},clip,width=.38\textwidth]{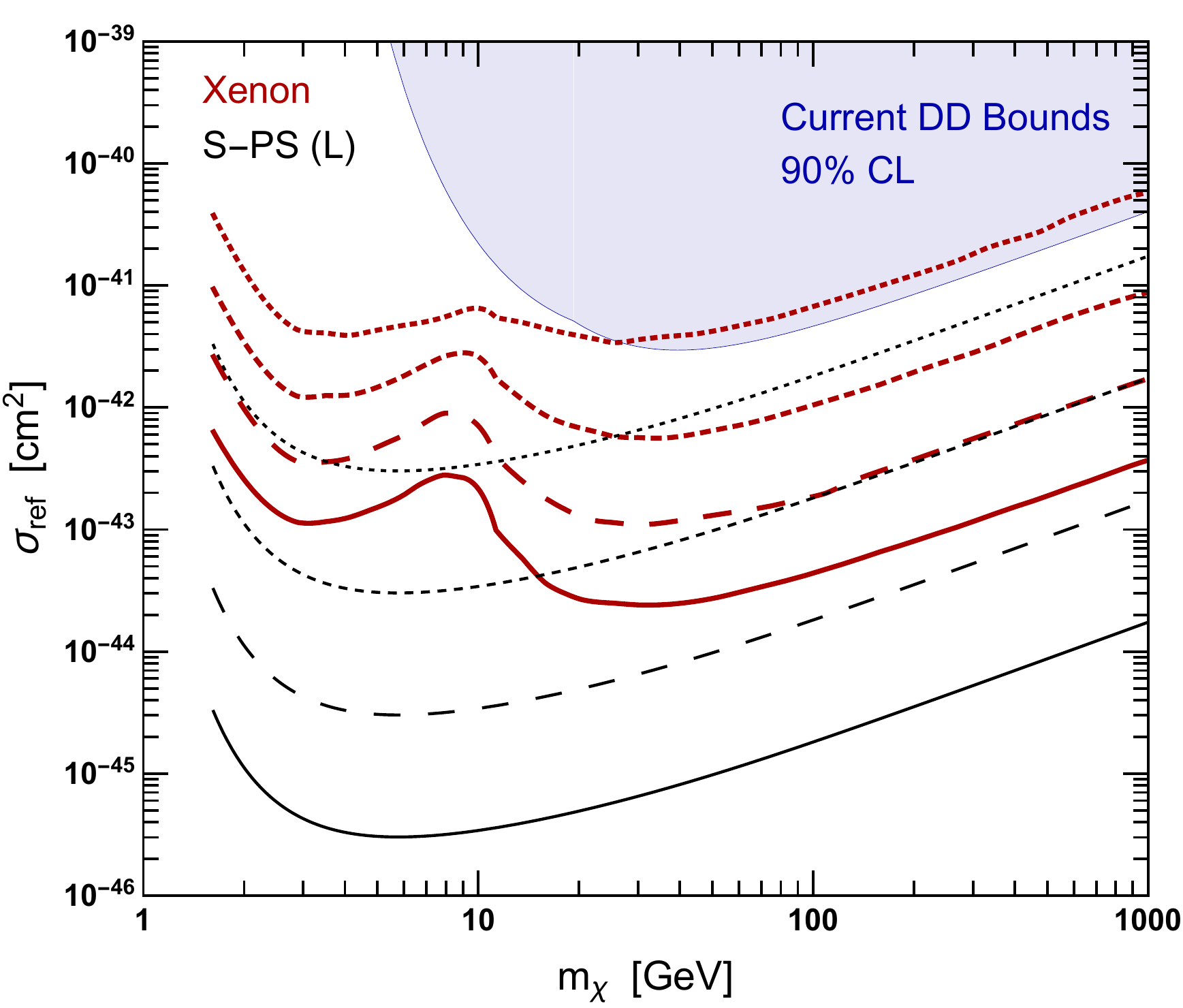}
 \includegraphics[trim={9mm 0mm 0 0},clip,width=.38\textwidth]{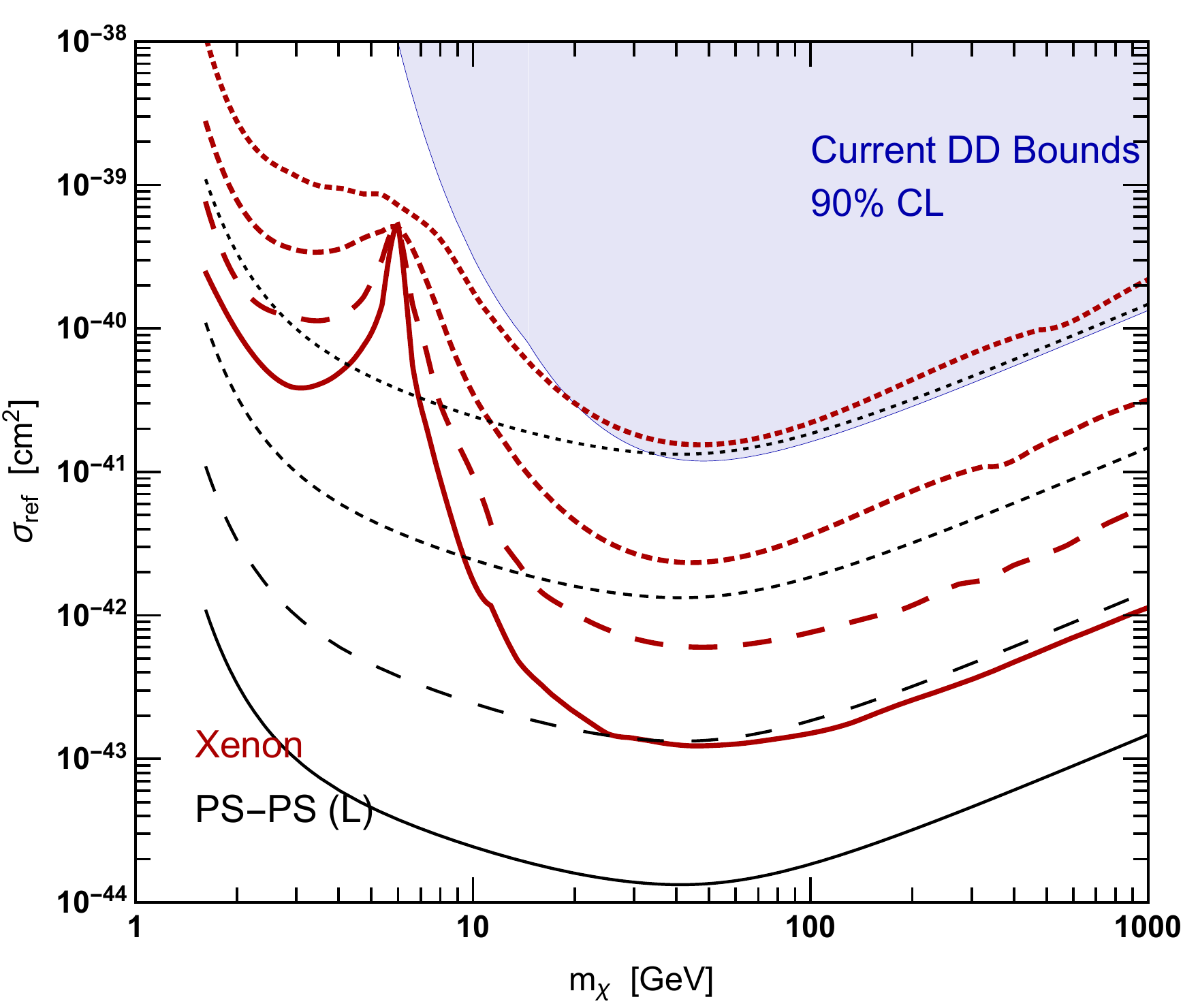}}

 \caption{\label{fig:xenon_floors_massless} Same as \Fig{fig:xenon_floors_massive} but assuming the mass of the mediator is negligible with respect to the momentum transfer.}
 \end{figure}

Fig.~\ref{fig:xenon_floors_massive} and Fig.~\ref{fig:xenon_floors_massless} show the $3\sigma$ discovery reach for xenon, considering exposures of $0.1$ ton-year (dotted), $1$ ton-year (short dashed), $10$ ton-years (long dashed), and $100$ ton-years (solid). The results for other target elements are deferred to Appendix~\ref{app:discpot} (Fig.~8 to Fig.~15). The results (denoted by colored lines) are compared to discovery limits that would be obtained in the absence of background (black lines). For each DM model, $90\%$ CL upper limits are calculated for Xenon1T \cite{Aprile:2017iyp}, LUX \cite{Akerib:2016vxi}, PandaX-II \cite{Cui:2017nnn} and PICO \cite{Amole:2017dex} data. For PICO, LUX and PandaX-II, the experimental configurations (i.e. resolution functions, efficiencies, etc.) are as defined in~\cite{Witte:2017qsy}, but for PandaX-II we employ the updated results from~\cite{Cui:2017nnn}. For Xenon1T, the energy efficiency is taken from Fig.~1 of ~\cite{Aprile:2017iyp}, the fiducial mass and runtime are taken to be 1042 kg and 34.2 days, and a limit is set using Poisson statistics with zero observed events. We have verified that this procedure reproduces the published limit to a high degree of accuracy. The combined strongest upper limit for each mass and model defines the blue region labeled `Current DD Bounds'. Note that these upper limits cannot be directly translated to our calculated discovery potential results.
\subsection{Inelastic scattering}

\begin{figure}
\mycenter{
\includegraphics[width=.42\textwidth]{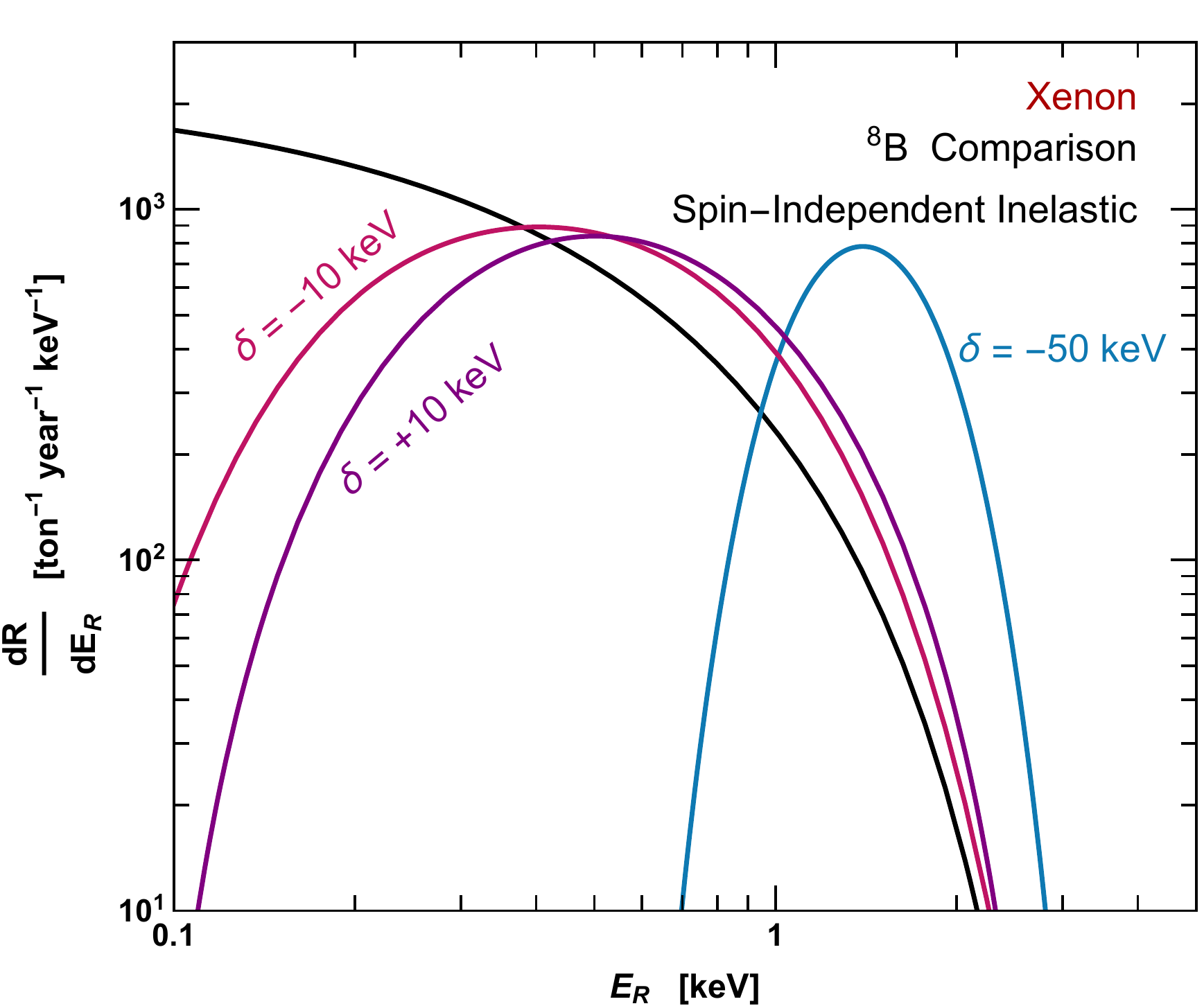}
\includegraphics[trim={15mm 0 0 0},clip,width=.3875\textwidth]{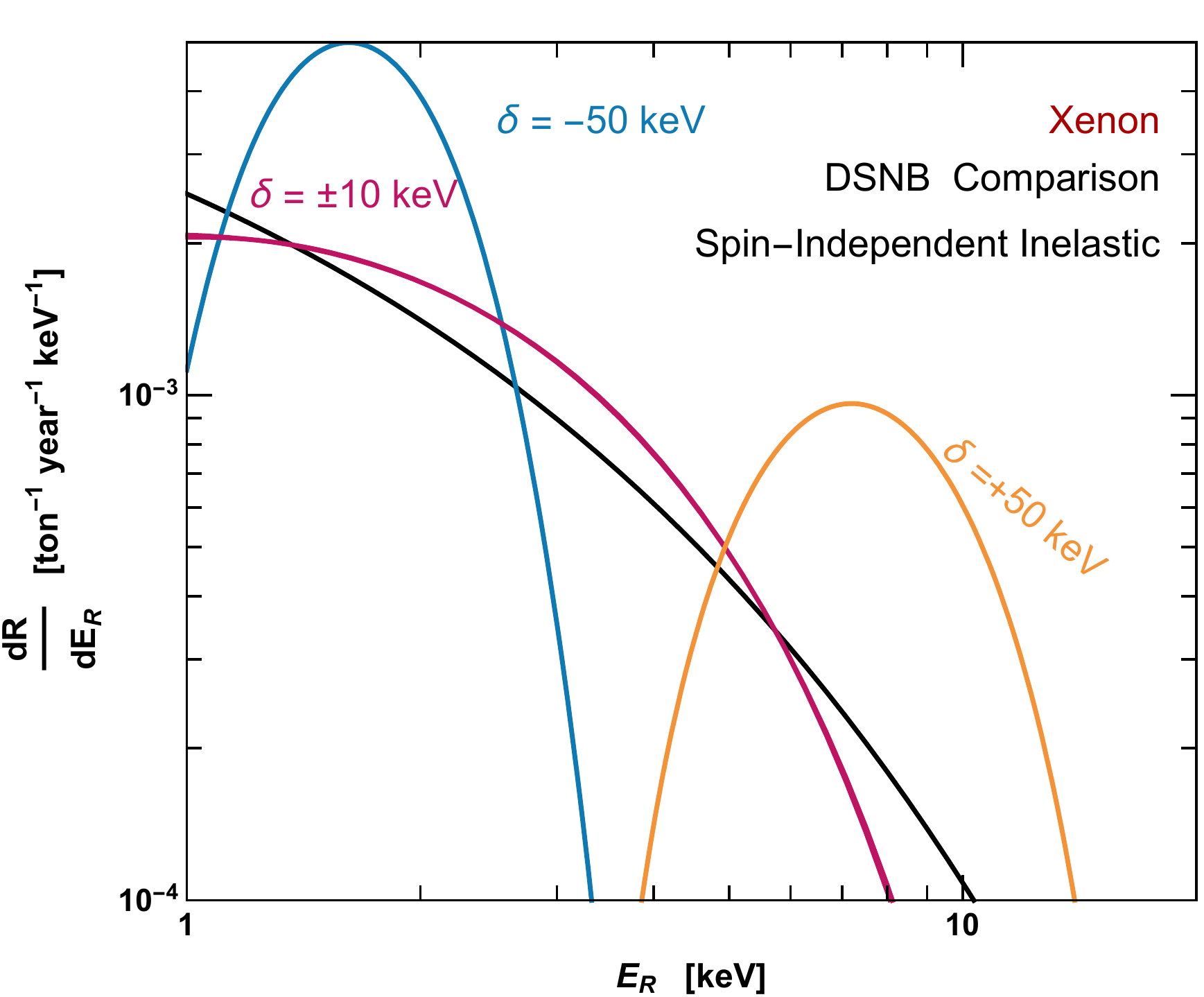}
\includegraphics[trim={15mm 0 0 0},clip,width=.3875\textwidth]{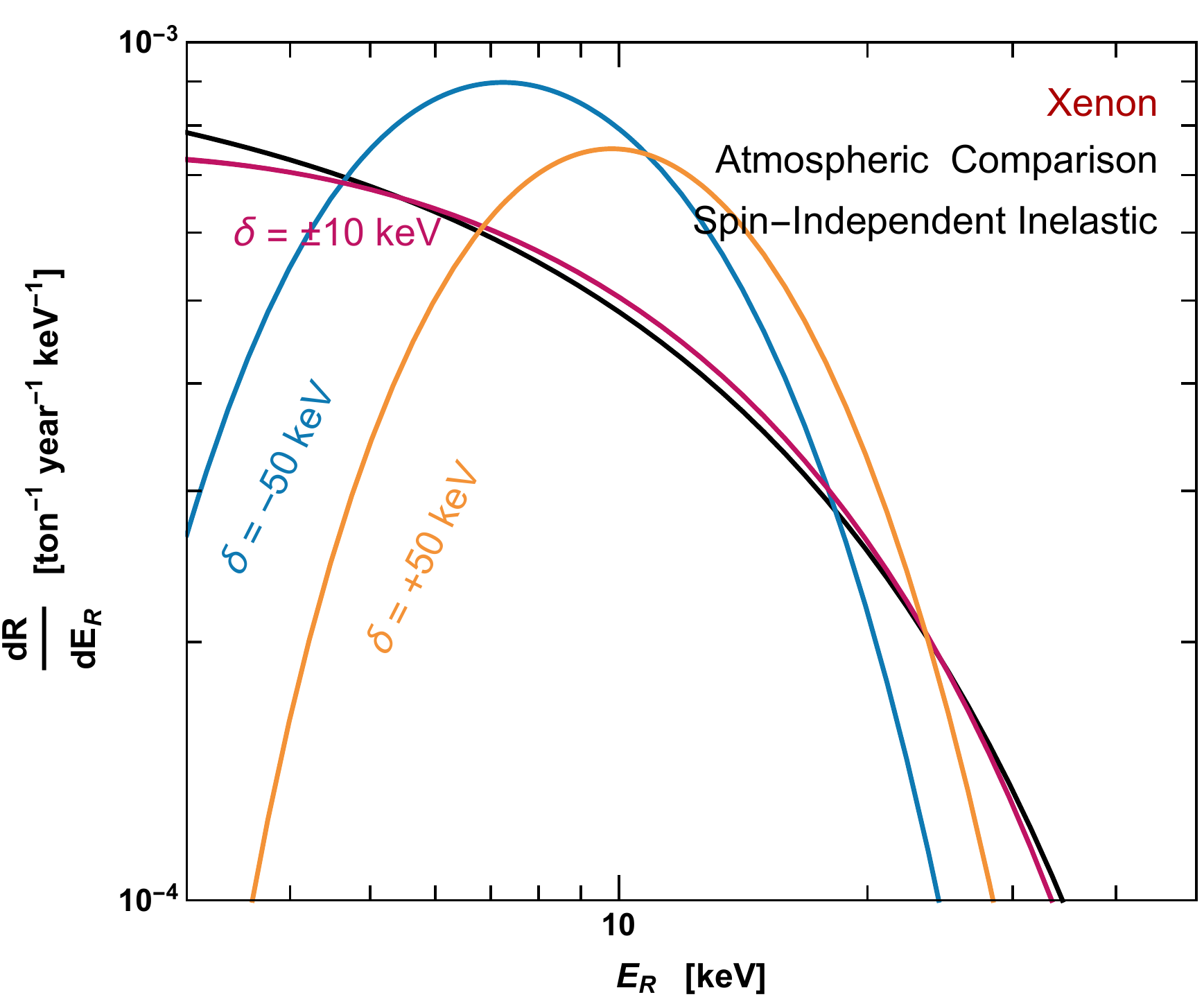}}
 \caption{\label{fig:xenon_degeneracy_inelastic}Comparison of recoil spectra induced in a xenon target from solar $^8$B (left), DSNB (middle) and atmospheric (right) neutrinos (black) along with the recoil spectrum arising from spin-independent heavy-mediator inelastic scattering. Results for DM mass splitting $\delta = - 50$ keV (blue), $\delta = + 10$ keV (magenta), and $\delta = + 50$ keV (orange). The best fit DM masses are given in \Tab{tab:bestfitDM_inelastic}.}
\end{figure}

The above analysis can be also applied to inelastic scattering. In \Fig{fig:xenon_degeneracy_inelastic} we display the recoil spectrum induced in xenon from solar $^8$B, DSNB, and atmospheric neutrinos, as well as the recoil spectrum arising from a SI interaction with a heavy mediator and inelastic exothermic ($\delta > 0$), endothermic ($\delta < 0$) scattering. As can be seen, as $|\delta|$ increases the DM spectrum becomes progressively more distinguishable from the neutrino spectrum. The DM masses for the spectra in \Fig{fig:xenon_degeneracy_inelastic} (i.e. those that give the best-fit to the respective neutrino background for fixed mass differences $\delta = + 10$ keV and $\delta \pm 50$ keV between the final and initial DM particles are given in \Tab{tab:bestfitDM_inelastic}). We note that the minimum recoil energy in xenon for a DM candidate with $\delta = +50$ keV in the DM mass range scanned is larger than the maximum recoil energy produced by the $^8$B neutrinos and is thus omitted in the left panel of~\Fig{fig:xenon_degeneracy_inelastic}.

\begin{table*}[tbp]
  \setlength{\extrarowheight}{2pt}
  \begin{center}
     \begin{threeparttable}
	\begin{tabular}{|c|c|c|c|}  \hline
	 \multirow{2}{*}{Model}	      & \multicolumn{3}{c|}{Best-fit DM mass (GeV)}      \\ \cline{2-4}
  & Solar ($^8$B) & DSNB & Atmospheric  \\ \hline
	\hline
	SI ($\delta = 0$)\tnote{1} 		& 5.8 	& 15.9	& 172.4 \\  \hline
    SI ($\delta = +10$ keV)			& 6.6	& 15.2	& 136.6 \\  \hline
    SI ($\delta = -50$ keV)		 	& 2.8	& 4.1	& 26.2\tnote{2}  \\  \hline
    SI ($\delta = +50$ keV)			&   -- 	& 21.9	& 39.0 \\  \hline
	\end{tabular}
    \begin{tablenotes}
 \item[1] These values correspond to the best-fit DM masses of elastic scattering.
  \item[2] With $\delta = -50$ keV the scattering off xenon is purely exothermic for $m_{\chi} \leq 17$ GeV. The actual best fit would depend on the fraction of lighter and heavier DM states.
 \end{tablenotes}
\caption{\label{tab:bestfitDM_inelastic}  DM masses for which the DM spectrum best fits the neutrino recoil spectrum assuming SI scattering with a heavy mediator and inelastic scattering with DM mass splitting $\delta = + 10$ keV and $\delta = \pm 50$ keV (see \Fig{fig:xenon_degeneracy_inelastic}).} 
  \end{threeparttable}
    \end{center}
\end{table*} 

\begin{figure}[h]
\mycenter{
\includegraphics[width=.42\textwidth]{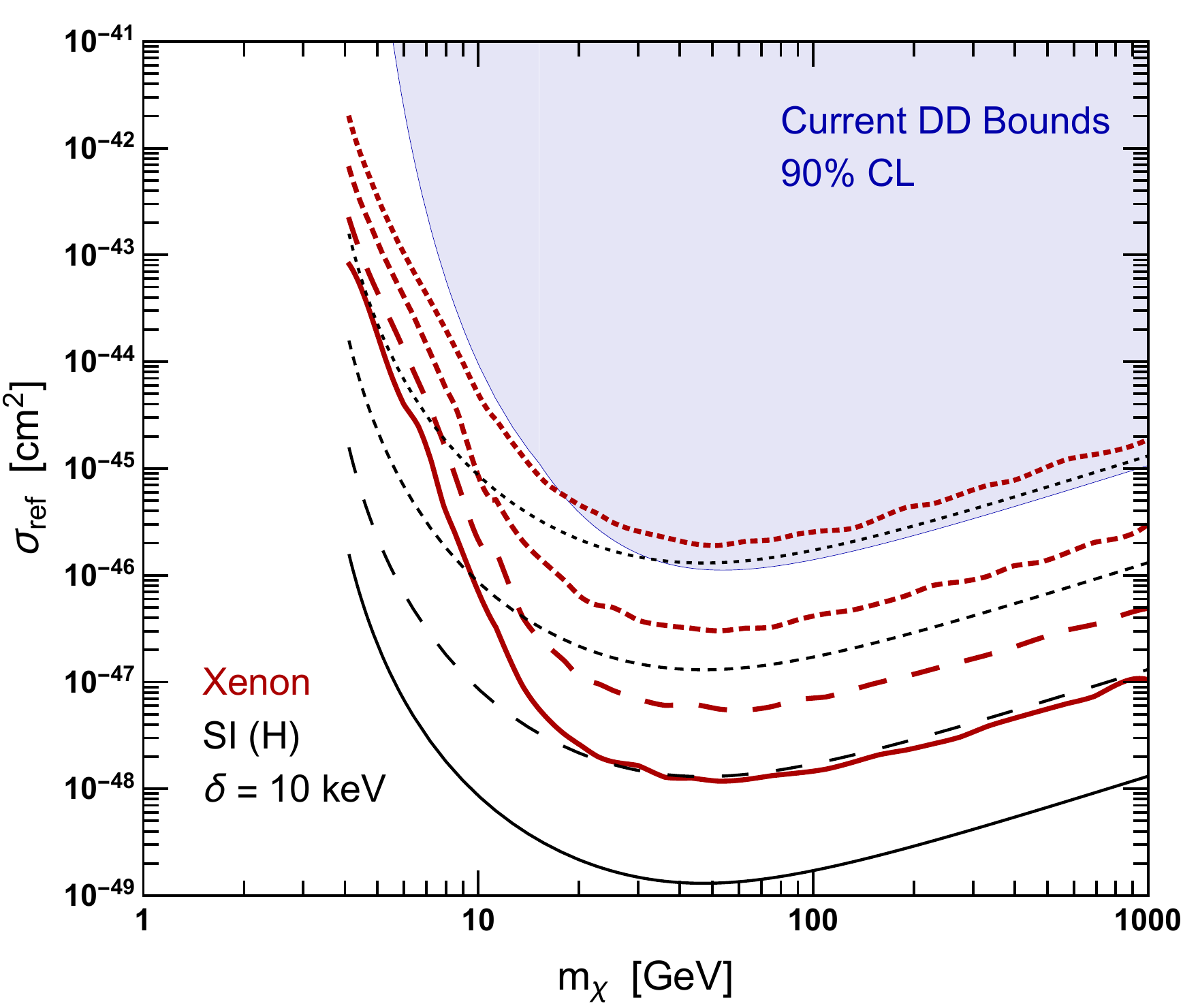}
\includegraphics[trim={15mm 0 0 0},clip,width=.3875\textwidth]{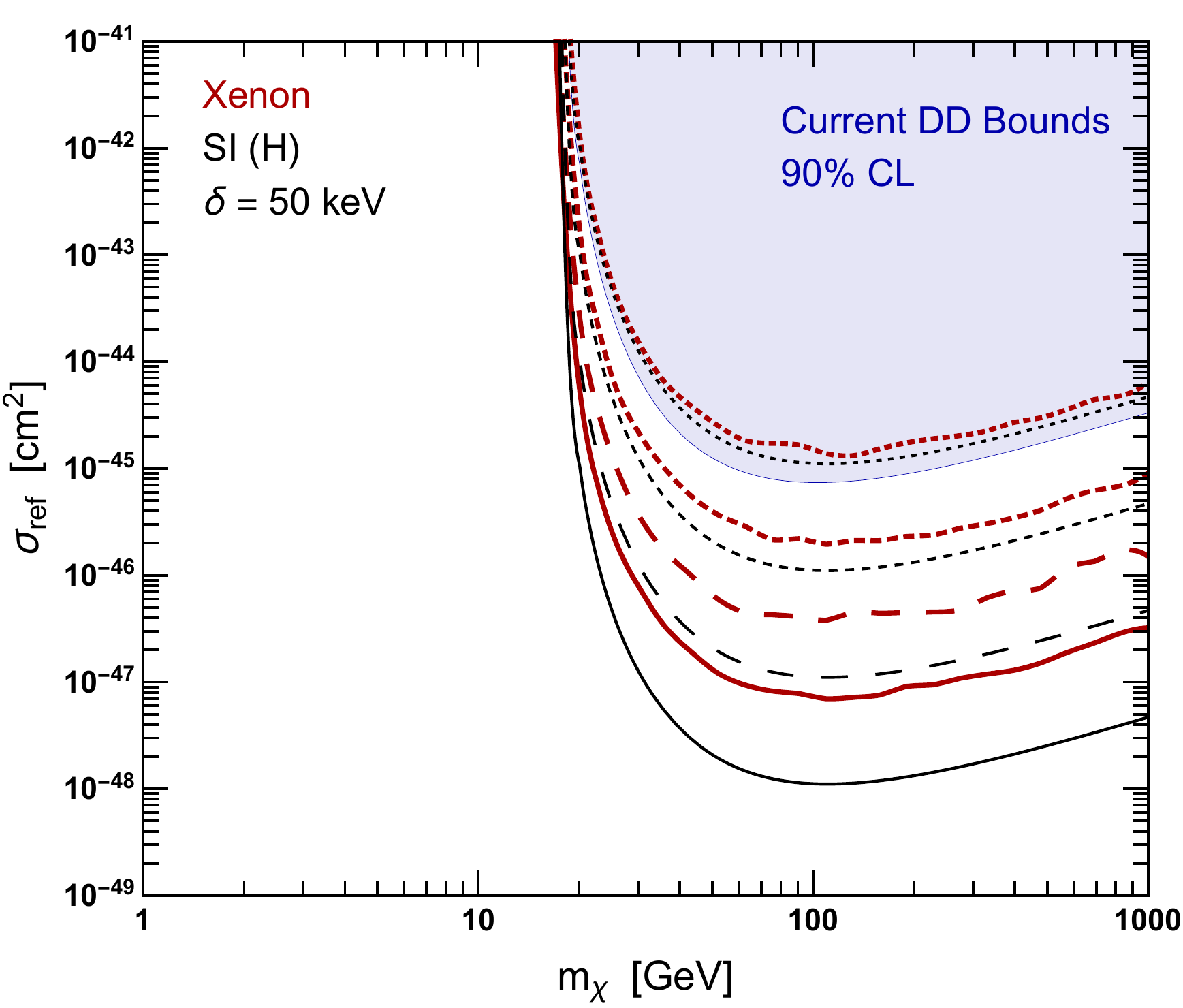}
\includegraphics[trim={15mm 0 0 0},clip,width=.38\textwidth]{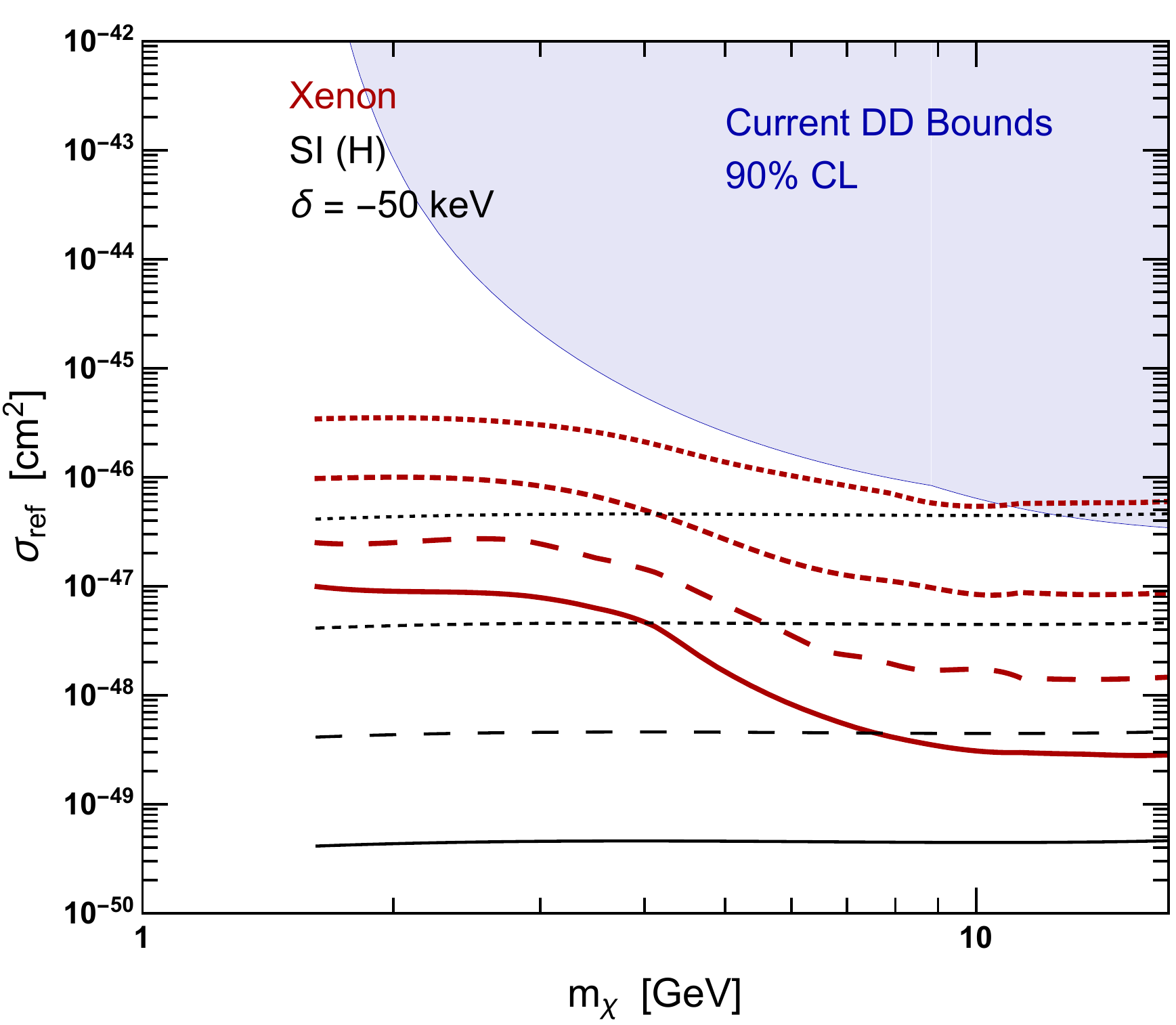}}
\caption{\label{fig:xenon_floors_inelastic} Same as \Fig{fig:xenon_floors_massive} but for a spin-independent interaction with exothermic and endothermic scattering, assuming $\delta = + 10$ keV and $\delta = \pm 50$ keV. The range of masses for $\delta = -50$ keV is cut at $m_{\chi} = 17$ GeV, since the scattering is not purely exothermic for larger masses.}
\end{figure} 

Exothermic scattering assumes that the DM at present consists of two different states of slightly different masses, the lightest being stable and the heaviest meta-stable. Then, the heavier state may down-scatter off the nuclei. However, this type of scattering dominates only if the scattering of each state to itself is suppressed or impossible (due to the couplings of the mediator) and the up-scattering (i.e. the endothermic scattering) off the lighter state is kinematically forbidden.
Thus, when considering exothermic scattering ($\delta<0$), one must consider the kinematic constraints on the endothermic process. If the endothermic scatter is allowed one must add both processes to obtain the complete DM scattering rate. Hence, studying a pure exothermic scattering process is only of interest if the minimum speed required for an endothermic scatter is larger than the galactic escape speed in the Earth's rest frame (i.e. when $\sqrt{2|\delta|/\mu} > |\vec{v}_{\rm esc}| + |\vec{v}_{\odot}|$). Otherwise, one must consider a joint analysis containing both contributions. For example, for $\delta = -10$ keV (-50 keV) both exothermic and endothermic scattering contributions to the rate should be added for $m_{\chi} > 1$ GeV (17 GeV) in a proportion that depends on the fraction of light and heavy states in the halo. Due to this uncertainty we do not include this case.The fits shown in \Fig{fig:xenon_degeneracy_inelastic} do not enforce the constraint on the DM mass to have only exothermic scattering, however only the best-fit mass for the atmospheric neutrino background for $\delta = -50$ keV in \Tab{tab:bestfitDM_inelastic} violates this condition. With $\delta = -50$ keV the scattering is purely exothermic only for $m_{\chi} \leq 17$ GeV, thus we cut the range of DM masses for the figure shown in the right panel of \Fig{fig:xenon_floors_inelastic}.

Interestingly, the kinematic features of inelastic scattering could provide a distinctive signature in future direct detection experiments if at least one target elements could uniquely isolate the purely exothermic scattering process while another target element maintained sensitivity to both endo- and exothermic scatters. A purely endothermic scattering helps to lift the neutrino-DM degeneracy even for SI interactions.

\section{Summary and Outlook\label{sec:summary}}

As DM direct detection experiments continue to improve their sensitivity they will inevitably encounter an irreducible background, the so-called ``neutrino-floor'', arising from coherent neutrino scattering. The neutrino floor has a detrimental effect on the capability of experiments to detect DM. In particular, assuming the conventional SI or SD interactions, the recoil spectrum of the neutrino background mimics the DM signal associated with particular WIMP masses, resulting in a degeneracy between the two. In this case, an increase in the experiment's exposure by multiple orders of magnitude may not yield any significant increase in sensitivity. The SI and SD interactions, however, do not adequately reflect the whole landscape of the well-motivated DM models, which include other interactions. In this work we have studied the relevance of the neutrino background for the discovery sensitivity of a variety of DM-nucleus interactions and several potential futuristic experimental settings, considering different target elemental compositions. This study could assist with mapping out the optimal DM detection strategy for the next generation of experiments.

For realistic exposures (i.e. $M T \sim \mathcal{O}(10^2)$ ton$\cdot$years), as expected of next generation direct detection experiments, we find that the degeneracy of DM and neutrino spectra primarily originates from solar $^8$B neutrinos. Due to the much lower flux levels of DSNB and atmospheric neutrinos, the spectral degeneracies with these neutrinos that happen for some DM mass values and interactions (see central and left panels of \Fig{fig:Dark_Matter_Degeneracy}, Appendix~\ref{app:recoilspec} \Fig{fig:recoilspec_full} as well as Appendix~\ref{app:recoilspec} \Tab{tab:bestfitDM_elastic_xenon} to  \Tab{tab:bestfitDM_elastic_sodium}) would become relevant only for exposure levels many orders of magnitude beyond those expected of experiments within the foreseeable future. This is consistent with findings of previous studies, which primarily focused on SD and SI interactions (e.g. see \cite{Ruppin:2014bra}, Fig. 3).

Several key features of the presented DM discovery potential results can be readily understood from the momentum transfer $|\vec{q}|$ dependence of the differential DM-nucleus scattering cross-sections $d\sigma_{\chi}/d E_R$ (see \Tab{tab:fermop} and the cross-sections of Sec. \ref{sssec:models}). As shown in Sec. 4.2, for cross-sections with no dependence on the momentum exchange, i.e. proportional to $|\vec{q}|^0$, the degeneracy with $^8$B neutrinos happens at masses close to 10 GeV. For heavy mediators, these correspond to the following interactions: SI, SD, mC and AV-V (only if the AV-V cross-section is dominated by the charge $Z_T^2 v^2$ term, e.g. in Xe \Fig{fig:xenon_floors_massive} because the charge of Xe is large and its magnetic moment is relatively small, see Eq.~(2.33); also Ar Fig.~10, since Ar has no spin). For light (or $M = 0$) mediator the cross-sections with no dependence on $|\vec{q}|$ correspond to the interactions: PS-PS, MD (if the MD cross-section in Eq.~(2.21) is dominated by the spin coupling, e.g. in Na Fig.~13) and Anapole (if dominated by the charge $Z_T^2 v^2$ term, e.g. in Xe Fig.~4, Ar Fig.~11). The neutrino-DM spectral degeneracy severely undermines the experimental sensitivity for $m_{\chi} \approx 10$ GeV. However, for masses above roughly $50$ GeV, such cross-sections have less than one order of magnitude suppression of the discovery limits for exposures as large as MT $\sim (10)^2$ ton$\cdot$years, making the neutrino background significantly less detrimental (see e.g. top left and middle panels of \Fig{fig:xenon_floors_massive} and bottom right panel of Fig.~4).

We note that for elastic scattering with $d \sigma_{\chi}/d E_R \sim |\vec{q}|^0$ there exists a general scaling relation for the DM signal-neutrino floor degeneracy with the WIMP mass. After having obtained from numerical calculations the DM mass for which the DM induced recoil spectrum is degenerate with the neutrinos for one target element, the DM mass for which spectral degeneracy occurs for another target element can be approximately found from kinematic relations as follows. The shape of the nuclear recoil spectrum depends on a common function of $v_{\rm min}$ for all targets. Two recoil spectra have the same shape only if they correspond to the same $v_{\rm min}$ range. Since degeneracy implies the recoil spectra are the same, one can equate the $E_{\rm R}$ of each target element. Hence, knowing that for a particular element (denoted as $T_1$) the DM spectrum is degenerate for a particular DM particle mass (denoted $m_1$) we can find the value of the respective mass (i.e. $m_2$) for the degeneracy in another element (denoted as $T_2$) by requiring that the respective nuclear recoil energies for the same $v_{\rm min}$ value coincide as
\begin{equation}
E_R^{T_1} (v_{\rm min}, m_1) = E_R^{T_2} (v_{\rm min}, m_2)~,
\end{equation}
which can be rewritten as
\begin{equation}
\dfrac{2 \mu_{T_1}^2 v_{\rm min}^2}{M_{T_1}} = \dfrac{2 \mu_{T_2}^2 v_{\rm min}^2}{M_{T_2}}~.
\end{equation}
Thus, 
\begin{equation}
\dfrac{\mu_{T_1}^2}{M_{T_1}} = \dfrac{\mu_{T_2}^2}{M_{T_2}}~.
\end{equation}
For masses close to 10 GeV (e.g. for Xe one observes the degeneracy with $^8$B neutrinos at $m_1 \approx 6$ GeV) we have that for all the considered nuclei $\mu_{T_1} \simeq m_1$ and $\mu_{T_2} \simeq m_2$, resulting in 
\begin{equation}
m_2 \simeq \sqrt{\dfrac{M_{T_1}}{M_{T_2}}} m_1~.
\end{equation}
This approximate scaling relation provides insight into the degeneracy behavior in different elements as identified in Fig. 7 of \cite{Ruppin:2014bra} for SI and SD interactions.

For cross-sections inversely dependent on powers of $|\vec{q}|$ (i.e. $1/|\vec{q}|^4$ or $1/|\vec{q}|^2$), there is no spectral degeneracy (see e.g. top left and middle panels of \Fig{fig:xenon_floors_massless}). However, there is an enhancement of the DM recoil spectrum at low recoil energies where the flux from solar neutrinos is large. Hence, the respective discovery limits are affected already for exposures drastically lower than for cross-sections independent of $|\vec{q}|$, except for masses near the degeneracy. This can be understood by considering that for a fixed  DM particle mass, cross sections independent of $|\vec{q}|$ extend to larger recoil energies where the neutrino flux is reduced, making this part of the spectra more differentiable from that of the background neutrinos. The effect of the concentration of the DM signal at low $E_R$ is exacerbated for low mass WIMPs, whose recoil spectrum already tends to be rather steep and concentrated at low energies. For interactions with $|\vec{q}|^{-4}$, i.e. light mediator ($M=0$) SI, SD, mC, and AV-V (assuming the AV-V cross-section is dominated by the $Z_T^2 v^2$ term, e.g. in Xe Fig.~4 and Ar Fig.~11), the discovery limits are suppressed by more than 3 orders of magnitude for exposures $MT \geq (10)^2$ ton$\cdot$years. The effect is less pronounced, leading to a suppression of around 1.5 orders of magnitude in the discovery limits for $m > 50$ GeV, for cross-sections proportional to $1/|\vec{q}|^2$. The interactions with these cross-sections are, always for light (or $M = 0$) mediators: MD (dominated by the $Z_T^2 v^2$ term, e.g. Xe Fig.~4 or Ar Fig.~11), ED, PS-S and S-PS (see e.g. central middle panel of Fig.~4)

The effect of $^8$B neutrinos is least pronounced for recoil momentum suppressed  DM cross-sections, i.e. cross sections proportional to $|\vec{q}|^b$ with $b > 0$. These correspond to interactions with heavy mediators. The interactions with $d \sigma_{\chi}/ d E_R \sim |\vec{q}|^2$ are MD (dominated by the $Z_T^2 v^2$ term, e.g. Xe Fig.~3 and Ar Fig.~10), ED, PS-S and S-PS. Those with $d \sigma_{\chi}/ d E_R \sim |\vec{q}|^4$ are MD (dominated by the magnetic moment coupling, e.g. Na Fig.~12), Ana (dominated by the $Z_T^2 v^2$ term) and PS-PS. The interaction with $d \sigma_{\chi}/ d E_R \sim |\vec{q}|^6$ is Ana (dominated by the magnetic term as in Na Fig.~12).

For momentum suppressed cross-sections there is no  neutrino-DM spectra degeneracy and, moreover, the DM recoil spectrum peaks at some non-zero value of the momentum transfer (note that this is contrary to interactions that are either independent of the momentum transfer or inversely dependent on powers of the momentum transfer, which both have a maximum differential rate for $|\vec{q}| \rightarrow 0$). There is an enhancement in sensitivity with respect to momentum independent cross-sections. This enhancement is modest  for light DM masses (except for the region of degeneracy where the enhancement is significant), for which $^8$B neutrinos are the main background. This can be seen e.g. for Xe in the bottom row, the left and middle panels of the central row and the top right panel of Fig.~3. This is so because larger powers of $|\vec{q}|$ produce increasingly peaked recoil spectra which thus become more differentiable from background.  The larger the power $b$ of $|\vec{q}|^b$ the smaller the effect of $^8$B neutrinos.

For large DM masses and heavier nuclei the enhancement in sensitivity with respect to momentum independent cross-sections is large, because a considerable portion of the recoils for momentum suppressed cross sections occur at energies above the recoil energies produced by atmospheric neutrinos. The suppression of the atmospheric background is a consequence of the nuclear form factor behavior (the resulting recoil spectra from atmospheric neutrinos in different targets can be seen e.g. in Fig.~5 of \cite{Gutlein:2010tq}). For large recoil energies there is no ambiguity as to whether an event arose from a DM candidate or a background neutrino, at least for all of the exposures considered here.

Recall that for momentum suppressed interactions, we perform our analysis over two energy regions, one extending to 50 keV and the other to 300 keV. The derived experimental sensitivity is then taken to be the stronger of the two. For large DM masses and heavier target nuclei the broader energy interval provides the stronger limits. For a Xe target for example, we can see in Fig.~3 that for DM masses above a few 100's GeV the discovery limit shows no suppression at all due to the neutrino background for Ana, MD and PS-PS interactions. All of these interactions have the strongest momentum suppression of $|\vec{q}|^4$ or $|\vec{q}|^6$ in their cross sections. For interactions with a weaker dependence  $|\vec{q}|^2$ in their cross sections (for ED, AV-V, PS-S and S-PS) the discovery limits in Xe at large masses and in the larger recoil energy interval also improve considerably with respect to  those derived from the low energy interval,  although only for S-PS interactions we find no suppression of discovery limits for large DM masses (and this is due to the nuclear form factor for this interaction decreasing very slowly with energy).

The same situation we just described for Xe (see Fig.~3) holds for Ge and I (see Figs.~8 and 14), as well as potentially for other heavy nuclei. The same is not found in Ar or Na (see Figs.~10 and 12).  This can be understood using kinematic arguments. The recoil energy for DM particles heavy with respect to the nuclear mass is simply $E_R \simeq 2 m_T v_{\rm min}^2$, thus heavier nuclei have larger recoil energies, while for lighter nuclei most recoils are below 50 keV -- thus considering larger recoil energies does not improve the discovery limits.

Thus far in this section our discussion has exclusively been limited to elastic DM-nuclei scattering. In this paper, we demonstrated explicitly that inelastic scattering, in which an initial DM particle of mass $m_{\chi}$ scatters into another of mass $(m_{\chi} + \delta)$, produces a distinctive recoil spectra that allows for easy discrimination with the neutrino background. This was explicitly shown for an SI interaction with  positive and negative values of $\delta$ (with $|\delta| \ll m_{\chi}$), but it should be emphasized that different interactions (e.g. the anapole or magnetic dipole) may produce more distinctive recoil spectra.

Target material complementarity can both mitigate the effect of the neutrino background and help discriminate the various DM models. This can be qualitatively understood by inspecting the specific dependence on target properties of the cross-sections (see Sec. 2.2.2). One can thus identify target elements with vastly different predicted event rates. For differential cross-sections consisting of a single term, the overall predicted rate 
 would often be sufficient to determine if a particular DM model can account for the results of two different experiments with different target materials. For the neutrino background, the ratio of the scattering rates in two target elements $T_1$ and $T_2$ should be proportional to the ratio of the square of their neutron numbers $(A_{T_1} - Z_{T_1})^2/(A_{T_2} - Z_{T_2})^2 = N_{T_1}^2 / N_{T_2}^2$. However, for DM interactions other than the SI with coupling predominantly to neutrons the ratio of the scattering event rates would be different. In the context of discriminating between DM models, interpreting a putative signal in a xenon target experiment in terms of any spin-dependent interaction inherently implies non-observation of the signal in an argon detector (since argon has no nuclear spin). Additionally, should the DM differential cross section contain multiple non-negligible terms with differing dependence on the nuclear properties of the target (e.g. the magnetic dipole interaction), not only can the overall scattering rate change but also the spectral shape can change with different targets, enhancing the discriminating power when observations are made in multiple experiments. Target complementarity would also be a particularly useful tool for inelastic DM, as endothermic interactions favor heavier targets and exothermic interactions favor lighter targets. Thus, this could be a helpful probe of models with multicomponent dark sector.

Until a convincing DM signal appears, it is of paramount importance that experiments maintain the broadest possible sensitivity to the wide array of possible DM interactions, including the inelastic scattering that is an often neglected but a viable possibility. Therefore it is important to maintain a multi-pronged approach to DM direct detection in the coming generation of experiments. As previously discussed, various target elements provide complimentary sensitivities that may help both differentiate a putative signal from background, and differentiate viable DM models from each other. Xenon has sensitivity to all the  possible DM interactions we have explored, and is easily scalable to large mass experiments. However, the spin and nuclear magnetic moment of the various xenon isotopes are relatively low. Argon is easily scalable to large mass experiments and has an extremely low background. Hence, it is ideal for interactions proportional to the number of nucleons. On the other hand, argon has no spin or a magnetic moment, and is thus insensitive to a variety of interactions. Therefore, argon experiments could be a powerful discriminating tool for a spin or magnetic-moment-dependent DM interaction. Germanium (and silicon, which to some degree is similar) has a broad sensitivity to many interactions, although like xenon it has a small spin and a small nuclear magnetic moment. In silicon the spin and nuclear magnetic moment is even smaller. Germanium (and silicon) experiments can reach a low threshold, implying a strong sensitivity to low mass DM candidates. Sodium and iodine provide spin-dependent interactions with protons (note that this is distinctive from both germanium and xenon which have spin-dependent couplings predominantly with  neutrons) and large nuclear magnetic moments\footnote{Fluorine also has a large nuclear magnetic moment and a large spin-dependent proton coupling, however this element has not been studied here since experiments using F only measure the energy-integrated rate, making neutrino background discrimination difficult.}, however these elements have larger backgrounds and require binning, implying a much lower sensitivity than would otherwise be suggested by this study. 

We reiterate that the fundamental goal of this work is to highlight the future sensitivity of direct detection experiments employing various target elements to a wide array of possible DM-nucleus interactions, particularly focusing on the extent to which background neutrinos will inhibit the ability to probe interesting parameter space. We identify a large number of interactions and parameter space for which the effect of the neutrino floor is significantly reduced, and a sizable amount of parameter space for which it is strengthened, relative to the conventionally studied SI and SD interactions. We also identify strategies that could be exploited to optimize the sensitivities of future DM direct detection experiments to a wide-range of possible DM candidates.


\acknowledgments
\addcontentsline{toc}{section}{Acknowledgments}

We would like to thank Robert Cousins and Michail Bachtis for helpful discussions regarding the statistical analysis. The work of GG and VT was supported, in part, by the
U.S. Department of Energy (DOE) under Grant No. DE-SC0009937.
SJW is supported by the European Union's Horizon 2020 research and innovation program under the Marie Sklodowska-Curie grant agreement No. 674896. The work of GG and VT was also supported, in part, by the National Science Foundation under Grant No. NSF PHY-1748958 due to their stay at the Kavli Institute for Theoretical Physics (KITP) at the University of California, Santa Barbara (UCSB), while completing this paper.

\appendix

\section{Recoil spectrum degeneracy}
\label{app:recoilspec}

\begin{figure}[H]
\mycenter{
\includegraphics[trim={0mm 15mm 0 0},clip,width=.41\textwidth]{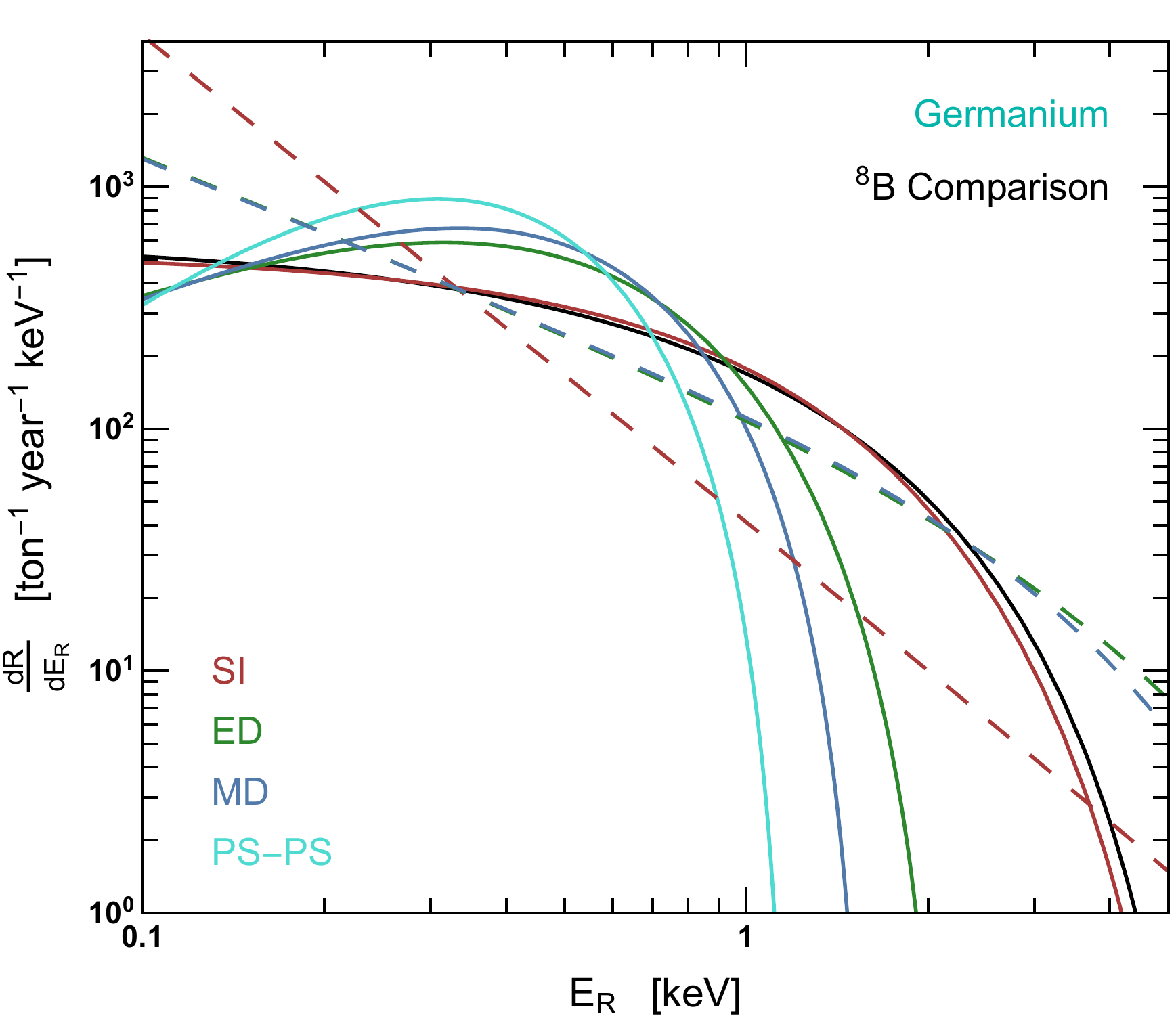}
\includegraphics[trim={12.5mm 15mm 0 0},clip,width=.38\textwidth]{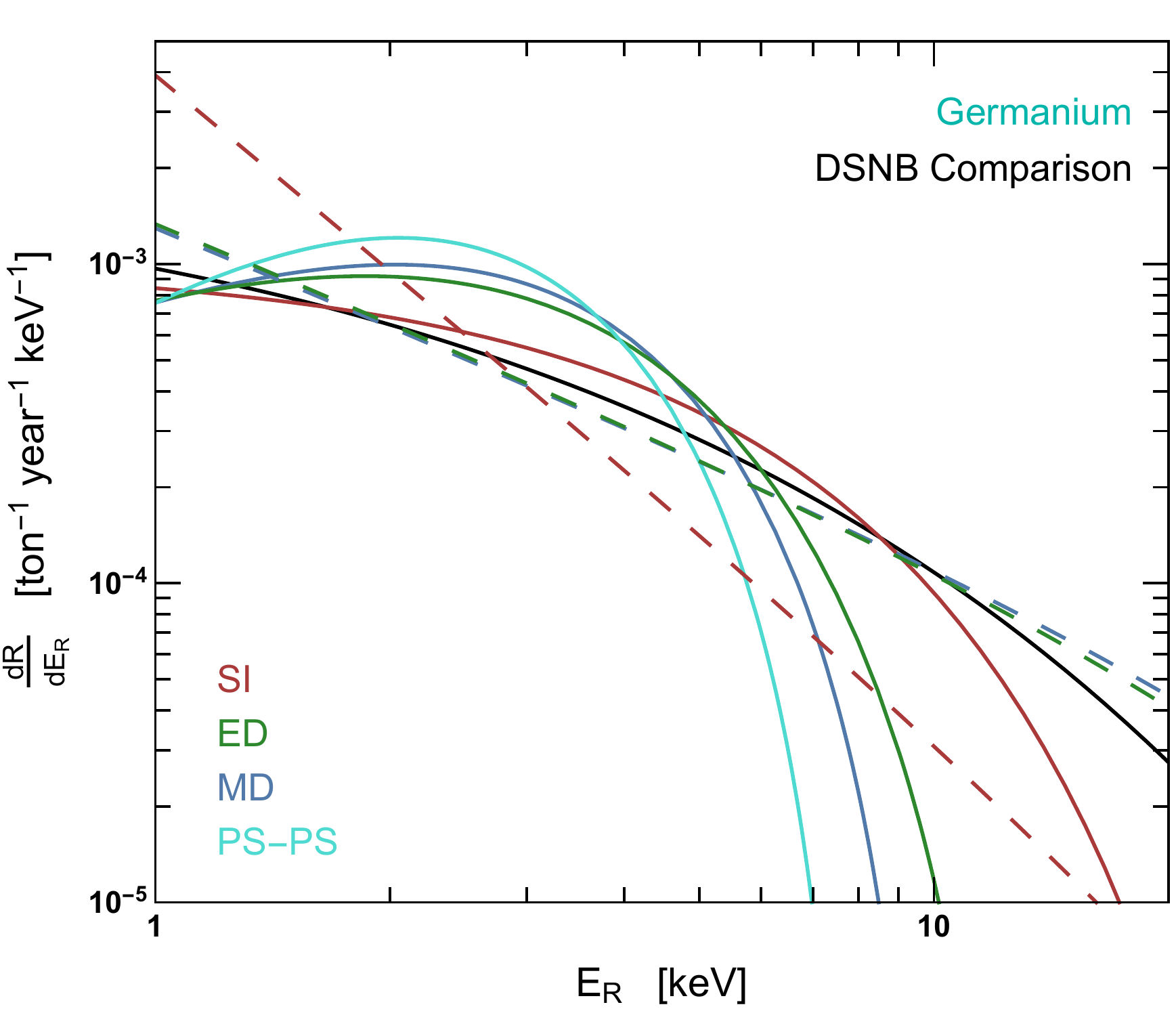}
\includegraphics[trim={12.5mm 15mm 0 0},clip,width=.38\textwidth]{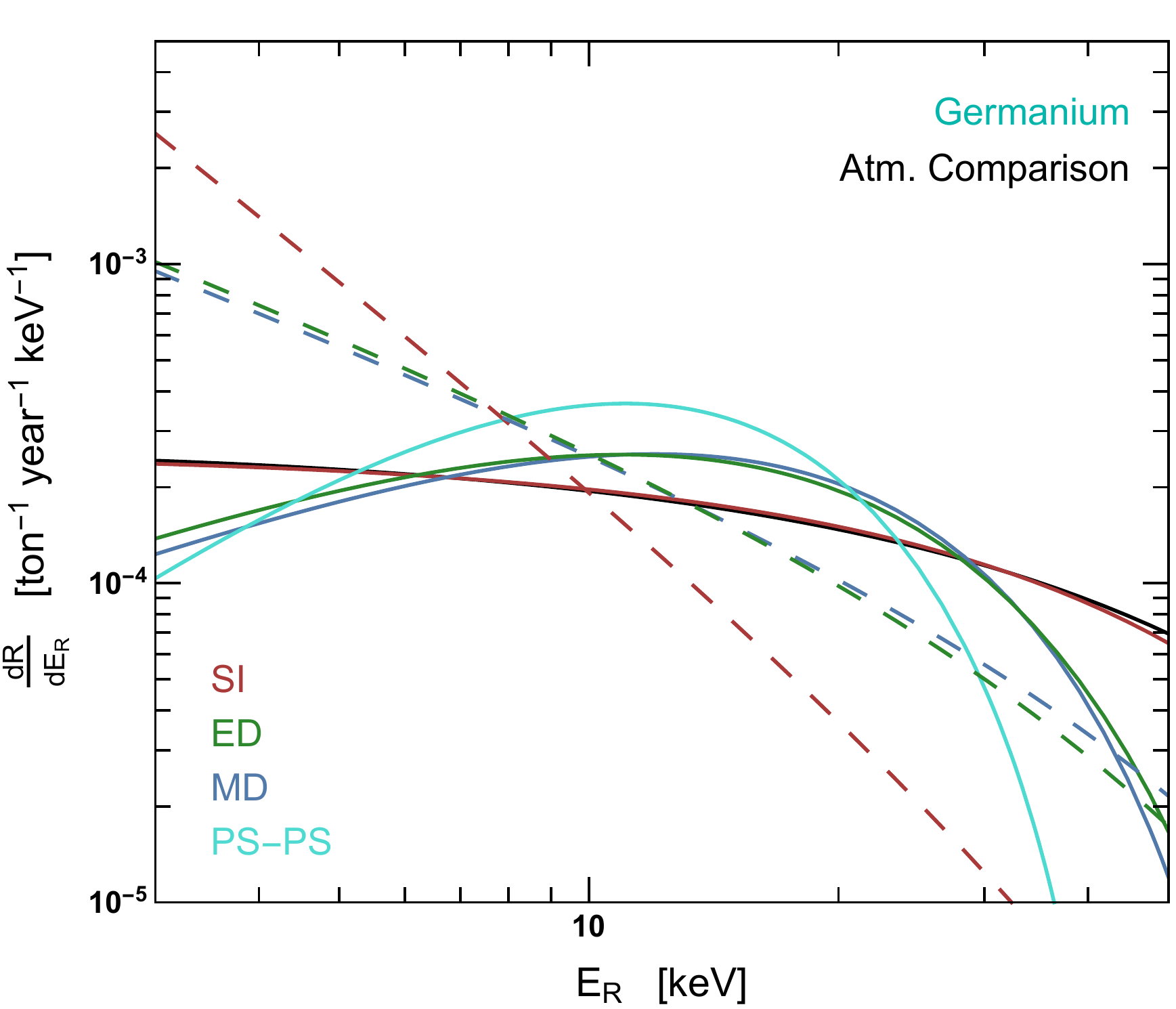}}
\mycenter{
\includegraphics[trim={0mm 15mm 0 0},clip,width=.41\textwidth]{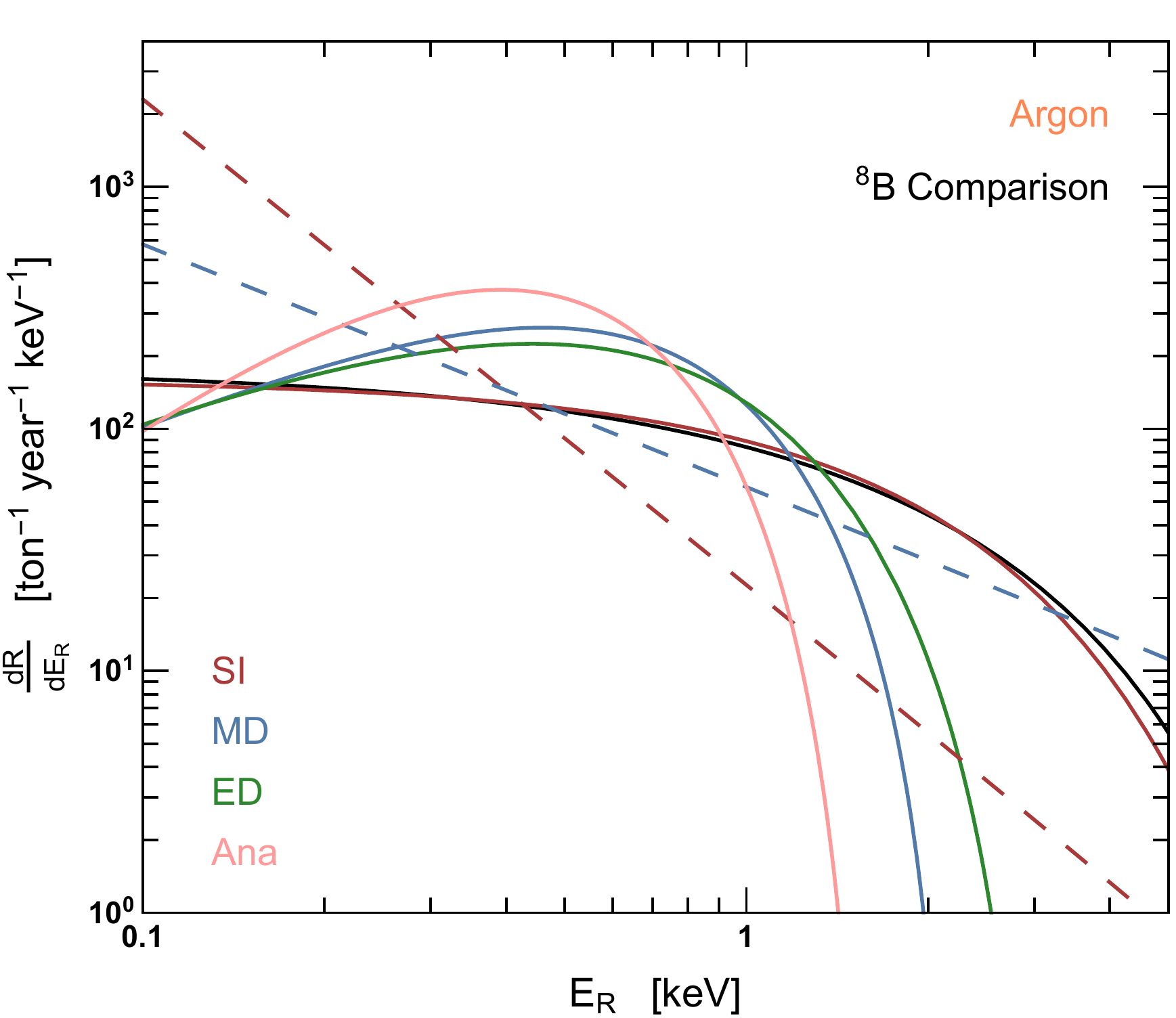}
\includegraphics[trim={12.5mm 15mm 0 0},clip,width=.38\textwidth]{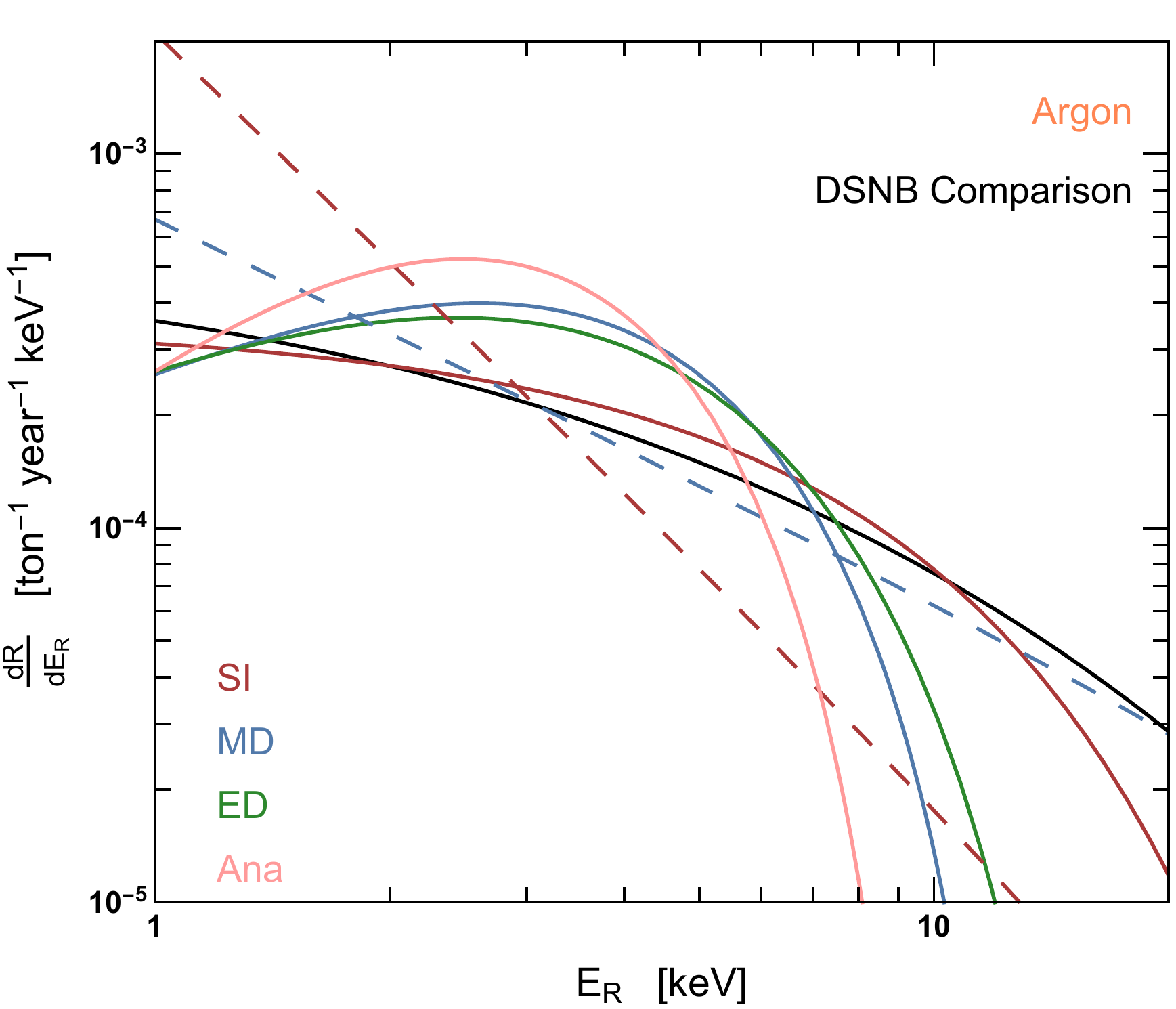}
\includegraphics[trim={12.5mm 15mm 0 0},clip,width=.38\textwidth]{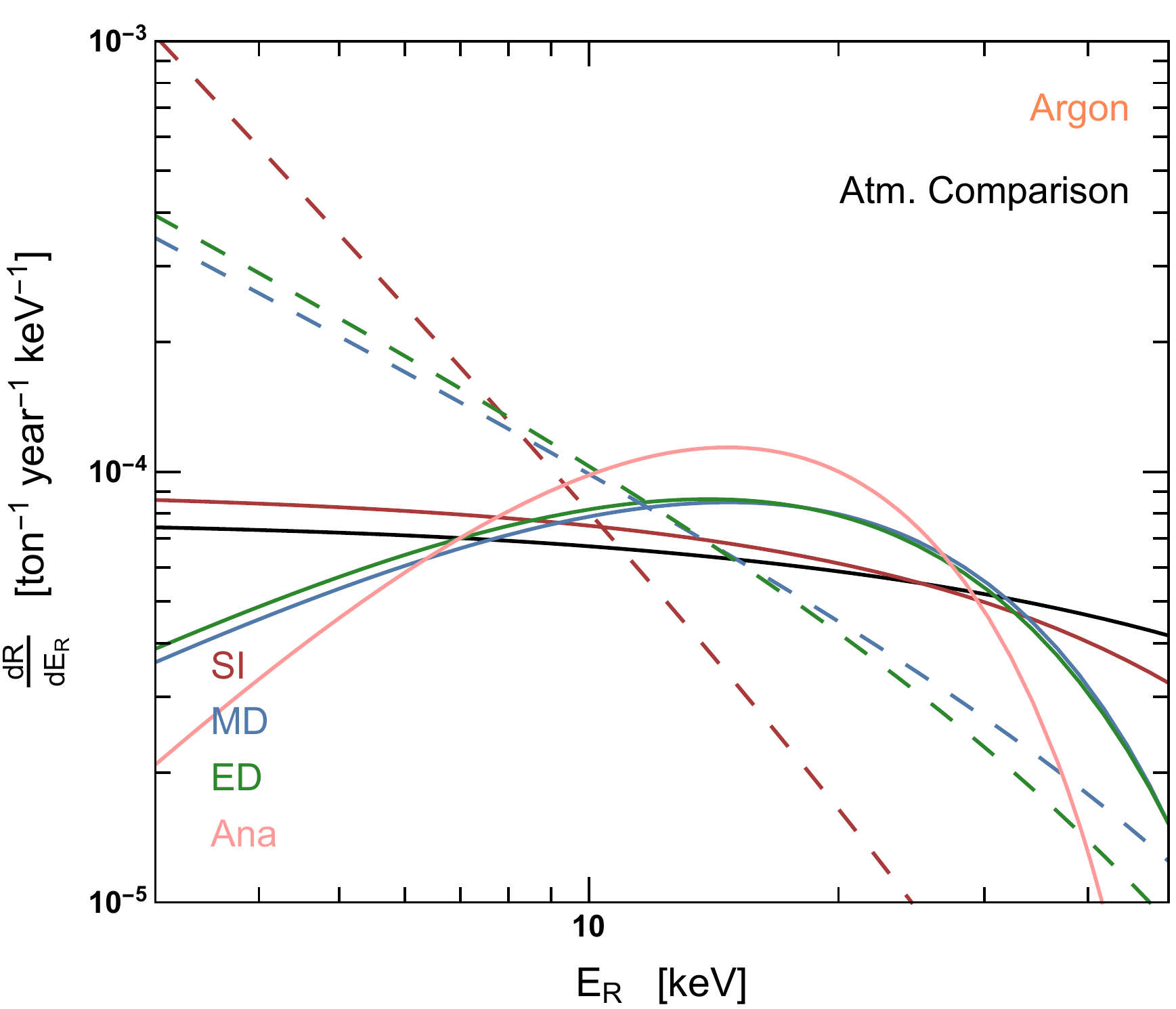}}
\mycenter{
\includegraphics[trim={0mm 15mm 0 0},clip,width=.41\textwidth]{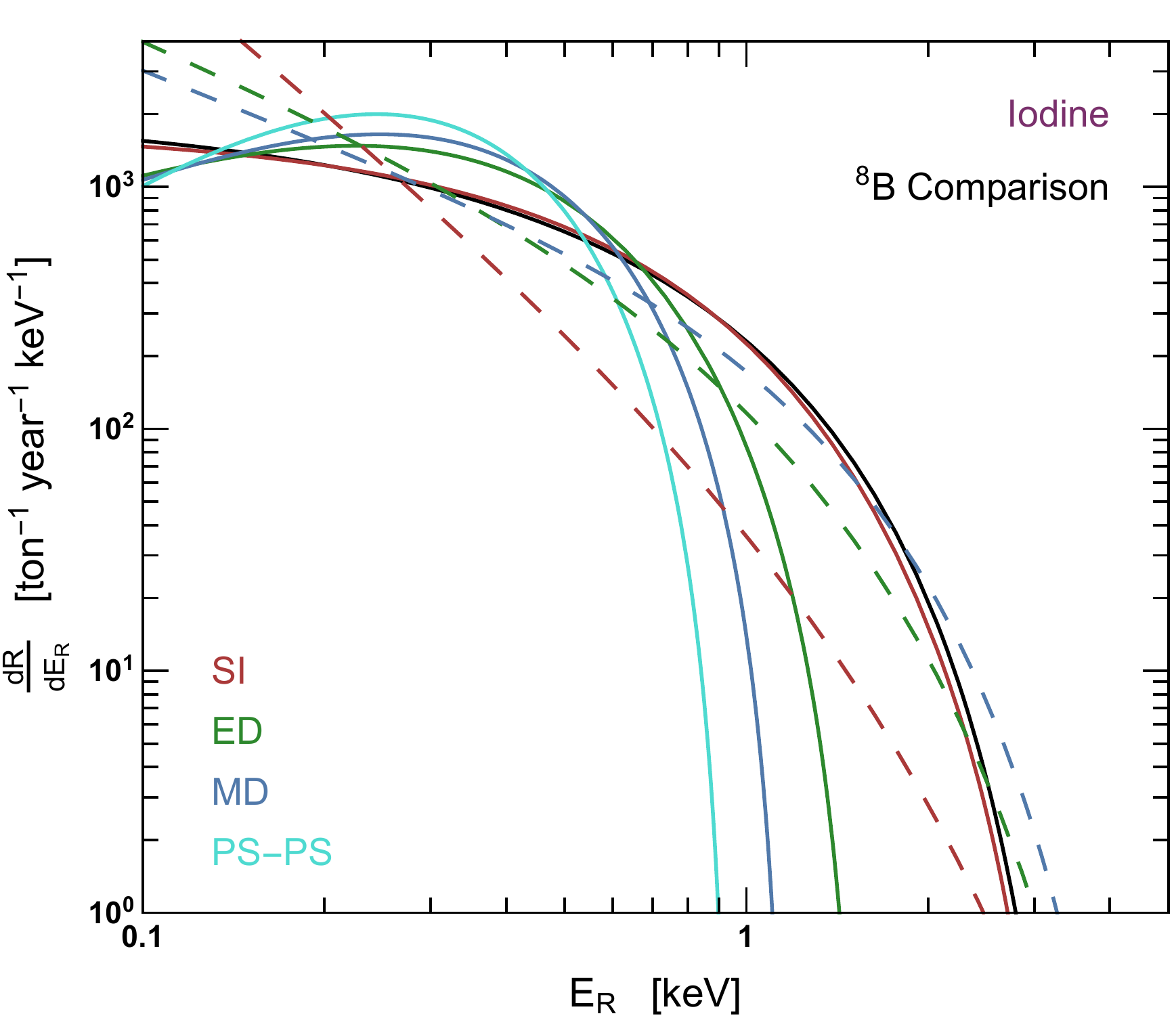}
\includegraphics[trim={12.5mm 15mm 0 0},clip,width=.38\textwidth]{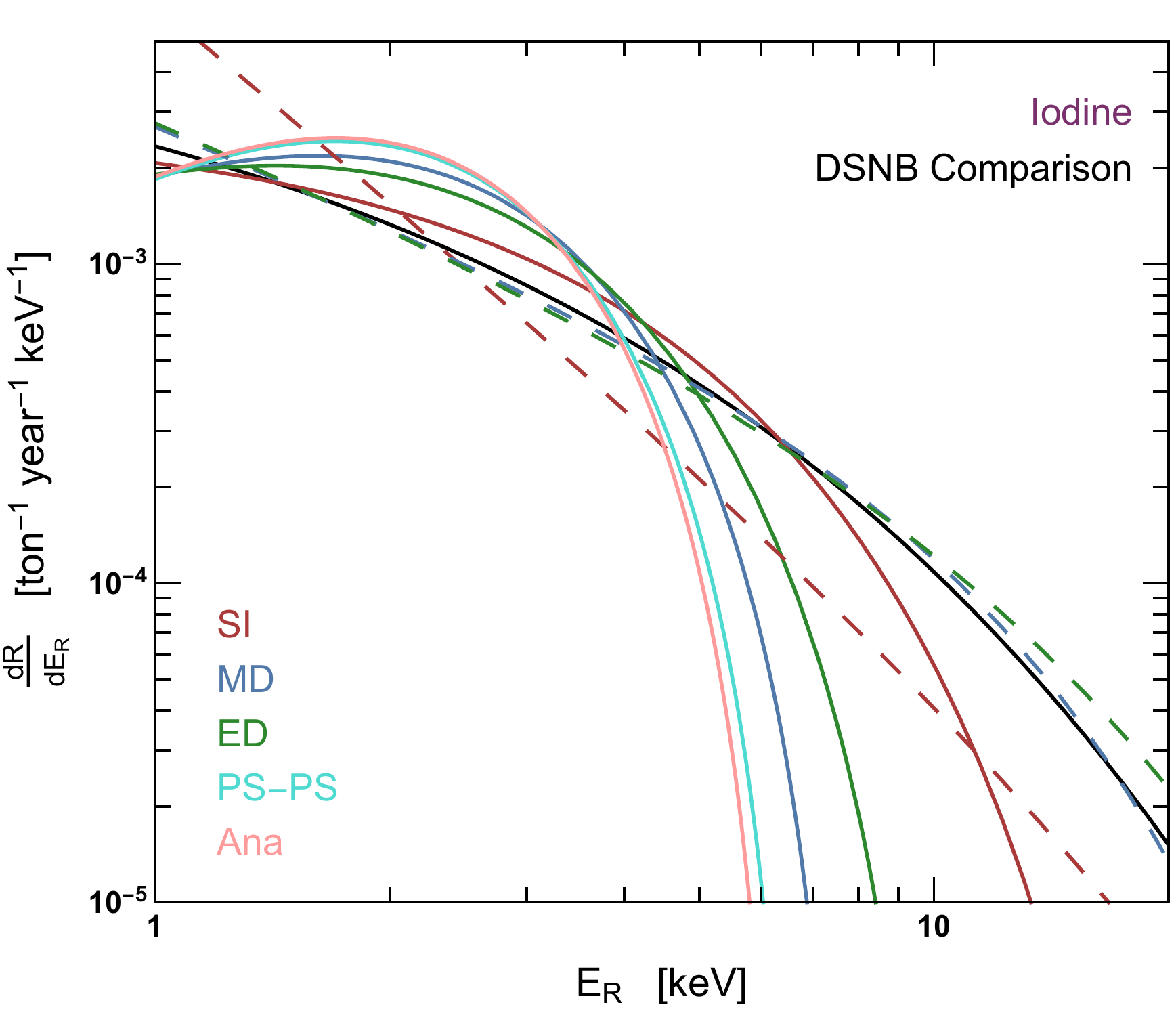}
\includegraphics[trim={12.5mm 15mm 0 0},clip,width=.38\textwidth]{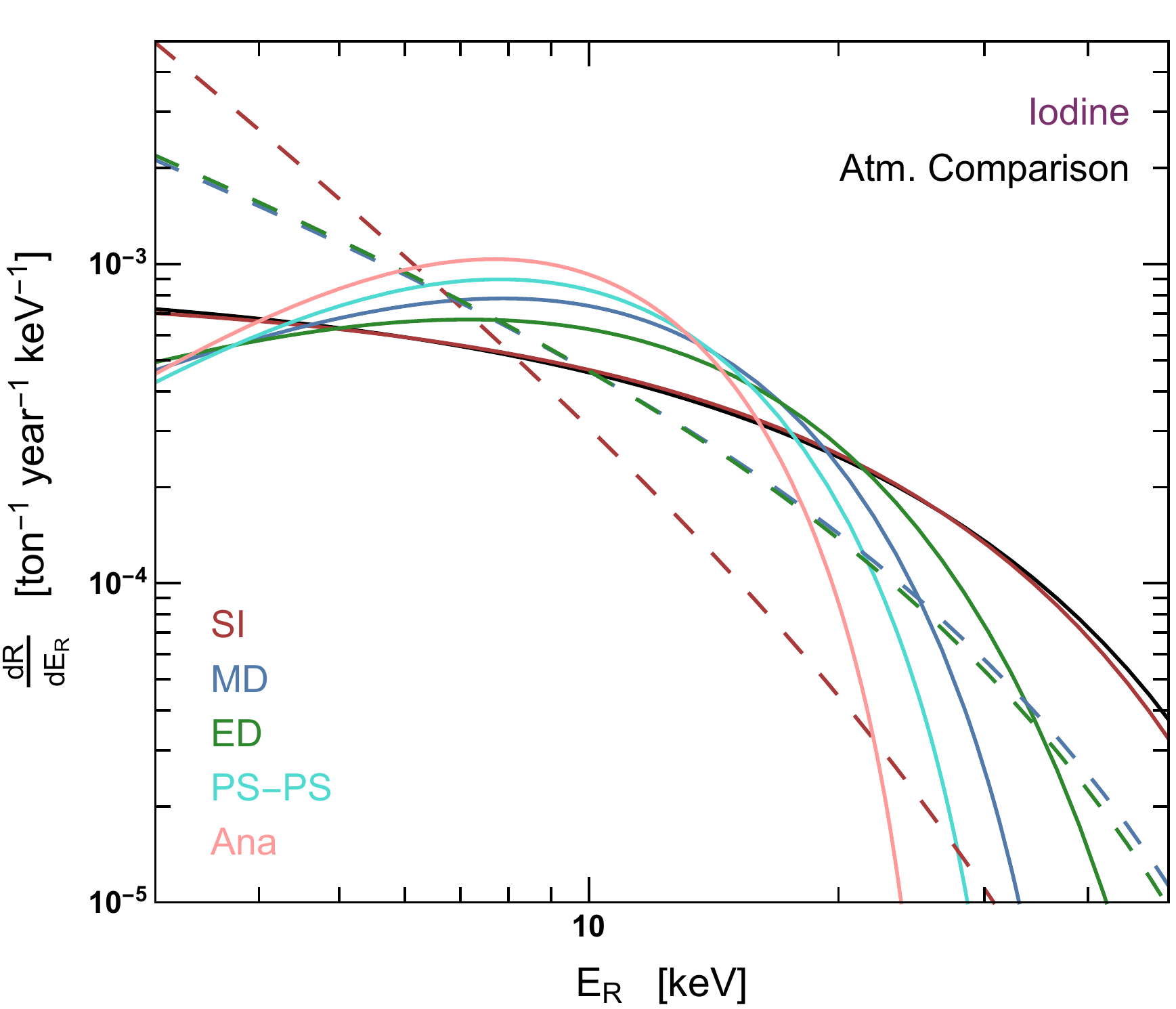}}
\mycenter{
\includegraphics[trim={0mm 0mm 0 0},clip,width=.41\textwidth]{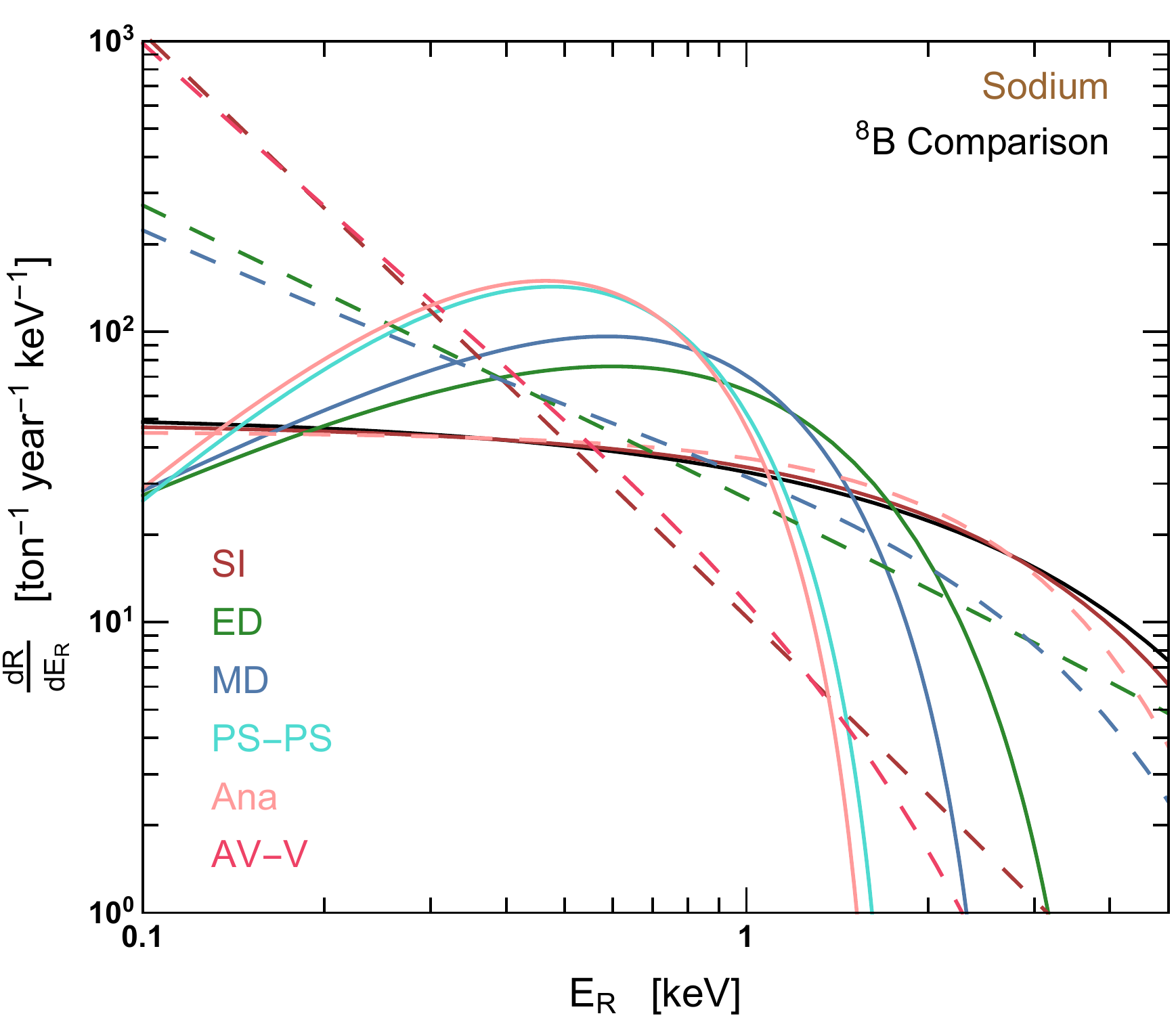}
\includegraphics[trim={12.5mm 0mm 0 0},clip,width=.38\textwidth]{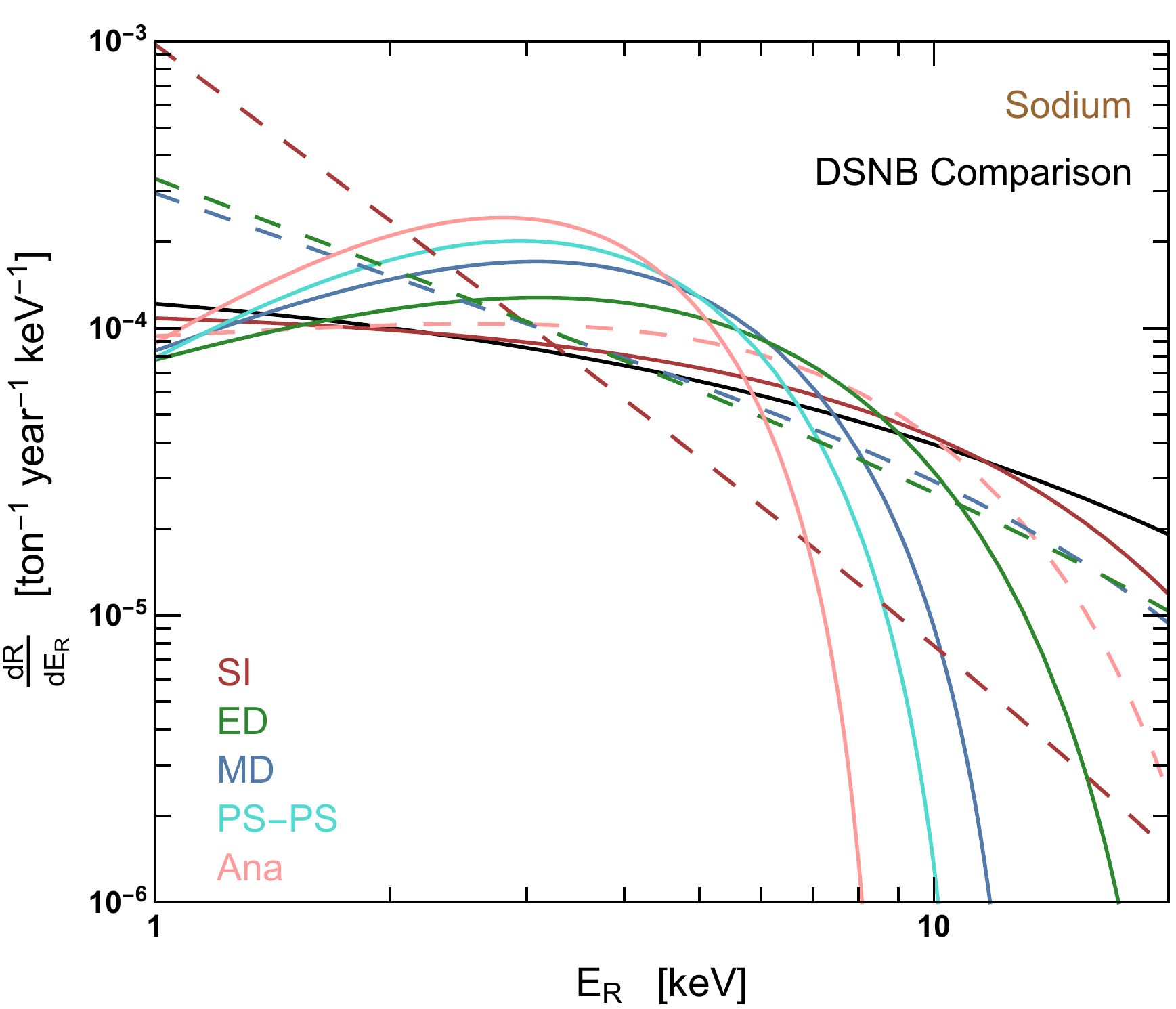}
\includegraphics[trim={12.5mm 0mm 0 0},clip,width=.38\textwidth]{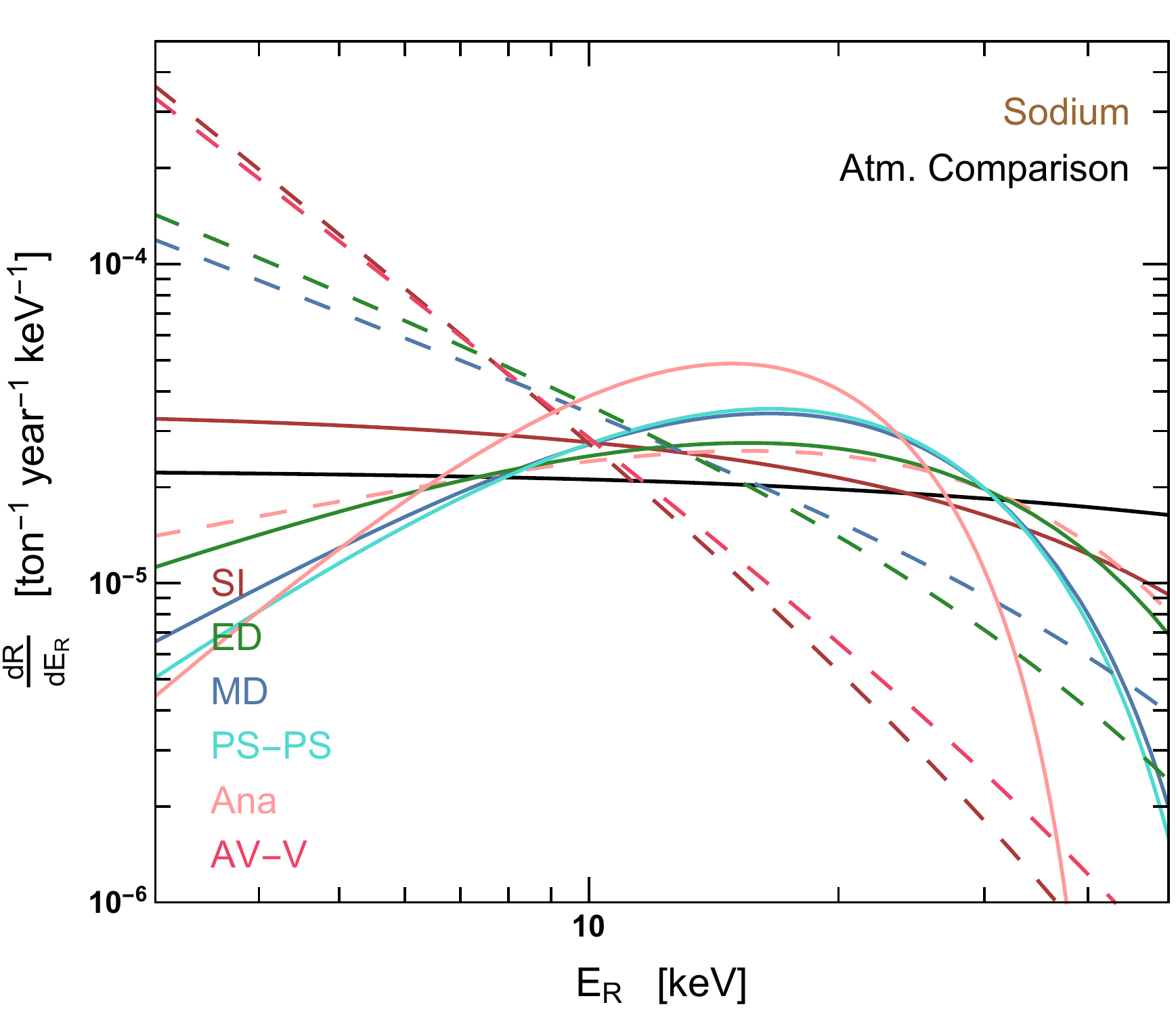}}

\caption{\label{fig:recoilspec_full} Same as Fig. 2 but for Ge, Ar, I and Na (see Tables 11 to 14 for best-fit masses).}
\end{figure}

\begin{table*}[h]
  \setlength{\extrarowheight}{2pt}
  \begin{center}
     \begin{threeparttable}
	\begin{tabular}{|c|c|c|c||c|c|c|}  \hline
	 \multirow{3}{*}{Model}	      & \multicolumn{6}{c|}{Best-fit DM mass (GeV)}      \\ \cline{2-7}
          	      & \multicolumn{3}{c||}{Heavy mediator}  & \multicolumn{3}{c|}{Light mediator}    \\ \cline{2-7}
  & Solar ($^8$B) & DSNB & Atmospheric & Solar ($^8$B) & DSNB & Atmospheric \\ \hline
	\hline
	SI   			& 5.8 	& 15.9	& 172.4 & 7.5 & $\gtrsim \mathcal{O}(10^3)$ & $\gtrsim \mathcal{O}(10^3)$  \\  \hline
    SD		 		& 5.8	& 16.1 & 147.0 & 7.5 & $\gtrsim \mathcal{O}(10^3)$ & $\gtrsim \mathcal{O}(10^3)$   \\  \hline
    Anapole			& 3.2 	& 8.8 & 21.8 & 6.1  & 16.5 & 134.9    \\  \hline
    ED		 		& 3.9  & 10.7 & 31.7 & 7.0  & 46.2 & $\gtrsim \mathcal{O}(10^3)$  \\  \hline
    MD			  	& 3.4  & 9.3 & 26.9 & 6.8  & 29.0	& $\gtrsim \mathcal{O}(10^3)$  \\  \hline
    PS-S   			& 3.9  & 10.7 & 32.3 & 7.0  & 48.3	& $\gtrsim \mathcal{O}(10^3)$   \\  \hline
    S-PS		 	& 3.9  & 10.6 & 29.2 & 7.0  & 38.7 & $\gtrsim \mathcal{O}(10^3)$  \\  \hline
    PS-PS			& 3.0  & 8.4 & 20.3 & 5.8 	& 15.5 & 60.4  \\  \hline
    AV-V			& 6.1  & 16.5 & 134.9& 8.0 & $\gtrsim \mathcal{O}(10^3)$ &  $\gtrsim \mathcal{O}(10^3)$ \\  \hline
	\end{tabular}
\caption{\label{tab:bestfitDM_elastic_xenon} Best-fit DM masses obtained by fitting the three mentioned neutrino backgrounds (see Sec.~4.2) with the DM-nucleus interactions given in the first column and either a heavy or a light mediator (see Sec.~2.2.2), assuming a Xe target. Some of the corresponding DM spectra are shown in Fig.~2.} 
  \end{threeparttable}
    \end{center}
\end{table*}

\begin{table*}[h]
  \setlength{\extrarowheight}{2pt}
  \begin{center}
     \begin{threeparttable}
	\begin{tabular}{|c|c|c|c||c|c|c|}  \hline
	 \multirow{3}{*}{Model}	      & \multicolumn{6}{c|}{Best-fit DM mass (GeV)}      \\ \cline{2-7}
          	      & \multicolumn{3}{c||}{Heavy mediator}  & \multicolumn{3}{c|}{Light mediator}    \\ \cline{2-7}
  & Solar ($^8$B) & DSNB & Atmospheric & Solar ($^8$B) & DSNB & Atmospheric \\ \hline
	\hline
	SI   			& 6.0	& 15.7 & $\gtrsim \mathcal{O}(10^3)$ & $\gtrsim \mathcal{O}(10^3)$ & $\gtrsim \mathcal{O}(10^3)$  	&  $\gtrsim \mathcal{O}(10^3)$ \\  \hline
    SD		 		& 6.0	& 16.1 & $\gtrsim \mathcal{O}(10^3)$ & $\gtrsim \mathcal{O}(10^3)$ & $\gtrsim \mathcal{O}(10^3)$   	&  $\gtrsim \mathcal{O}(10^3)$  \\  \hline
    Anapole			& 2.7	& 7.4 & 20.9 	&  6.3  & 16.4 & 377.4     \\  \hline
    ED		 		& 3.5	& 9.4 & 32.2  & 14.4 & $\gtrsim \mathcal{O}(10^3)$ 	&  $\gtrsim \mathcal{O}(10^3)$ \\  \hline
    MD			  	& 3.0	& 8.1 & 28.3 & 10.1 & 283.9  	&  $\gtrsim \mathcal{O}(10^3)$ \\  \hline
    PS-S   			& 3.5	& 9.4 & 32.5 & 14.5 & $\gtrsim \mathcal{O}(10^3)$  	&  $\gtrsim \mathcal{O}(10^3)$  \\  \hline
    S-PS		 	& 3.5	& 9.4 & 32.5 & 14.5 & $\gtrsim \mathcal{O}(10^3)$  	&  $\gtrsim \mathcal{O}(10^3)$ \\  \hline
    PS-PS			& 2.6	& 7.1 & 20.3 & 6.0 & 15.8  	&  866.2 \\  \hline
    AV-V			& 6.3  & 16.5 & 377.4 & $\gtrsim \mathcal{O}(10^3)$ & $\gtrsim \mathcal{O}(10^3)$  	&  $\gtrsim \mathcal{O}(10^3)$ \\  \hline
	\end{tabular}
\caption{\label{tab:bestfitDM_elastic_germanium} Same as Table~10 but assuming a Ge target and with some of the corresponding DM spectra shown in the top row of Fig.~7.} 
  \end{threeparttable}
    \end{center}
\end{table*}

\begin{table*}[h]
  \setlength{\extrarowheight}{2pt}
  \begin{center}
     \begin{threeparttable}
	\begin{tabular}{|c|c|c|c||c|c|c|}  \hline
	 \multirow{3}{*}{Model}	      & \multicolumn{6}{c|}{Best-fit DM mass (GeV)}      \\ \cline{2-7}
          	      & \multicolumn{3}{c||}{Heavy mediator}  & \multicolumn{3}{c|}{Light mediator}    \\ \cline{2-7}
  & Solar ($^8$B) & DSNB & Atmospheric & Solar ($^8$B) & DSNB & Atmospheric \\ \hline
	\hline
	SI   			&  6.3	&  16.3 & $\gtrsim \mathcal{O}(10^3)$  &  $\gtrsim \mathcal{O}(10^3)$ 	& $\gtrsim \mathcal{O}(10^3)$ & $\gtrsim \mathcal{O}(10^3)$ \\  \hline
    SD		 		&  NA & NA & NA & NA  & NA &  NA  \\  \hline
    Anapole			&  2.3 & 6.4 & 21.3  &  6.8 & 17.7 & $\gtrsim \mathcal{O}(10^3)$  \\  \hline
    ED		 		&  3.2 & 8.6 & 32.2 & $\gtrsim \mathcal{O}(10^3)$ & $\gtrsim \mathcal{O}(10^3)$ &  $\gtrsim \mathcal{O}(10^3)$ \\  \hline
    MD			  	&  2.7 & 7.4 & 30.0 & $\gtrsim \mathcal{O}(10^3)$ & $\gtrsim \mathcal{O}(10^3)$ &  $\gtrsim \mathcal{O}(10^3)$ \\  \hline
    PS-S   			&  3.2 & 8.6 & 32.3 & $\gtrsim \mathcal{O}(10^3)$ &  $\gtrsim \mathcal{O}(10^3)$ &  $\gtrsim \mathcal{O}(10^3)$  \\  \hline
    S-PS		 	&  NA & NA & NA & NA  & NA &  NA \\  \hline
    PS-PS			&  NA & NA & NA & NA & NA &  NA \\  \hline
    AV-V			&  6.8 & 17.7 & $\gtrsim \mathcal{O}(10^3)$ & $\gtrsim \mathcal{O}(10^3)$  	& $\gtrsim \mathcal{O}(10^3)$ &  $\gtrsim \mathcal{O}(10^3)$ \\  \hline
  
	\end{tabular}
\caption{\label{tab:bestfitDM_elastic_argon} Same as Table~10 but assuming an Ar target and with some of the corresponding DM spectra shown in the second row of Fig.~7.} 
  \end{threeparttable}
    \end{center}
\end{table*}

\begin{table*}[h]
  \setlength{\extrarowheight}{2pt}
  \begin{center}
     \begin{threeparttable}
	\begin{tabular}{|c|c|c|c||c|c|c|}  \hline
	 \multirow{3}{*}{Model}	      & \multicolumn{6}{c|}{Best-fit DM mass (GeV)}      \\ \cline{2-7}
          	      & \multicolumn{3}{c||}{Heavy mediator}  & \multicolumn{3}{c|}{Light mediator}    \\ \cline{2-7}
  & Solar ($^8$B) & DSNB & Atmospheric & Solar ($^8$B) & DSNB & Atmospheric \\ \hline
	\hline
	SI   			& 5.9 & 15.9 & 176.2 & 8.3 & $\gtrsim \mathcal{O}(10^3)$ & $\gtrsim \mathcal{O}(10^3)$ \\  \hline
    SD		 		& 5.9 & 16.2 & 107.0 & 8.3 & $\gtrsim \mathcal{O}(10^3)$ &  $\gtrsim \mathcal{O}(10^3)$ \\  \hline
    Anapole			& 3.09 & 8.2 & 18.2 &  5.9 & 14.0 & 40.4  \\  \hline
    ED		 		& 3.9 & 10.6 & 31.9 & 7.1 & 48.3 & $\gtrsim \mathcal{O}(10^3)$  \\  \hline
    MD			  	& 3.35 & 9.1 & 23.6 & 6.8 & 26.9 & $\gtrsim \mathcal{O}(10^3)$  \\  \hline
    PS-S   			& 5.9 & 10.6 & 32.4 & 7.12 & 50.3 &  $\gtrsim \mathcal{O}(10^3)$  \\  \hline
    S-PS		 	& 3.9 & 10.6 & 30.8 & 7.12 & 45.5 &  $\gtrsim \mathcal{O}(10^3)$ \\  \hline
    PS-PS			& 3.0 & 8.4 & 20.9 & 5.8 & 15.7 & 67.2 \\  \hline
    AV-V			& 5.9 & 14.0 & 40.4 & 8.0 & $\gtrsim \mathcal{O}(10^3)$ & $\gtrsim \mathcal{O}(10^3)$ \\  \hline
 
	\end{tabular}
\caption{\label{tab:bestfitDM_elastic_iodine} Same as Table~10 but assuming an I target and with some of the corresponding DM spectra shown in the third row of Fig.~7.} 
  \end{threeparttable}
    \end{center}
\end{table*}

\begin{table*}[h]
  \setlength{\extrarowheight}{2pt}
  \begin{center}
     \begin{threeparttable}
	\begin{tabular}{|c|c|c|c||c|c|c|}  \hline
	 \multirow{3}{*}{Model}	      & \multicolumn{6}{c|}{Best-fit DM mass (GeV)}      \\ \cline{2-7}
          	      & \multicolumn{3}{c||}{Heavy mediator}  & \multicolumn{3}{c|}{Light mediator}    \\ \cline{2-7}
  & Solar ($^8$B) & DSNB & Atmospheric & Solar ($^8$B) & DSNB & Atmospheric \\ \hline
	\hline
	SI   			& 6.9 & 19.0 & $\gtrsim \mathcal{O}(10^3)$  & $\gtrsim \mathcal{O}(10^3)$ &   $\gtrsim \mathcal{O}(10^3)$	& $\gtrsim \mathcal{O}(10^3)$ \\  \hline
    SD		 		& 6.9 & 19.1	&  $\gtrsim \mathcal{O}(10^3)$	& $\gtrsim \mathcal{O}(10^3)$ &   $\gtrsim \mathcal{O}(10^3)$	&  $\gtrsim \mathcal{O}(10^3)$  \\  \hline
    Anapole			& 1.8 & 4.6 &  13.9	&  5.04 & 10.2 & 42.6   \\  \hline
    ED		 		& 3.0 & 8.2 &  36.5 & 7.66 &  $\gtrsim \mathcal{O}(10^3)$ 	& $\gtrsim \mathcal{O}(10^3)$  \\  \hline
    MD			  	& 2.4 & 6.0 & 20.7 & 6.1 &  25.5	&  $\gtrsim \mathcal{O}(10^3)$ \\  \hline
    PS-S   			& 3.0 & 8.3 & 36.6 & $\gtrsim \mathcal{O}(10^3)$ &  $\gtrsim \mathcal{O}(10^3)$ 	&  $\gtrsim \mathcal{O}(10^3)$  \\  \hline
    S-PS		 	& 3.0 & 8.3 & 36.9 & $\gtrsim \mathcal{O}(10^3)$ & $\gtrsim \mathcal{O}(10^3)$ 	&  $\gtrsim \mathcal{O}(10^3)$ \\  \hline
    PS-PS			& 1.9 & 5.4 & 20.0 & 6.9 &  19.0 	&  $\gtrsim \mathcal{O}(10^3)$ \\  \hline
    AV-V			& 5.04 & 10.2 & 42.6 & 4.31 &  18.0 & $\gtrsim \mathcal{O}(10^3)$  \\  \hline
  
	\end{tabular}
\caption{\label{tab:bestfitDM_elastic_sodium} Same as Table~10 but assuming a Na target and with some of the corresponding DM spectra shown in the bottom row of Fig.~7.} 
  \end{threeparttable}
    \end{center}
\end{table*}

~\clearpage
\section{Discovery potential}
\label{app:discpot}
\begin{figure}[h]
\mycenter{
\includegraphics[trim={0mm 16mm 0 0},clip,width=.38\textwidth]{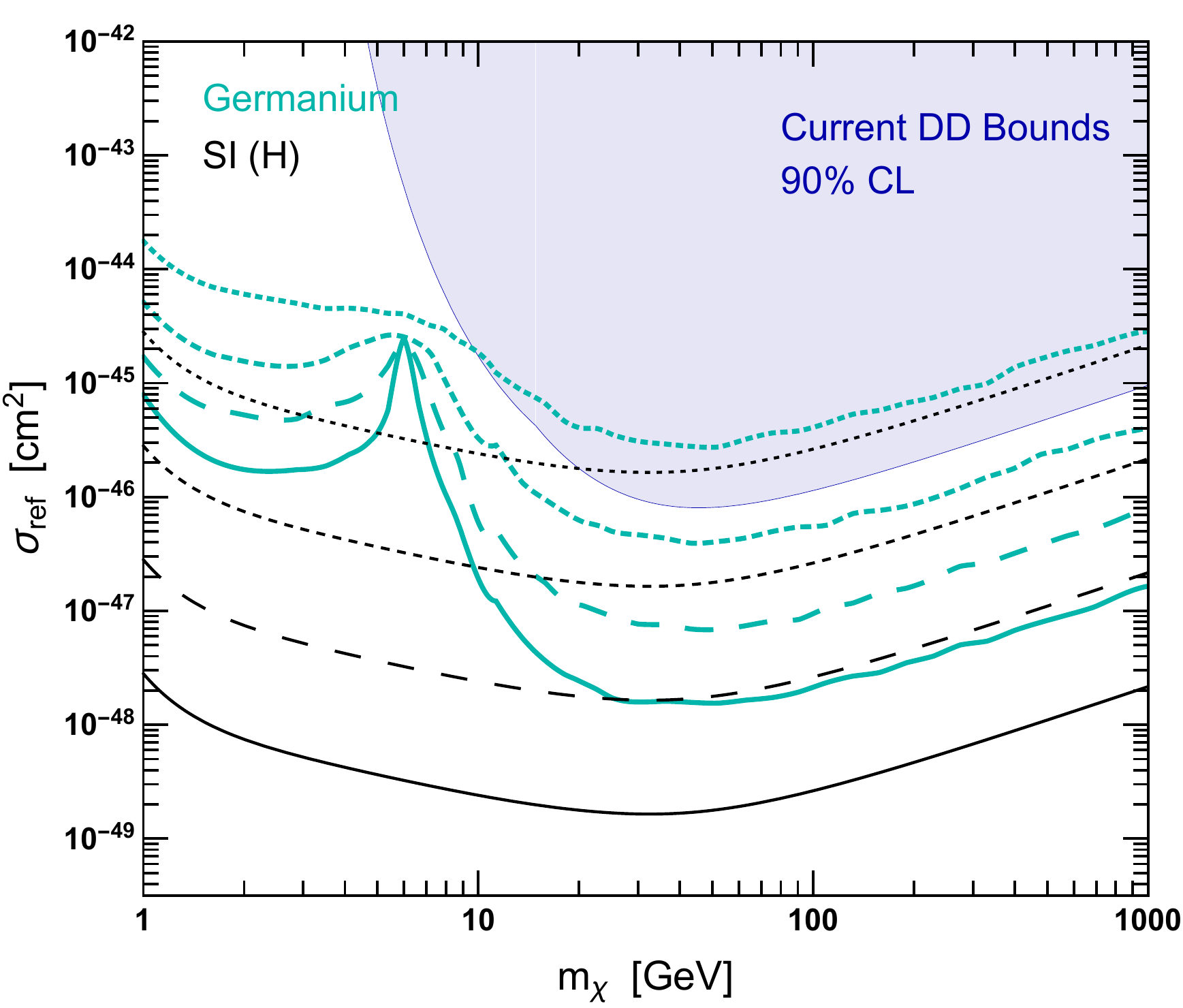}
\hspace{3mm}
\adjustbox{padding =0.5ex 0.425ex 0.5ex 0.425ex,margin*=1.5mm,  width=5 cm, max height = 4 cm,frame,raise=4mm }
{\maxsizebox*{0.7cm}{3ex}{\shortstack{\textcolor{teal}{Germanium} \\SD (H)\\  \\  \\  \\  \\ \textit{similar to} \\  SI(H), PS-PS(L), \\ Ana (L), AV-V(H)  }}}
 \hspace{5.75mm}
 \adjustbox{padding =0.85ex 0.4ex 0.85ex 0.4ex,margin*=2.0mm,  width=5 cm, max height = 4 cm,frame,raise=3mm }
{\maxsizebox*{0.7cm}{3ex}{\shortstack{\textcolor{teal}{Germanium} \\Ana (H)\\  \\  \\  \\  \\ \textit{similar to} \\  PS-PS(H)  \\ ~ }}}
}
\mycenter{
\includegraphics[trim={0mm 0mm 0 0},clip,width=.38\textwidth]{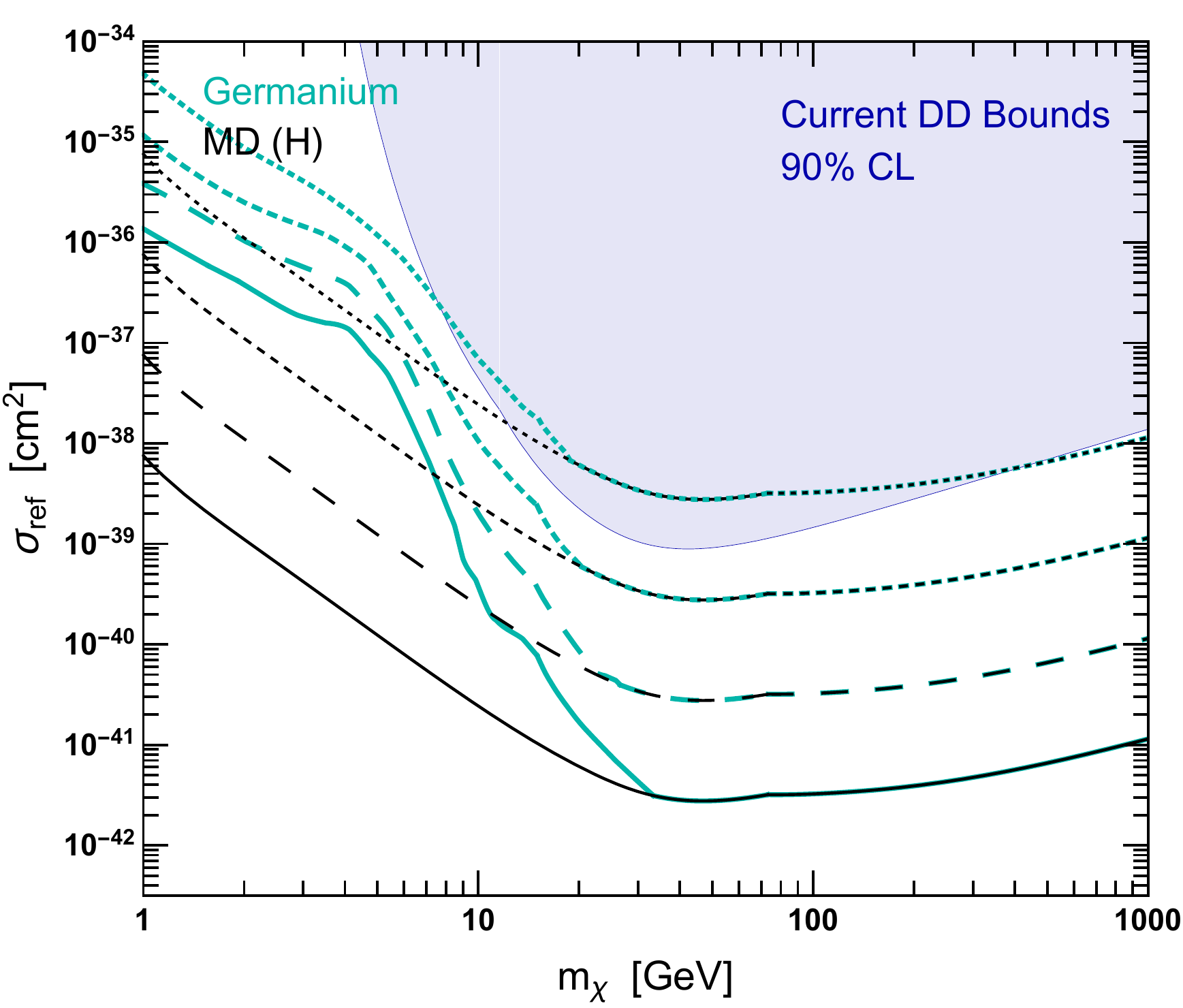}
\includegraphics[trim={ 9mm 0mm 0 0},clip,width=.3625\textwidth]{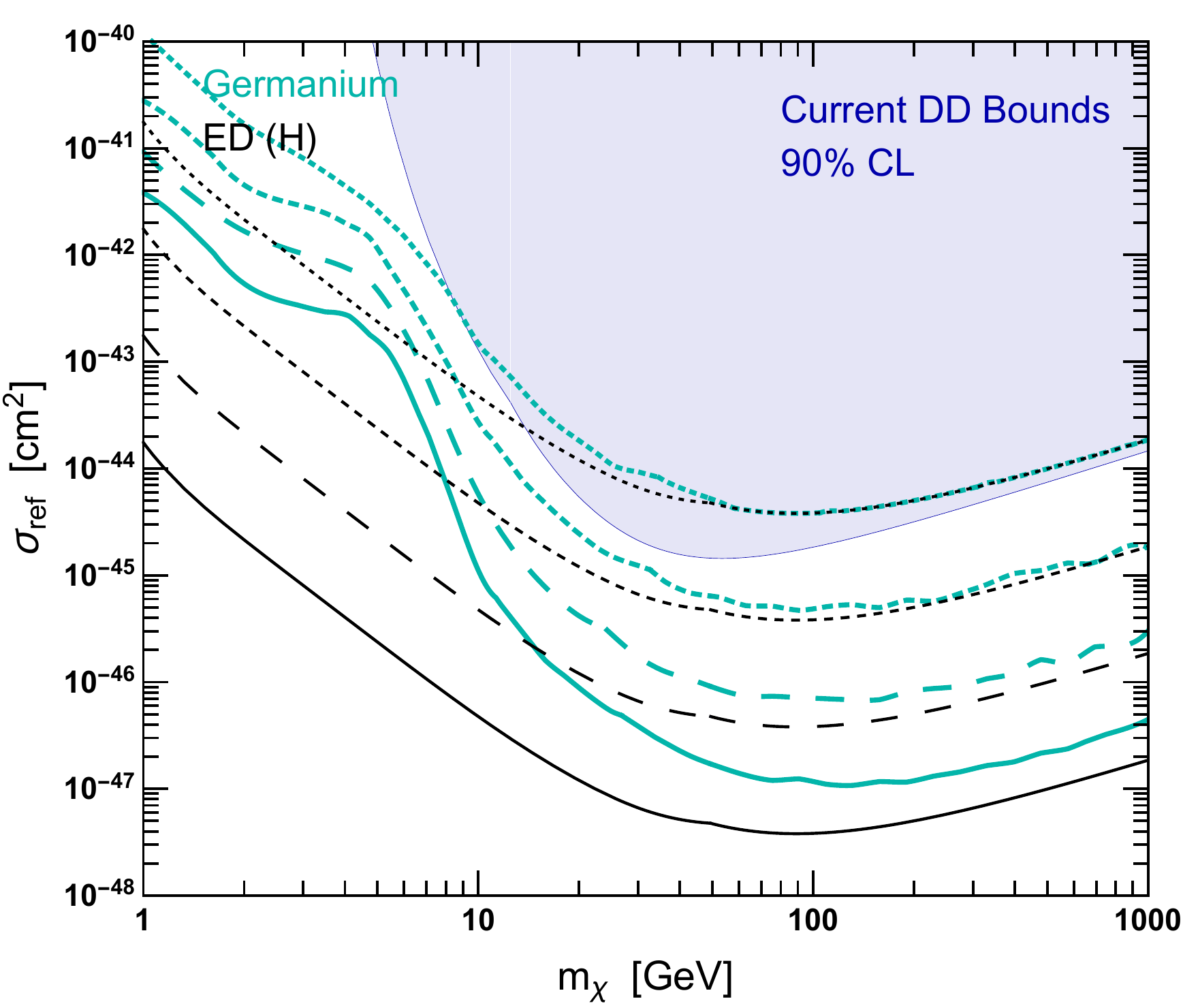}
 
  \hspace{4mm}
\adjustbox{padding =0.4ex 0.3ex 0.4ex 0.3ex,margin*=1.5mm,  width=5 cm, max height = 4 cm,frame,raise=8.5mm }
{\maxsizebox*{0.7cm}{3ex}{\shortstack{\textcolor{teal}{Germanium} \\AV-V (H) \\  \\  \\  \\  \\ \textit{similar to} \\  SI(H), PS-PS(L)  \\ SD(H), Ana(L) }}}
}
\mycenter{
\hspace{6mm}
\adjustbox{padding =0.5ex 0.3ex 0.5ex 0.3ex,margin*=2mm,  width=5 cm, max height = 4 cm,frame,raise=7.5mm }
{\maxsizebox*{0.7cm}{3ex}{\shortstack{\textcolor{teal}{Germanium} \\PS-S (H) \\  \\  \\  \\  \\ \textit{similar to} \\  S-PS(H), ED(H)  \\ ~ }}}
\hspace{4.5mm}
\adjustbox{padding =0.4ex 0.3ex 0.4ex 0.3ex,margin*=2mm,  width=5 cm, max height = 4 cm,frame,raise=7mm }
{\maxsizebox*{0.7cm}{3ex}{\shortstack{\textcolor{teal}{Germanium} \\S-PS (H)\\  \\  \\  \\  \\ \textit{similar to} \\ PS-P(H), ED(H) \\ ~  }}}
\includegraphics[trim={1mm 0mm 0 4mm},clip,width=.3625\textwidth]{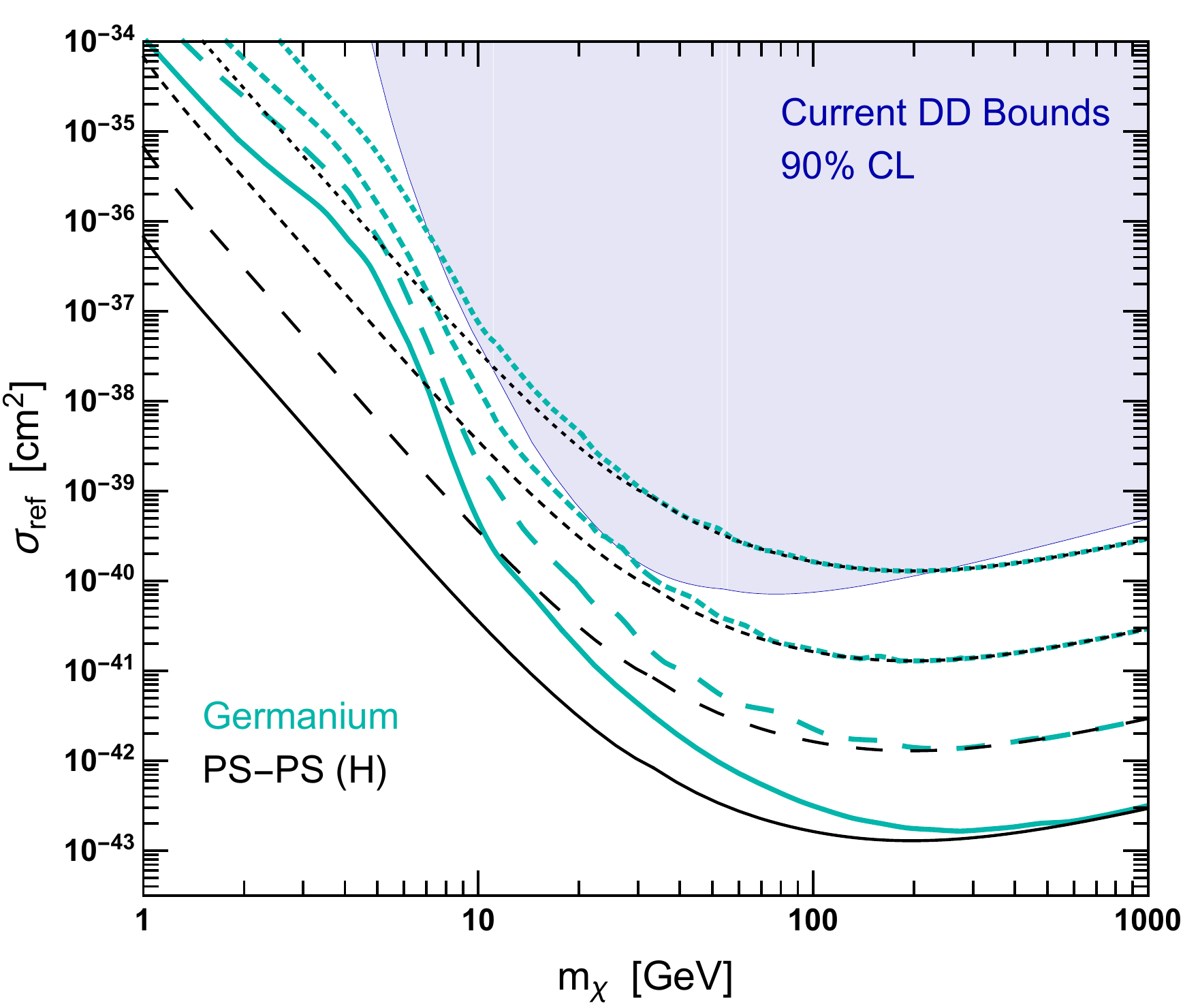}}

\caption{\label{fig:germanium_floors} 3$\sigma$ discovery limit for a germanium-based experiment for a 0.1 ton-year exposure (dotted), 1 ton-year exposure (short dashed), 10 ton-year exposure (long dashed), and 100 ton-year exposure (solid), including (teal) and neglecting (black) the neutrino background. Shown for comparison is the current 90\% upper limits from XENON1T and LUX (shaded blue).}
\end{figure}

\begin{figure}[H]
\mycenter{
\includegraphics[trim={0mm 16mm 0 0},clip,width=.38\textwidth]{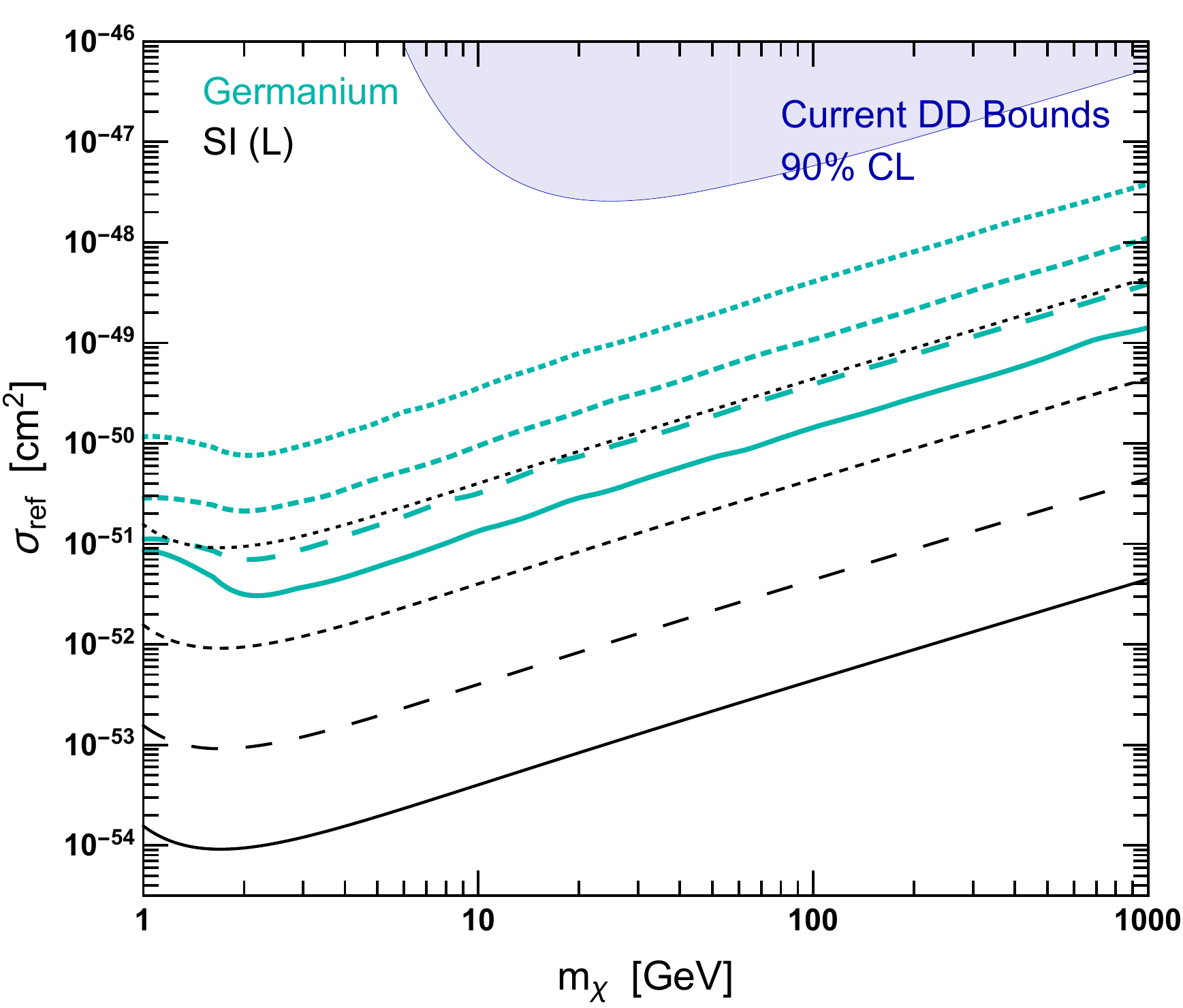}
\hspace{2.5mm}
\adjustbox{padding =0.6ex 0.425ex 0.6ex 0.425ex,margin*=2mm,  width=5 cm, max height = 4 cm,frame,raise=3mm}
{\maxsizebox*{0.7cm}{3ex}{\shortstack{\textcolor{teal}{Germanium} \\SD (L)\\  \\  \\  \\  \\ \textit{similar to} \\  SI(L), AV-V(L) \\ ~  }}}
 \hspace{3.75mm}
 \adjustbox{padding =0.85ex 0.4ex 0.85ex 0.4ex,margin*=2mm,  width=5 cm, max height = 4 cm,frame,raise=3mm }
{\maxsizebox*{0.7cm}{3ex}{\shortstack{\textcolor{teal}{Germanium} \\Ana (L)\\  \\  \\  \\  \\ \textit{similar to} \\  PS-PS(L) \\ ~  }}}
}
\mycenter{
\includegraphics[trim={0mm 0mm 0 1mm},clip,width=.38\textwidth]{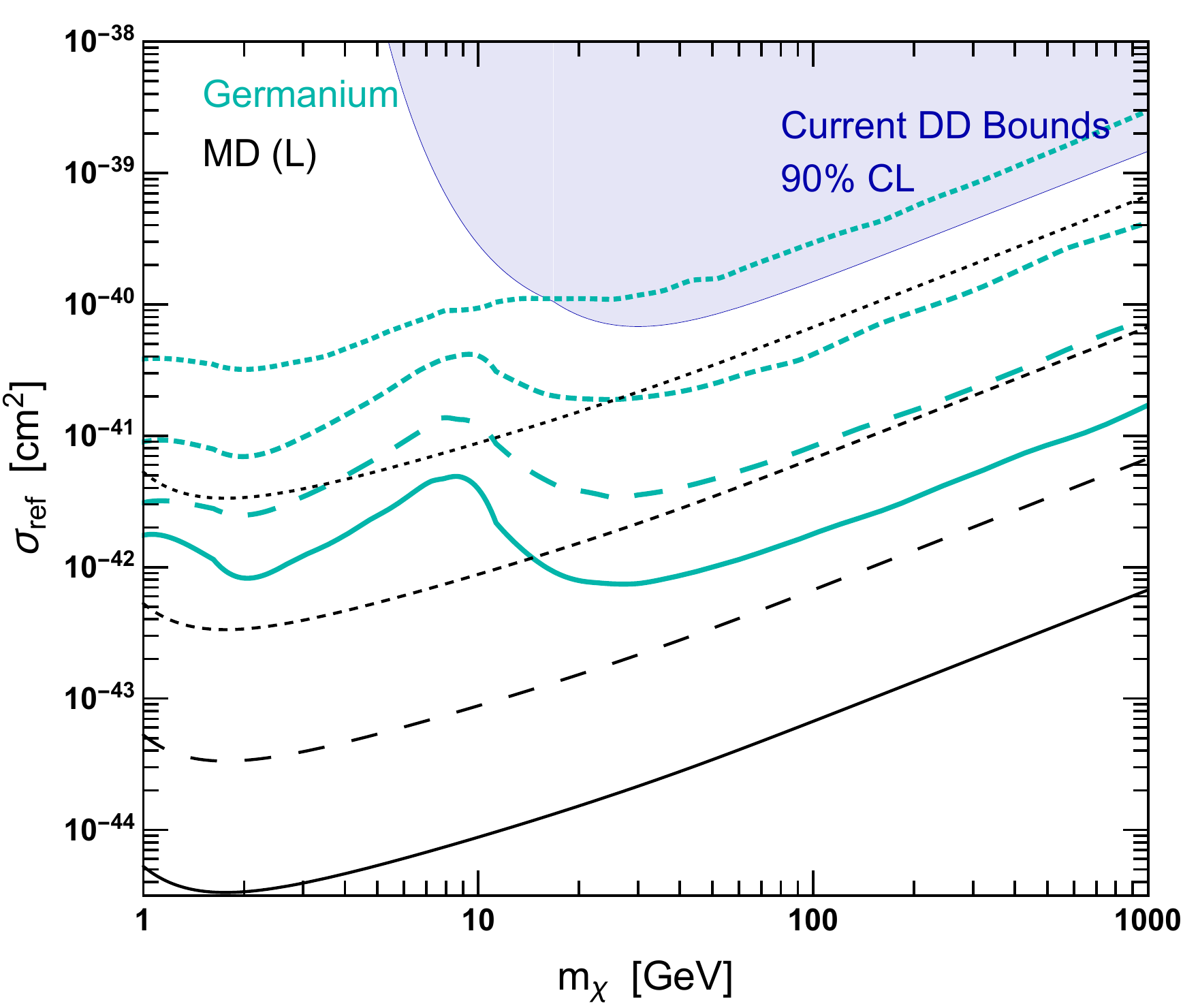}
 
\includegraphics[trim={ 9mm 0mm 0 1mm},clip,width=.36 \textwidth]{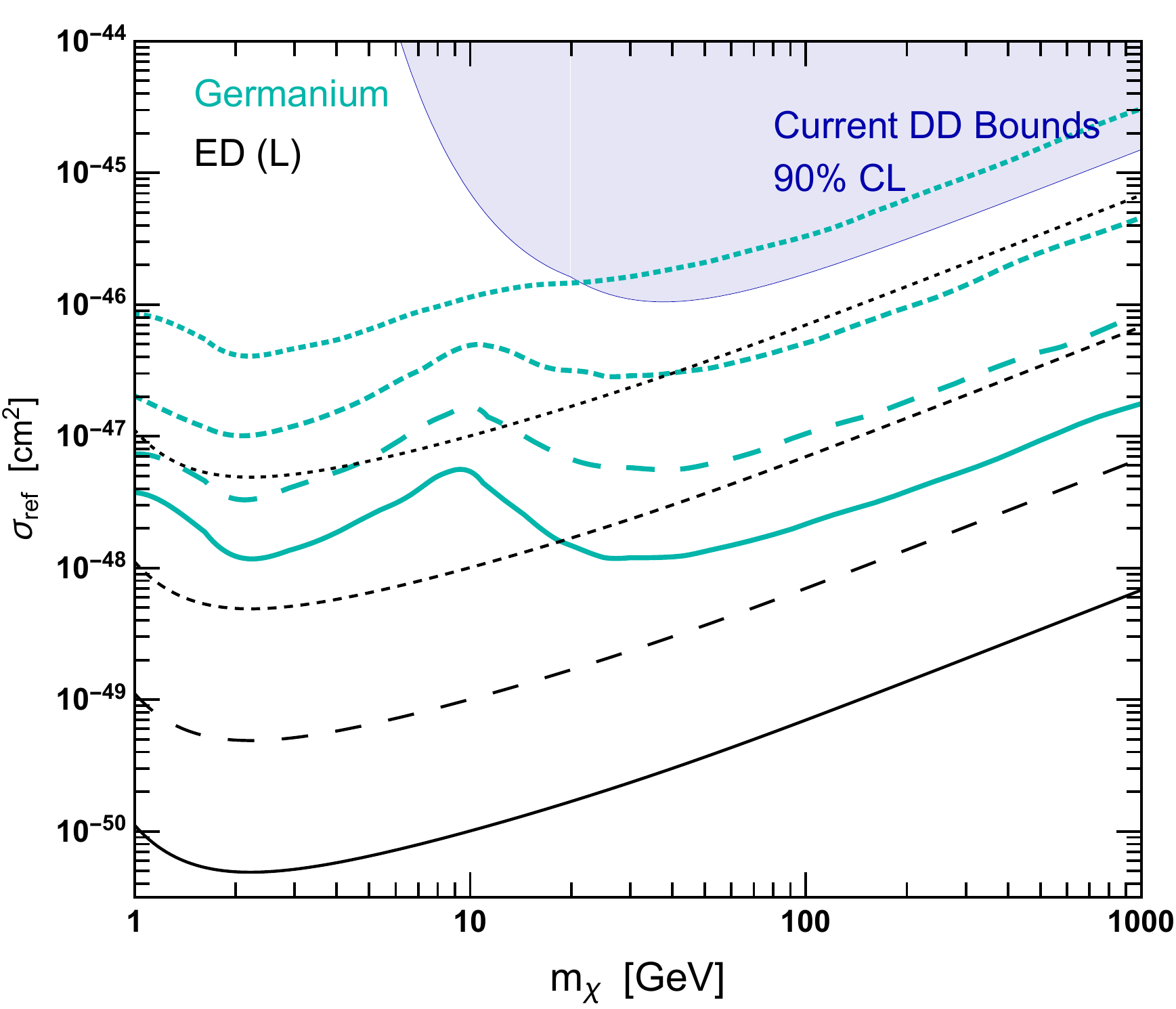}
 
  \hspace{2mm}
\adjustbox{padding =0.7ex 0.4ex 0.7ex 0.4ex,margin*=2.0mm,  width=5 cm, max height = 4 cm,frame,raise=8mm }
{\maxsizebox*{0.7cm}{3ex}{\shortstack{\textcolor{teal}{Germanium} \\AV-V (L) \\  \\  \\  \\  \\ \textit{similar to} \\  SI(L), SD(L)  \\ ~ }}}
}
\mycenter{
\hspace{6mm}
\adjustbox{padding =0.5ex 0.3ex 0.5ex 0.3ex,margin*=2.0mm,  width=5 cm, max height = 4 cm,frame,raise= 7mm }
{\maxsizebox*{0.7cm}{3ex}{\shortstack{\textcolor{teal}{Germanium} \\PS-S (L) \\  \\  \\  \\  \\ \textit{similar to} \\  S-PS(L), ED(L) \\ ~ }}}
\hspace{4.5mm}
\adjustbox{padding =0.4ex 0.3ex 0.4ex 0.3ex,margin*=2mm,  width=5 cm, max height = 4 cm,frame,raise= 7mm }
{\maxsizebox*{0.7cm}{3ex}{\shortstack{\textcolor{teal}{Germanium} \\S-PS (L)\\  \\  \\  \\  \\ \textit{similar to} \\ PS-P(L), ED(L) \\ ~  }}}
\hspace{4.25mm}
\adjustbox{padding =0.25ex 0.25ex 0.25ex 0.25ex,margin*=2mm,  width=5 cm, max height = 4 cm,frame,raise=7.5mm }
{\maxsizebox*{0.7cm}{3ex}{\shortstack{\textcolor{teal}{Germanium} \\PS-PS (L)\\  \\  \\  \\  \\ \textit{similar to} \\ SI(H), SD(H) \\ Ana(L), AV-V(H)  }}}}

\caption{\label{fig:germanium_floors} 3$\sigma$ discovery limit for a germanium-based experiment for a 0.1 ton-year exposure (dotted), 1 ton-year exposure (short dashed), 10 ton-year exposure (long dashed), and 100 ton-year exposure (solid), including (teal) and neglecting (black) the neutrino background. Shown for comparison is the current 90\% upper limits from XENON1T and LUX (shaded blue). }
\end{figure}

\begin{figure}[H]
\mycenter{
\includegraphics[trim={ 0mm 10mm 0 0},clip,width=.38\textwidth]{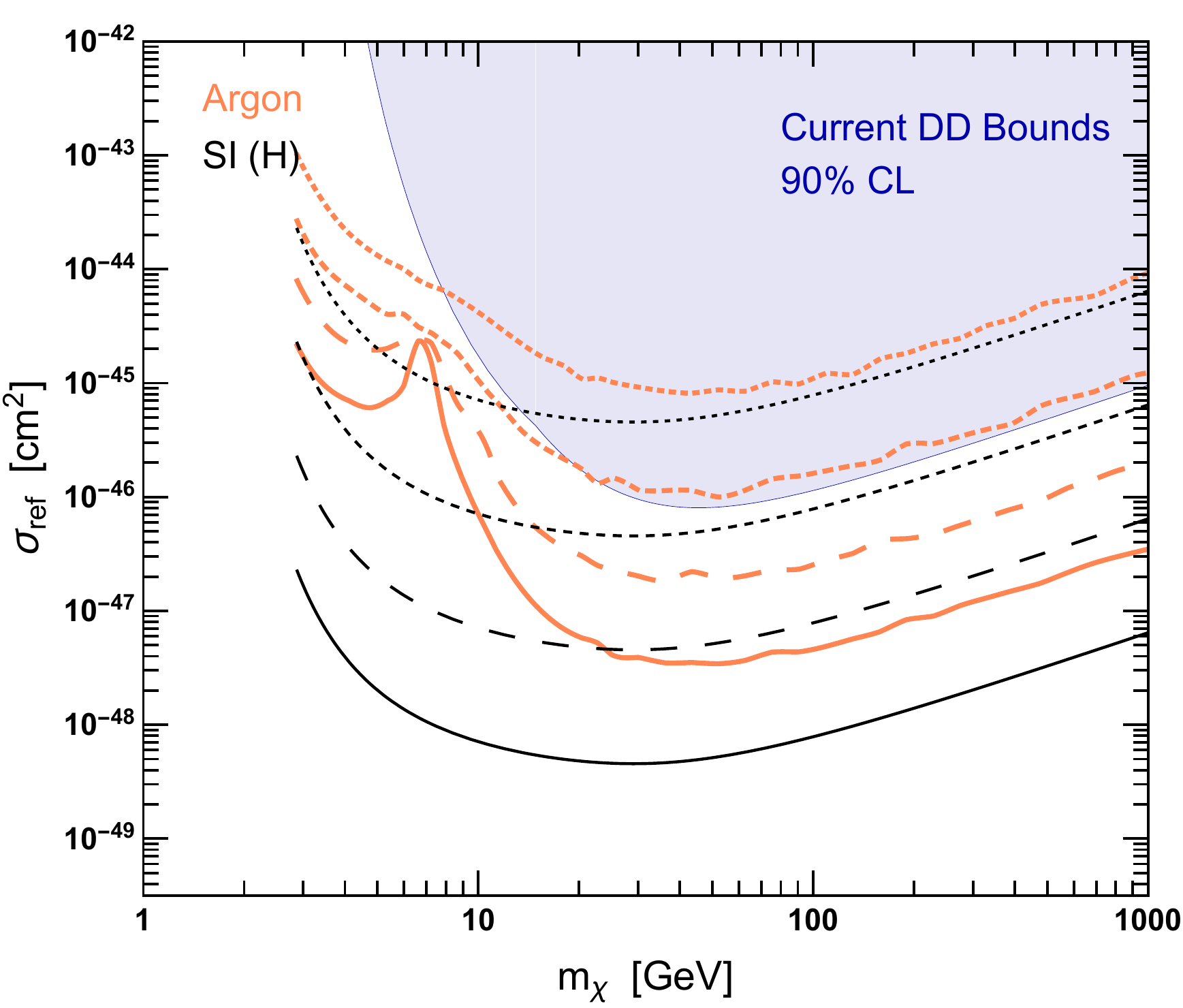}
\hspace{4mm}
\adjustbox{padding =0.6ex 0.5ex 0.6ex 0.5ex,margin*=2.5mm,  width=5 cm, max height = 4 cm,frame,raise=5mm }
{\maxsizebox*{0.7cm}{3ex}{\shortstack{\textcolor{orange}{Argon} \\SD (H) \\  \\  \\  \\  \\ \textit{no interaction} \\  ~ \\ ~  }}}
\hspace{1mm}
\includegraphics[trim={ 9mm 10mm 0 0},clip,width=.3625\textwidth]{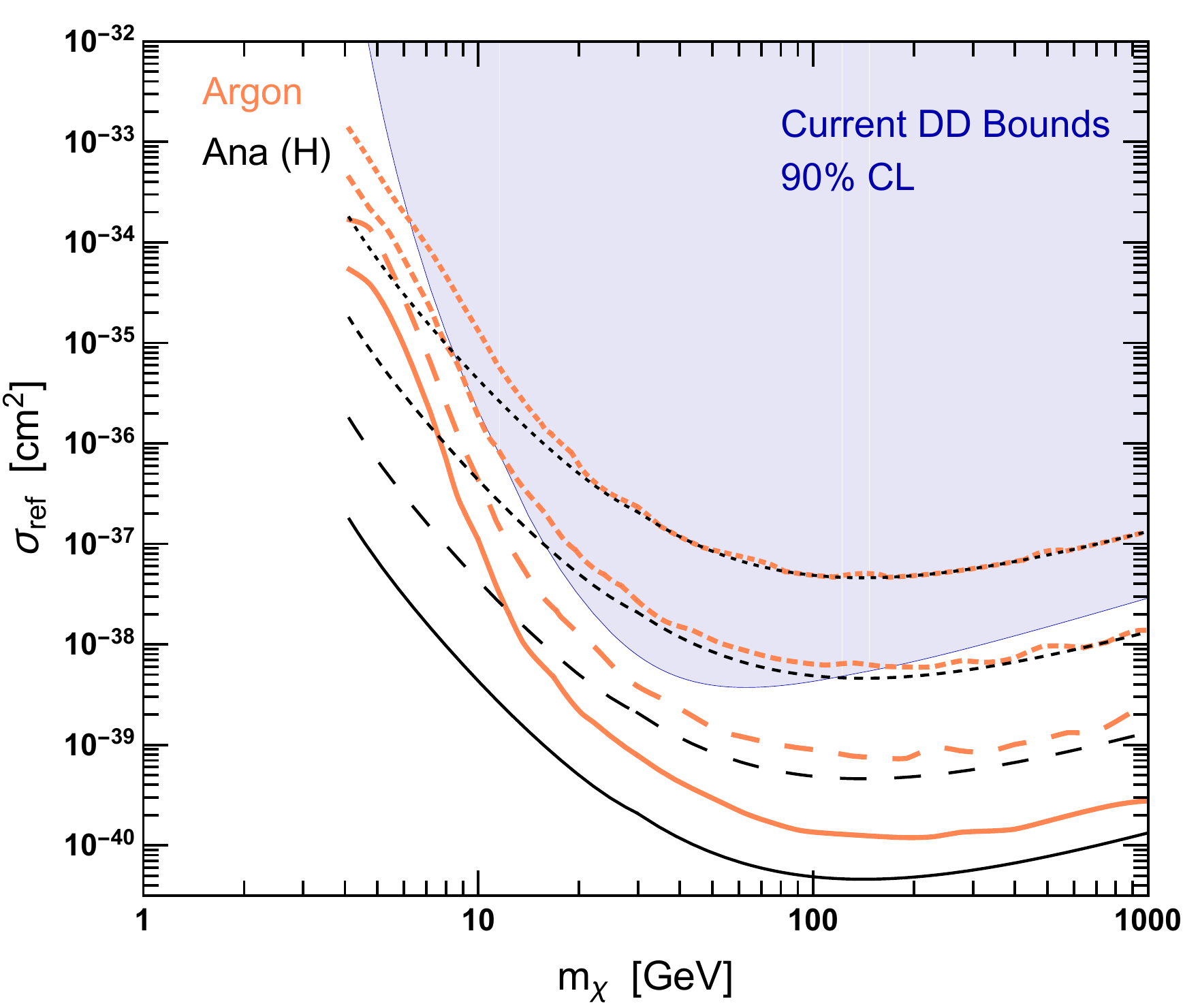}
}
\mycenter{
\includegraphics[trim={ 0mm 0mm 0 0},clip,width=.38\textwidth]{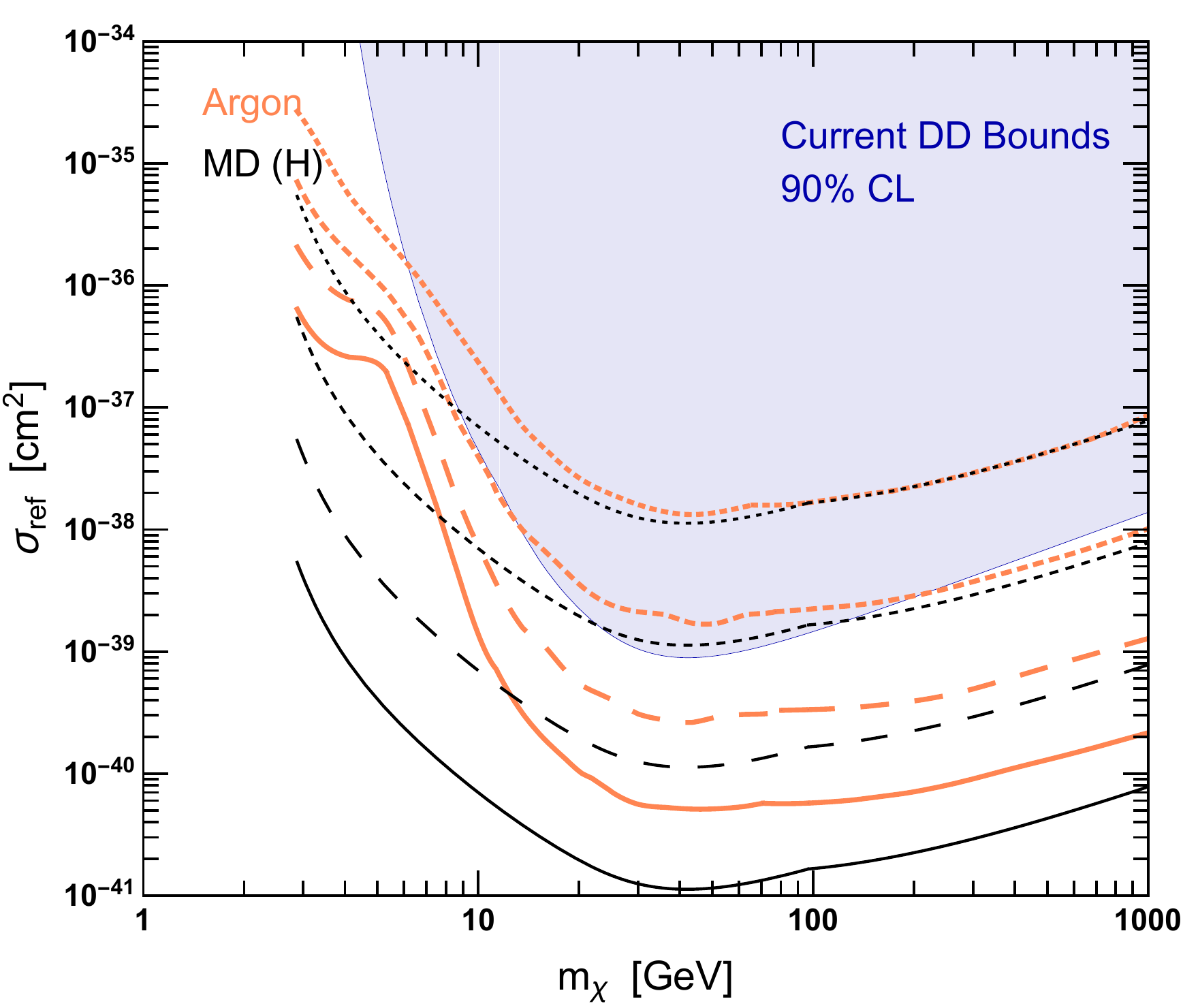}
\hspace{0 mm}
\includegraphics[trim={ 9mm 0mm 0 0},clip,width=.3625\textwidth]{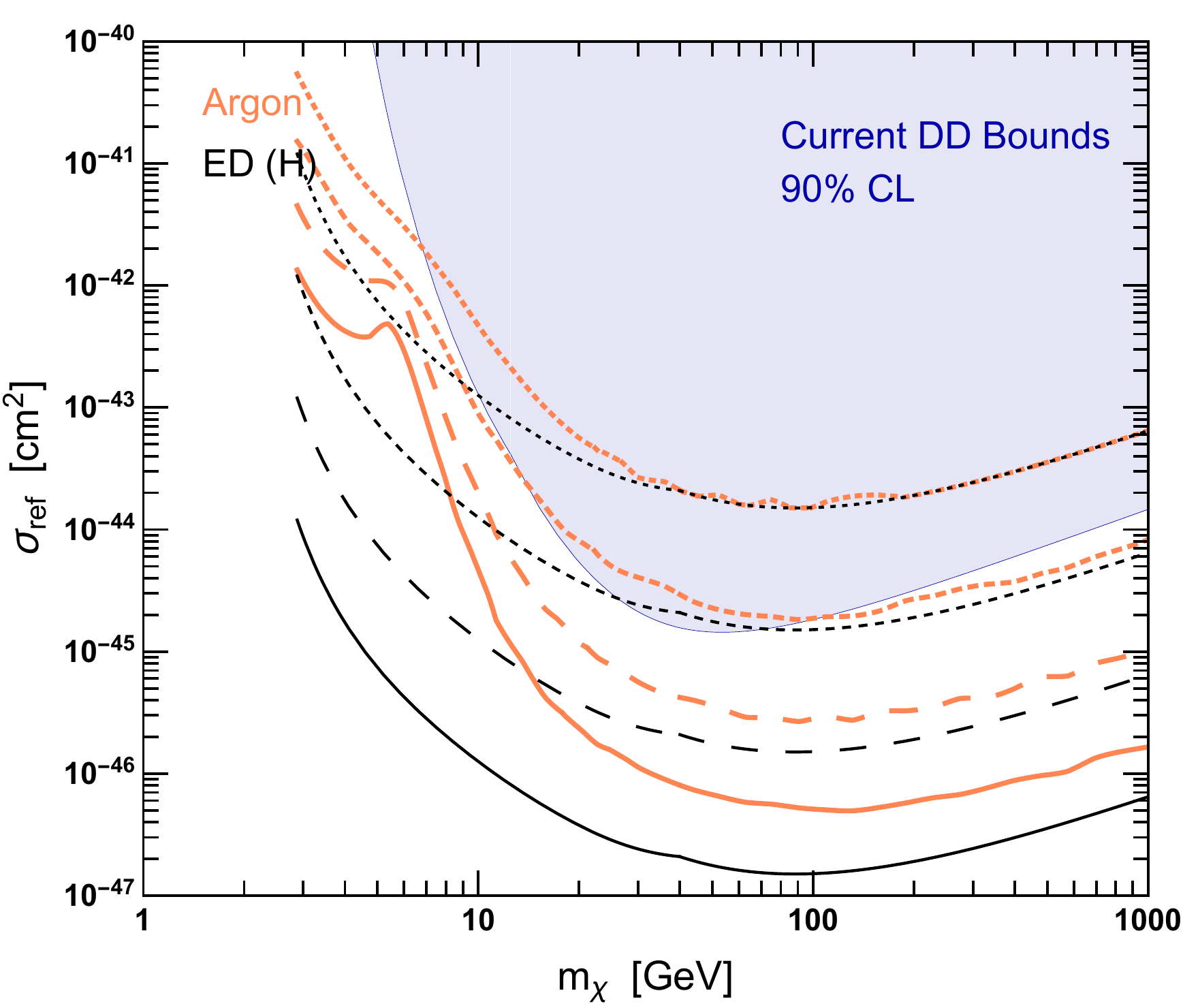}
  \hspace{3mm}
\adjustbox{padding =1.5ex 0.65ex 1.5ex 0.65ex,margin*=1.5mm,  width=5 cm, max height = 4 cm,frame,raise=10mm }
{\maxsizebox*{0.7cm}{3ex}{\shortstack{\textcolor{orange}{Argon} \\AV-V (H) \\  \\  \\  \\  \\ \textit{similar to} \\  SI(H)  \\ ~ }}}
}
\mycenter{
  \hspace{6mm}
\adjustbox{padding =1.4ex 0.6ex 1.4ex 0.6ex,margin*=1.5mm,  width=5 cm, max height = 4 cm,frame,raise=1mm }
{\maxsizebox*{0.7cm}{3ex}{\shortstack{\textcolor{orange}{Argon} \\PS-S (H) \\  \\  \\  \\  \\ \textit{similar to} \\  ED(H) \\ ~  }}}
  \hspace{6mm}
\adjustbox{padding =1.4ex 1.4ex 1.4ex 1.4ex,margin*=1.5mm,  width=5 cm, max height = 4 cm,frame,raise=4.5mm }
{\maxsizebox*{0.7cm}{3ex}{\shortstack{\textcolor{orange}{Argon} \\S-PS (H) \\  \\  \\  \\  \\ \textit{no interaction} \\  ~ \\ ~  }}}
  \hspace{4.5mm}
\adjustbox{padding =1.4ex 1.4ex 1.4ex 1.4ex,margin*=1.5mm,  width=5 cm, max height = 4 cm,frame,raise=4.5mm }
{\maxsizebox*{0.7cm}{3ex}{\shortstack{\textcolor{orange}{Argon} \\PS-PS (H) \\  \\  \\  \\  \\ \textit{no interaction} \\  ~ \\ ~  }}}
}

\caption{\label{fig:argon_floors} 3$\sigma$ discovery limit for a argon-based experiment for a 0.1 ton-year exposure (dotted), 1 ton-year exposure (short dashed), 10 ton-year exposure (long dashed), and 100 ton-year exposure (solid), including (orange) and neglecting (black) the neutrino background. Shown for comparison is the current 90\% upper limits from XENON1T and LUX (shaded blue).}
\end{figure}

\begin{figure}[H]
\mycenter{
\includegraphics[trim={ 0mm 16mm 0 0},clip,width=.38\textwidth]{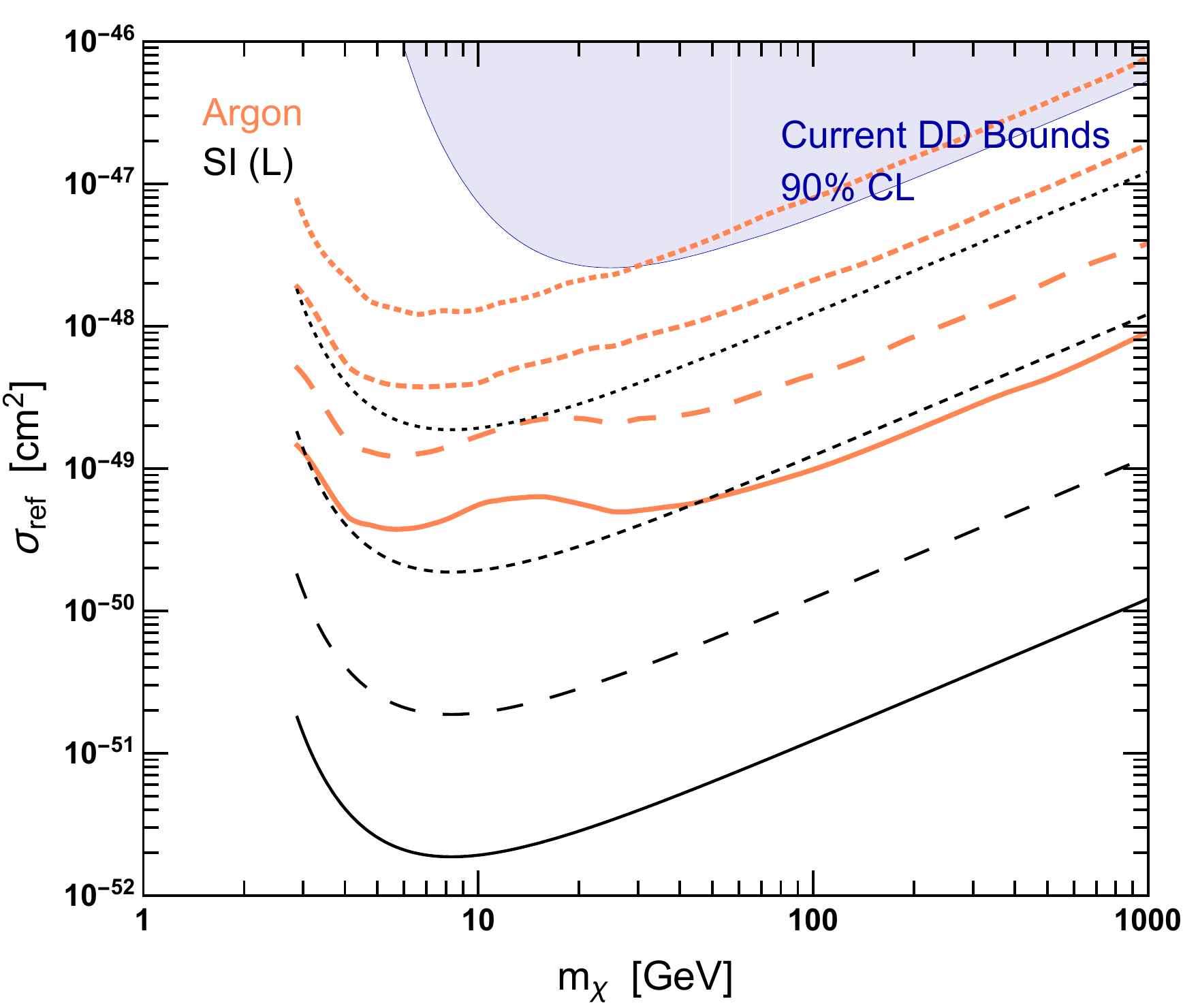}
 \hspace{3.0mm}
 \adjustbox{padding =0.85ex 0.75ex 0.85ex 0.75ex,margin*=1.5mm,  width=5 cm, max height = 4 cm,frame,raise=5.5mm }
{\maxsizebox*{0.7cm}{3ex}{\shortstack{\textcolor{orange}{Argon} \\SD(L) \\  \\  \\  \\  \\ \textit{no interaction} \\  ~ \\ ~  }}}
 \hspace{4.0mm}
 \adjustbox{padding =1.4ex 0.5ex 1.4ex 0.5ex,margin*=1.5mm,  width=5 cm, max height = 4 cm,frame,raise=4mm }
{\maxsizebox*{0.7cm}{3ex}{\shortstack{\textcolor{orange}{Argon} \\Ana(L) \\  \\  \\  \\  \\ \textit{similar to} \\ SI(H)    \\ ~  }}}

}
\mycenter{
\includegraphics[trim={ 0mm 0mm 0 0},clip,width=.38\textwidth]{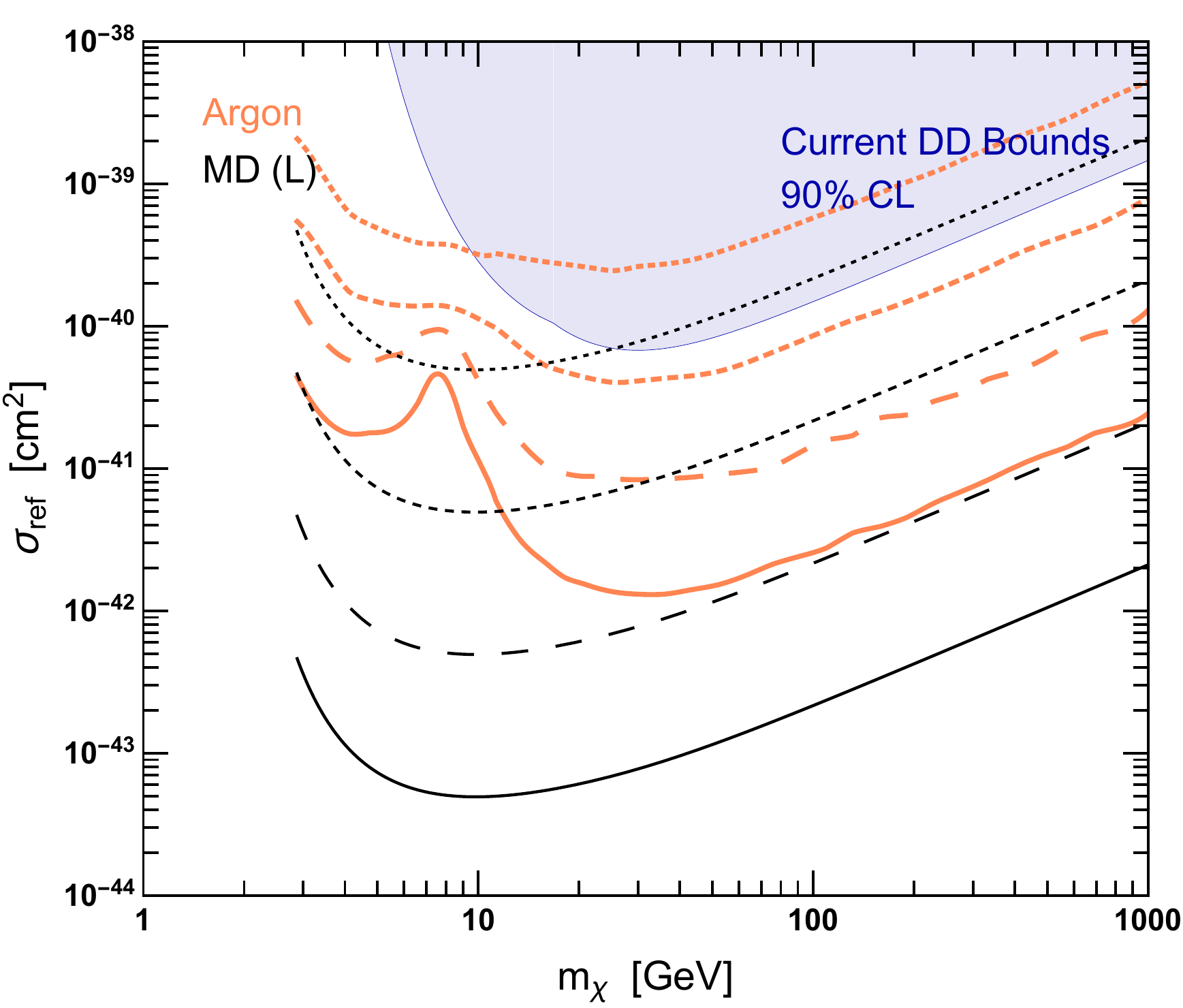}
\includegraphics[trim={ 9mm 0mm 0 0},clip,width=.3625\textwidth]{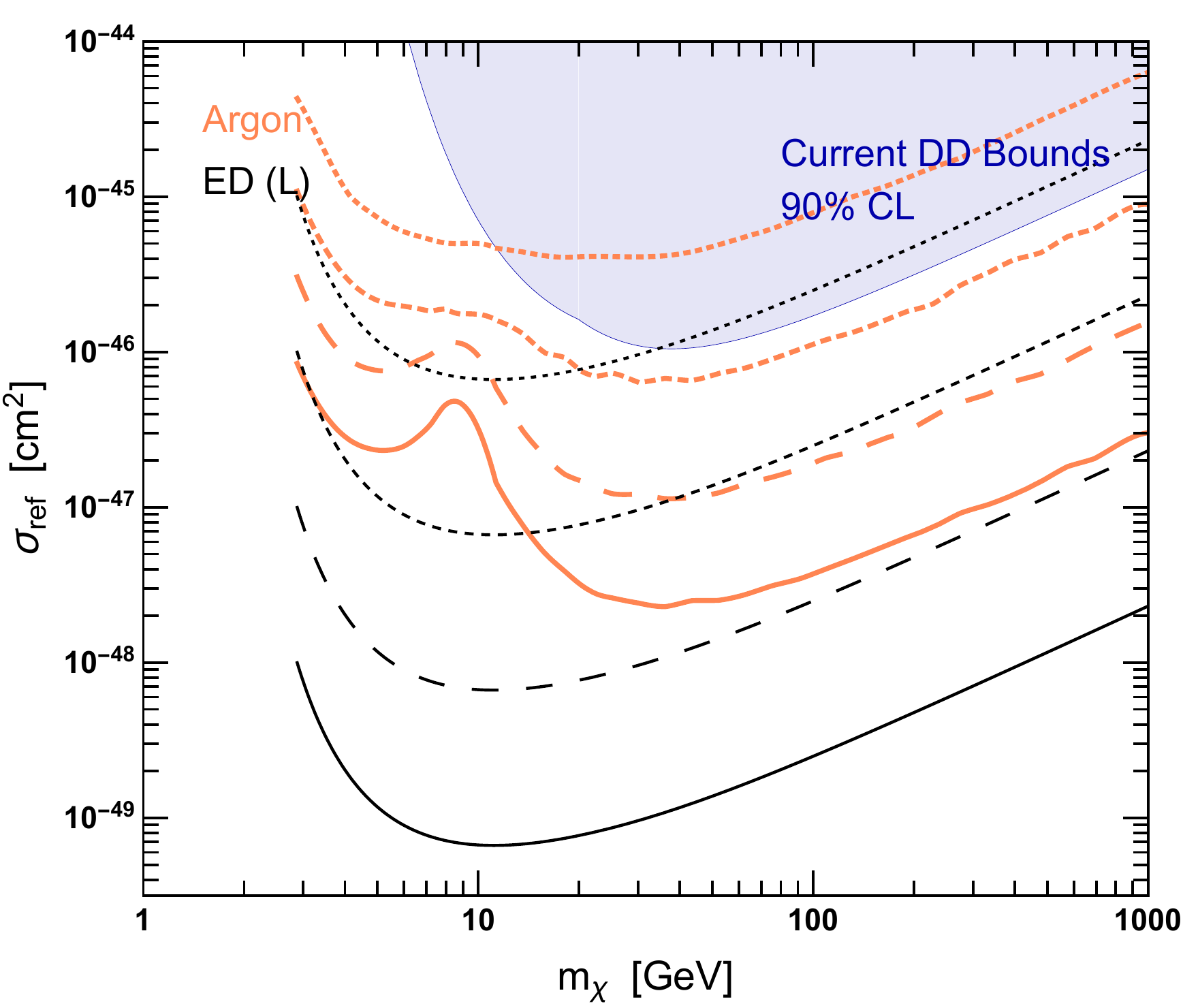}
 \hspace{2.25mm}
 \adjustbox{padding =1.4ex 0.5ex 1.4ex 0.5ex,margin*=1.5mm,  width=5 cm, max height = 4 cm,frame,raise=9.5mm }
{\maxsizebox*{0.7cm}{3ex}{\shortstack{\textcolor{orange}{Argon} \\AV-V (L) \\  \\  \\  \\  \\ \textit{similar to} \\  SI(L)  \\ ~ }}}
}
\mycenter{
 \hspace{6mm}
 \adjustbox{padding =1.4ex 0.5ex 1.4ex 0.5ex,margin*=1.5mm,  width=5 cm, max height = 4 cm,frame,raise=2mm }
{\maxsizebox*{0.7cm}{3ex}{\shortstack{\textcolor{orange}{Argon} \\PS-S (L) \\  \\  \\  \\  \\ \textit{similar to} \\  ED(L) \\ ~  }}}
 \hspace{4.5mm}
 \adjustbox{padding =1.4ex 1.3ex 1.4ex 1.3ex,margin*=1.5mm,  width=5 cm, max height = 4 cm,frame,raise=5.5mm }
{\maxsizebox*{0.7cm}{3ex}{\shortstack{\textcolor{orange}{Argon} \\S-PS (L) \\  \\  \\  \\  \\ \textit{no interaction} \\  ~ \\ ~  }}}
 \hspace{4.0mm}
 \adjustbox{padding =1.4ex 1.3ex 1.4ex 1.3ex,margin*=1.5mm,  width=5 cm, max height = 4 cm,frame,raise=5.5mm }
{\maxsizebox*{0.7cm}{3ex}{\shortstack{\textcolor{orange}{Argon} \\PS-PS (L) \\  \\  \\  \\  \\ \textit{no interaction} \\  ~ \\ ~  }}}
}

\caption{\label{fig:argon_floors} 3$\sigma$ discovery limit for a argon-based experiment for a 0.1 ton-year exposure (dotted), 1 ton-year exposure (short dashed), 10 ton-year exposure (long dashed), and 100 ton-year exposure (solid), including (orange) and neglecting (black) the neutrino background. Shown for comparison is the current 90\% upper limits from XENON1T and LUX (shaded blue).}
\end{figure}

\begin{figure}[H]
\mycenter{
 \hspace{3.0mm}
\includegraphics[trim={ 0mm 15.5mm 0 0},clip,width=.38\textwidth]{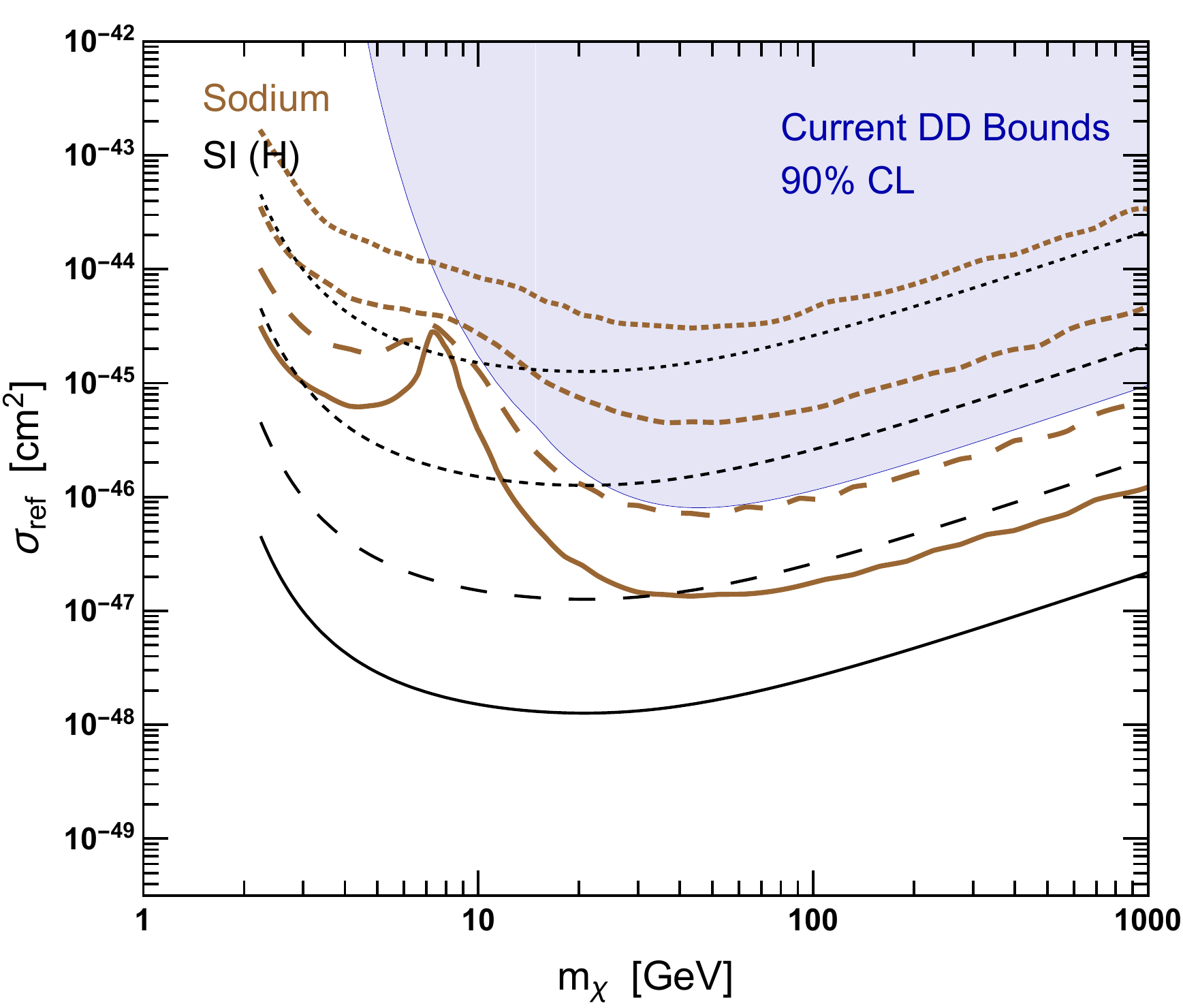}
 \hspace{2.75mm}
 \adjustbox{padding =0.4ex 0.45ex 0.4ex 0.45ex,margin*=2mm,  width=5 cm, max height = 4 cm,frame,raise=4.5mm }
{\maxsizebox*{0.7cm}{3ex}{\shortstack{\textcolor{brown}{Sodium} \\SD (H) \\  \\  \\  \\  \\ \textit{similar to} \\  SI(H), PS-PS(L) }}}
 \hspace{2.5mm}

\includegraphics[trim={9mm 15.5mm 0 0},clip,width=.3625\textwidth]{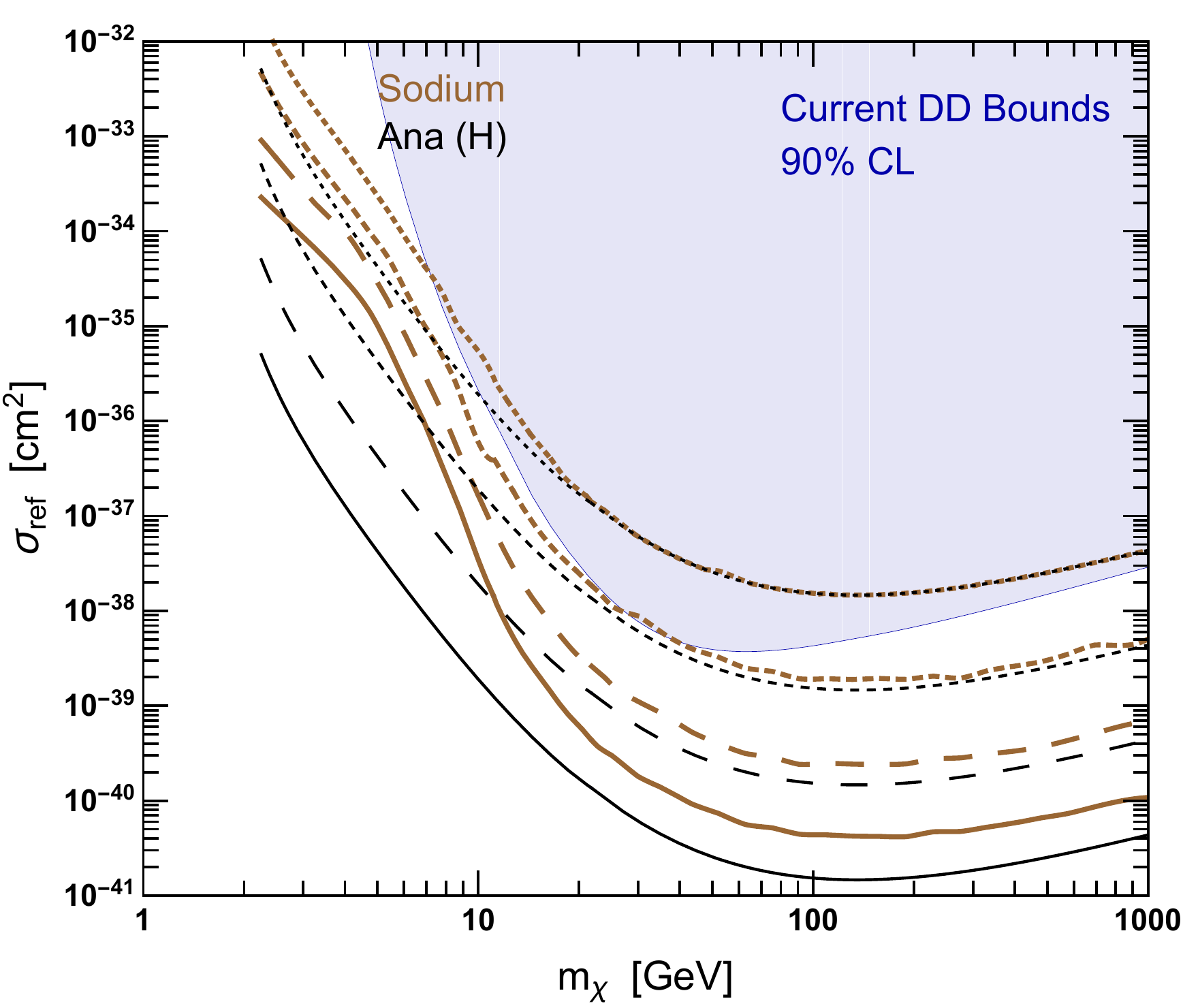}
}

\mycenter{
 \hspace{3mm}
\includegraphics[trim={ 0mm 15.5mm 0 0},clip,width=.38\textwidth]{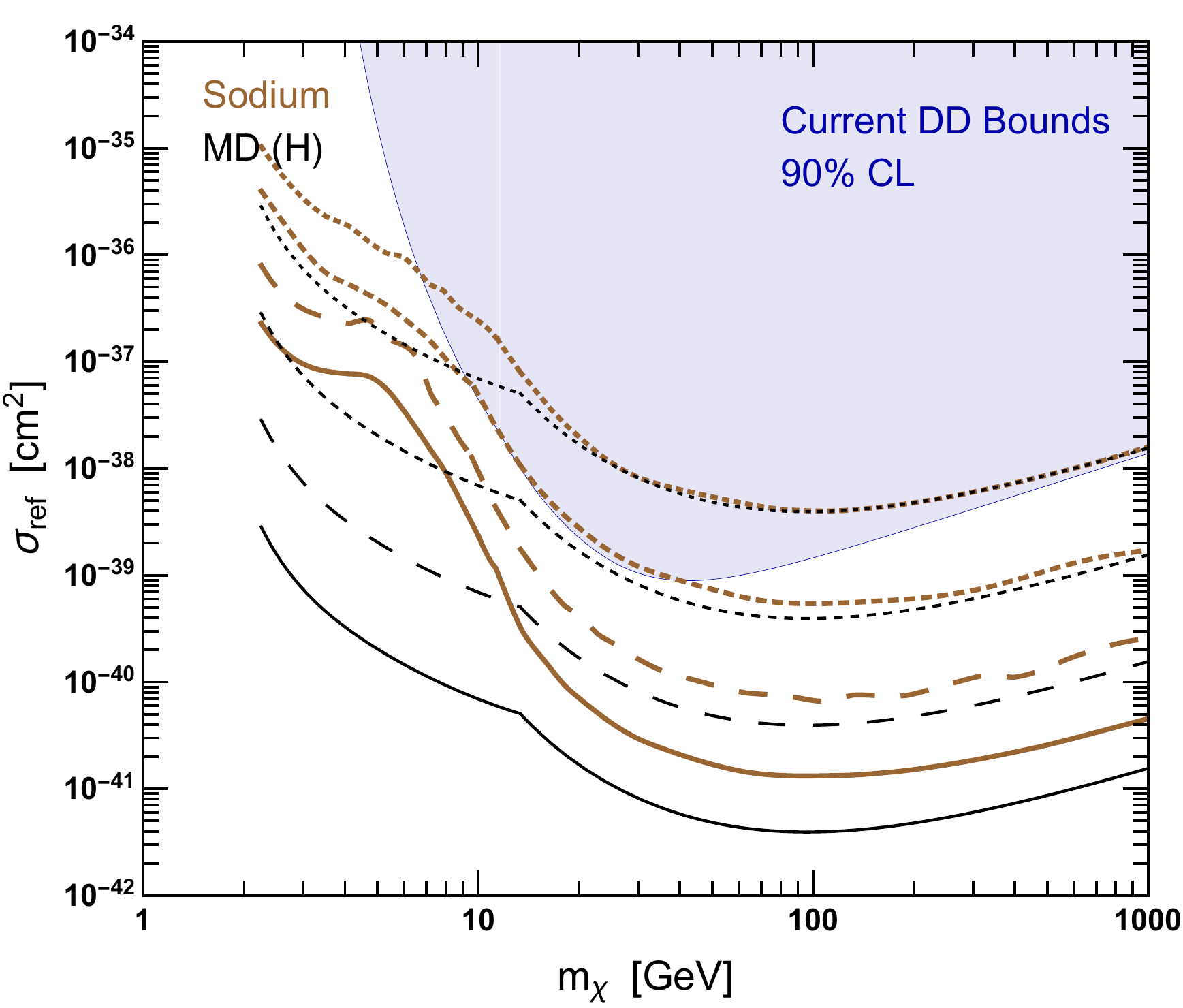}
\includegraphics[trim={ 9mm 15.5mm 0 0},clip,width=.3625\textwidth]{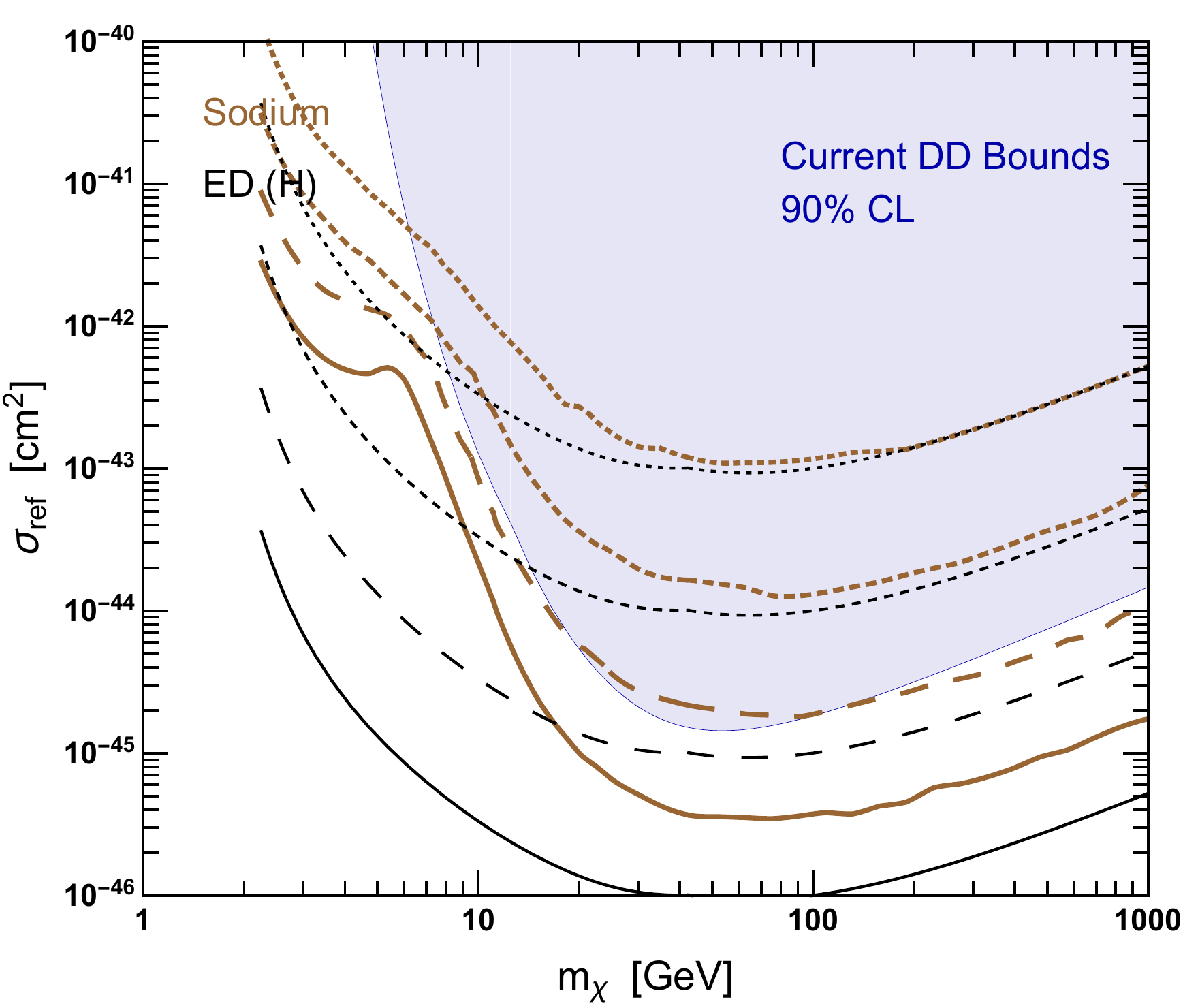}
\hspace{0.25mm}
\includegraphics[trim={ 9mm 15.5mm 0 0},clip,width=.3625\textwidth]{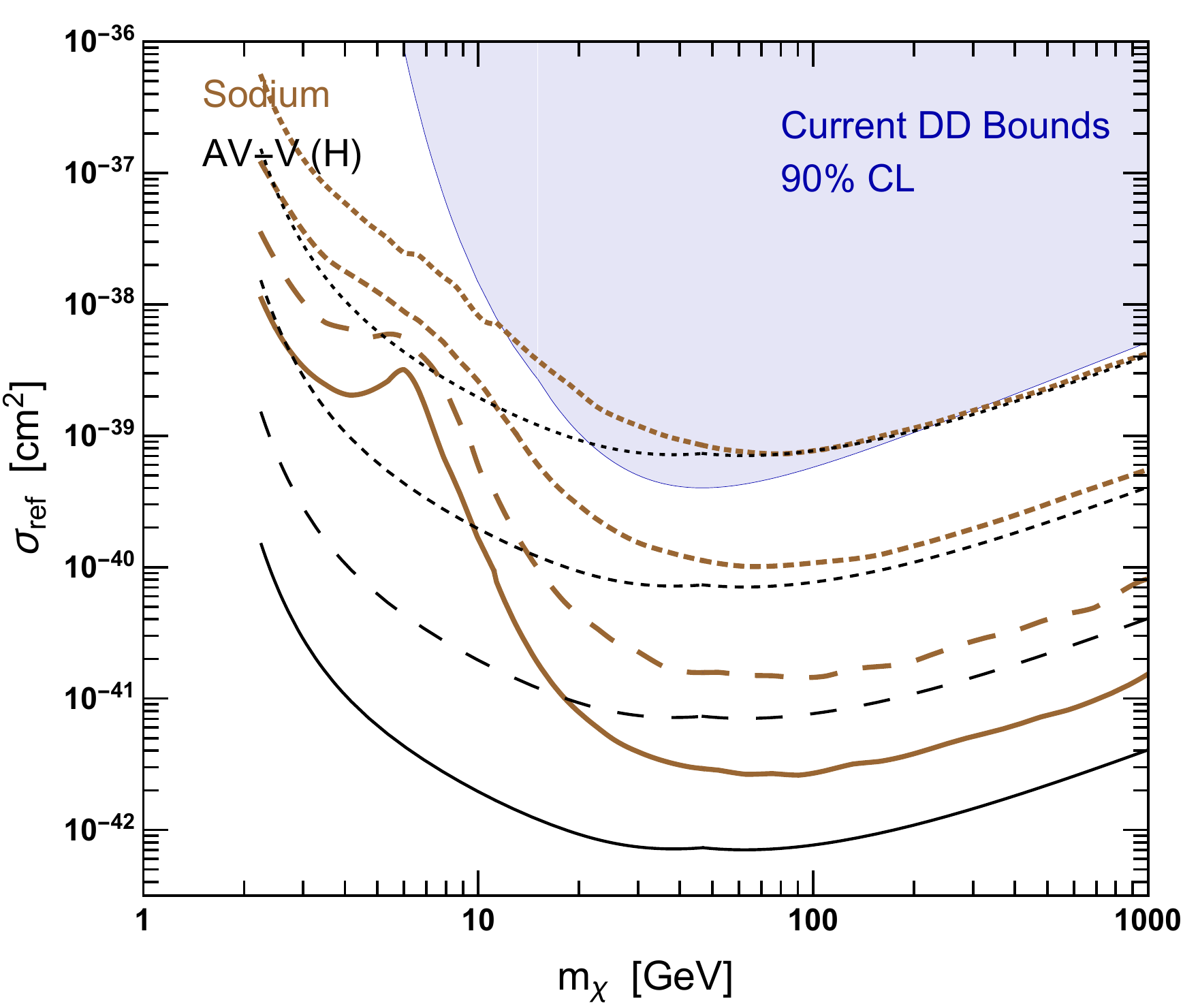}
}
\mycenter{
 \hspace{10mm}
 \adjustbox{padding =1.4ex 0.45ex 1.4ex 0.45ex,margin*=1.5mm,  width=5 cm, max height = 4 cm,frame,raise=9mm }
{\maxsizebox*{0.7cm}{3ex}{\shortstack{\textcolor{brown}{Sodium} \\PS-S (H) \\  \\  \\  \\  \\ \textit{similar to} \\  ED(H) \\ ~  }}}
 \hspace{5mm}
 \adjustbox{padding =1.4ex 1.3ex 1.4ex 1.3ex,margin*=1.5mm,  width=5 cm, max height = 4 cm,frame,raise=13.5mm}
{\maxsizebox*{0.7cm}{3ex}{\shortstack{\textcolor{brown}{Sodium} \\S-PS (H) \\  \\  \\  \\  \\ \textit{similar to} \\  PS-S(H), ED(H) }}}
\includegraphics[trim={0mm 0mm 0 0},clip,width=.3825\textwidth]{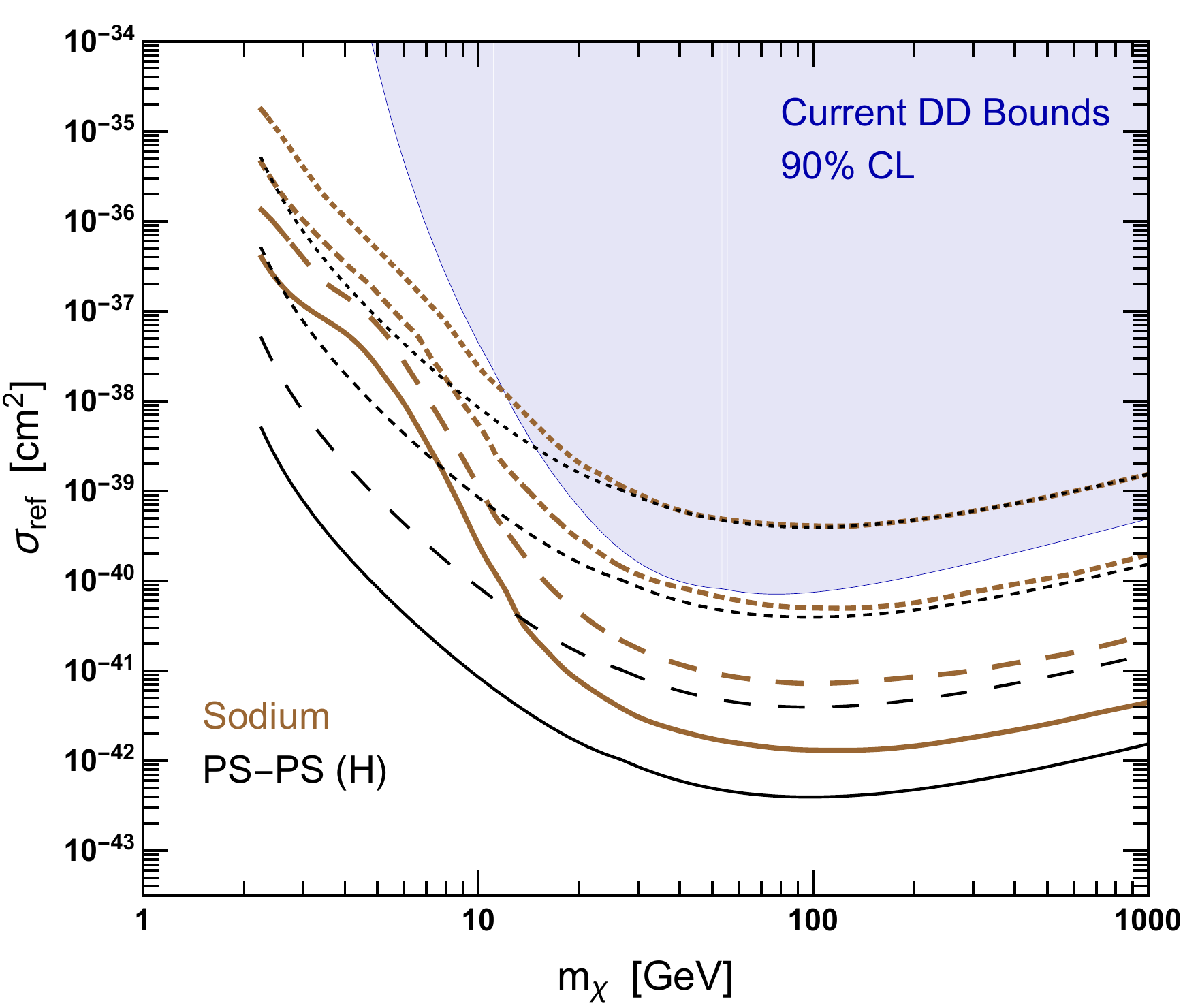} 
}

\caption{\label{fig:sodium_floors} 3$\sigma$ discovery limit for a sodium-based experiment for a 0.1 ton-year exposure (dotted), 1 ton-year exposure (short dashed), 10 ton-year exposure (long dashed), and 100 ton-year exposure (solid), including (yellow) and neglecting (black) the neutrino background. Shown for comparison is the current 90\% upper limits from XENON1T and LUX (shaded blue).}
\end{figure}

\begin{figure}[H]
\mycenter{
\includegraphics[trim={0mm 15.5mm 0 0},clip,width=.38\textwidth]{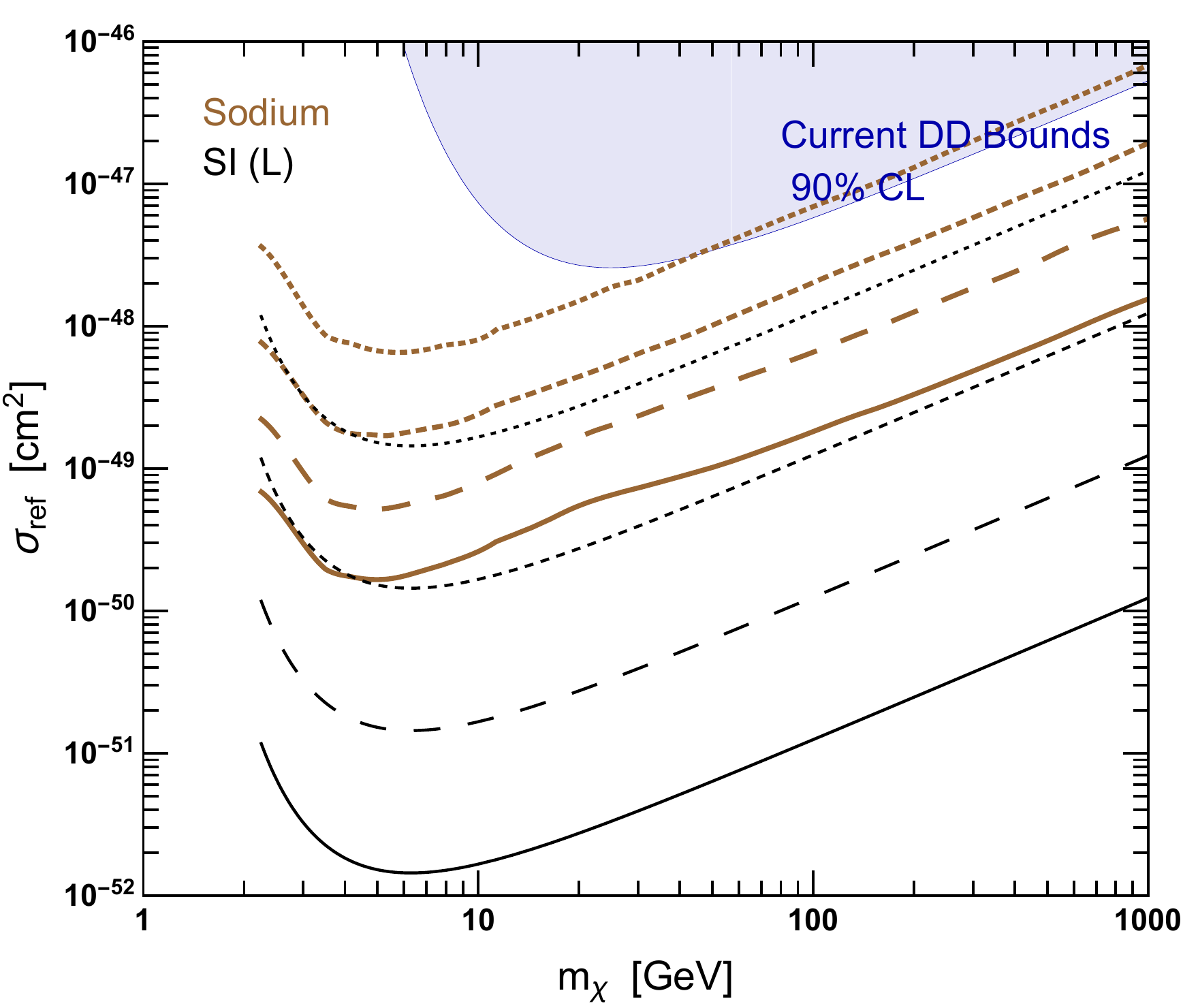}
 \hspace{3mm}
 \adjustbox{padding =1.4ex 0.55ex 1.4ex 0.55ex,margin*=1.5mm,  width=5 cm, max height = 4 cm,frame,raise=4mm }
{\maxsizebox*{0.7cm}{3ex}{\shortstack{\textcolor{brown}{Sodium} \\SD (L) \\  \\  \\  \\  \\ \textit{similar to} \\  SI(L) \\ ~  }}}
 \hspace{4.5mm}
 \adjustbox{padding =1.4ex 0.55ex 1.4ex 0.55ex,margin*=1.5mm,  width=5 cm, max height = 4 cm,frame,raise=4mm}
{\maxsizebox*{0.7cm}{3ex}{\shortstack{\textcolor{brown}{Sodium} \\Ava(L) \\  \\  \\  \\  \\ \textit{similar to} \\  AV-V(H) \\ ~ }}}
}
\mycenter{
\includegraphics[trim={ 0mm 0mm 0 0},clip,width=.38\textwidth]{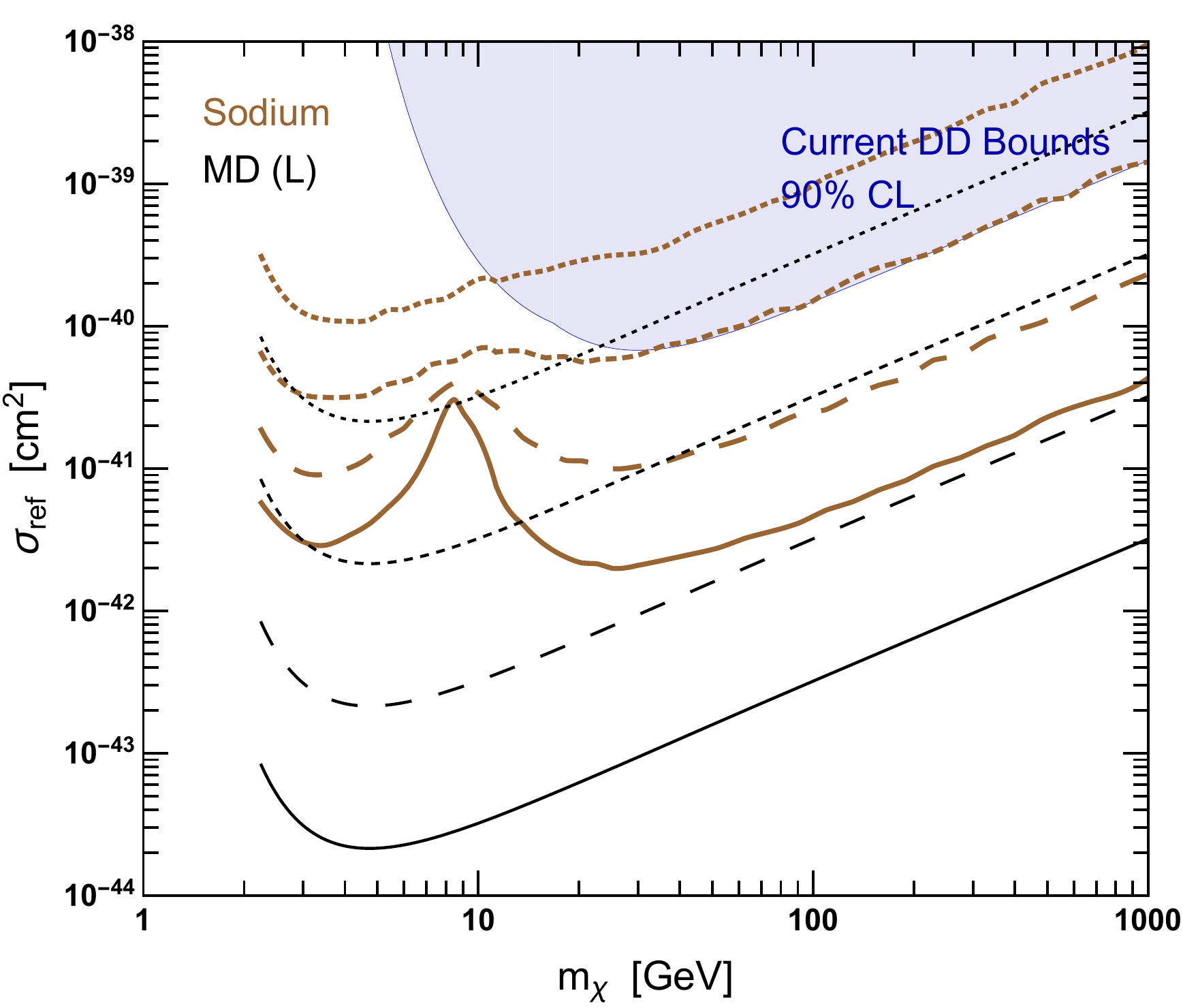} 
\includegraphics[trim={ 9mm 0mm 0 0},clip,width=.3625\textwidth]{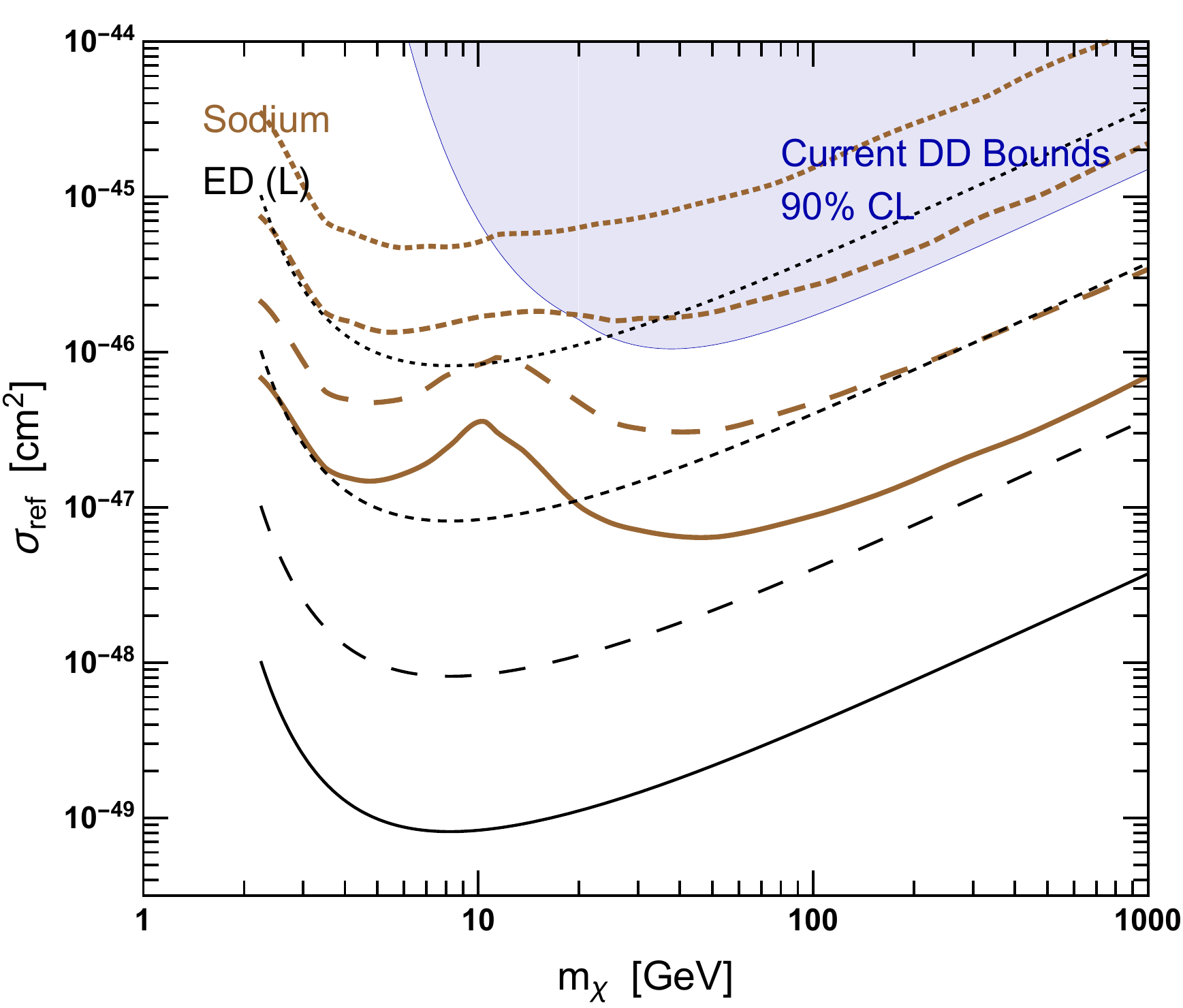}
\includegraphics[trim={9mm 0mm 0 0},clip,width=.3625\textwidth]{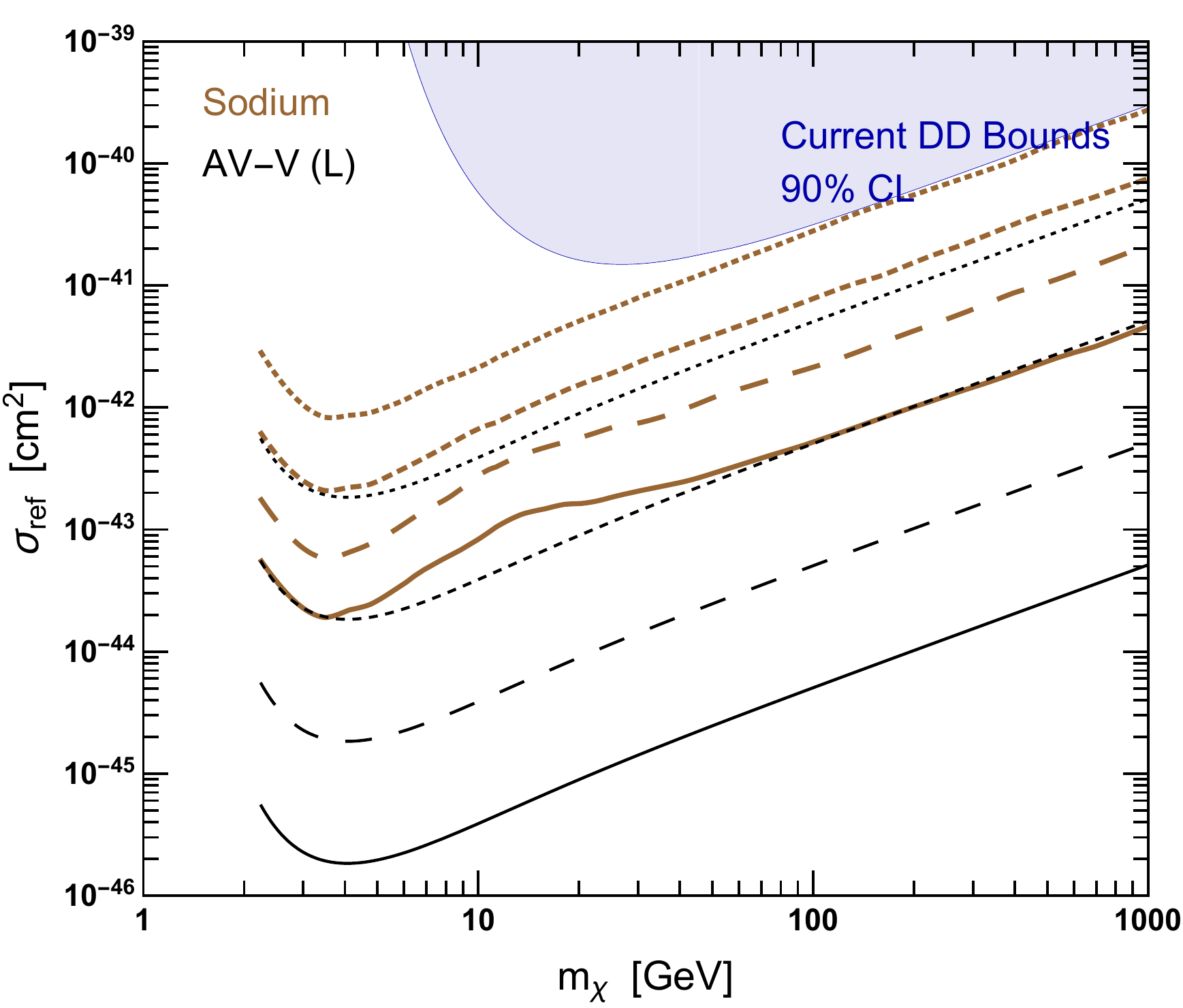}
}
\mycenter{
 \hspace{6mm}
 \adjustbox{padding =0.7ex 0.6ex 0.7ex 0.6ex,margin*=1.5mm,  width=5 cm, max height = 4 cm,frame,raise=9mm}
{\maxsizebox*{0.7cm}{3ex}{\shortstack{\textcolor{brown}{Sodium} \\PS-S (L) \\  \\  \\  \\  \\ \textit{similar to} \\   S-PS(L), ED(L) \\ ~ }}}
 \hspace{5mm}
 \adjustbox{padding =0.7ex 0.7ex 0.7ex 0.7ex,margin*=1.5mm,  width=5 cm, max height = 4 cm,frame,raise=10.5mm  }
{\maxsizebox*{0.7cm}{3ex}{\shortstack{\textcolor{brown}{Sodium} \\S-PS (L) \\  \\  \\  \\  \\ \textit{similar to} \\  PS-S(L), ED(L)  }}}
 \hspace{4.5mm}
 \adjustbox{padding =0.8ex 0.4ex 0.8ex 0.4ex,margin*=2.5mm,  width=5 cm, max height = 4 cm,frame,raise=8mm  }
{\maxsizebox*{0.7cm}{3ex}{\shortstack{\textcolor{brown}{Sodium} \\PS-PS (L) \\  \\  \\  \\  \\ \textit{similar to} \\  SI(H),SD(H)  }}}
}
 
\caption{\label{fig:sodium_floors} 3$\sigma$ discovery limit for a sodium-based experiment for a 0.1 ton-year exposure (dotted), 1 ton-year exposure (short dashed), 10 ton-year exposure (long dashed), and 100 ton-year exposure (solid), including (yellow) and neglecting (black) the neutrino background. Shown for comparison is the current 90\% upper limits from XENON1T and LUX (shaded blue).}
\end{figure}

\begin{figure}[H]
\mycenter{
\includegraphics[trim={0mm 16mm 0 0},clip,width=.38\textwidth]{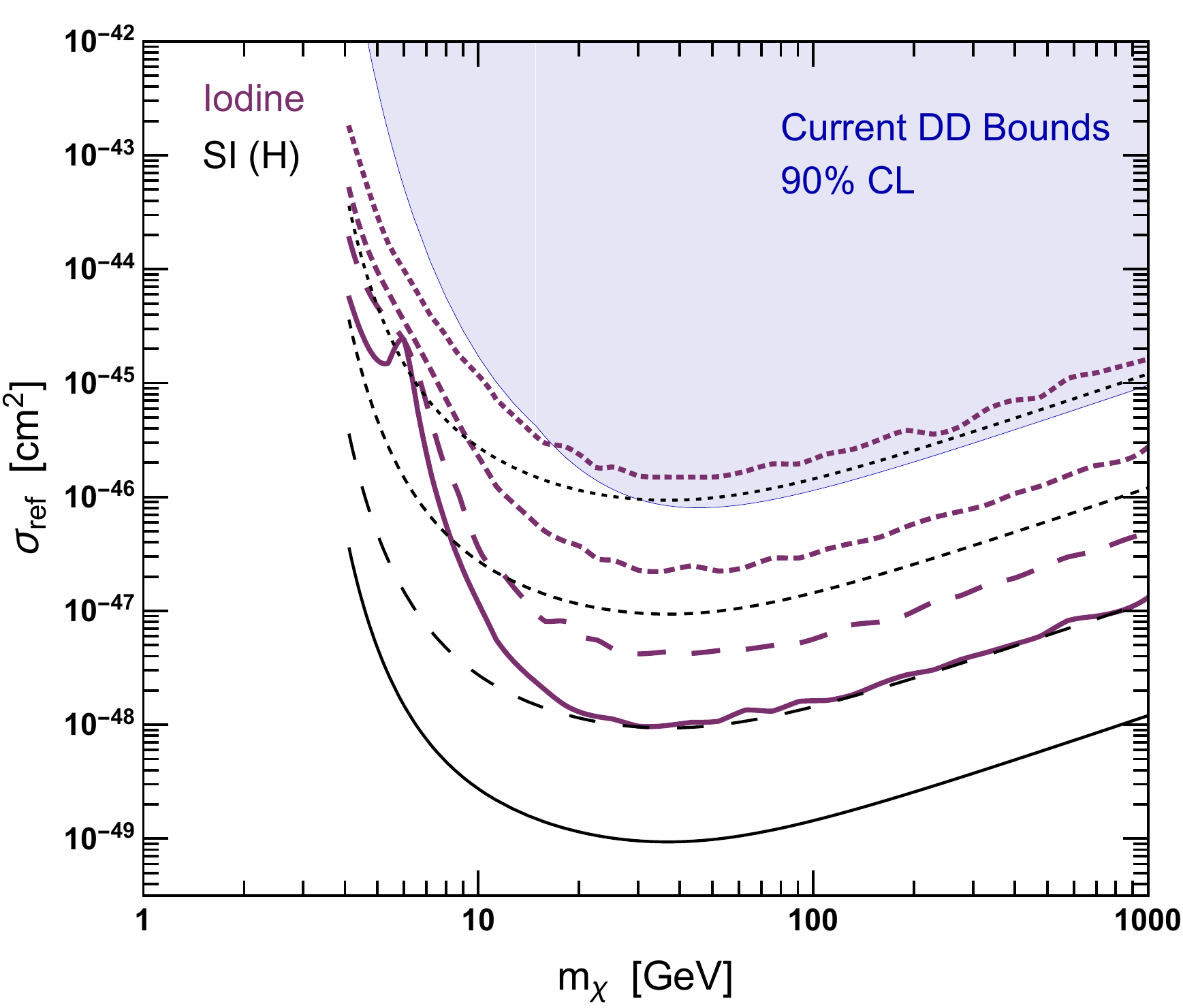}
 \hspace{3.75mm}
 \adjustbox{padding =0.6ex 0.5ex 0.6ex 0.5ex,margin*=1.5mm,  width=5.1 cm, max height = 4 cm,frame,raise=4mm }
{\maxsizebox*{0.7cm}{3ex}{\shortstack{\textcolor{purple}{Iodine} \\SD (H) \\  \\  \\  \\  \\ \textit{similar to} \\  SI(H), PS-PS(L) \\ ~  }}} 
\includegraphics[trim={0mm 16mm 0 0},clip,width=.38\textwidth]{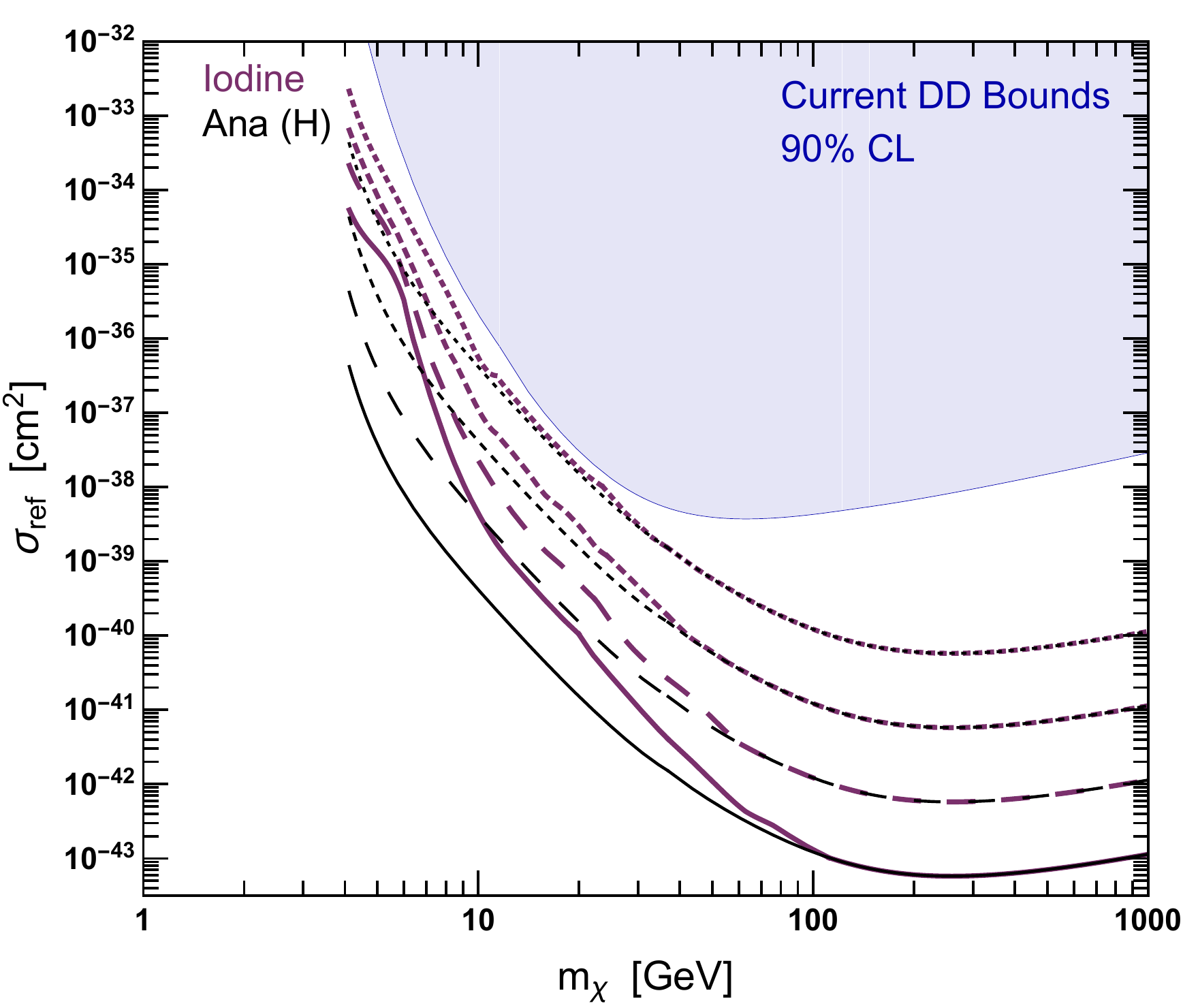}
}
\mycenter{
\includegraphics[trim={0mm 1mm 0 0},clip,width=.38\textwidth]{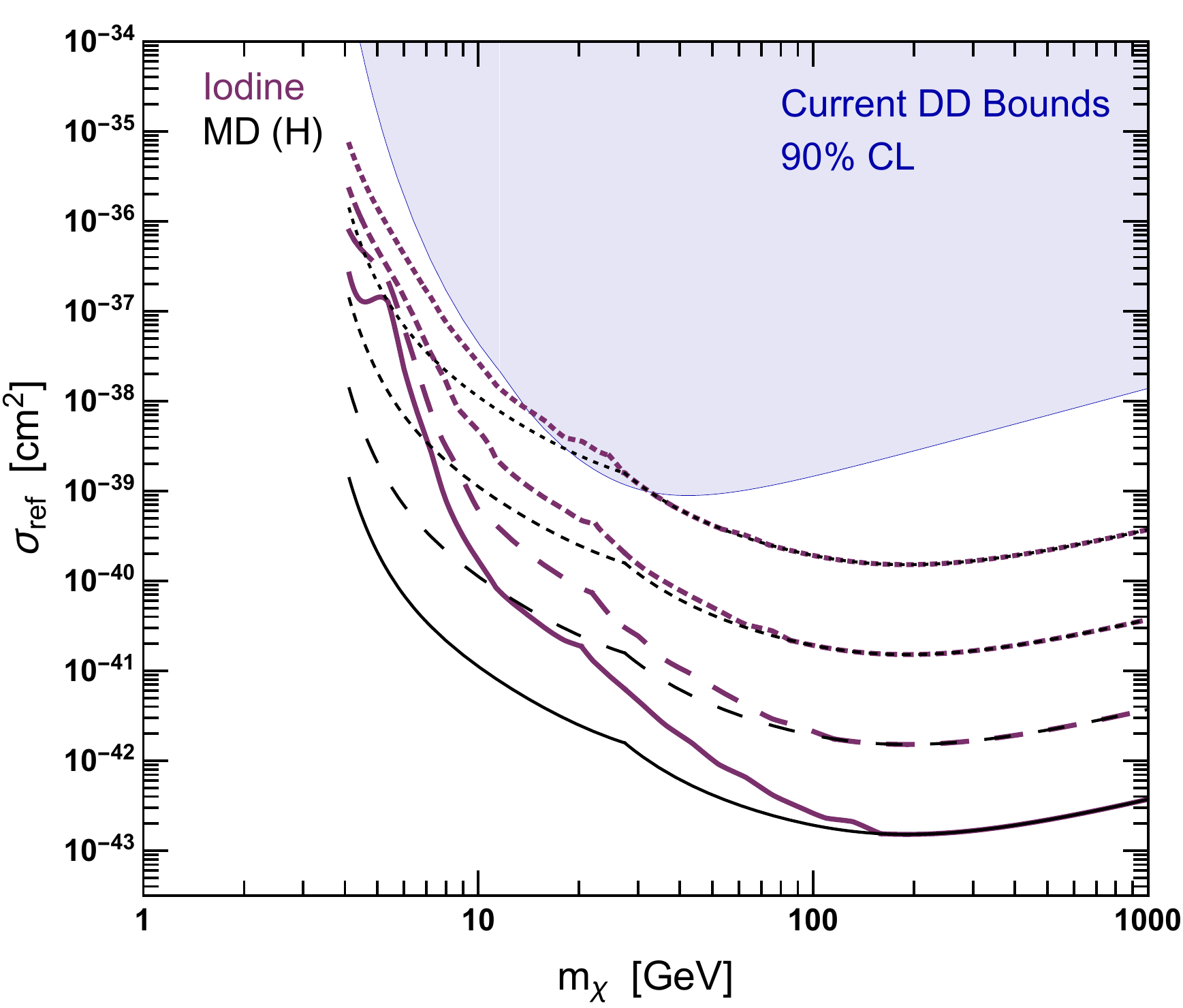}
\hspace{0mm}
\includegraphics[trim={9mm 1mm 0 0},clip,width=.365\textwidth]{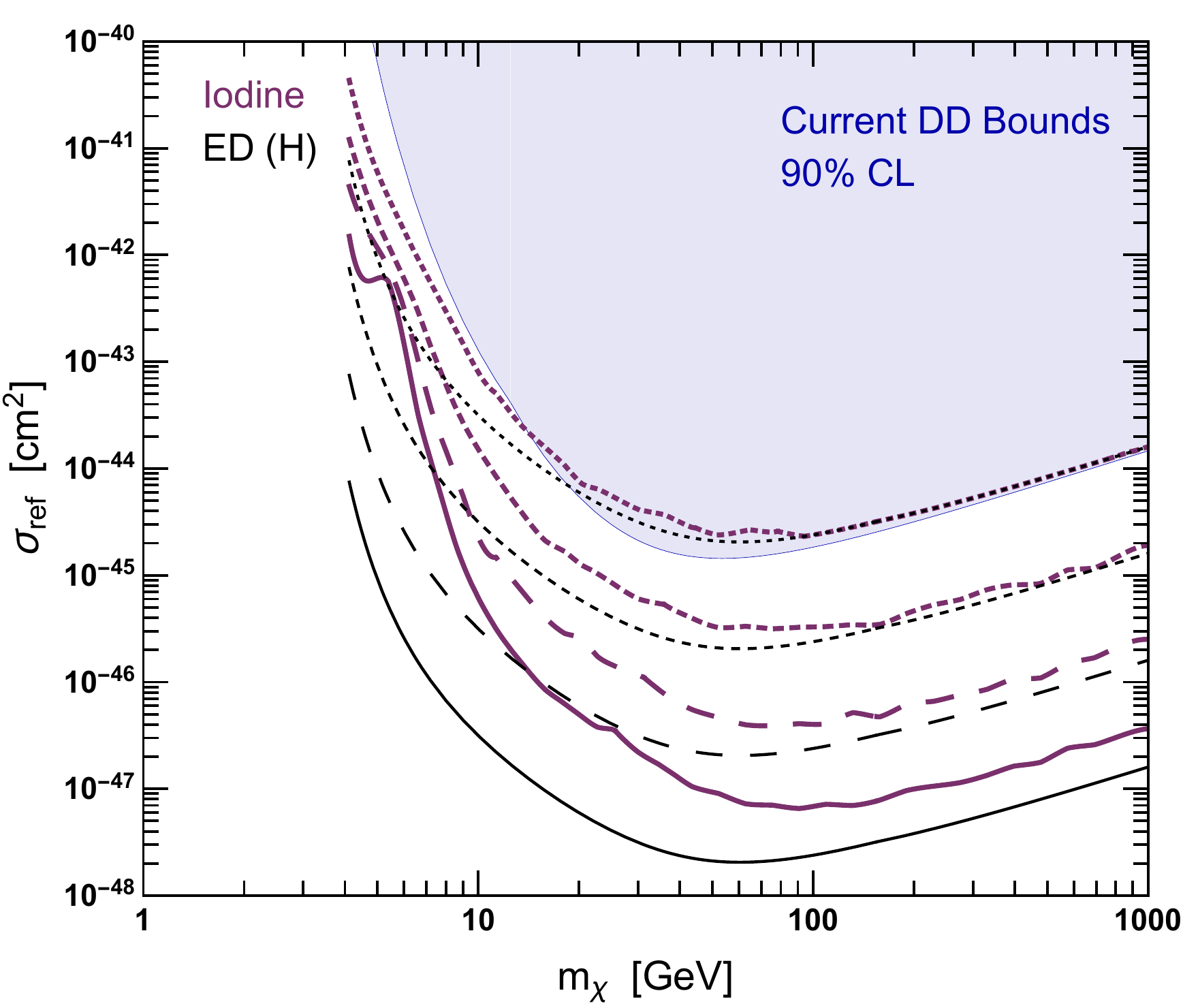}
 \hspace{4mm}
 \adjustbox{padding =1.2ex 0.5ex 1.2ex 0.5ex,margin*=1.5mm,  width=5 cm, max height = 4 cm,frame,raise=9.5mm }
{\maxsizebox*{0.7cm}{3ex}{\shortstack{\textcolor{purple}{Iodine} \\AV-V (H) \\  \\  \\  \\  \\ \textit{similar to} \\  SI(H) }}}
}
\mycenter{
 \hspace{6mm}
 \adjustbox{padding =0.6ex 0.5ex 0.6ex 0.5ex,margin*=1.5mm,  width=5 cm, max height = 4 cm,frame,raise=9mm }
{\maxsizebox*{0.7cm}{3ex}{\shortstack{\textcolor{purple}{Iodine} \\PS-S (H) \\  \\  \\  \\  \\ \textit{similar to} \\  S-PS(H), ED(H)  \\ ~   }}} 
 \hspace{6mm}
 \adjustbox{padding =0.6ex 0.5ex 0.6ex 0.5ex,margin*=1.5mm,  width=5 cm, max height = 4 cm,frame,raise=9mm }
{\maxsizebox*{0.7cm}{3ex}{\shortstack{\textcolor{purple}{Iodine} \\S-PS (H) \\  \\  \\  \\  \\ \textit{similar to} \\  PS-S(H), ED(H) \\ ~  }}} 
\includegraphics[trim={0mm 0mm 0 0},clip,width=.38\textwidth]{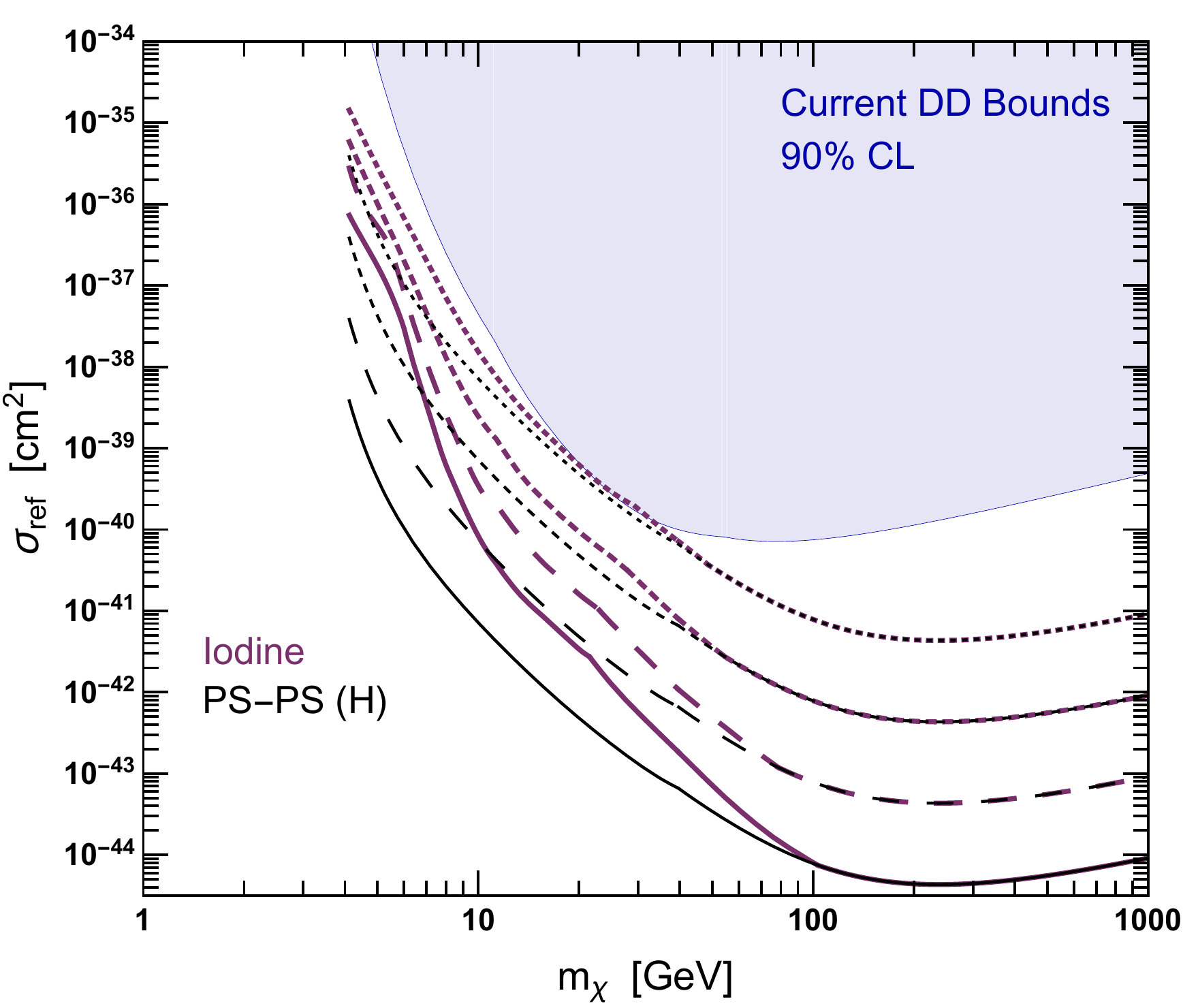}
}

\caption{\label{fig:iodine_floors} 3$\sigma$ discovery limit for a iodine-based experiment for a 0.1 ton-year exposure (dotted), 1 ton-year exposure (short dashed), 10 ton-year exposure (long dashed), and 100 ton-year exposure (solid), including (purple) and neglecting (black) the neutrino background. Shown for comparison is the current 90\% upper limits from XENON1T and LUX (shaded blue).}
\end{figure}

\begin{figure}[H]
\mycenter{
\includegraphics[trim={0mm 16mm 0 0},clip,width=.38\textwidth]{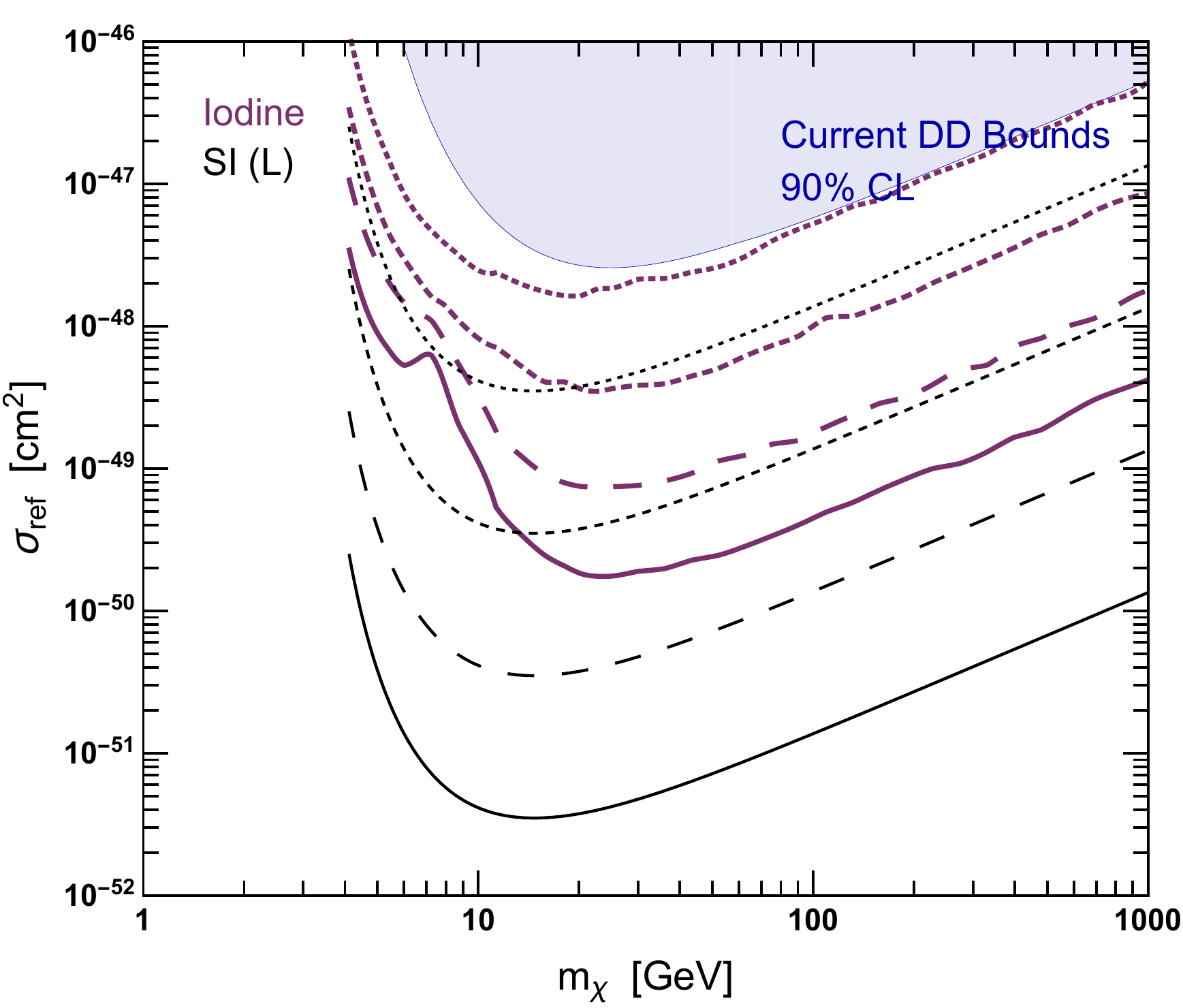}
 \hspace{3.0mm}
 \adjustbox{padding =1.2ex 0.45ex 1.2ex 0.45ex,margin*=1.5mm,  width=5 cm, max height = 4 cm,frame,raise=3.5mm }
{\maxsizebox*{0.7cm}{3ex}{\shortstack{\textcolor{purple}{Iodine} \\SD (L) \\  \\  \\  \\  \\ \textit{similar to} \\  SI(L)  \\ ~  }}} 
 \hspace{3.75mm}
 \adjustbox{padding =0.5ex 0.40ex 0.5ex 0.40ex,margin*=1.5mm,  width=5 cm, max height = 4 cm,frame,raise=3.75mm }
{\maxsizebox*{0.7cm}{3ex}{\shortstack{\textcolor{purple}{Iodine} \\Ana (L) \\  \\  \\  \\  \\ \textit{similar to} \\  SI(H), PS-PS(L) \\ SD(H)   }}} 
}
\mycenter{
\includegraphics[width=\triplt]{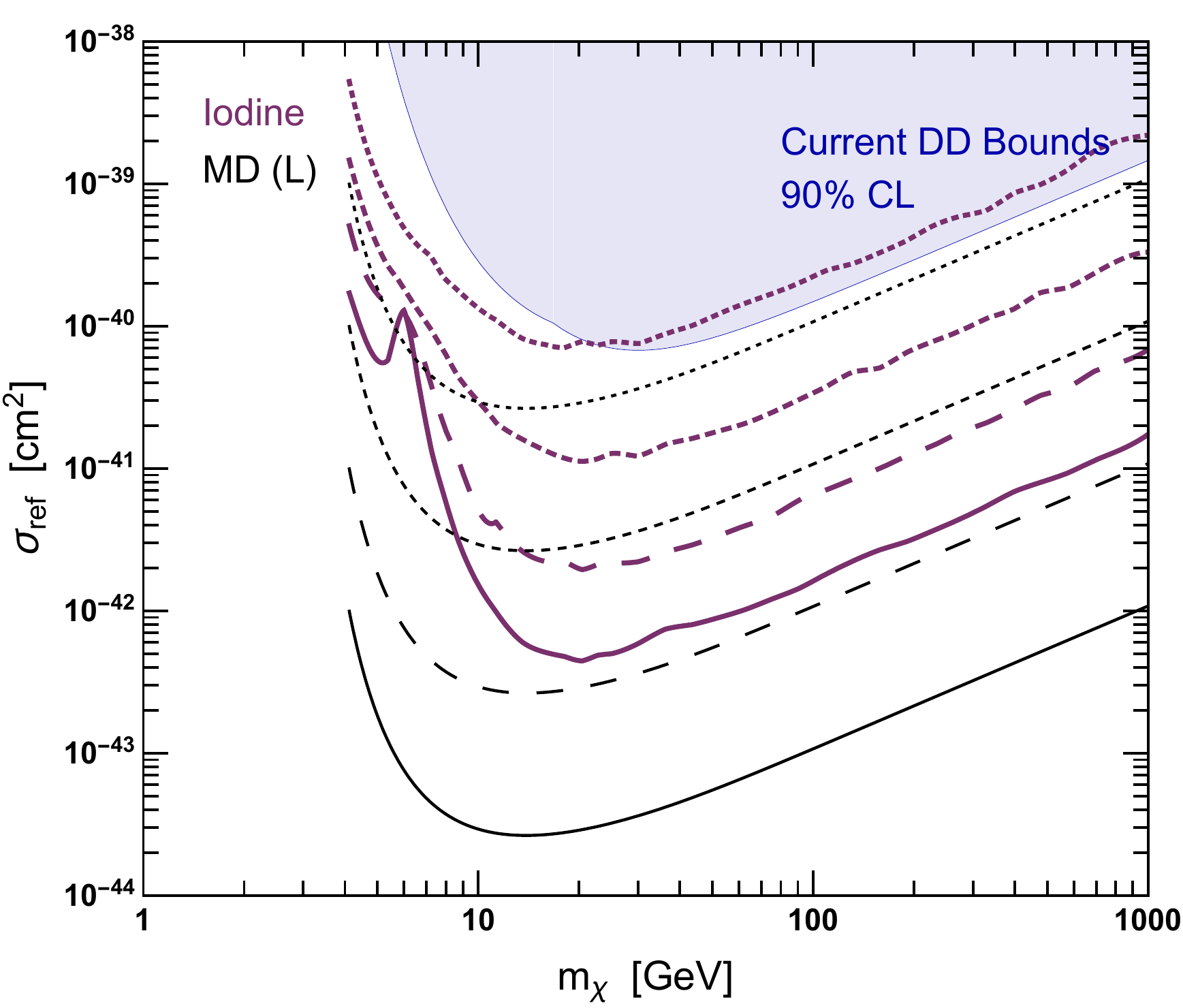}
\includegraphics[trim={9mm 0mm 0 0},clip,width=.3625\textwidth]{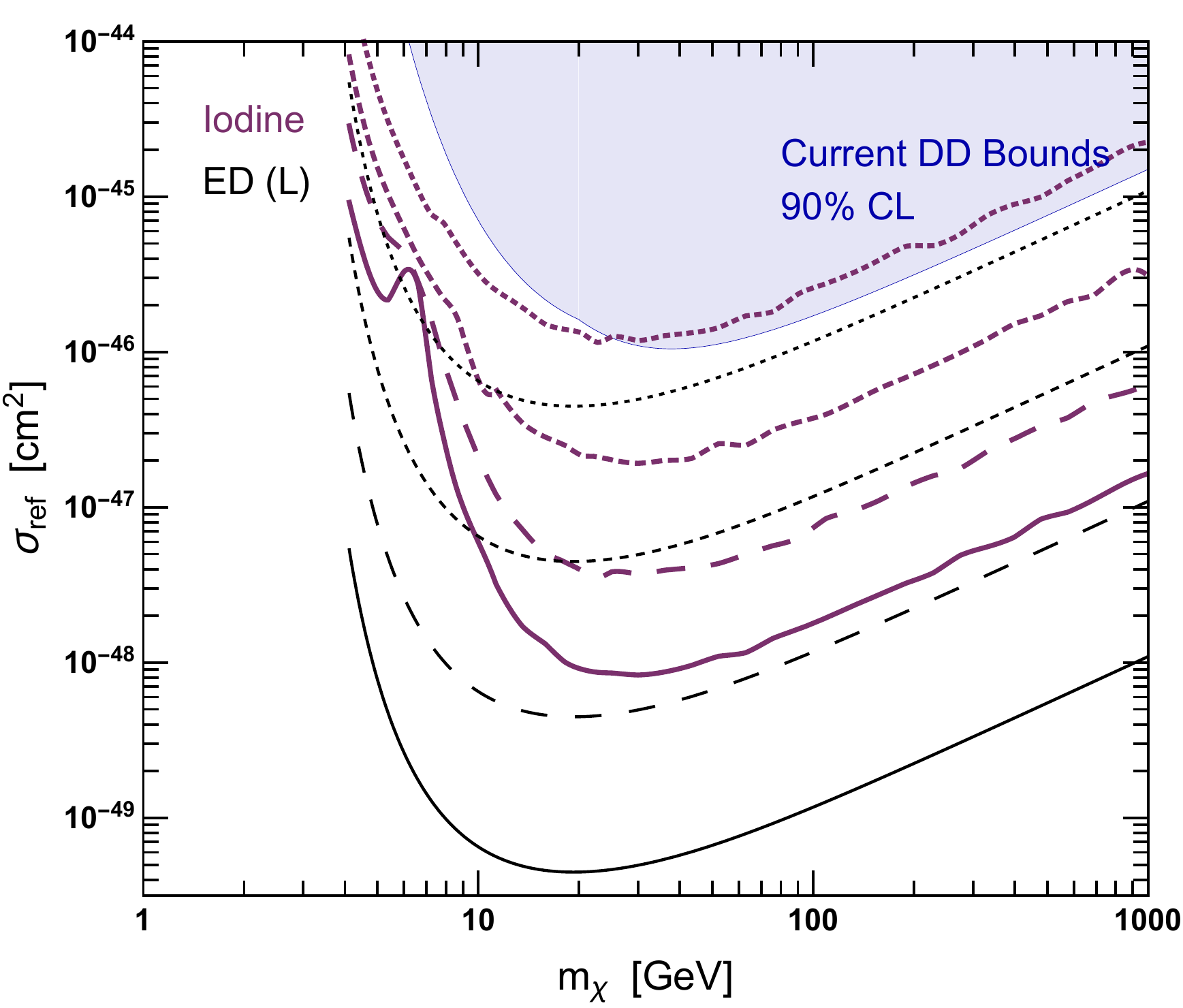}
 \hspace{1.75mm}
 \adjustbox{padding =1.1ex 0.45ex 1.1ex 0.45ex,margin*=1.5mm,  width=5 cm, max height = 4 cm,frame,raise=8.5mm }
{\maxsizebox*{0.7cm}{3ex}{\shortstack{\textcolor{purple}{Iodine} \\AV-V(L) \\  \\  \\  \\  \\ \textit{similar to} \\  SI(L) \\~  }}} 
}
\mycenter{
 \hspace{6mm}
 \adjustbox{padding =0.6ex 0.5ex 0.6ex 0.5ex,margin*=1.5mm,  width=5 cm, max height = 4 cm,frame,raise=6mm }
{\maxsizebox*{0.7cm}{3ex}{\shortstack{\textcolor{purple}{Iodine} \\PS-S (L) \\  \\  \\  \\  \\ \textit{similar to} \\  S-PS(L), ED (L) \\ ~  }}} 
 \hspace{4.5mm}
 \adjustbox{padding =0.6ex 0.5ex 0.6ex 0.5ex,margin*=1.5mm,  width=5 cm, max height = 4 cm,frame,raise=5.75mm }
{\maxsizebox*{0.7cm}{3ex}{\shortstack{\textcolor{magenta}{Iodine} \\S-PS (L) \\  \\  \\  \\  \\ \textit{similar to} \\   PS-S(L), ED (L) \\ ~  }}}
 \hspace{3.75mm}
 \adjustbox{padding =0.6ex 0.4ex 0.6ex 0.4ex,margin*=1.5mm,  width=5 cm, max height = 4 cm,frame,raise=5mm }
{\maxsizebox*{0.7cm}{3ex}{\shortstack{\textcolor{magenta}{Iodine} \\PS-PS (L) \\  \\  \\  \\  \\ \textit{similar to} \\   SI(H), SD(H) \\ ~  }}} 
}

\caption{\label{fig:iodine_floors} 3$\sigma$ discovery limit for a iodine-based experiment for a 0.1 ton-year exposure (dotted), 1 ton-year exposure (short dashed), 10 ton-year exposure (long dashed), and 100 ton-year exposure (solid), including (purple) and neglecting (black) the neutrino background. Shown for comparison is the current 90\% upper limits from XENON1T and LUX (shaded blue).}
\end{figure}

\clearpage
\section{Erratum}
\label{app:erratum}

This is an erratum  for  the original article above, published separately in Journal of Cosmology and Astroparticle Physics (JCAP).

Some of the entries in \Tab{tab:reactor_fit} of our original article contain transcription errors, although correct numbers were used in our calculations. Corrected entries are presented in  \Tab{tab:correct-reactor_fit}.
\begin{table*}[hbp]
  \setlength{\extrarowheight}{2pt}
  \setlength{\tabcolsep}{10pt}
  \begin{center}
	\begin{tabular}{c|c|c|c|c}  \hline\hline
	$i$  & $^{235}U$ & $^{238}U$ & $^{239}P$ & $^{241}P$ \\  
	\hline
	1 & 3.217 & 0.4833 & 6.413 & 3.251\\  \hline
	2 & -3.111 & 0.1927 & -7.432 & -3.204 \\  \hline
    3 & 1.395 & -0.1283 & 3.535 & 1.428 \\  \hline
    4 & -0.3690 & -0.006762 & -0.8820 & -0.3675 \\  \hline
    5 & 0.04445 & 0.002233 & 0.1025 & 0.04254 \\  \hline
	6 &  -0.002053 & -0.0001536 & -0.004550 & -0.001896 \\  \hline\hline
	\end{tabular}
  \end{center}
\caption{\label{tab:correct-reactor_fit} Corrected version of \Tab{tab:reactor_fit}. Fitted values of the reactor neutrino spectrum $\alpha_{i, k}$ coefficients, used in Eq.~\eqref{eq:reactor_nuspec}, are displayed for dominant nuclear isotopes, taken from \cite{Mueller:2011nm}.}
\end{table*}

\begin{figure}
\centering
\includegraphics[width=.6\textwidth]{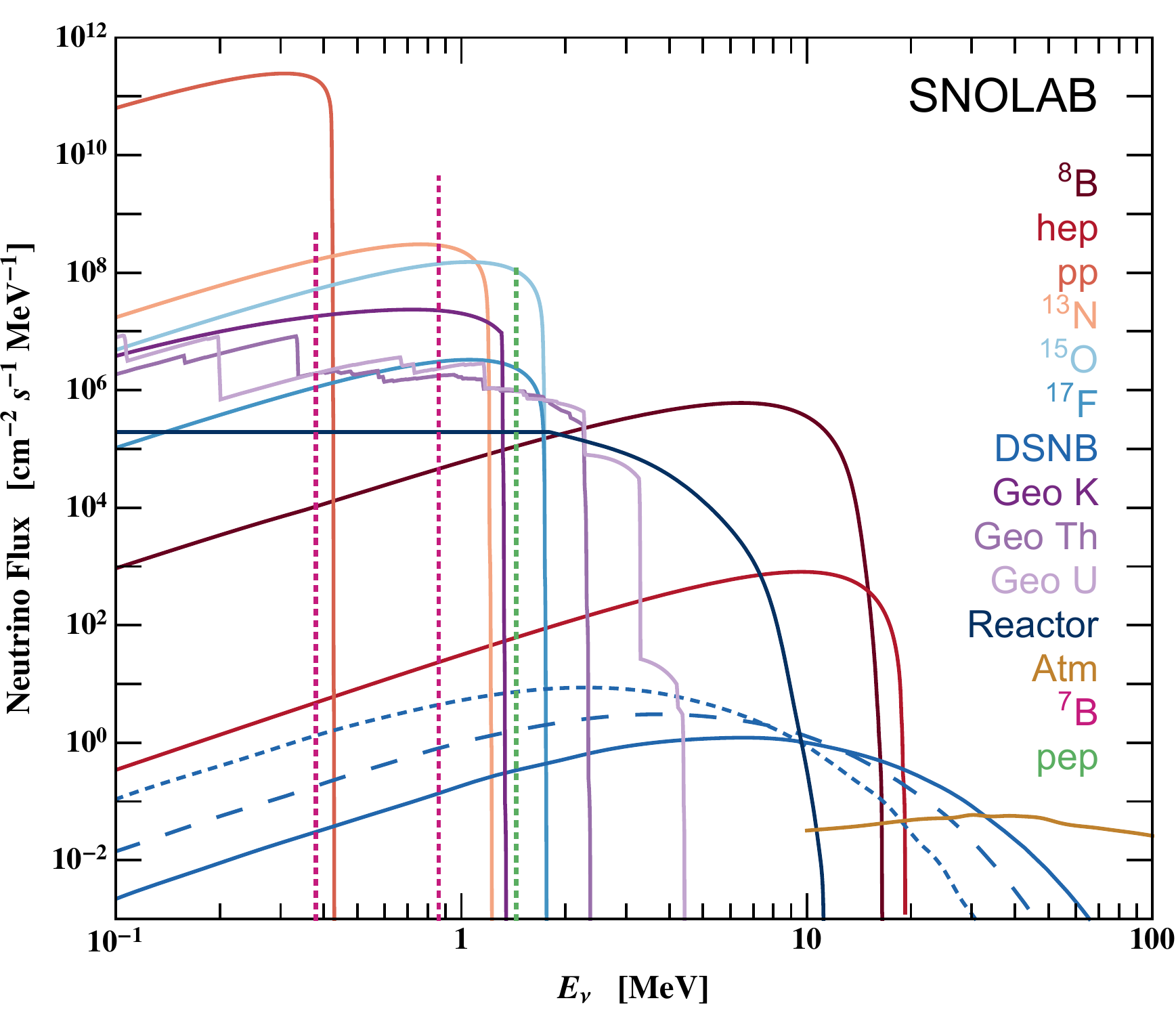}
\caption{\label{fig:correct-neutrino_flux} Corrected version of Fig.~\ref{fig:neutrino_flux}. Neutrino flux components comprising the ``neutrino floor'' at the SNOLAB location. The atmospheric neutrino contribution has been summed over all the neutrino flavors.}
\end{figure}

In  our earlier calculations of reactor neutrino fluxes we have used the electrical reactor power output in \Tab{tab:reactors}, instead of the thermal reactor output that is larger and is directly related to the intensity of the neutrino emission. The corrected reactor power outputs are given in \Tab{tab:correct-reactors}.  The corresponding  corrected reactor neutrino fluxes are about a factor of $\sim 3$ larger than before, as shown in Fig.~\ref{fig:correct-neutrino_flux} (which supersedes the original Fig.~\ref{fig:neutrino_flux}). However, the updated reactor neutrino flux still remains a sub-dominant neutrino background contribution for our analysis. We have verified that even in the cases in which the background of low energy neutrinos has the largest effect, namely for small dark matter particle masses, low experimental thresholds and either $q$-independent or $q$-enhanced cross-sections, the increase of reactor neutrinos had a negligible effect on our results. 
The updated total reactor anti-neutrino flux of \Tab{tab:nucomponents} is 5.96(1$\pm$ 0.080) $\times 10^{5}$ cm$^{-2}$ s$^{-1}$  and includes  all three main types of reactor anti-neutrinos (while the previous quoted value of 1.88(1$\pm$ 0.080) $\times 10^{5}$ cm$^{-2}$ s$^{-1}$ corresponded only to contribution from $^{235}$U).

We further clarify a minor inaccuracy in the explanation of the statistical analysis Section~\ref{sec:like_fit}, which states that the  PDF in Eq.~\eqref{eq:pval0} follows a $\chi^2$ distribution with one degree of freedom. Namely, we note that in the large sample limit, Wilks' theorem ensures that the mentioned PDF is given by one half times a delta function at $q_0 = 0$ plus one half times a $\chi^2$ distribution with one degree of freedom.
The original statistical analysis in was performed using the correct distribution, and is thus not affected.

\begin{table}[tbp]
  \setlength{\extrarowheight}{2pt}
  \setlength{\tabcolsep}{10pt}
  \begin{center}
	\begin{tabular}{|l|m{2.3cm}|c|c|c|}  \hline
	Nuclear Reactor Name & Location & Nearest Lab & Distance (km) & Output (MW) \\ \hline \hline
    Cooper  & 40$^\circ$21$'$43$''$\,N 95$^\circ$38$'$29$''$\,W & SURF & 801 & 2419  \\ \hline
    Monticello  & 45$^\circ$20$'$01$''$\,N 93$^\circ$50$'$57$''$\,W& SURF & 788 & 2004 \\ \hline
    Prarie Island  & 44$^\circ$37$'$18$''$\,N 92$^\circ$73$'$59$''$\,W & SURF & 835 & 3334 \\ \hline
    Nine Mile Point  & 43$^\circ$31$'$15$''$\,N 76$^\circ$24$'$25$''$\,W & SNOLAB & 498 & 5838 \\ \hline
    R.E. Ginna  & 43$^\circ$16$'$40$''$\,N 77$^\circ$18$'$36$''$\,W & SNOLAB & 468 & 1775 \\ \hline
    James A. Fitzpatrick  & 43$^\circ$31$'$04$''$\,N 76$^\circ$23$'$09$''$\,W & SNOLAB & 500 & 2536 \\ \hline
    Point Beach  & 44$^\circ$16$'$52$''$\,N 87$^\circ$32$'$12$''$\,W & SNOLAB & 552 & 3600 \\ \hline
    Enrico Fermi  & 41$^\circ$57$'$46$''$\,N 83$^\circ$15$'$27$''$\,W & SNOLAB & 527 & 3486 \\ \hline
    Davis Besse & 41$^\circ$35$'$48$''$\,N 83$^\circ$05$'$11$''$\,W & SNOLAB & 563 & 2817 \\ \hline
    Perry  & 41$^\circ$48$'$03$''$\,N 81$^\circ$08$'$36$''$\,W & SNOLAB & 519 & 3758 \\ \hline    
    Bruce  & 44$^\circ$19$'$31$''$\,N 81$^\circ$35$'$58$''$ W & SNOLAB & 240 & 21384 \\ \hline
    Darlington  & 43$^\circ$55$'$22$''$\,N 78$^\circ$43$'$11$''$\,W & SNOLAB & 343 & 11104 \\ \hline
    Tricastin  & 44$^\circ$19$'$47$''$\,N 04$^\circ$43$'$56$''$\,E & LNGS & 744 & 11140 \\ \hline
    Cruas & 44$^\circ$37$'$59$''$\,N 04$^\circ$45$'$29$''$\,E & LNGS & 750 & 11140 \\ \hline
    Saint-Alban & 45$^\circ$24$'$16$''$\,N 04$^\circ$45$'$19$''$\,E & LNGS & 778 & 7634 \\ \hline
    Bugey  & 45$^\circ$47$'$54$''$\,N 05$^\circ$16$'$15$''$\,E & LNGS & 760 & 13094 \\ \hline
	\end{tabular}
  \end{center}
\caption{\label{tab:correct-reactors} Corrected version of \Tab{tab:reactors}. List of most relevant nuclear reactors for SURF, SNOLAB, and LNGS. Columns contain, from left to right, the name of the reactor, the GPS location, the laboratory for which the reactor is relevant, the distance to the laboratory in kilometers, and the output of the reactor in MW~\cite{USNuclear,CanadaNuclear,EUNuclear}}  
\end{table}

The original analysis was done only at tree level. Thus, we would also like to clarify that conclusion, which states that since argon has no spin or a magnetic moment  it is insensitive to a variety of interactions, is only valid at tree level. In general, there could be contributions of the same interactions at loop level that may be independent of spin or magnetic moment and would thus be detectable with an argon based experiment.

\clearpage
\bibliography{nufloor}
\addcontentsline{toc}{section}{Bibliography}
\bibliographystyle{JHEP}

\end{document}